\documentclass[10pt,aps,rmp,twocolumn,showpacs,superscriptaddress,amsmath,amssymb]{revtex4-1}
\usepackage[english]{babel}
\usepackage{graphicx}
\usepackage{hyperref}
\usepackage{enumerate}
\usepackage{bm}
\usepackage{color}
\usepackage[utf8]{inputenc} 

\usepackage{cmap}
\definecolor{rltred}{rgb}{0.75,0,0}
\definecolor{rltgreen}{rgb}{0,0.5,0}
\definecolor{rltblue}{rgb}{0,0,0.75}
\hypersetup{colorlinks,%
hypertexnames=true,%
pdfauthor={xxx},%
pdftitle={xxx},%
pdfview=FitH,%
pdfstartview=FitH,%
pdfpagemode=UseNone,%
bookmarksopen=true,%
bookmarksnumbered=true,%
pdfhighlight=/I,%
linkcolor=rltred,%
citecolor=rltgreen,%
linktocpage=true}
\newcommand{\fig}[1]{Fig.~\ref{fig:#1}}
\newcommand{\eq}[1]{Eq.~(\ref{eq:#1})}

\newcommand{\bfr}{{\bf r}}

\usepackage{SIunits}

\usepackage{letltxmacro}

\LetLtxMacro{\ORIGselectlanguage}{\selectlanguage}
\makeatletter
\DeclareRobustCommand{\selectlanguage}[1]{%
  \@ifundefined{alias@\string#1}
    {\ORIGselectlanguage{#1}}
    {\begingroup\edef\x{\endgroup
       \noexpand\ORIGselectlanguage{\@nameuse{alias@#1}}}\x}%
}
\newcommand{\definelanguagealias}[2]{%
  \@namedef{alias@#1}{#2}%
}
\makeatother

\definelanguagealias{en}{english}

\begin{document}
\title{Light fields in complex media: mesoscopic scattering meets wave control}\author{Stefan Rotter}\email{stefan.rotter@tuwien.ac.at}\affiliation{Institute for Theoretical Physics, Vienna University of Technology (TU Wien), Wiedner Hauptstra\ss e 8--10/136, 1040 Vienna, Austria, EU}\author{Sylvain Gigan}\email{sylvain.gigan@lkb.ens.fr}\affiliation{Laboratoire Kastler Brossel, UMR8552 Université Pierre et Marie Curie, Ecole Normale Sup{\'e}rieure, Collège de France, CNRS, 24 rue Lhomond, 75005 Paris, France, EU}
\date{\today}

\begin{abstract}
The newly emerging field of wave front shaping in complex media has recently seen enormous progress. The driving force behind these advances has been the experimental accessibility of the information stored in the scattering matrix of a disordered medium, which can nowadays routinely be exploited to focus light as well as to image or to transmit information even across highly turbid scattering samples. We will provide an overview of these new techniques, of their experimental implementations as well as of the underlying theoretical concepts following from mesoscopic scattering theory. In particular, we will highlight the intimate connections between quantum transport phenomena and the scattering of light fields in disordered media, which can both be described by the same theoretical concepts. We also put particular emphasis on how the above topics relate to application-oriented research fields such as optical imaging, sensing and communication. 
\end{abstract}

\maketitle
\tableofcontents

\section{Introduction}\label{section1}
Recent years have witnessed enormous conceptual and experimental progress in the ability to manipulate  light fields both spatially and temporally. On the experimental side these advances have largely been enabled by the availability of highly tunable digital arrays, known also as spatial light modulators \cite{savage_digital_2009}, which are meanwhile being used to create arbitrarily complex light fields. In this sense the current state-of-the-art in the field of optical wave front shaping is reminiscent of the situation in related areas like acoustics and seismology, which were similarly promoted by antenna or transducer arrays that can retrieve information from a complex environment. The availability of such versatile tools now also in optics opens up the way to address a topic where conventional optical techniques are hard to apply, like the control of light propagation in turbid media such as in amorphous or disordered materials, in biological tissues, complex photonic structures, plasmonic systems, multimode fibers 
etc. Starting points for these activities were a number of proof-of-principle experiments that have recently demonstrated that a disordered material can be used to focus light \cite{vellekoop_focusing_2007,van_putten_scattering_2011} and that its transmission matrix can be measured in detail \cite{popoff_measuring_2010,popoff_exploiting_2011} such as to reconstruct the transmission of images across  highly scattering samples  \cite{popoff_image_2010}. Beyond spatial control, explicitly time-dependent measurements were able to show that a wave scattered in a disorder region can not only be focused in space but also in time \cite{mounaix_spatiotemporal_2016,mccabe_spatio-temporal_2011,katz_focusing_2011,aulbach_control_2011}. Following the pioneering concepts introduced in \cite{freund_looking_1990}, further work successfully demonstrated that the information stored in the scattering matrix of a disordered system can be used for turning a disordered sample into a perfect mirror \cite{katz_looking_2012} or into a high resolution spectral filter or spectrometer \cite{small_spectral_2012,redding_compact_2013}. These insights can be expected to have impact on a very broad range of fields like on biology and medicine \cite{cox_optical_2012}, where imaging through disorder is a major challenge; 
on nanophotonics \cite{kawata_nano-optics_2002}, where the challenge is to address and control quantum systems in a disordered environment \cite{sapienza_long-tail_2011}; on quantum information \cite{wolterink_programmable_2016,defienne_two-photon_2016,ott_quantum_2010}, where entangled states could be guided and transformed; as well as on communication technology \cite{tse_fundamentals_2005}, where the principal goal is to secure that information sent through a complex environment ends up at a desired receiver.
 
A sound theoretical basis required to describe all of the above phenomena is given in terms of scattering theory. In the specific context of disorder scattering it is mostly the work in mesoscopic physics \cite{stockmann_quantum_2006,akkermans_mesoscopic_2007,imry_introduction_2002,sebbah_waves_2001}, quantum transport \cite{datta_electronic_1997,ferry_transport_1997,mello_quantum_2004} and Random Matrix Theory \cite{alhassid_statistical_2000,mitchell_random_2010,beenakker_random-matrix_1997} that has been the principal driving force behind theoretical progress. This is because electron scattering through disordered or chaotic systems has been and continues to be one of the paradigms in the mesoscopic physics community. In spite of the progress made, many of the results obtained for the situation on the mesoscopic scale do, however, remain unknown to the newly emerging scientific communities working on wave front shaping in complex media. The reason why many insights penetrated only weakly outside the community of mesoscopic physics is probably due to the vastness of the field, which makes it difficult to overlook, and due to a specific scientific jargon which scientists working outside this community are typically not familiar with.

The intended goal of our review article will  be to bridge this knowledge gap. Our strategy will be to demonstrate how theoretical insights from mesoscopic scattering theory have direct relevance for the recent wave control experiments and vice versa. We will start in section \ref{section2} with a brief review of mesoscopic transport theory in which basic concepts like the scattering matrix and its statistical properties following from Random Matrix Theory or related approaches are introduced. We will discuss here the concept of transmission eigenchannels 
as well as their connection to electronic shot noise, which provides indirect access to the distribution of transmission eigenvalues in measurements of electronic current. 
A particular emphasis will also be put on the concept of time-delay in scattering and 
its relation to the density of states as well as to coherent wave absorption and to the quantum-to-classical crossover. 
With such a solid theoretical basis being established, we move on in section
\ref{section3} to review a number of mesoscopic transport effects that have meanwhile been observed in optical experiments without the help of any wavefront shaping tools. As we will show, quite a number 
of theoretical concepts first studied in a mesoscopic context could be successfully transferred and, indeed, observed with light fields in complex media. Examples which we highlight here are those 
related to the quantization of the conductance and its universal fluctuations, weak localization, the memory effect etc. 
While these observations are very encouraging for the applicability of the theoretical
tools introduced in \ref{section2}, much more can be done in these experiments with the tools of wavefront shaping. These tools will then be reviewed in section \ref{section4},
starting with a historical perspective on where such experimental techniques were first employed such as in adaptive optics. Once spatial light modulators are introduced, we will discuss how they can be used for measuring and modulating the light transmitted through the paradigmatic case of a thin disordered slab. As will be discussed in our review, such an ``opaque lens'' can be used for focusing, imaging as well as for controlling the polarization and the temporal shape of the transmitted light. 
In section \ref{section5} we will review how the predictions from mesoscopic transport theory can be fully brought to bear using the wavefront shaping tools introduced earlier. First, we will focus on effects that were already realized in corresponding experiments such as those related to open and closed transmission channels, the memory effect etc. In a next step we will provide a collection of many interesting predictions that still await an implementation in the laboratory. This outlook also serves the purpose of indicating future directions of research and of demonstrating how much ``uncharted territory'' is yet to be developed in this increasingly active field of research. Our review is rounded off with a summary in section \ref{section6}.

We have intentionally restricted the scope of this review to the interface between mesoscopic scattering theory and the recent advances in wavefront shaping. With this focus we hope to provide some added value to both of the corresponding two communities that we are trying to better connect with our piece. At this point we emphasize that excellent reviews and books are already available for each of these two separate fields: Regarding mesoscopic scattering, our readers may find a wealth of information on specific topics, like on random matrix theory  \cite{beenakker_random-matrix_1997,akemann_oxford_2011,guhr_random-matrix_1998}, or on the maximum entropy approach \cite{mello_quantum_2004}, as well as on the many interesting connections between electronic transport and light scattering \cite{dragoman_quantum-classical_2004,lagendijk_resonant_1996,akkermans_mesoscopic_2007}. Also for wavefront shaping first short reviews have 
recently become available \cite{vos_light_2014,mosk_controlling_2012, vellekoop_feedback-based_2015, shi_statistics_2015}.  
The niche we intend to fill with our own contribution is to highlight the underappreciated connection between the above topical areas and its potential for future research.

\section{Scattering theory for complex media}\label{section2}
\begin{figure}
  \centering
  \includegraphics[width=0.76\linewidth]{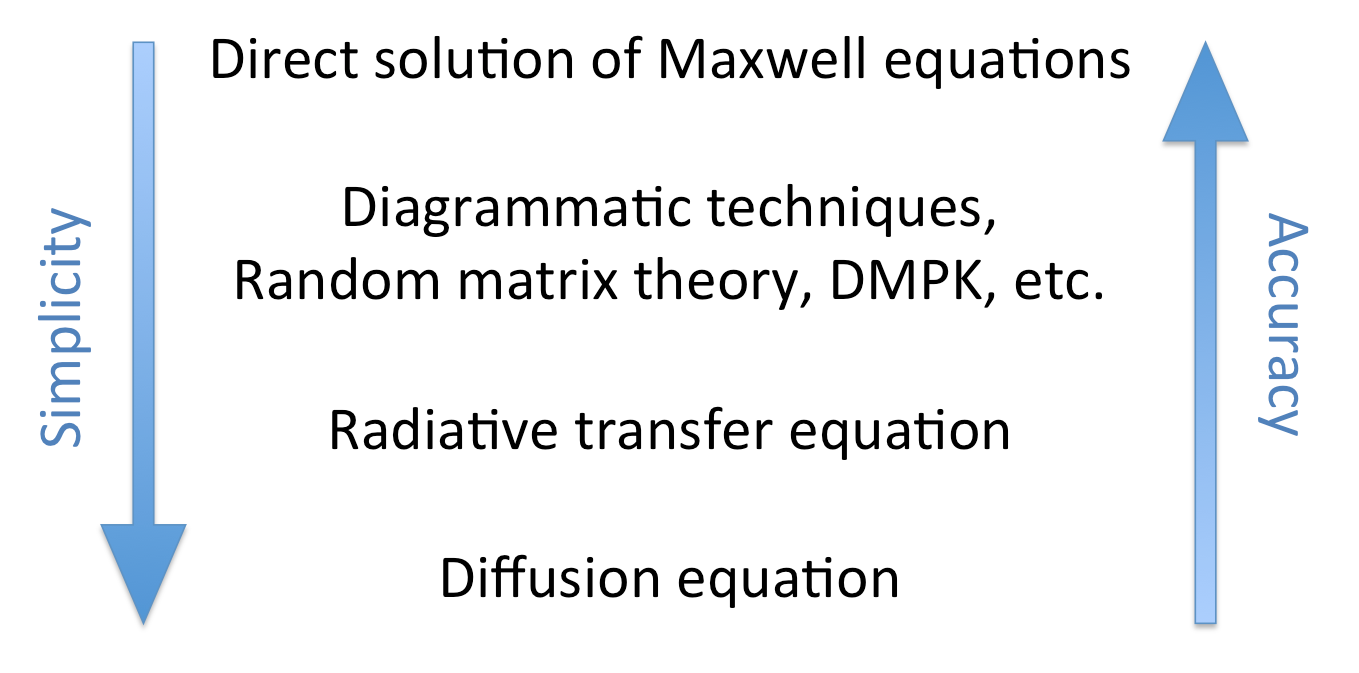}
  \caption{(color online). Schematic 
  of different methods used for describing wave scattering in disordered media
  on different levels of accuracy and simplicty,
  adapted from \cite{bertolotti_light_2011}.}
  \label{fig:methods_figure}
\end{figure}

The scattering of waves through disordered or otherwise complex media is a problem that 
can be approached from many different angles. In particular, a whole hierarchy of different 
methods have been developed that provide insights on different levels of accuracy,
typically anti-correlated with the complexity of a specific method (see \fig{methods_figure}).
For the purpose of this review, we will mostly be interested in those approaches, which 
retain the wave nature of the scattering process, such as to incorporate effects due to 
interference. A full solution of the corresponding wave equation is, however, very costly  numerically
and often does not provide much insight into the general features underlying a whole class
of related problems. To overcome such limitations, much work has been invested into 
``mesoscopic scattering theory'', which we will provide a short review of. This term  
refers to a set of theoretical tools that were largely developed in the context of mesoscopic electron transport, in which the phase coherence of electrons, the finite number of modes through which they can scatter as well as the correlations between these modes play a significant role.

Whatever the formalism chosen to describe  wave propagation in disordered media, there are a few common parameters to quantify the scattering properties of the medium. The most important one is probably the transport mean free path, usually referred to as $\ell_{\rm tr}$ or $\ell^\star$ in the literature (we will use the notation $\ell^\star$ in this review). This length scale measures after which distance the propagation direction of an incoming photon is randomized and thus governs the macroscopic transport properties of the medium. At a more microscopic level, the scattering mean free path $\ell$ (or $\ell_{\rm s}$ in the literature) measures the average distance traveled between two scattering events and thus quantifies the scattering strength of the medium. The link between $\ell^\star$ and $\ell$ is through the anisotropy of scattering, $\ell^\star={\ell}/{(1-g)}$, quantified by the anisotropy factor $g= \langle \cos{\theta} \rangle$, i.e., the average of the cosine of the scattering angle $\theta$. An important case is the one of isotropic scattering, for which $\langle \cos \theta \rangle =0$, $g=0$ and $\ell^\star=\ell$.

\subsection{Basic formalism}\label{subsection2.1}

\subsubsection{Wave equations}\label{subsubsection2.1.1}
A good starting point for setting up the formalism for mesoscopic scattering is the observation that electromagnetic waves in a dielectric medium behave similarly as electrons in a potential (see here also the corresponding references listed in the last paragraph of the introduction).
Since this analogy will also be the bridge across which many of the results from
mesoscopic transport theory can be carried over to the domain of optics, we shall start here by elucidating this connection. 

Consider first the Schr\"odinger equation for the evolution of a particle of mass $m$ in a potential $V({\bf r})$,
\begin{equation}
\left[\frac{{\bf p}^2}{2m}+V({\bf r})\right] \psi({\bf r},t)=i\hbar \partial_t \psi({\bf r},t)\,.
\label{eq:schrodinger}
\end{equation}
For stationary states with a well-defined real energy $E=\hbar\omega $ and a corresponding time-evolution 
$\exp(-i\omega t)$ the Schr\"odinger equation reduces to the following form
without a time-derivative,
\begin{equation}
\left\{\Delta- \frac{2m}{\hbar^2}\left[ V({\bf r})-E\right]\right\} \psi_E({\bf r})=0\,,
\label{eq:schrodinger2}
\end{equation}
where we have used the standard definition for the momentum operator ${\bf p}=-i\hbar {\bf \nabla}$.
For appropriate boundary conditions the differential operator
in \eq{schrodinger2} will be Hermitian such that the eigenstates, $\psi_{m}({\bf r})$ (labeled by their mode index $m$) satisfy the conventional orthogonality relations
\begin{equation}
\int d\bfr \, \psi_{m}({\bf r})^*\psi_{n}({\bf r})=\delta_{mn}\,,
\label{eq:orthosch}
\end{equation}
and form a complete basis of states.

To find equivalent relations also for light scattering, consider first the wave equation for 
the electric field ${\bf E}({\bf r},t)$
in a source-free, linear and frequency-independent dielectric medium 
with dielectric function $\varepsilon(\bf r)$, which is 
directly derived from Maxwell's equations 
\cite{jackson_classical_1998},
\begin{equation}
-{\bf \nabla}\times{\bf \nabla}\times {\bf E}({\bf r},t)=\frac{\varepsilon({\bf r})}{c^2}\partial_t^2 {\bf E}({\bf r},t)\,.
\label{eq:helmholtz}
\end{equation}
To make the analogy to the Schr\"odinger equation, we first restrict ourselves to 
monochromatic states, which, in perfect analogy to the 
stationary states of the Schr\"odinger equation, feature a harmonic time-dependence
${\bf E}({\bf r},t)={\bf E}_\omega({\bf r})\exp(-i\omega t)$. Unless explicitly stated otherwise, we will 
work with this complex notation in the following with the understanding that the 
real (physical) electric field 
is extracted as the real part of this complex quantity, Re$[{\bf E}({\bf r},t)]$. 
As it turns out, even for appropriate boundary conditions, the curl operator above is not a Hermitian operator when the dielectric function $\varepsilon(\bf r)$ is spatially varying and when
the conventional inner product is used
\cite{viviescas_field_2003}. We thus rewrite the electric field through 
the vector-valued function ${\boldsymbol \phi}_\omega(\bf r)=\sqrt{\varepsilon(\bfr)}\,
{\bf E}_\omega({\bf r})$, such that \eq{helmholtz} can be rewritten with the
Hermitian differential operator $\bm{\mathcal L}$,
\begin{equation}
\bm{\mathcal L}\,{\boldsymbol \phi}_\omega({\bf r})\equiv\frac{1}{\sqrt{\varepsilon(\bfr)}}\nabla\times
\left[\nabla\times \frac{{\boldsymbol \phi}_\omega(\bf r)}
{\sqrt{\varepsilon(\bfr)}} \right]=\frac{\omega^2}{c^2}\,{\boldsymbol \phi}_\omega({\bf r})\,.
\label{eq:curlnew}
\end{equation}
With appropriate boundary conditions, the eigenstates, ${\boldsymbol \phi}_{m}(\bf r)$ 
are then orthogonal
to each other and provide a complete basis of states. As a consequence, the 
electric field satisfies the following orthogonality relation,
\begin{align}
\int d\bfr \, {\boldsymbol \phi}_{m}({\bf r})\,{\boldsymbol \phi}_{n}({\bf r})=&\int d\bfr \,\varepsilon(\bfr)\, {\bf E}_{m}({\bf r})\,{\bf E}_{n}({\bf r})=\delta_{mn}\,.&
\label{eq:orthoelectric}
\end{align}
To make the analogy to the stationary Schr\"odinger equation even more apparent, we simplify 
the curl operator by the following vector identity
$\nabla\times(\nabla\times {\bf A})=\nabla(\nabla\cdot{\bf A})-\Delta {\bf A}$, 
such that we arrive at what is termed the vectorial Helmholtz equation,
\begin{equation}
\left\{\Delta -\frac{\omega^2}{c^2}[1-\varepsilon({\bf r})]+\frac{\omega^2}{c^2}\right\}{\bf E}_\omega(\bf r)=0\,.
\label{eq:helmholtz2}
\end{equation}
We need to emphasize at this point, however, that the above equation holds only 
under the assumption that $\nabla\cdot {\bf E}=0$ as for 
for source-free media which are  linear, homogeneous and isotropic (in the 
linear regime gain and loss in the medium may be included through a complex 
dielectric $\varepsilon({\bf r})$). For inhomogeneous media, where the dielectric permittivity is 
position dependent, the wave equation Eq.~(\ref{eq:helmholtz}) as well as the above Helmholtz equation
Eq.~(\ref{eq:helmholtz2}) do not hold. One may just consider the approximation of a locally homogeneous
medium for which the variation of $\varepsilon({\bf r})$ is slow as compared to the wavelength $\lambda$
\cite{lifante_integrated_2003}. Alternatively, when the medium consists of piecewise homogeneous 
constituents, one may use the Helmholtz equation for each sub-part, but different field-components get
mixed at the boundaries between them. We also mention here, that it may computationally be more efficient
to consider the magnetic field, rather than the electric  field in a nonmagnetic medium \cite{joannopoulos_photonic_2008}. 

Even the vector Helmholtz equation itself is very difficult to solve for most cases and closed
solutions only exist in very special limits. In practice, one therefore often reduces the 
Helmholtz equation to a scalar form, in which the scalar quantity $\psi_\omega({\bf r})$
stands for one of the three components of the electric or magnetic field. 
Implicit in this strategy is the assumption that the coupling of different vectorial components does not contain important physics -- a point which in many publications remains open [see, e.g., a corresponding analysis in \citep{bittner_experimental_2009}]. Certainly, the description of light as a scalar field may lead to quite different results than those based on a full solution of the Maxwell equations \cite{lagendijk_resonant_1996,skipetrov_absence_2014},
such that a careful analysis for each individual case at hand must be recommended.

In all cases where the scalar Helmholtz equation provides a good approximation of the 
real physics \cite{kragl_cases_1992},
\begin{equation}
\left\{\Delta - \frac{\omega^2}{c^2} [1-\varepsilon({\bf r})]+\frac{\omega^2}{c^2}\right\}\psi_\omega({\bf r})=0\,,
\label{eq:helmholtz3}
\end{equation} 
all the quantities in this scalar equation for light fields can be directly compared with those of the Schr\"odinger equation for electrons, \eq{schrodinger2}. For the case of 
the dielectric constant of vacuum, $\varepsilon({\bf r})=1$, or, equivalently, for
the case of vanishing potential $V({\bf r})$ in the Schr\"odinger case, we
can see immediately, that the resulting two equations, \eq{schrodinger2} and \eq{helmholtz3}, 
are the same if we identify the ``light energy'' as follows, $E_{\rm light}=(\hbar\omega)^2/(2mc^2)$
\cite{lagendijk_resonant_1996}. 
For the case of free space propagation, both equations also have the same fundamental 
plane wave solutions,   
$\psi_E({\bf r})=\psi_\omega({\bf r})=\psi_{{\bf k},\omega}\exp(i{\bf k r}-i\omega t)$,
characterized by a single frequency $\omega$ and a single wave vector ${\bf k}$, where 
$|{\bf k}|=k=\omega\sqrt{\varepsilon\mu_0}=n k_0$ with $n$ the refractive index and $k_0=\omega/c$. 
In the case of a spatially non-uniform  
dielectric function, $\varepsilon({\bf r})$, or potential landscape, $V({\bf r})$, 
the scalar Helmholtz and the Schr\"odinger equation can still be mapped onto each 
other for any given frequency $\omega$, when we identify the following  relation for the ``light potential'' 
$V_{\rm light}({\bf r})=E_{\rm light}[1-\varepsilon({\bf r})]$ and keep in mind that
the Helmholtz equation is valid for locally homogeneous media only \cite{lifante_integrated_2003}. 

The equivalence between the fundamental 
equations, which describe electronic and light scattering, will be essential for many of the
effects discussed in this review and for their occurrence in both of the different research fields. Note,
however, that this analogy also has well-defined limits, as, e.g., when attempting to describe the
microscopic details of the scattering field in a disordered medium, which goes beyond the capacity of the Helmholtz equation and requires a full treatment based on  Maxwell's equations. As we will see below, for many other quantities, in particular for those related to the statistical properties of scattering amplitudes, many common aspects in electron and light scattering can be identified. 

Fundamental differences between the scattering of electrons and light do, however, remain: These become apparent, e.g, when going away from the stationary picture at a given scattering energy $E$ or frequency $\omega$. Due to the difference between the linear dispersion relation for light ($\omega=k c$, or, equivalently, $E\propto p$) and the quadratic dispersion for matter ($E\propto p^2$), 
the temporal dynamics in scattering will be very different for
these two cases. Consider here, e.g., the free motion of a wave packet in one dimension
which satisfies \eq{helmholtz} with the linear dispersion relation $\omega=kc/n$. In free space,
where $n=1$,  
both the group velocity $v_g=\partial \omega/(\partial k)=c$ and the phase velocity 
$v_\phi=\omega/ k=c$ are independent of $\omega$ or $k$ such that wave packets of 
light preserve their
shape in vacuum. In contrast, for electronic matter waves the corresponding velocities $v_g=\hbar k/m$ and $v_\phi=\hbar k/(2m)$ do depend on $k$, such that different frequency components of the wave packet travel with different speeds, leading to
wave packet spreading even in vacuum.  Furthermore, since
the relation $\varepsilon({\bf r})>1$ implies that the light potential can never exceed the light energy, 
$V_{\rm light}<E_{\rm light}$, a dielectric medium can never form a tunneling barrier for light in the
same way as an electrostatic potential can do  for electrons. Also any effects related to
the vectorial character of the electric field (such as the polarization of light) have no 
simple analogy to the electronic case.  When considering stationary scattering problems in which
the polarization does not play an important role,  the analogy between
electron and light scattering can, however, be used extensively. At points where this analogy breaks down, this will be mentioned explicitly. 

\subsubsection{Continuity equation and flux}\label{subsubsection2.1.3}
The scattering of electrons and the scattering of light in a  lossless, static and linear dielectric medium  
have in common that a conservation relation applies,
\begin{equation}
\partial_t W(\bfr,t) +\nabla\cdot {\bf J}(\bfr,t)=0\,.
\label{eq:continuity}
\end{equation}
This so-called ``continuity equation'' states that any temporal change of the density $W$ must
be compensated by a corresponding flux ${\bf J}$. For the electronic case these two quantities
are given by the probability density $W(\bfr,t)=|\psi(\bfr,t)|^2$ and by the probability current
density ${\bf J}(\bfr,t)={\rm Re}\left[\psi(\bfr,t)^*\,{\bf p}\,\psi(\bfr,t)\right]/m$, respectively. The corresponding
quantities for light are the electromagnetic energy density $u(\bfr,t)$ and the Poynting vector ${\bf S}(\bfr,t)$, which fulfill the relation in 
Eq.~(\ref{eq:continuity}), now termed ``Poynting theorem'', when the following replacements are made: 
$W(\bfr,t)\to u(\bfr,t)=\frac{1}{2}\left[\varepsilon{\bf E}(\bfr,t)^2+\mu^{-1}
{\bf B}(\bfr,t)^2\right]$ and ${\bf J}(\bfr,t)\to {\bf S}(\bfr,t)={\bf E}(\bfr,t)\times {\bf B}(\bfr,t)/\mu$ (for which definitions we used real-valued fields ${\bf E},{\bf B}$)
\cite{griffiths_introduction_1999}.
Note that these quantities are of particular importance in experiments, since what detectors typically measure 
is the integrated flux, counted in terms of the number of electrons or of photons that hit the detector surface \cite{lagendijk_resonant_1996,van_tiggelen_analogies_1994}.

\subsubsection{Green's function}\label{subsubsection2.1.2}
A central issue that we will address in this review is the question of how the radiation emitted by a set
of given sources is scattered to a set of receivers. 
For this purpose it is convenient to resort to the concept of the Green's function \cite{morse_methods_1953}. 
We start here again with Maxwell's equations for  a non-magnetic
medium described by a dielectric function $\varepsilon_\omega({\bf r})=\varepsilon_\omega^r+\varepsilon_\omega^s({\bf r})$
that is embedded in an infinite homogeneous reference medium $\varepsilon_\omega^r$. In the presence of external current sources  ${\bf J}_\omega(\bfr)$
we end up with an inhomogeneous vector Helmholtz equation of the following form \cite{tsang_scattering_2004}
\begin{equation}
-{\bf \nabla}\times{\bf \nabla}\times {\bf E}_\omega({\bf r})+\left(\frac{\omega}{c}\right)^2\varepsilon_\omega({\bf r}){\bf E}_\omega({\bf r})=i\mu_0\omega\,{\bf J}_\omega(\bfr)\,,
\label{eq:helmholtznabla}
\end{equation}
with $\omega/c=k_0$ being the vacuum wavenumber. 
When simplifying the notation in the following way $-{\bf \nabla}\times{\bf \nabla}\times\to\bm{\mathcal D}$,
$k_0^2\varepsilon^r_\omega({\bf r})\to{\bf e}^r$, $k_0^2\varepsilon^s_\omega({\bf r})\to{\bf e}^s$, the above Eq.~(\ref{eq:helmholtznabla}) is written as
$(\bm{\mathcal D}+{\bf e}^r+{\bf e}^s){\bf E}=i\mu_0\omega\,{\bf J}_\omega$. The desired Green's function ${\bf G}$ (which is actually a dyadic tensor) satisfies the corresponding equation 
$(\bm{\mathcal D}+{\bf e}^r+{\bf e}^s){\bf G}=\delta({\bf r}-{\bf r}'){\bf 1}$, where we have used the following simplified notation: ${\bf G}({\bf r},{\bf r}',\omega)\to{\bf G}$ and ${\bf 1}$ is the unit tensor.
With the help of the tensorial Green's function, we may relate vectorial current sources with vectorial electric fields through a convolution, 
\begin{equation}
{\bf E}_\omega({\bf r})=i\mu_0\omega\int d{\bf r}'\,{\bf G}({\bf r},{\bf r}',\omega){\bf J}_\omega({\bf r}')\,.
\label{eq:convolution}
\end{equation}
In the case that no current sources are present in a medium, the inhomogeneity in Eq.~(\ref{eq:helmholtznabla}) vanishes. An electric field can still be present, however,
when an incident field is considered. Such a scattering problem can be treated by setting up equivalent relations for the
incident field ${\bf E}^0$ that satisfies 
$(\bm{\mathcal D}+{\bf e}^r){\bf E}^0=0$ in the homogeneous and source-free reference system. With the corresponding Green's function ${\bf G}^0$ satisfying
$(\bm{\mathcal D}+{\bf e}^r){\bf G}^0=\delta({\bf r}-{\bf r}'){\bf 1}$ one finds the so-called Dyson equation ${\bf G}={\bf G}^0-{\bf G}^0{\bf e}_s{\bf G}$  and 
${\bf E}=({\bf 1}-{\bf G}{\bf e}_s){\bf E}^0$ \cite{martin_generalized_1995}. One thus has a generalized field propagator ${\bf K}=({\bf 1}-{\bf G}{\bf e}_s)$ at hand 
that connects the incident field with the full field distribution (including the scattered part) again through a convolution,
\begin{equation}
{\bf E}_\omega({\bf r})=\int d{\bf r}'\,{\bf K}({\bf r},{\bf r}',\omega){\bf E}^0_\omega({\bf r}')\,.
\label{eq:propagator}
\end{equation}
Note that both Eq.~(\ref{eq:convolution}) and Eq.~(\ref{eq:propagator}) are valid independently of the form of the current sources or of the incident field. 
The central piece of information necessary to solve these equations is the system response encapsulated in the Green's tensor ${\bf G}$. To obtain this
quantity, one may proceed through direct inversion of the Dyson equation (${\bf G}^0$ is known analytically \cite{morse_methods_1953}), or through iteration (corresponding iteration schemes 
have been put forward both for electromagnetic wave propagation \cite{martin_generalized_1995} as well as for mesoscopic electron scattering \cite{datta_electronic_1997,ferry_transport_1997,rotter_modular_2000}).

In the general case of a non-uniform medium, which may also change the polarization of
the electric field, the tensorial character of the Green's function is essential.  The reduction to a scalar Green's function is 
allowed, however, when considering the emission and detection in well-defined polarization
directions only \cite{papas_theory_2011}.

\begin{figure}
  \centering
  \includegraphics[width=0.9\linewidth]{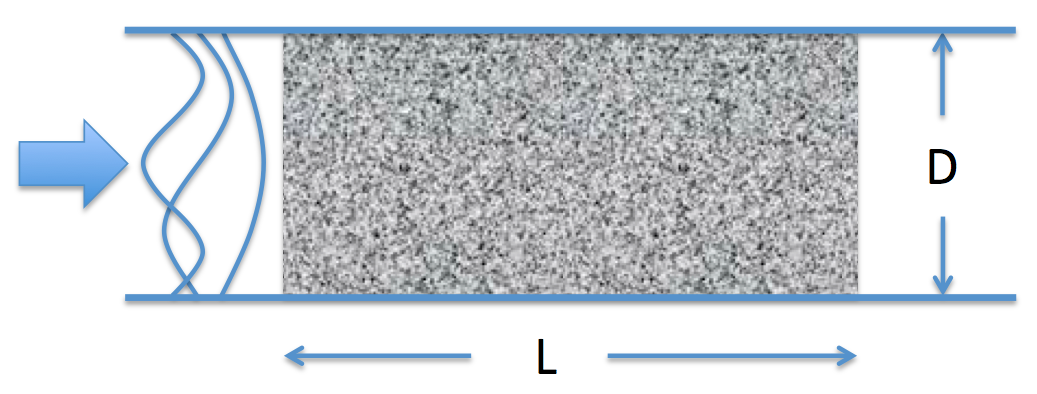}
  \caption{(color online).
  Illustration of the scattering system considered in the text: A rectangular disordered region 
  of length $L$  fills the middle part of an infinite wave of height $D$. Here the flux injected from the left
  through transverse waveguide modes can be transmitted (to the right) or reflected (to the left).}
  \label{fig:slab_figure}
\end{figure}

\subsubsection{Scattering matrix}\label{subsubsection2.1.4}
A primary goal in scattering theory is to connect the incoming flux that is impinging on the
system of interest to the outgoing flux 
scattered away from this system. 
A convenient tool for carrying out the corresponding book-keeping is the scattering matrix
which connects all the incoming and outgoing flux ``channels'' to be defined in detail below. The scattering matrix, in turn,
is intimately related to the Green's function since the latter connects all points in space with each other. 

To illustrate this in detail, we introduce 
as a model system a two-dimensional slab geometry of uniform height $D$ with a disordered dielectric medium of length $L$ in the middle and lossless semi-infinite waveguides of the same height
attached on the left and right (see \fig{slab_figure}). This model system will serve as a convenient tool
to many of the features which we want to explain below for scattering through waveguides, 
fibres and disordered media in
general. To simplify matters, we will assume that we can 
use the scalar Helmholtz equation [see \eq{helmholtz3} above], 
as for a transverse magnetic polarized electro-magnetic field mode in a three-dimensional medium 
which is invariant in $z$-direction. The relevant scalar field component which
we thus describe is the $z$-component of the electric field, $E_z$, assuming hard-wall (Dirichlet)-boundary conditions at the upper and lower boundaries of the scattering domain. 

In the asympotic regions (far away from the disordered part), the field in the left ($\alpha=l$) and right ($\alpha=r$) 
wave guide  will naturally be decomposed into
different waveguide modes $\chi_n(y)=\sqrt{2/D}\sin(n\pi y/D)$, as determined by the boundary conditions in transverse direction,
\begin{equation}
\psi_\omega({\bf x})=\sum_{n=1}^N c_{\alpha,n}^+\,\chi_n(y)\frac{e^{ik_n^x x}}{\sqrt{k_n^x}} +c_{\alpha,n}^- \,\chi_n(y)\frac{e^{-ik_n^x x}}{\sqrt{k_n^x}}\,.
\label{eq:modes}
\end{equation}
For fully defining the above scattering state in the asymptotic region we have summed over all 
$N=\lfloor\omega D/(c\pi)\rfloor$ flux-carrying modes for which the propagation constant 
$k_n^x=\sqrt{\omega^2/c^2-(n\pi/D)^2}$ is real (evanescent modes with an imaginary propagation constant have died out
asymptotically). 
The complex expansion coefficients $c_n^\pm$ correspond to right-moving (+) and left-moving waves (--), respectively. The terms in the denominators
$\sim \sqrt{k_n^x}$ are required to make sure that all the basis states on which we expand the field, have the same flux in longitudinal direction 
$J_\parallel$ (see definitions in section \ref{subsubsection2.1.3}). Based on the above representation of the field in the asymptotic region, we can define the
scattering matrix as the complex matrix which connects the incoming expansion coefficients with the outgoing coefficients,
\begin{equation}
{\bf c}_{\rm out}=\textbf{S}\,{\bf c}_{\rm in}\quad{\rm with}\quad{\bf c}_{\rm in}\equiv\left( {\bf c}_l^+\atop {\bf c}_r^-\right),\,
{\bf c}_{\rm out}\equiv\left( {\bf c}_l^-\atop {\bf c}_r^+\right).
\label{eq:smatrix}
\end{equation}
The $2N\times 2N$ complex coefficients in the scattering matrix can be subdivided into four block matrices,
\begin{equation}
{\bf S}=\left( \begin{array}{cc}
{\bf r} & {\bf t}'   \\
{\bf t}  & {\bf r}'  \end{array} \right)\,,
\label{eq:smatrixblock}
\end{equation}
where the quadratic blocks on the diagonal contain the reflection amplitudes for incoming modes from the left ($r_{mn}$) and from the right
($r'_{mn}$), respectively. The off-diagonal blocks contain the transmission amplitudes for scattering from left to right ($t_{mn}$) and from 
right to left ($t'_{mn}$), respectively. Note that in the case that the number of modes is different on the left ($N$) and right side 
($M$), the reflection matrices  ${\bf r,r'}$ remain quadratic (of size $N\times N$ and $M\times M$, respectively), whereas the
transmission matrices ${\bf t,t'}$ are then just rectangular (of size $M\times N$ and $N\times M$, respectively). 
In this general case the total transmission $T_n$ and reflection $R_n$ associated with a given incoming mode $n$ on the left
read as follows, $T_n=\sum_{m=1}^M |t_{mn}|^2$ and $R_n=\sum_{m=1}^N |r_{mn}|^2$ 
(equivalent relations also hold for the quantities $T'_n,\,R'_n,\,T',\,R'$ with incoming 
modes from the right). We choose here the convention to label the incoming (outgoing) mode with the second (first) subindex in the matrices such as to coincide with the convention of matrix multiplication, i.e., ${\bf c}_{\rm out}=\textbf{S}\,{\bf c}_{\rm in}$ is $c_{{\rm out},m}=\sum_{n}S_{mn}\,c_{{\rm in},n}$.

In electronic scattering the modes cannot be addressed individually, such that the relevant quantity 
in this context is the total transmission $T=\sum^N_{n=1} T_n$, corresponding to the transmission through all
equally populated incoming modes from the left (similarly the total reflection 
$R=\sum^N_{n=1} R_n$). Neglecting the smearing effect of a finite temperature and counting
each of the spin polarizations separately, this
total transmission $T$ can be directly related to the electronic conductance 
\begin{equation}
G=(2e^2/h)\,T=(2e^2/h)\,\sum_{m,n}|t_{mn}|^2\,,
\label{eq:landauer}
\end{equation}
a quantity which is directly measurable in the experiment.  
The above connection between the conductance and the transmission was first derived in 
\cite{economou_static_1981,fisher_relation_1981}, and is commonly 
known as the ``Landauer formula" \cite{landauer_spatial_1957}. Its generalization to multi-terminal systems \cite{buttiker_four-terminal_1986} is referred to as the ``Landauer-B{\"u}ttiker formalism''. 

Since in a scattering process without gain and loss the combined value of the transmission and reflection 
for each mode must be one, we can write $T_n+R_n=1$ and $T'_n+R'_n=1$ or, more generally, $T+R=N$ and 
$T'+R'=M$. These relations, together with the flux-normalization of modes in \eq{modes}, demonstrate the
conservation of flux in systems without sources or sinks. In other words, the incoming flux in
a scattering process, $\sum_n |c_{{\rm in},n}|^2=|{\bf c}_{\rm in}|^2$, must be equal to the outgoing flux 
$\sum_n |c_{{\rm out},n}|^2=|{\bf c}_{\rm out}|^2$, such that,
\begin{equation}
{\bf c}^\dagger_{\rm out}{\bf c}_{\rm out}={\bf c}^\dagger_{\rm in}{\bf c}_{\rm in}\quad\rightarrow\quad
{\bf c}^\dagger_{\rm in}({\bf S}^\dagger{\bf S}-\textbf{1}){\bf c}_{\rm in}=0\,,
\label{eq:unitarity}
\end{equation}
which relation can only be fulfilled if the scattering matrix is unitary, 
${\bf S}^\dagger{\bf S}=\textbf{1}$. Inserting the block-matrix form of the scattering matrix, \eq{smatrixblock},
into this unitarity condition, we arrive at the corresponding relations which the transmission and reflection matrices
have to satisfy: 
\begin{equation}
{\bf t}^\dagger {\bf t}+{\bf r}^\dagger {\bf r}={\bf t}'^\dagger {\bf t}'+{\bf r}'^\dagger {\bf r}'={\bf 1}
,\quad\!\!\!\!{\bf r}^\dagger {\bf t}'+{\bf t}^\dagger {\bf r}'={\bf t}'^\dagger {\bf r}+{\bf r}'^\dagger {\bf t}={\bf 0}
\label{eq:tdaggert1}
\end{equation}
as well as 
\begin{equation}
{\bf t} {\bf t}^\dagger+{\bf r}' {\bf r}'^\dagger={\bf t}' {\bf t}'^\dagger+{\bf r} {\bf r}^\dagger={\bf 1},
\,\quad\!\!\!\! 
{\bf r} {\bf t}^\dagger+{\bf t}' {\bf r}'^\dagger={\bf t} {\bf r}^\dagger+{\bf r}' {\bf t}'^\dagger={\bf 0}
\label{eq:tdaggert2}
\end{equation}
following from the alternative formulation of the unitarity condition, 
${\bf S}{\bf S}^\dagger=\textbf{1}$. 

The above Hermitian matrices ${\bf t}^\dagger {\bf t},\,{\bf r}^\dagger {\bf r},\,{\bf t} {\bf t}^\dagger,\,{\bf r} {\bf r}^\dagger$
and their primed counterparts play an important role in the theoretical description of multi-mode scattering problems.
This is because they can be used to conveniently express several of the scattering quantities of interest. Consider, e.g., 
the total transmission $T$ and reflection $R$ for all incoming modes from the left (see above), which can also be written 
as $T={\rm Tr}({\bf t}^\dagger {\bf t})=\sum_n\tau_n$ and $R={\rm Tr}({\bf r}^\dagger {\bf r})=\sum_n \rho_n$, where $\tau_n$ 
and $\rho_n$ are the real eigenvalues of ${\bf t}^\dagger {\bf t}$ and ${\bf r}^\dagger {\bf r}$, respectively. Also the 
transmission $T_n$ or reflection $R_n$ of a given mode $n$ on the left can be expressed as follows: $T_n=({\bf t}^\dagger 
{\bf t})_{nn}$, $R_n=({\bf r}^\dagger {\bf r})_{nn}$. Note that the transmission of a lead mode $T_n$ is different from that
of a ``transmission eigenchannel'' $\tau_n$, only their sum is the same $T=\sum_n T_n=\sum_n \tau_n$, as required by the
invariance of the trace. From the requirement that ${\bf t}^\dagger {\bf t}+{\bf r}^\dagger {\bf r}={\bf 1}$ we can further
deduce that the quadratic matrices ${\bf t}^\dagger {\bf t}$ and ${\bf r}^\dagger {\bf r}$ are simultaneously
diagonalizable and that their eigenvalues are related as follows: $\tau_n=1-\rho_n$.  

To better understand the relation between the matrices ${\bf t}^\dagger {\bf t}$ and ${\bf t} {\bf t}^\dagger$ we
can write the non-Hermitian and not necessarily quadratic transmission matrix ${\bf t}$ in its singular value decomposition, 
${\bf t}={\bf U}\,{\boldsymbol \Sigma}\,{\bf V}^\dagger$,
where the unitary matrices ${\bf U}$ (of size $M\times M$) and ${\bf V}$ (of size $N\times N$) contain, as their columns, the \textit{left} and 
\textit{right} singular vectors of ${\bf t}$, respectively. The matrix ${\boldsymbol \Sigma}$ in the center contains the real and non-negative
singular values $\sigma_i$ on its diagonal. These quantities are now elegantly connected to the quadratic and Hermitian matrices 
${\bf t}^\dagger {\bf t}$ and ${\bf t} {\bf t}^\dagger$: The left singular vectors of ${\bf t}$ (contained in ${\bf U}$) 
are the orthogonal eigenvectors of ${\bf t} {\bf t}^\dagger$ and the right singular vectors of ${\bf t}$ (contained in ${\bf V}$) 
are the orthogonal eigenvectors of ${\bf t}^\dagger {\bf t}$. The non-zero singular values $\sigma_i$ of ${\bf t}$  (contained in ${\boldsymbol \Sigma}$)
are the square roots of the non-zero eigenvalues of ${\bf t}^\dagger {\bf t}$ and ${\bf t} {\bf t}^\dagger$, i.e., $\sigma_i=\sqrt{\tau_i}$ (if these two matrices are 
different in size, $M\neq N$, the larger matrix has at least $|M-N|$ zero eigenvalues). With the help of these identities, we can write
${\bf t}^\dagger {\bf t}={\bf V}\,{\boldsymbol \Sigma}^2\,{\bf V}^\dagger={\bf V}\,{\boldsymbol \tau}
\,{\bf V}^\dagger$ and ${\bf t} {\bf t}^\dagger={\bf U}\,{\boldsymbol \Sigma}^2\,{\bf U}^\dagger={\bf U}\,{\boldsymbol
\tau}\,{\bf U}^\dagger$, where the diagonal matrix ${\boldsymbol
\tau}$ contains the transmission eigenvalues from above on its diagonal ${\boldsymbol
\tau}={\rm \textbf{diag}}(\tau_1,\ldots,\tau_M)$ . 
From these identities we can conclude that ${\bf t}^\dagger {\bf t}$ and ${\bf t} {\bf t}^\dagger$ share the same
eigenvalues (except for $|M-N|$ zero eigenvalues) and due to the identities \eq{tdaggert1} and \eq{tdaggert2} these 
eigenvalues are also the same as those of the matrices ${\bf t}'^\dagger {\bf t}'$, ${\bf t}' {\bf t}'^\dagger$,
${\bf 1}-{\bf r}^\dagger {\bf r}$, ${\bf 1}-{\bf r} {\bf r}^\dagger$, and 
${\bf 1}-{\bf r}'^\dagger {\bf r}'$, ${\bf 1}-{\bf r}' {\bf r}'^\dagger$.

Additional, so-called reciprocity relations (also called Onsager relations) can 
be obtained for the scattering matrix \cite{jalas_what_2013}.  
In terms of the transmission and reflection matrix elements from above, reciprocity translates into an identity between 
the amplitude for scattering from mode $m$ to another mode $n$ and the amplitude for the reverse process (i.e., from mode $n$ to mode $m$):
 $r_{nm}=r_{mn},\,r'_{nm}= r'_{mn},\,t_{nm}= t'_{mn}$, corresponding to a transposition-symmetric scattering matrix, $\textbf{S}=\textbf{S}^{\rm T}$.
Similar reciprocity relations can also be derived for generalized transmission and reflection coefficients  \cite{nieto-vesperinas_generalized_1986}.
Generally speaking, these relations tell us that if one can scatter from a mode $m$ to another mode $n$ then the reverse 
process also happens with the same amplitude. One may be tempted to associate this property with time-reversal symmetry, which
is, however, misleading. Time-reversal symmetry implies reciprocity, but not the other way around. The best example to illustrate this fact
is a medium with absorption, for which time-reversal symmetry is obviously broken but the above reciprocity relations may still hold 
\cite{van_tiggelen_reciprocity_1997}. Breaking the reciprocity of a medium typically requires a time-dependent dielectric function, non-linear effects or a magnetic field \cite{jalas_what_2013}.

Our choice to evaluate the scattering matrix ${\bf S}$ in the lead-mode basis $\chi_n$ is arbitrary and other
basis sets can be more useful for addressing particular problems. A natural basis, in which the
scattering matrix is diagonal is, of course, its eigenbasis,
\begin{equation}
{\bf S}={\boldsymbol \Omega} \,{\rm \bf diag}(e^{i\phi_1},\ldots,e^{i\phi_{2N}})\,{\boldsymbol \Omega}^\dagger\,.
\label{eq:eigenbasis}
\end{equation}  
Being a unitary matrix, the eigenvalues of $\textbf{S}$ lie on the unit circle in the complex plane and 
can be parametrized as above, using the so-called scattering phase shifts $\phi_n$. 
The transformation to the eigenbasis is mediated by the unitary matrix ${\boldsymbol \Omega}$ which contains 
the eigenvectors of ${\bf S}$. In the presence of time-reversal symmetry where the scattering matrices 
are transposition-symmetric, ${\boldsymbol \Omega}$ can be chosen real and is then an orthogonal matrix
${\boldsymbol \Omega}^{\rm T}{\boldsymbol \Omega}={\bf 1}$. 

The above parametrization of the scattering matrix has the disadvantage that the modes on the left and right of the
sample are strongly mixed as eigenvectors of ${\bf S}$ typically feature components from
all modes, irrespective of their asymptotic behavior. An alternative parametrization 
which disentangles the modes
on the left and right side was proposed in \cite{mello_macroscopic_1988,martin_wave-packet_1992}. This so-called ``polar decomposition'' is based on the singular value 
decomposition of the scattering matrix blocks discussed above and reads as follows
\begin{equation}
{\bf S}=\left( \begin{array}{cc}
{\bf V} & {\bf 0}   \\
{\bf 0}  & {\bf U}  \end{array} \right)\,
\left( \begin{array}{cc}
{\bf -\sqrt{1-{\boldsymbol\tau}}} & {\bf \sqrt{\boldsymbol \tau}}   \\
{\bf \sqrt{\boldsymbol \tau}}  & {\bf \sqrt{1-{\boldsymbol\tau}}}  \end{array} \right)\,
\left( \begin{array}{cc}
{\bf V}' & {\bf 0}   \\
{\bf 0}  & {\bf U}'  \end{array} \right)\,.
\label{eq:polardecomposition}
\end{equation}
In the general case the primed matrices satisfy ${\bf U}'={\bf U}^\dagger,\,{\bf V}'={\bf V}^\dagger$
and in the presence of time-reversal symmetry one 
has ${\bf U}'={\bf U}^{\rm T},\,{\bf V}'={\bf V}^{\rm T}$.
The above transformation from the lead modes to the transmission eigenchannels of ${\bf t}^\dagger {\bf t}$ 
and ${\bf t} {\bf t}^\dagger$ has the advantage that the scattering amplitudes on either side of the medium 
stay well separated but their interrelation becomes  maximally transparent. 

The scattering matrix $\textbf{S}$ which relates the flux-amplitudes of incoming to outgoing modes 
at a fixed scattering frequency $\omega$ shares a very close relationship (Fisher-Lee relation) 
 with the corresponding retarded Green's function $G^+$ at the same value of $\omega$
\cite{fisher_relation_1981} (see \cite{ferry_transport_1997,datta_electronic_1997} for a review). 
This close connection is well 
exemplified by considering the scattering matrix elements corresponding to incoming modes
from the left,
\begin{eqnarray}\label{eq:greensmat0}
t_{nm}(\omega)\!&=&\!-i\sqrt{k_m^x k_n^x}\times\\\nonumber
&&\times\int_0^D\!dy_l \int_0^D\! dy_r \,\chi_m(y_l)G^+(y_l,y_r,\omega)\chi_n(y_r)\nonumber\\
r_{nm}(\omega)\!&=&\!\delta_{nm}-i\sqrt{k_m^x k_n^x}\times\label{eq:greensmat}\\
&&\times\int_0^D\!dy'_l \int_0^D\! dy_l \,\chi_m(y'_l)G^+(y'_l,y_l,\omega)\chi_n(y_l)\nonumber\,
\end{eqnarray}
where the appearing integrals are evaluated along a transverse section in the left ($y_l$) and 
in the right ($y_r$) lead. The corresponding relations for incoming modes from the right 
lead is fully equivalent. 
The flux normalization factors $\sqrt{k_n^x}$, which are necessary to convert field amplitudes into flux amplitudes, correspond to the direction cosines in Fresnel-Kirchhoff diffraction theory \cite{born_principles_1999}. 

Based on the above, it is interesting to take note of the difference
in the information content between the Green's function and the scattering matrix: Whereas the
Green's function along the considered sections contains an infinite set of propagation amplitudes from any 
point on the transverse section to any other point, the scattering matrix relates only the 
finite and discrete set of flux-carrying modes to one another. The reduced information content 
in the scattering matrix is due to the neglect of evanescent modes which carry no flux and thus
also do not contribute to transport. Note, however, that evanescent modes play a crucial role
in the near field of the scattering region where they need to be taken into account if the field distribution close to the scattering region is of interest. Correspondingly, extended definitions of the scattering matrix as well as of their unitarity and reciprocity relations that also include 
evanescent modes have been put forward in \cite{carminati_reciprocity_2000}. 

Since \eq{greensmat0} and \eq{greensmat} provide a relation between the scattering matrix ${\bf S}$ and the scalar 
Green's function $G^+(\bfr,\bfr',\omega)$ and the latter is, in turn, related to the Helmholtz operator 
$\bm{\mathcal L}$ defined in \eq{curlnew}, there should also exist a direct link between ${\bf S}$
and $\bm{\mathcal L}$. To uncover this relation one first subdivides space into an interior (scattering)
region ${\mathcal Q}$, where $\varepsilon(\bfr)$ may vary in space, and an exterior (asymptotic) 
region ${\mathcal P}$ with a constant $\varepsilon(\bfr)$, where the scattering matrix ${\bf S}$ is  
evaluated. Following the so-called Feshbach projection operator technique 
\cite{feshbach_unified_1958,feshbach_unified_1962} (see \cite{zaitsev_recent_2010} for a review),
such a sub-division is carried out with corresponding projection operators,
\begin{equation}
\bm{\mathcal Q}=\int_{\bfr \in {\mathcal Q}} |\bfr\rangle\langle\bfr|,\quad \bm{\mathcal P}=\int_{\bfr \in {\mathcal P}}
|\bfr\rangle\langle\bfr|,
\label{eq:projection}
\end{equation}
which project onto the corresponding regions and satisfy $[\bm{\mathcal P},\bm{\mathcal Q}]=0,\,\bm{\mathcal P}+\bm{\mathcal Q}={\bf 1}$.
With these operators, we can write \eq{curlnew} in the equivalent form as,
\begin{equation}
\left( \begin{array}{cc}
\bm{\mathcal L}_\mathcal{QQ} & \bm{\mathcal L}_\mathcal{QP}  \\
\bm{\mathcal L}_\mathcal{PQ}  &\bm{\mathcal L}_\mathcal{PP} \end{array} \right)
\left({\boldsymbol \mu}_\omega \atop {\boldsymbol \nu}_\omega\right)=\frac{\omega^2}{c^2} \left({\boldsymbol \mu}_\omega \atop 
{\boldsymbol \nu}_\omega\right),\,
\label{eq:projop}
\end{equation}
where the Hermitian diagonal matrix blocks $\bm{\mathcal L}_\mathcal{QQ}=\bm{\mathcal L}_\mathcal{QQ}^\dagger,\,
\bm{\mathcal L}_\mathcal{PP}=\bm{\mathcal L}_\mathcal{PP}^\dagger$ are the projections of 
$\bm{\mathcal L}$ into the 
scattering and asymptotic region, respectively, and the non-Hermitian off-diagonal blocks
$\bm{\mathcal L}_\mathcal{QP}^\dagger=\bm{\mathcal L}_\mathcal{PQ}$ are the coupling
operators between these two regions. The reduced Hermitian operators are now used to define
eigenvalue problems in the spaces ${\mathcal Q}$ and ${\mathcal P}$,
\begin{equation}
\bm{\mathcal L}_\mathcal{QQ} \,{\boldsymbol \mu}_m=\frac{\omega_m^2 }{c^2}\,{\boldsymbol \mu}_m,\quad
{\rm and }\quad\bm{\mathcal L}_\mathcal{PP}\, {\boldsymbol \nu}_{n,\omega}=\frac{\omega^2 }{c^2}\, {\boldsymbol \nu}_{n,\omega}\,.
\label{eq:reducedop}
\end{equation}
Due to the confinement of states ${\boldsymbol \mu}_m$ in $\mathcal{Q}$ the corresponding eigenvalues $\omega_m$
are discrete, whereas the eigenvalues $\omega$ in the unconfined asymptotic regions are continuous ($n$ is just a channel index in this case).
In both regions ${\mathcal P},\,{\mathcal Q}$ the eigenfunctions form a complete set and can thus be used to expand modes of arbitrary complexity in the respective subspaces. 
The interface between ${\mathcal P}$ and ${\mathcal Q}$ can
be chosen anywhere in the asymptotic region and also the boundary conditions on ${\mathcal Q}$ are arbitrary but 
should be such that the operator $\bm{\mathcal{L}}_\mathcal{QQ}$ is Hermitian (with Dirichlet or Neumann boundary 
conditions being the standard choices). If needed, the boundary between ${\mathcal P}$ and ${\mathcal Q}$ can
also be placed in the direct vicinity of the scattering region, in which case the coupling to evanescent modes 
needs to be properly taken into account \cite{viviescas_field_2003}.

To make the connection with the scattering matrix ${\bf S}$ we place the boundary between ${\mathcal P}$ 
and ${\mathcal Q}$ outside the sections where the scattering matrix  is being evaluated in \eq{greensmat}.
It can be shown in this case \cite{mahaux_shell-model_1969} (see \cite{rotter_non-hermitian_2009,guhr_random-matrix_1998,datta_electronic_1997} 
for a review) that the retarded Green's function $G^+(\bfr,\bfr',\omega)$
appearing in \eq{greensmat} with $\bfr,\,\bfr'\in{\mathcal Q}$ can then again be written as a resolvent, 
${\bf G}_\mathcal{QQ}=[\omega^2-\bm{\mathcal{L}}_{\rm eff}]^{-1}$, with the help of an \textit{effective} non-Hermitian 
operator $\bm{\mathcal L}_{\rm eff}=\bm{\mathcal L}_\mathcal{QQ}+\bm{\Sigma}(\omega)$. The so-called \textit{self-energy} 
$\bm{\Sigma}(\omega)$ can be written as follows 
\begin{equation}
\bm{\Sigma}(\omega)=\bm{\mathcal L}_\mathcal{QP}(\omega)\frac{1}{\omega^2-\bm{\mathcal L}_\mathcal{PP}(\omega)+i\varepsilon}
\bm{\mathcal L}_\mathcal{PQ}(\omega)\,,
\label{eq:selfenergy}
\end{equation}
with $\varepsilon$ being here an infinitesimal positive number. Through this self-energy the Green's function
${\bf G}^0_\mathcal{QQ}=[\omega^2-\bm{\mathcal{L}}_\mathcal{QQ}]^{-1}$ 
of the closed region $\mathcal{Q}$ 
(the superscript 0 denotes the absence of coupling to $\mathcal{P}$) turns into the Green's function of the corresponding open system 
${\bf G}_\mathcal{QQ}=[\omega^2-\bm{\mathcal L}_\mathcal{QQ}+\bm{\Sigma}(\omega)]^{-1}$ (where the coupling to 
$\mathcal{P}$ is included). Due to its restriction to the interface between $\mathcal{P}$ and 
$\mathcal{Q}$ the self-energy is nothing else than a non-Hermitian boundary condition which
parametrically depends on the real scattering frequency $\omega$ in the outside domain $\mathcal{P}$. 
This boundary condition is known under the name of Kapur-Peierls or constant-flux boundary condition 
\cite{kapur_dispersion_1938,tureci_self-consistent_2006} with the latter terminology being motivated by the fact that the outgoing flux in $\mathcal{P}$
is conserved (corresponding to a real value of $\omega$). 
In the context of quantum scattering, the operator $\bm{\mathcal L}_{\rm eff}$ is also known as the 
``effective'' or non-Hermitian Hamiltonian. Being non-Hermitian, the operator $\bm{\mathcal L}_{\rm eff}$
has complex eigenvalues, corresponding to eigenstates which decay through the system boundaries
and thus have only a finite lifetime. 

To establish the link between $\bm{\mathcal L}_{\rm eff}$ and the scattering matrix ${\bf S}$ 
we need to express the above operators in the basis vectors of the regions $\mathcal{P}$ and $\mathcal{Q}$, respectively.
For the closed system operator $\bm{\mathcal{L}}_\mathcal{QQ}$ the matrix elements are $H_{\lambda m}=
\langle{\boldsymbol \mu}_\lambda|{\mathcal L}_\mathcal{QQ}|{\boldsymbol \mu}_m\rangle=\omega_m^2 
\delta_{\lambda m}$ and for the self-energy we get $\Sigma_{\lambda m}(\omega)=
-2\Delta_{\lambda m}(\omega)-2i\pi ({\bf WW}^\dagger)_{\lambda m}(\omega)$, 
where the real matrix elements $\Delta_{\lambda m}$ contain frequency shifts and the Hermitian matrix
${\bf WW}^\dagger$ contains damping terms resulting from the coupling between the bounded region  $\mathcal{Q}$ with the continuum region $\mathcal{P}$, i.e.,  
$W_{\lambda m}(\omega)=\langle{\boldsymbol \mu}_\lambda|{\mathcal L}_\mathcal{QP}|{\boldsymbol \nu}_m 
(\omega)\rangle\sqrt{k_m^x}$. These matrix elements which can be calculated analytically for simple systems \cite{viviescas_field_2003}
or numerically for complex geometries \cite{sadreev_s-matrix_2003,stockmann_effective_2002},
allow us to write the effective operator $\bm{\mathcal{L}}_{\rm eff}=\bm{\mathcal{L}}_\mathcal{QQ}-2\bm{\Delta}-2i\pi \bm{W W}^\dagger$ 
and with it the scattering matrix in mode representation \cite{mahaux_shell-model_1969}
\begin{equation}
S_{mn}(\omega)=\delta_{mn}-2i \,\left[\bm{W}^\dagger(\omega) \frac{1}{\omega^2-\bm{\mathcal{L}}_{\rm eff}(\omega)}\bm{W}(\omega)\right]_{mn}\,,
\label{eq:sgreenmat}
\end{equation}
which represents the desired relation between the scattering matrix and the differential operator $\bm{\mathcal{L}}$ introduced at the
beginning of this section, see \eq{curlnew}.

An interesting correspondence that can be established based on Eq.~(\ref{eq:sgreenmat}) is that between the poles of the Green's function ${\bf G}_{\mathcal{Q}\mathcal{Q}}$
and the resonances in the transmission and reflection amplitudes in the scattering matrix ${\bf S}(\omega)$. The complex frequency values $\omega$ at 
which these poles are located are implicitly defined through the eigenvalues $\Omega_k(\omega)$ of $\bm{\mathcal{L}}_{\rm eff}(\omega)$, which have to satisfy
the relation $\omega^2-\Omega^2_k(\omega)=0$. To find the solutions of this equation one 
can iteratively track the values of $\omega$ from the real axis to the desired fixed point for each specific eigenvalue $\Omega_k$. 
The complex resonance values found in this way play an important role
for scattering problems, as their real parts specify the positions of scattering resonances and their imaginary parts fix the corresponding resonance widths, which, in turn, are inversely proportional 
to the decay time of a resonant state in this open system. Due to their finite life-time the resonances are also often referred to as quasi-bound states or quasi-modes of the system and 
starting from the original work by Gamow \citep{gamow_zur_1928} many theoretical studies are based on these states \cite{moiseyev_non-hermitian_2011}. In contrast to 
the constant-flux states mentioned above, the quasi-bound states do, however, have the problem that they diverge to infinity outside of the system boundaries, which requires much care when 
using them to expand a field in this set of states \cite{ching_quasinormal-mode_1998}. On the other hand, quasi-bound states do not feature a parametric dependence on the 
frequency outside the system (as the constant-flux states do), since for the quasi-bound states both of 
the involved frequencies are equal: $\omega_k=\Omega_k$.

\subsubsection{Random Matrix Theory (RMT)}\label{subsubsection2.1.5}
A very convenient tool to understand the statistical rather than the system-specific 
properties of scattering processes in disordered media is Random Matrix Theory (RMT). 
The basic assumption of RMT is that the statistical properties of a sufficiently chaotic 
or disordered system are the same as those of those of a suitably 
chosen ensemble of random matrices.  
This idea, which was originally introduced by Wigner
to model the distribution of energy spacings in nuclei  \cite{wigner_characteristic_1955}, has 
meanwhile found a broad range of applications, not only in nuclear physics (see \cite{weidenmuller_random_2009,mitchell_random_2010} for a review), but also in mesoscopic physics 
(see \cite{beenakker_random-matrix_1997} for a review)
and increasingly so in disordered photonics (see \cite{beenakker_applications_2011} for a review). 
The broad applicability of RMT (see \cite{stockmann_quantum_2006} for a review) is strongly linked
to the so-called Bohigas-Giannoni-Schmitt (BGS) conjecture \cite{bohigas_characterization_1984}
 according to which RMT describes well  
 the spectral statistics of any ``quantum'' or ``wave'' system (governed by a wave equation) whose ``classical'' counterpart (governed by a corresponding Hamiltonian equation of motion) 
is \textit{chaotic}. Classically, such chaotic systems are characterized by having 
more degrees of freedom than constants of motion.
Quantum mechanically this translates into having more degrees of freedom than 
``good'' quantum numbers. Finding a proof for the BGS conjecture has
turned out to be very difficult (proofs in certain limits have meanwhile been proposed 
\citep{muller_semiclassical_2004}). 
Extensive theoretical and experimental work (see \citep{beenakker_random-matrix_1997}  for
a review) has, however, shown that the BGS 
conjecture is very well satisfied in many different physical scenarios 
not only  for mesoscopic quantum systems and the corresponding matter waves, but for many other types of waves as well
(like optical, acoustic and micro-waves etc.) \cite{guhr_random-matrix_1998,graf_distribution_1992,stockmann_quantum_2006,ellegaard_spectral_1995,dietz_quantum_2015}. Furthermore, based on the analogies (see section \ref{subsubsection2.1.1}) between quantum 
systems (described by a Schr\"odinger equation) and optical scattering systems (described by a Helmholtz equation) 
many of the results that have been explored in the field of mesoscopic physics can now be carried
over to the domain of optical scattering. Before demonstrating this explicitly by means of concrete 
examples (see section \ref{section3}), we will first review the basic theoretical concepts of RMT.

\begin{figure}
  \centering
  \includegraphics[width=1\linewidth]{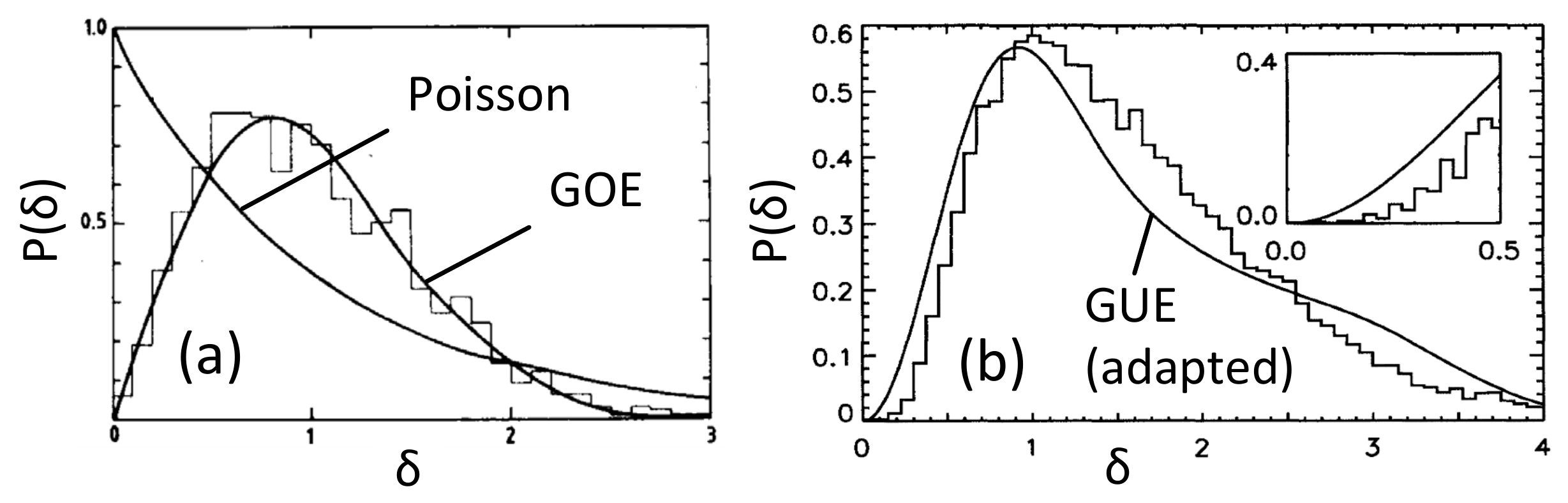}
  \caption{(a) Comparison of the nearest neighbor level spacing distribution  
  $P(\delta)$ in  a set of nuclear scattering resonances (histogram) with the corresponding RMT prediction from the Gaussian Orthogonal Ensemble (GOE) for time-reversal invariant systems. The very good agreement found confirms that the statistical property of nuclei can be approached through RMT. For comparison also the prediction from a Poisson distribution is shown, corresponding to uncorrelated levels. Image adapted from \cite{bohigas_nuclear_1983}. (b) Level spacing distribution in a quasi two-dimensional microwave billiard with an isolator attached. Here the data is well described by the Gaussian Unitary Ensemble (GUE). The solid line is adapted to account for missing levels and the inset shows the quadratic increase of the level repulsion for small values as characteristic for GUE. Image adapted from \cite{stoffregen_microwave_1995}.}
  \label{fig:spacingdist}
\end{figure}

Our starting point for applying RMT to the systems considered in this review will be
the approach by Wigner and Dyson \cite{wigner_characteristic_1955}, which
consists in replacing
the matrix representation of the differential operator ${\mathcal{L}}_\mathcal{QQ}$ in 
Eqs.~(\ref{eq:projop},\ref{eq:reducedop}) for a specific 
closed system $\mathcal{Q}$ by a random matrix ${\bf H}$. The latter contains as each of its elements 
$H_{mn}$ a randomly generated number from an ensemble with the following Gaussian 
distribution $P(H_{mn})=(w\sqrt{2\pi})^{-1}\exp[-H_{mn}^2/(2w^2)]$
and zero average $\langle H_{mn}\rangle=0$ (the value of $w$ determines the mean level spacing
of the corresponding eigenvalues). Since the matrix elements $H_{mn}$ can, in general, be complex, we may
 choose both the real and imaginary parts from this Gaussian ensemble independently.
The only additional constraint that is imposed on the matrix elements $H_{mn}$
is that they are those of a Hermitian matrix ${\bf H}^\dagger={\bf H}$, i.e., $H^*_{mn}=H_{nm}$.
For a time-reversal-symmetric system, the matrix elements are real and symmetric, i.e., $H_{mn}=
H_{nm}\in{\mathbb R}$. Having replaced the  differential operator $\bm{\mathcal{L}}_\mathcal{QQ}$ by a random
matrix ${\bf H}$, all system-specific information about $\mathcal{Q}$ is lost and only statements
on the statistical properties of a whole class of 
systems can be made that can be associated with the same Gaussian ensemble of matrices
 \cite{porter_statistical_1965} (see \cite{mehta_random_2004} for a review). For the Hermitian
matrices with complex, Gaussian-distributed elements this ensemble is called the Gaussian Unitary 
Ensemble (GUE). The term \textit{unitary} refers here to the unitary matrices containing the eigenvectors 
of these Hermitian matrices. Similarly, the
symmetric matrices with real, Gaussian-distributed elements are referred to as the Gaussian
Orthogonal Ensemble (GOE), where \textit{orthogonal} refers to the corresponding orthogonal eigenvector matrix.
It can now be shown 
that each of these matrix ensembles has a very specific distribution
of eigenvalues. In particular, when we take Gaussian random matrices of very large size ($N\to\infty$)
and compute the set of eigenvalues $E_\alpha$, then their distribution will be universal in the
average over many matrix realizations. The corresponding distribution function for the
eigenvalues, 
\begin{equation}
P(\{E_n\})\approx const.\times\prod_{m<n}^N|E_n-E_m|^\beta\times\prod_{n}^N\exp[-E_n^2/(2w^2)]\,
\label{eq:eigenvaluedistribution}
\end{equation}
is known as the Wigner-Dyson distribution, following the original work by these two authors \cite{dyson__1962,wigner__1957,wigner_random_1967}.
The parameter $\beta$ here is assigned the value $\beta=1$ for GOE and $\beta=2$ for GUE.
An interesting result contained in this distribution is the repulsion of nearby levels, i.e., 
$P(\delta)\to 0$ for $\delta\to 0$. More specifically, when considering the normalized spacing between nearest 
eigenvalues $\delta=(E_{\alpha+1}-E_\alpha)/\Delta$ with the mean level spacing $\Delta=\langle E_{\alpha+1}-E_\alpha\rangle$ 
then one finds that the level repulsion scales like $P(\delta\ll 1)\propto \delta^\beta$.
Comparing this result with experimental data both for time-reversal invariant systems with $\beta=1$ 
(see Fig.~\ref{fig:spacingdist}(a))
as well as for systems without time-reversal symmetry and $\beta=2$ (see 
Fig.~\ref{fig:spacingdist}(b)), shows good agreement. Note that the results shown in
Fig.~\ref{fig:spacingdist} stem from very different physical systems, like an atomic nucleus in (a) and 
a microwave billiard with an attached isolator in (b).

To investigate how these results for bounded systems carry over to the case of unbounded scattering systems, we follow the so-called ``Heidelberg approach''
 \cite{mahaux_shell-model_1969} (see \cite{guhr_random-matrix_1998} for a review). 
Here, the random Hamiltonian matrix ${\bf H}$ describing the bounded region ($\mathcal{Q}$) 
is coupled to the unbounded outside
domain ($\mathcal{P}$) by way of the frequency-dependent coupling matrices $W_{\lambda,m}(\omega)$ introduced in section \ref{subsubsection2.1.4}.
Using Eq.~(\ref{eq:sgreenmat}) then yields the corresponding scattering matrix for transmission and reflection through
the region described by the Hamiltonian ${\bf H}$. Note that in this approach 
no approximation is introduced by the subdivision of space into 
$\mathcal{P}$ and $\mathcal{Q}$. 
When one is interested in the statistical
properties of the scattering matrix, a less rigorous calculation is usually sufficient. A common approximation is, e.g., to 
neglect the frequency-dependence of the coupling matrix elements $W_{\lambda,m}(\omega)$, which are then drawn from an 
ensemble of random numbers, just like the matrix elements of ${\bf H}$. 

In this approach, which has found interesting applications in nuclear scattering 
\cite{iida_wave_1990,iida_statistical_1990,verbaarschot_grassmann_1985}, the only frequency dependence in the scattering matrix
comes from the $\omega^2$-term in Eq.~(\ref{eq:sgreenmat}). This explicit frequency dependence is essential as it can be used
to study frequency correlations in the scattering matrix elements \cite{guhr_random-matrix_1998}.
Consider here, in particular,
that the bound eigenstates of the Hamiltonian ${\bf H}$ are coupled by the matrix elements $W_{\lambda,m}(\omega)$ to the waveguide modes, which
turns these states into quasi-bound resonances (as discussed in the last paragraph of section
\ref{subsubsection2.1.4}).
Depending on whether the coupling strength (as determined by the matrix elements $W_{\lambda,m}(\omega)$) is smaller or larger than the mean
level spacing of the Hamiltonian eigenstates (as determined by the variance of the matrix elements $H_{mn}$) these
resonances will be isolated (weak coupling) or overlapping (strong coupling), resulting in very different frequency correlations
in the scattering matrix elements (see also \cite{brouwer_generalized_1995} for more details
on the situation with non-ideal waveguide coupling).

A different strategy to set up a random matrix theory for coherent scattering, also known
as the ``Mexico approach'', starts not with the Hamiltonian ${\bf H}$, but with the
scattering matrix ${\bf S}$ as the fundamental quantity \cite{friedman_marginal_1985,friedman_information_1985,mello_information_1985,baranger_mesoscopic_1994,jalabert_universal_1994}. In this approach (see \cite{alhassid_statistical_2000} for a review), 
which was developed independently of the ``Heidelberg approach'', 
one replaces the scattering matrix elements by random complex numbers. In analogy to the random Hamiltonian matrix elements from
above, which had to be chosen such as to respect the Hermiticity of the Hamiltonian, the random scattering matrix  elements 
have to respect the unitarity of the scattering matrix. In the case of time-reversal symmetry, the scattering matrix additionally 
has to be symmetric (see discussion in section \ref{subsubsection2.1.4}). 
The corresponding matrix ensembles are 
referred to as Dyson's circular ensemble 
\cite{dyson_statistical_1962} with the parameter $\beta=1$ assigned to
unitary symmetric and $\beta=2$ for general unitary matrices. Assuming such a distribution leads to very specific
correlations in the scattering phase shifts in Eq.~(\ref{eq:eigenbasis}),
\begin{equation}
P(\{\phi_n\})\propto\prod_{n<m}\left|\exp(i\phi_n)-\exp(i\phi_m)\right|^\beta\,,
\label{eq:phasecorrelation}
\end{equation}
which were found by Bl\"umel and Smilansky \cite{blumel_random-matrix_1990} 
to describe the phase shifts 
in chaotic scattering very well. To link the circular ensemble to experimentally more accessible quantities
like the statistics of transmission and reflection,  the
 corresponding distribution of the
transmission eigenvalues $\tau_n$ of the matrices ${\bf t}^\dagger {\bf t}$ and ${\bf t} {\bf t}^\dagger$
needs to be evaluated \cite{baranger_mesoscopic_1994,jalabert_universal_1994}.
The corresponding joint probability
density of transmission eigenvalues is given as follows,
\begin{equation}
P(\{\tau_n\})\propto \prod_{n<m}\left|\tau_n-\tau_m\right|^\beta
\times\prod_p\tau_p^{-1+\beta/2}\,.
\label{eq:transcorrelation}
\end{equation}
The product between neighboring transmission eigenvalues leads to 
a level repulsion similar to the repulsion of energy eigenvalues of the random Hamiltonian
${\bf H}$. Note, however, that 
the above Eq.~(\ref{eq:transcorrelation}) also applies for the case of just a few scattering channels $N$ down to the 
single channel case where $N=1$. 

\begin{figure}
  \centering
  \includegraphics[width=0.85\linewidth]{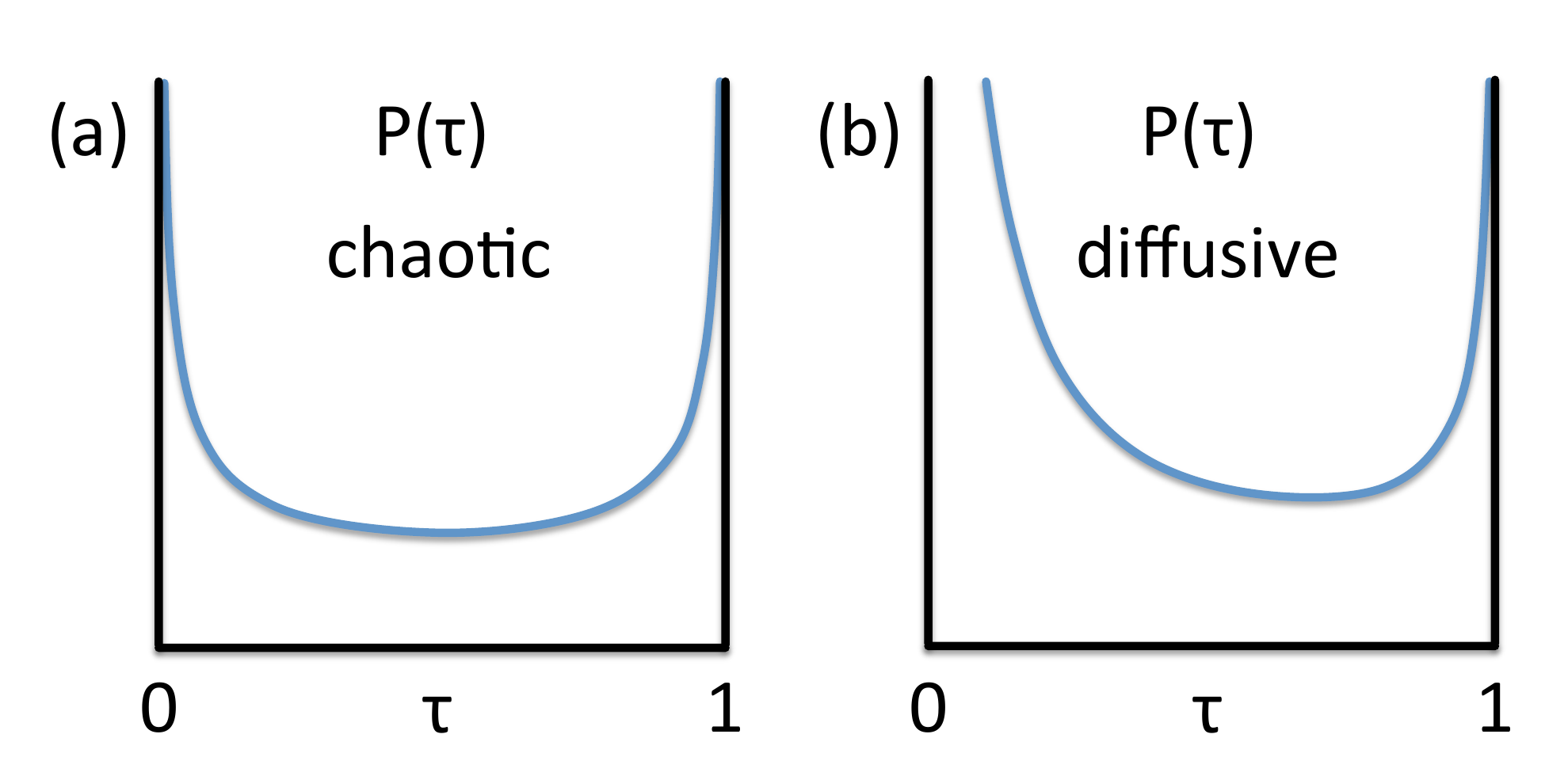}
  \caption{(color online). Distribution of transmission eigenvalues $\tau$ for the case (a) of chaotic scattering and (b) of diffusive scattering. The corresponding  analytical expressions for these functions (following from random matrix theory) are spelled out in Eq.~(\ref{eq:onepoint}) and in Eq.~(\ref{eq:bimodal}), respectively. }
  \label{fig:bimodaldist}
\end{figure}

The above result already contains very interesting physics. Consider, e.g., the limiting
case of the above distribution for a very large number of scattering channels ($N\to\infty$), broken down to the
one-point probability density of transmission eigenvalues $P(\tau)$. The latter is given as the mean value of the 
microscopic density $\rho(\tau)=\sum_n^N\delta(\tau-\tau_n)$ with respect to the ensemble average
according to the probability distribution from above,
\begin{equation}
P(\tau)\equiv\langle\rho(\tau)\rangle=\int_0^1 d\tau_1\ldots\int_0^1 d\tau_N P(\{\tau_n\})\rho(\tau)\,,
\label{eq:densityaverage}
\end{equation}
resulting in the following bi-modal distribution 
for the limit $N\gg 1$ illustrated in Fig.~\ref{fig:bimodaldist}(a)
\cite{baranger_mesoscopic_1994,jalabert_universal_1994},
\begin{equation}
P(\tau)=\frac{1}{\pi\sqrt{\tau(1-\tau)}}\,.
\label{eq:onepoint}
\end{equation} 
Results for small $N$ or for an asymmetric number of channels on the left and right are
provided in \cite{savin_shot_2006}.
Note that in the case of broken time-reversal symmetry ($\beta=2$) for which 
Eq.~(\ref{eq:onepoint}) was derived, the
distribution $P(\tau)$ is symmetric around $\tau=1/2$. For the time-reversal symmetric case ($\beta=1$), however, 
the second product in Eq.~(\ref{eq:transcorrelation})
induces an asymmetry into this distribution which 
leads to the following results for the average transmission \cite{baranger_mesoscopic_1994,jalabert_universal_1994}
\begin{equation}
\langle T\rangle=\frac{N}{2}+\left(\frac{1}{4}- \frac{1}{2\beta}\right)+\mathcal{O}(1/N)\,.
\label{eq:avtransm}
\end{equation}
The reduction in transmission for $\beta=1$ is called the {\it weak localization} correction, which 
in a semi-classical picture can
be partially associated with the presence of time-reversed path pairs that enhance the reflection $R$ (hence the name
``localization''). The term ``weak'' refers here
to the fact that the correction is of order $\mathcal{O}(1)$, which is much smaller than the leading term $\mathcal{O}(N)$. Experimental demonstrations of this effect will be discussed in section \ref{subsubsection3.1.3}.

The first product in Eq.~(\ref{eq:transcorrelation}) involving pairs of transmission
eigenvalues suppresses the likelihood of neighboring transmission eigenvalues approaching each 
other very closely. This eigenvalue repulsion leads also to a spacing distribution between nearest neighbor transmission 
eigenvalues which scales like $P(\delta \ll 1)\propto s^\beta$ where the normalized spacing is given by
$\delta=(\tau_{m+1}-\tau_m)/\Delta$, with $\Delta=\langle\tau_{m+1}-\tau_m\rangle$. Note the similarity here
with the spacing distribution obtained earlier for the eigenvalues of a 
random Hamiltonian, see Eq.~(\ref{eq:eigenvaluedistribution}). 
 Due to this ``spectral rigidity'' the transmission eigenvalues
only fluctuate between the limits imposed by their neighboring values, which is obviously much 
less than for uncorrelated transmission eigenvalues that would be described by a Poisson distribution 
$P(\delta \ll 1)\propto \exp(-\delta/\Delta)$. This suppression of fluctuations is so strong that 
the variance of the fluctuations in the total transmission (as defined below Eq.~(\ref{eq:smatrixblock})) 
approach a universal, but $\beta$-specfic value \cite{lee_universal_1985},
which is of order $\mathcal{O}(1)$ and thus independent of $N$,
\begin{equation}
{\rm var}\,T=\frac{1}{8 \beta}\,.
\label{eq:}
\end{equation}
This prediction for universal conductance fluctuations in 
 a chaotic cavity was obtained with the ``Heidelberg approach'' for the Hamiltonian
by \cite{iida_wave_1990,iida_statistical_1990,verbaarschot_grassmann_1985} and with the ``Mexico approach'' for the scattering matrix by \cite{baranger_mesoscopic_1994,jalabert_universal_1994}.
An extended discussion of universal conductance fluctuations can be found in section \ref{subsubsection3.1.2}.

Another highly non-trivial aspect of coherent chaotic scattering is contained in the shape of the distribution function
of transmission eigenvalues (see Fig.~\ref{fig:bimodaldist}a): Contrary to what one would
naively expect, the transmission eigenvalues $\tau_n$ are not uniformly distributed between the limiting
values 0 and 1; instead, Eq.~(\ref{eq:onepoint}) predicts that the $\tau_n$ are peaked near 0 and 1, corresponding 
to transmission channels that are almost closed (near 
$\tau\approx 0$) and others that are almost open (near $\tau\approx 1$) -- a phenomenon also known under the name ``maximal fluctuation theorem" \cite{pendry_maximal_1990}. These open and closed transmission eigenchannels (see Fig.~\ref{fig:bimodaldist})
which were first discovered by Dorokhov 
\cite{dorokhov_coexistence_1984} will play an important role for many of the
effects which we are going to discuss in this review. 
To understand the origins of the bi-modal distribution consider the repulsion of transmission eigenvalues inherent in Eq.~(\ref{eq:transcorrelation}). The closer a given transmission eigenvalue is to 0 (to 1), the more it is repelled by its higher (lower) neighbors (simply because there are more of them). Together with the restriction to the inverval $\tau\in[0,1]$, this leads to the clustering near the limiting values 0 and 1. It is also instructive 
to compare the RMT distribution for $P(\tau)$ with the distribution which one would get for the scattering of classical particles
rather than of waves. Since particles (like billiard balls) that enter a scattering region connected to a left and right port, can 
either be fully reflected or fully transmitted, but nothing in between, the corresponding classical distribution function 
has two delta peaks, $P_{\rm cl}(\tau)=\alpha\,\delta(\tau)+(1-\alpha)\,\delta(1-\tau)$, with $\alpha=1/2$ when assuming the probability for 
transmission and for reflection to be equal. We can thus conclude that the RMT-distribution for the transmission eigenvalues
is peaked at exactly those values of transmission which are classically allowed. In turn, only a comparatively small fraction of  
transmission eigenchannels features transmission in the classically forbidden region around $\tau\approx 0.5$. This analogy between
the scattering of waves and of particles is all the more remarkable as the open and closed transmission eigenchannels in
the RMT-distribution are an interference effect, in contrast to the classical case where evidently no interference occurs. 

When comparing the predictions of RMT with the results of a real experiment or a simulation for a specific scattering geometry
it is important to keep in mind that such a comparison can only be meaningful on a statistical level. This is because RMT does not
contain any information about non-universal, i.e., system-specific details. Also, it is important to ask which requirements a system 
needs to fulfill such that its statistical properties can meaningfully be compared with RMT. For the case considered right above, where
the scattering matrix was replaced by a random unitary matrix with the same number of channels on the left and right, we can think of
a scattering system that is connected to an incoming and an outgoing port of equal size. Following the BGS conjecture from above, the
scattering region in between the ports should be classically chaotic. In addition, the non-universal scattering contributions should 
be as weak as possible which is equivalent to the requirement that all incoming channels should get more or less equally randomized before
exiting again from the scattering region. 

\begin{figure}
  \centering
  \includegraphics[width=0.85\linewidth]{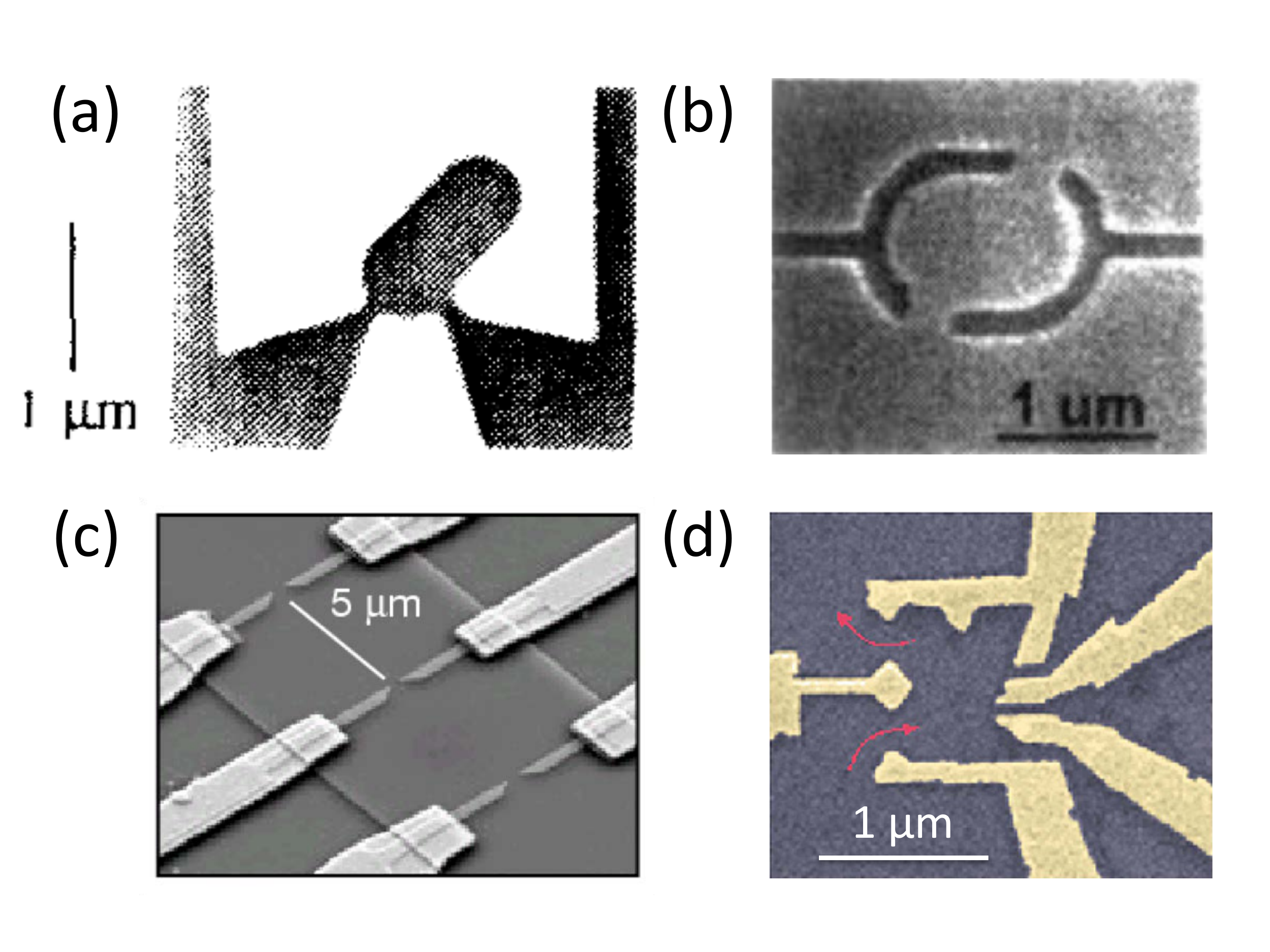}
  \caption{(color online). Different experimental realizations of chaotic quantum dots with figures adapted from \cite{marcus_conductance_1992} (a),
  \cite{chang_weak_1994} (b), \cite{oberholzer_crossover_2002} (c), and \cite{marcus_quantum_1997} (d). In all four cases these electronic billiards
  are fabricated based on a semi-conductor heterostructure with high mobility. The current between the source and drain enters and exits through 
  slits (quantum point contacts), which are small compared to the overall dimension of the chaotic scattering region in between.}
  \label{fig:quantum_dot_figure}
\end{figure}

\begin{figure}
  \centering
  \includegraphics[width=0.85\linewidth]{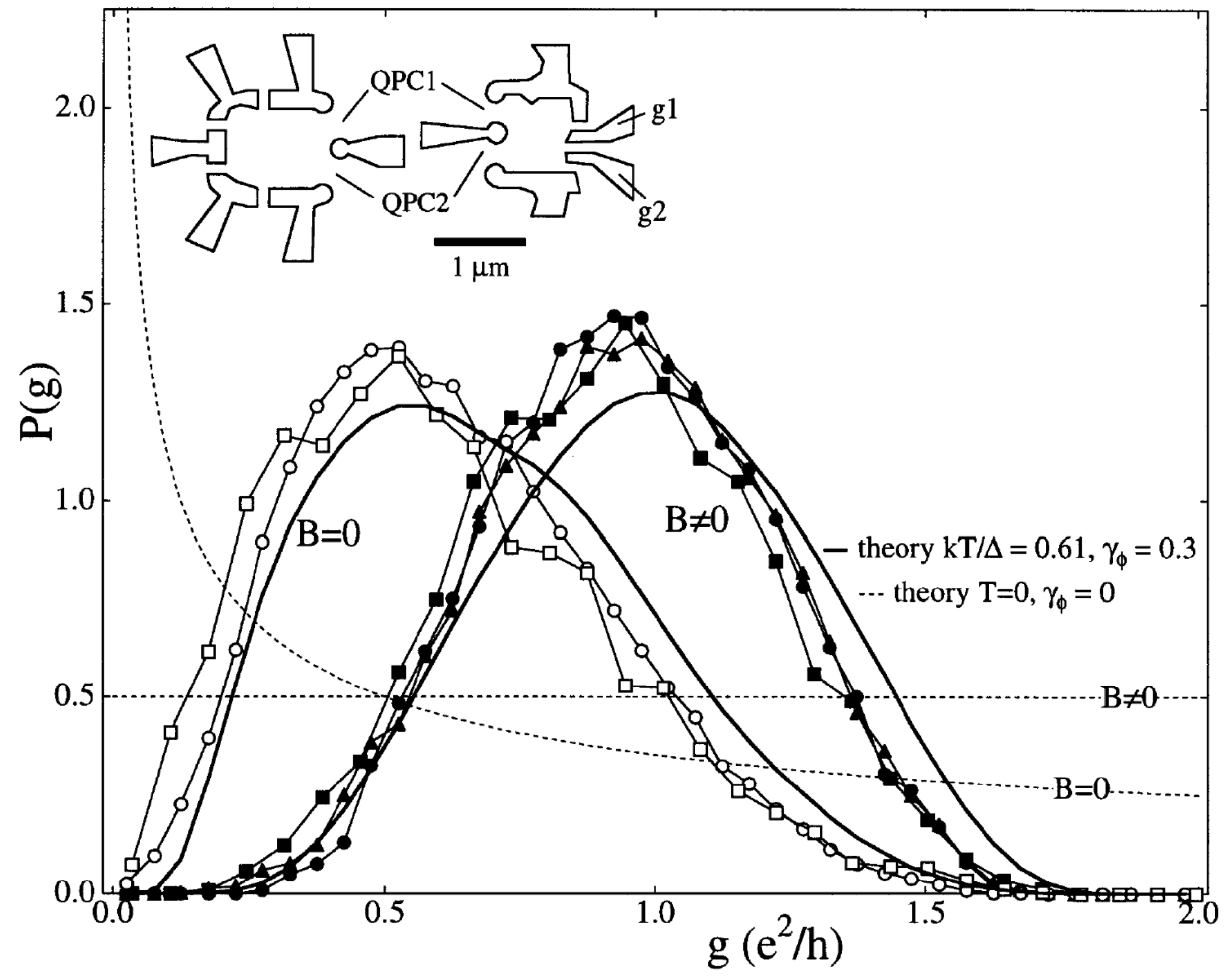}
  \caption{Experimental results from \cite{huibers_distributions_1998} on the distribution of the conductance $P(g)$ in electrostatically defined quantum dots (see schematic on the top). Both the input as well as the output point contacts only feature a single open transverse mode such that these distributions are equivalent to the distribution of transmission eigenvalues in the single-channel limit. The corresponding theoretical predictions from RMT (dashed lines) do not reproduce the experimental data (connected symbols). Only when effects due to finite temperature ($T$) and dephasing (through the dephasing rate $\gamma_\phi$) are taken into account (see solid lines) good agreement is found. Both the situation with and without time-reversal symmetry are considered with the latter case being realized through the application of a finite magnetic field ($B$) applied perpendicular to the scattering area.}
  \label{fig:cond_distribution}
\end{figure}

Figure \ref{fig:quantum_dot_figure} 
shows a few scattering setups (for electrons) that fulfill these requirements 
to a satisfactory degree. The special shape of these systems already suggests that typical scatterers from the real world are, however, generally not described by RMT (the world would, indeed, be a dull place if this was the case). We also show in Fig.~\ref{fig:cond_distribution} that mesoscopic 
transport experiments usually suffer from several imperfections (decoherence processes, 
finite temperature, etc.) which severely spoil the agreement between the
measurement data and an RMT prediction (even for well-engineered electron billiards as shown in 
Fig.~\ref{fig:quantum_dot_figure}). Figure \ref{fig:cond_distribution} also shows that a decent agreement can be
found when the influence of these real world effects is taken into account
in the corresponding RMT model. As it turns out, also the optical scattering
through a disordered medium (as discussed in the latter sections of this review) 
is not well described by the above simple models. 
The amendments to RMT that are necessary in these cases are, however, 
different from the ones employed for electron transport through quantum dots as in  Fig.~\ref{fig:cond_distribution}. 

\subsubsection{DMPK equation}\label{subsubsection2.1.5a}
\begin{figure}
  \centering
  \includegraphics[width=0.8\linewidth]{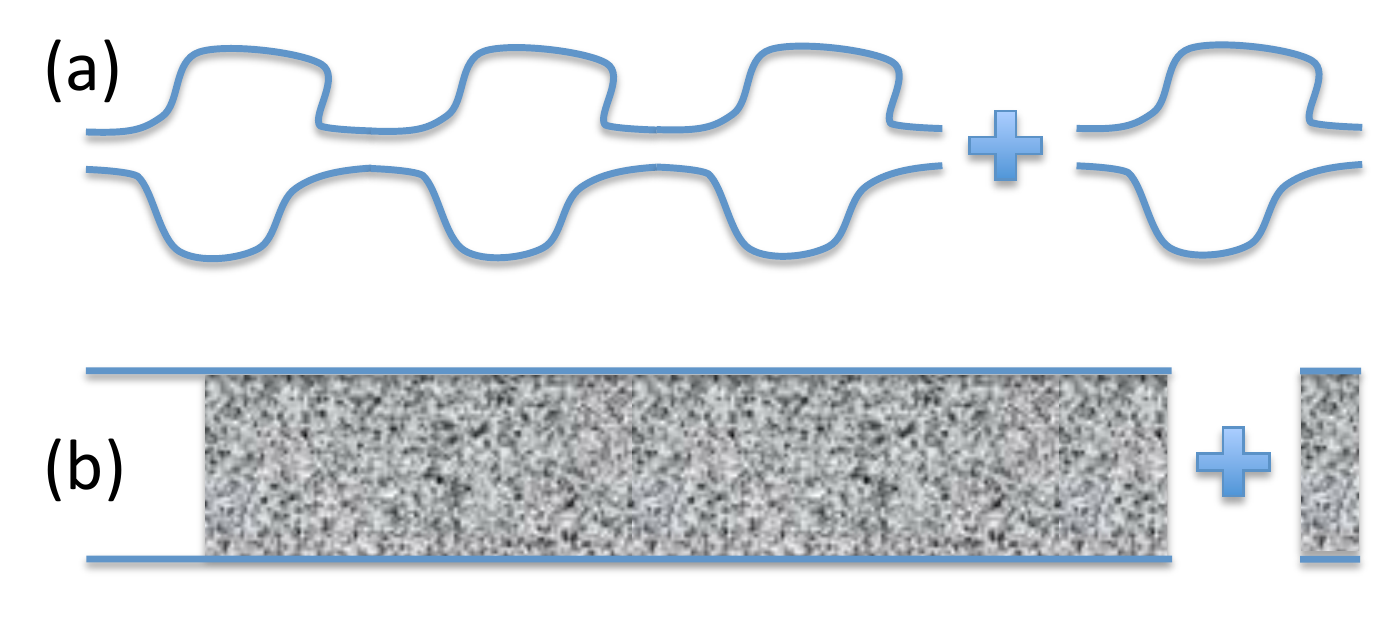}
  \caption{(color online). (a) Stacking chaotic quantum dots behind each other leads to an effective wire geometry with statistics that can be well described by a corresponding random matrix approach. (b) In the more refined DMPK approach weakly scattering segments are recursively added to the wire geometry. }
  \label{fig:dmpk_figure}
\end{figure}

The starting point for this section is the insight that a disordered system is inherently more complex than
a chaotic quantum dot and its RMT description. This is because a disordered medium does more than
just randomize all incoming waves in equal measure and letting them escape again symmetrically on either side. In particular, depending on how long the incoming waves remain inside the medium the degree of disorder scattering that they will suffer from will be very different. Also transmission and
reflection will of course depend on the thickness of this medium as compared to the transport
mean free path $\ell^\star$. To cope with this situation the simple RMT models from section \ref{subsubsection2.1.5} 
were extended by concatenating many random scattering matrices from the 
appropriate RMT ensemble in series. This approach has first been used in \cite{iida_wave_1990,iida_statistical_1990,altland_conductance_1991,weidenmuller_scattering_1990} 
to describe electronic scattering through a disordered
wire and corresponds to the description of a system consisting of a series of chaotic cavities,
see Fig.~\ref{fig:dmpk_figure}(a). Whereas this ansatz allows one 
to conveniently extend RMT to such more complicated scenarios, the approach also has several limitations, in particular, as the different transport regimes 
in a wire (ballistic, diffusive, localized) and their respective crossovers are hard to treat with it
(see \cite{dembowski_anderson_1999} for a microwave experiment on coupled cavities).  

To properly describe all of these regimes Dorokhov, Mello, Pereira and Kumar (DMPK) 
already earlier proposed a model \cite{dorokhov_transmission_1982,mello_macroscopic_1988},
which subdivides
the scattering region into a series of weakly scattering segments rather than using the
fully randomized matrices from RMT, see Fig.~\ref{fig:dmpk_figure}(b). 
Choosing each segment of length $\Delta z$ shorter than the transport mean free path, 
$\Delta z\ll \ell^\star$, but longer than the wavelength, $\Delta z\gg \lambda$, has the advantage that adding a new segment 
can be described as a perturbative correction. Assuming, in addition, that in each segment all 
incoming channels are scattered by the disorder into all of the available channels
isotropically (i.e., with equal weight), one can derive a Fokker-Planck equation for the ``Brownian motion'' 
of the transmission eigenvalues $\tau_n$ (with constant diffusion
coefficient). The corresponding evolution equation for the distribution of 
transmission eigenvalues $\tau_n$ as a function of the wire length $L$ is known as the DMPK equation,
\begin{equation}
\frac{\partial}{\partial s}P(\{x_n\},s)=\frac{1}{2\gamma}\sum_{m=1}^N
\frac{\partial}{\partial x_m}\left[\frac{\partial P}
{\partial x_m}+\beta P\frac{\partial}{\partial x_m}
\Omega(\{x_n\})\right]
\label{eq:dmpk}
\end{equation}
where we have substituted the transmission eigenvalues $\tau_n$ 
by new variables $x_n$ according to $\tau_n=1/\cosh^2x_n$ and used
$s=L/\ell^\star$, $\gamma=\beta N+2-\beta$ as well as
\begin{eqnarray}
\Omega(\{x_n\})&=&-\sum_{m<n}\ln|\sinh^2 x_n-\sinh^2x_m|
\nonumber\\&&-\frac{1}{\beta}\sum_m\ln|\sinh 2x_m|\,.
\label{eq:jacobian}
\end{eqnarray}
The variables $x$ used above for simplifying the equations can be 
interpreted such that $L/x_m$ is the channel-specific ``localization length'' for the 
transmission channel $m$ in the disordered region (we will see in a later part of this 
section what localization is).

Note that in real disordered wires the isotropy assumption,
which corresponds to an ergodicity assumption in the transverse direction,
is in general not well fulfilled for short lengths $L$. This is because 
any specific disorder profile typically features a
non-uniform differential scattering cross-section that
leads to preferential coupling between specific mode pairs.
Since the length scale for transverse diffusion is not taken into account in the DMPK equation,
its validity is restricted to ``quasi-one-dimensional'' (quasi-1D) wire geometries which are 
much longer than their transverse width,  $L\gg W$. For very long systems (with $L\gg \ell^\star$) 
the solutions to the DMPK
equation have equivalent statistics as those of the concatenated random scattering matrix
model from above. In a similar spirit, one
can also set up an alternative model where not the scattering matrix but the Hamiltonian is
the central quantity of interest. As demonstrated by Efetov and Larkin \cite{efetov_kinetics_1983} 
such an approach can be mapped onto a so-called supersymmetric nonlinear $\sigma$-model, which
was shown to be equivalent to the DMPK equation \cite{brouwer_quantum_1996} in the ``thick
wire limit'' with many scattering channels $N$. 
Since all of these models were extensively discussed already in 
several reviews and books 
\cite{beenakker_random-matrix_1997,brouwer_random-matrix_1997,janssen_fluctuations_2001}, 
we will not review them again here.
Rather, we will present in the following a summary of the main results of the DMPK model that will 
be useful for later chapters. For this purpose we will rely on the exact solutions of the DMPK equation.
As the case of broken time-reversal symmetry ($\beta=2$) is much easier to treat, we will discuss it first and then point out corrections for the case when time-reversal symmetry is restored ($\beta=1$).

Focusing on the case of a thick wire with many transverse channels ($N\gg 1$) the DMPK equation 
makes specific predictions for three characteristic regimes: 

(i) In the {\it ballistic} regime the 
scattering is very weak, such that the system length is smaller than the mean free path, $L\lesssim \ell$.
In this limit the waves can be thought of as travelling ballistically on straight lines, rather than being multiply 
scattered by the disorder. Correspondingly, the transmission eigenvalues are all very close to one and no reflection occurs. In optics the ballistic regime is very important since most imaging techniques only
work in the ballistic limit. Correspondingly, in systems through which X-rays propagate ballistically and visible light does not -- a situation commonly encountered in biomedical imaging -- the former type of radiation is better suited  for imaging purposes.

(ii) In the {\it diffusive} regime,
where the system length is in between the mean free path and the 
localization length, $\ell^\star\lesssim L \lesssim \xi$ (with $\xi\approx \beta N \ell^\star$ 
in quasi-1D systems), the transmitted waves have 
already undergone many scattering events. This translates into a randomization of the transmission eigenvalues $\tau_n\,$,
which is quantified by solving the DMPK equation starting with a ``ballistic initial condition'' $\tau_n=1$ for all 
$n$ imposed at $s=L/\ell^\star=0$ up to the length $L$ at which $s=L/\ell^\star\gg 1$. 
The solution for the  joint probability density of transmission eigenvalues
reads as follows \cite{beenakker_exact_1994}
\begin{eqnarray}
P(\{x_n\},s)&\propto&\prod_{i<j}\left[\left(\sinh^2 x_j-\sinh^2 x_i\right)(x_j^2-x_i^2)\right]\times\nonumber\\
&&\prod_i\left[\exp(-x_i^2N/s)(x_i\sinh 2x_i)^{1/2}\right]\,.
\label{eq:dmpkdiffusive}
\end{eqnarray}
Distilling out of this result the one-point probability density of transmission eigenvalues (by integration) one finds that in the regime of very long systems $s\gg 1$ 
the leading term of order $\mathcal{O}(N)$ (which
is independent of $\beta$) features a {\it uniform} distribution of the transformed transmission eigenvalues $x_n$,
\begin{equation}
P(x,s)\equiv\langle\rho(x)\rangle_s=\frac{N}{s}\Theta(s-x)\,.
\label{eq:uniformdistrib}
\end{equation}
Note that the only length dependence that remains here is the upper cut-off of this uniform
distribution introduced by the Heaviside-Theta function $\Theta$. This cut-off
sets the transmission of all modes to zero for which $x\gtrsim L/\ell^\star$ and it keeps the normalization
at $\int_0^\infty P(x,s)dx=N$. If we translate this result back to the transmission eigenvalues $\tau_n$
we find that out of
the $N$ transmission channels about $N\ell^\star/L$ have a finite 
transmission with $\tau> 4\exp(-2L/\ell^\star)$, which again follow a bi-modal but asymmetric distribution, see Fig.~\ref{fig:bimodaldist}(b) \cite{pendry_universality_1992,imry_active_1986,dorokhov_coexistence_1984},
\begin{equation}
P(\tau)=\frac{N\ell^\star}{2L}\frac{1}{\tau\sqrt{1-\tau}}\,.
\label{eq:bimodal}
\end{equation}  
The remaining $N(1-\ell^\star/L)$ channels are closed (i.e., $\tau\approx 0$ for them).
The normalization is here such that $\int_{\tau_0}^1
P(\tau)d\tau=N$ with the lower integration limit $\tau_0=4\exp(-2L/\ell^\star)$. 
In practice the closed transmission eigenvalues are smeared over several eigenvalue spacings; 
since, however, they only contribute weakly to transport this cut-off is usually not specified
in more detail. Most importantly, we have thus obtained the spectacular result that even in transmission
through a highly scattering quasi-1D system open transmission channels with $\tau\approx 1$ exist,
which are chiefly responsible for the transmission (i.e., conductance). For the typical situation 
encountered in optics it will be important to know that these open transmission channels are also 
present in the case of a disordered slab (see section \ref{subsubsection4.3}), which, contrary 
to the assumptions in the DMPK
model, is much shorter than its transverse width  $L\ll W$ \cite{goetschy_filtering_2013}. 

The above results obtained for $\beta=2$ are subject to corrections of next to leading order 
$\mathcal{O}(1)$ for the 
time-reversal symmetric case of $\beta=1$. These corrections were measured in the experiment 
\cite{mailly_sensitivity_1992} and also appear in the average transmission and the variance of 
the fluctuations
of the transmission as induced by changing the disorder configuration or an external parameter (like the 
scattering wavenumber  $k$),
\begin{equation}
\langle T\rangle=\frac{N l}{L}+ \frac{\beta-2}{3\beta}+\mathcal{O}(1/N),\quad{\rm var}\,T=\frac{2}{15\beta}+\mathcal{O}(1/N)
\label{eq:gandvarg}
\end{equation}
Note the interesting analogy of the leading order term in transmission $\propto 1/L$ to the Ohmic behavior of a classical
wire whose resistance ($\propto 1/T$) scales linearly with the length $L$. Given the fact that we have used here a wave picture of 
transport rather than a classical trajectory picture (as in the Drude model), this analogy is far from obvious, in particular, in view
of the bi-modal distribution of transmission eigenvalues. Another interesting observation based on Eq.~(\ref{eq:gandvarg}) is
that the application of a mechanism that breaks time-reversal symmetry (like a magnetic
field for electrons) leads to a slight increase of the average transmission (weak localization) and to a two-fold decrease of the transmission fluctuations ${\rm var}\,T$. The latter are apparently also universal in the case 
of diffusive scattering, with an $N$-independent value for the leading order term 
${\rm var}\,T=2/(15\beta)$.
Note that these results agree exactly with those from an independent calculation using a diagrammatic perturbation theory \cite{anderson_possible_1979,gorkov_particle_1979,altshuler_fluctuations_1985,lee_universal_1985}. Extensive reviews of the diagrammatric framework  can be found in \cite{dragoman_quantum-classical_2004,montambaux_coherence_2006,akkermans_mesoscopic_2007}. 
Diagrammatic techniques have the downside that they do not provide
access to the full distribution of transmission eigenvalues, which is why we review them only
very briefly in section \ref{subsubsection3.1.2} 

(iii) In the {\it localized }regime we are in the situation where the system length is larger than the so-called ``localization length'', $L\gtrsim\xi$, 
which, in quasi-1D systems is connected to the mean free path $\ell^\star$ by the relation, $\xi\approx\beta N \ell^\star$. The effect of localization, originally 
proposed by P.W.~Anderson in \cite{anderson_absence_1958}, 
exponentially suppresses transmission due to multiple interference and is thus entirely due to 
the wave nature 
of the scattered flux (the effect is nonexistent in a trajectory picture as for classical particles). 
As reviewed in  \cite{lagendijk_fifty_2009,abrahams_50_2010}), several experiments have meanwhile successfully demonstrated localization of different kinds of waves (as for sound, microwaves, light and cold atomic gases).
Solving the DMPK equation in the localized limit ($s\gg N$) 
\cite{dorokhov_electron_1983,pichard__1991,dorokhov_transmission_1982}
yields a 
joint probability density of the transmission eigenvalues  
which nicely factorizes into a product of Gaussian distributions,
\begin{equation}
P(\{x_n\},s)=\left(\frac{\pi s}{N}\right)^{-N/2}\prod_{n=1}^N\exp\left[-(N/s)
(x_n-\bar{x}_n)^2\right]
\label{eq:distr_loc}
\end{equation}
centered around the regularly spaced mean values $\bar{x}_n=(s/N)(2n-1)/2$. Since, in the limit $s\gg N$, the 
width of the Gaussians is much smaller than the 
spacing to the nearest neighbors, the transmission
eigenvalues  $1\ll x_1\ll x_2\ll 
\ldots \ll x_N$ ``crystallize'' on a regular lattice with 
a lattice spacing of $\delta x=N/(Ll)$ (for $N\gg 1$)
\cite{muttalib_random_1990,pichard_theory_1990,stone_random_1991,frahm_equivalence_1995}. 
It is interesting to compare this crystal-like
behavior in the localized regime with the liquid-like behavior found 
for the diffusive regime, where the $x_n$ are uniformly distributed, see Eq.~(\ref{eq:uniformdistrib}). 
In the transition region between these two regimes also the
distribution function is intermediate between a constant function
with a cut-off and a series of Gaussians as expected for a partially melted solid
(see also experimental data in Fig.~\ref{fig:crystallization} of 
section \ref{subsubsection3.1.5}).

If we translate this result for the $x_n$ to the conventional transmission
eigenvalues $\tau_n=1/\cosh^2 x_n$, we can use the fact that $x_n\gg 1$ 
to simplify $\tau_n\approx 4\exp(-2x_n)$. The transmission eigenvalues
thus have a log-normal distribution. Since for the total transmission $T$
the first transmission eigenvalue $\tau_1$ then dominates over all others, 
we also find that the total transmission, $T\approx 4\exp(-2x_1)$ takes on a
log-normal distribution with the following mean value and variance
\begin{equation}
\langle T\rangle=-s/N+\mathcal{O}(1)\,,\quad
{\rm var}\langle T\rangle=-2\langle \ln T\rangle=4L\xi\,.
\label{eq:localtandvart}
\end{equation}
These results were calculated for the case of broken time-reversal 
symmetry, $\beta=2$. The connection between the mean and the variance of the
conductance, however, stays valid for other values of $\beta$ \cite{beenakker_exactly_1994,pichard__1991}. Note, how, based on this 
relationship, the variance of the transmission increases as the transmission
itself is reduced, a result which nicely contrasts the constant value
of the transmission fluctuations in the diffusive case. 
To arrive at this result one uses the fact that the crystallization of transmission
eigenvalues in the localized regime reduces the multi-channel scattering problem
effectively to a one-channel problem. 

This single-channel regime of transport occurring in the deeply localized limit is also very instructive for relating the
transmission eigenchannels with the internal modes in the system -- a connection that is already inherent in the definition of the scattering matrix, see discussion below  Eq.~(\ref{eq:sgreenmat}). 
The internal ``quasi-bound states'' or ``resonances'' are responsible for mediating
the transmission from one side of the medium to the other. In most circumstances, like in the diffusive scattering
regime, these modes will have a resonance width $\delta\nu$ which exceeds their mean level spacing $\Delta\nu$, 
resulting in a ratio (called the Thouless number) \cite{edwards_numerical_1972,thouless_maximum_1977}
$\delta\equiv \delta\nu/\Delta\nu > 1$
for which many modes are strongly overlapping 
such that they are extractable from the transmission data only through special techniques \cite{kuhl_resonance_2008,persson_observation_2000}.
The opposite limit of well-resolved modes and with it the regime of Anderson localization
itself, is characterized by $\delta<1$ (Thouless criterion). It was  shown in 
\cite{thouless_maximum_1977,abrahams_scaling_1979} that in the localized limit the
transmission $T$ through the system (or, equivalently, the dimensionless conductance $g=G\,h/(2e^2)$)
and the Thouless number become the same $g=\delta$ (whereby the spectral and the 
transport properties become intimately connected). The Thouless number also turns out to
govern all statistical properties of Anderson localization \cite{abrahams_scaling_1979}.  

As was meanwhile demonstrated successfully 
in an experimental micro-wave study, the good resolution of modes in the localized limit 
($\delta<1$) allows to decompose a speckle
pattern of radiation transmitted through a disordered sampled  
into a sum of only a few individual mode patterns  \cite{wang_transport_2011}. 
Not only do very few localized modes dominate transmission in the localized regime, but also just a few transmission channels are open. As has meanwhile been demonstrated
explicitly, these modes and channels are not merely strongly correlated \cite{choi_perfect_2012}, 
but, in fact, directly linked with each other \cite{pena_single-channel_2014}: 
In the deeply localized regime the single dominant transmission eigenchannel 
is given either by a single localized mode or by a so-called ``necklace state'', which is a highly transmitting superposition of 
overlapping localized modes \cite{pendry_quasi-extended_1987}. 
Another curious observation in this deeply localized limit is that due to
the dominance of a single transmission transmission eigenchannel, the entire scattering system can be mapped onto a strictly
one-dimensional system with the same statistical properties, provided that its localization length is properly
renormalized \cite{pena_single-channel_2014}. 
Such a mapping can also be carried out for media with multiple open transmission 
channels that can then be mapped onto a sum of several one-dimensional systems, 
not only in terms of the transmission
statistics, but also in terms of the density of states in the medium and the corresponding time-delay \cite{davy_transmission_2015,davy_universal_2015}.

\subsection{Open transmission eigenchannels and shot noise}\label{subsection2.2}
One of the most spectacular predictions of RMT and of the DMPK equation is the existence of
``open transmission eigenchannels'' which have been discovered first by Oleg Dorokhov in 1984 
\cite{dorokhov_coexistence_1984} (see
the corresponding distribution of transmission eigenvalues in Fig.~\ref{fig:bimodaldist}). 
Due to the absence of wave front shaping tools for coherent electron scattering, directly probing these channels with electrons is, however, only possible for very simple geometries like quantum point contacts. 
As noted first by Yoseph Imry \cite{imry_active_1986}, 
the open transmission eigenchannels do, however,
leave very conspicuous statistical signatures on the transport properties of electrons. In particular, as we will explain
in the following, the presence of open transmission eigenchannels is detectable in the electronic shot noise (not
to be confused with photonic shot noise or conductance fluctuations). 

The term ``shot noise'' was originally introduced by Walter Schottky 
who was measuring the temporal fluctuations of the electric
current in a vacuum tube \cite{schottky_uber_1918}. 
As he first pointed out, these time-dependent fluctuations around the mean current value are due
to the granularity of the electronic charge. In other words, since electrons come in discrete charge packets (i.e., the
elementary charge) they don't produce a fluid-like flow of current but rather a random succession of discrete charge impact events.
Comparing this situation to the (acoustic) noise produced 
by the small metal pellets from a ``shot gun'' when impinging on a solid surface, Schottky predicted shot noise to be a convenient tool
to measure the value of the electron charge. Specifically, he proposed a relationship between the so-called shot noise spectral density 
$S(\omega)=\langle\delta I(\omega)^2\rangle/\delta\omega$ based on the frequency-dependent current fluctuations around
the mean current value, $\delta I(\omega)=I(\omega)-\langle I(\omega)\rangle$ and the mean current, $S(\omega)=2e\langle I(\omega)\rangle$.
(Note that the factor of 2 comes from the contribution of positive and negative frequencies and that the formula only holds in the
limit where contributions from thermal or $1/f$-noise can be disregarded.)
Since both $S(\omega)$ and $\langle I(\omega)\rangle$ can be measured in an experiment, one should be able to determine the elementary
charge $e$ according to Schottky, who assumed electrons to be completely uncorrelated (Poisson distributed) to derive this
relation. Due to the residual correlations among electrons (even in a vacuum tube), Schottky's prediction, however, failed to reach
the accuracy of the seminal Millikan experiment using oil droplets (see 
\cite{beenakker_quantum_2003} and \cite{blanter_shot_2000} for a review). 

Contrary to the expectation from the famous Franck-Hertz experiment, the shot noise produced in Schottky's vacuum tube can be understood completely classically \cite{schonenberger_shot_2001}. In the mescoscopic limit, however, where electrons behave as quantum matter waves (as in ultra-thin wires at a few milli-Kelvin) the correlations that lead to deviations from the Schottky formula become dominant.
These deviations are typically quantified in terms of the so-called ``Fano factor'', $F=S/S_P$, which is the ratio of the 
noise spectral density $S$ for a given system (with correlations), as compared to the uncorrelated value of Schottky 
for Poissonian statistics $S_P=2e\langle I\rangle$. In the mesoscopic limit these quantities can be conveniently 
evaluated using the Landauer-B\"uttiker framework (see section \ref{subsubsection2.1.4}) 
to estimate both the current \cite{buttiker_absence_1988}, 
$\langle I\rangle=(2 e^2/h)V\sum^N_{n=1}\tau_n$, as well as the noise spectral density
\cite{buttiker_scattering_1990,khlus_current_1987,lesovik_excess_1989}, $S=2e(2e^2/h)V\sum^N_{n=1}\tau_n(1-\tau_n)$.
Note that the latter prediction relates the magnitude of the time-dependent current fluctuations ($S$) with the
time-independent transmission eigenvalues $\tau_n$. An intuitive interpretation for this expression 
can be given as follows \cite{beenakker_quantum_2003}: Since according to the Pauli principle, at zero temperature (as considered here) all levels up to the Fermi
energy $E_F$ are filled with electrons and above $E_F$ all levels are empty, all thermal fluctuations are suppressed.
The quantum shot noise thus comes from the electrons in a given transmission eigenchannel, attempting to transmit 
from source to drain with transmission probability $\tau_n$. Since the electron in channel $n$ can either pass
or not pass, one gets binomial statistics as in a sequence of statistically independent yes/no experiments, each of which
has a probability of $\tau_n$ to give ``yes'' as an answer. Correspondingly, the fluctuations in the transmitted 
current will be proportional to $\tau_n(1-\tau_n)$ for channel $n$; since, furthermore, all channels are statistically
independent, the total fluctuations will be proportional to $\sum^N_{n=1}\tau_n(1-\tau_n)$, just as predicted in the above
formula for the noise spectral density. Note that in contrast to classical electronic shot noise which is due to the randomness
associated with the emission of electrons, for quantum electronic shot noise the randomness in emission is completely suppressed
by the Pauli principle. Instead, the noise is here due to the intrinsic indeterminism inherent in any quantum transmission problem
to which only a transmission ``probability'' can be assigned. Due to the different statistics (Bose-Einstein vs.~Fermi-Dirac)
the shot noise will also be different when replacing electrons with photons -- even when considering systems with the same
scattering matrix. Loosely speaking, photons are more ``noisy'' due to bunching, whereas electrons are more ``quiet'' 
due to anti-bunching. As a consequence, the results from above can not be directly mapped from the electronic to the photonic case,
where primarily amplification and absorption, rather than scattering, shift the Fano factor away from its Poissonian value \cite{beenakker_photon_1999}.
More recent work, however, has also found that mesoscopic fluctuations influence the photocount statistics of coherent light scattered in a random medium
\cite{balog_photocount_2006}.

\begin{figure}
  \centering
  \includegraphics[width=1\linewidth]{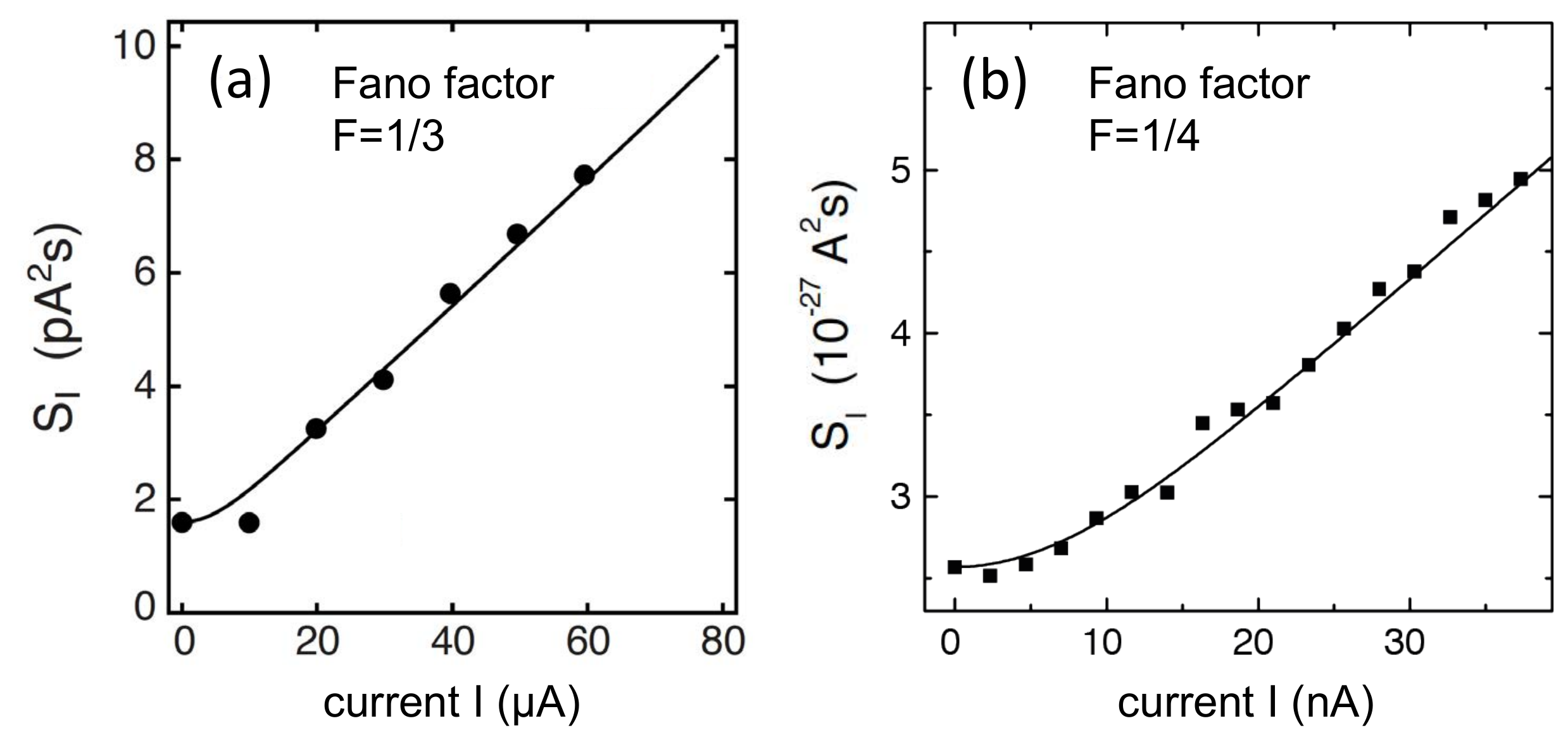}
  \caption{Shot noise power (a) in a metallic diffusive wire and (b) in a chaotic quantum cavity. Figures adapted from \cite{henny_1/3-shot-noise_1999} for (a) and \cite{oberholzer_crossover_2002} for (b). The linear rise of the experimentally obtained noise power with increasing current has a slope that follows the theoretical predictions for the universal Fano factors $F=1/3$ and $F=1/4$, respectively. For small currents the noise deviates from the linear increase due to finite temperature effects.}
  \label{fig:onethird}
\end{figure}

To make contact with the open and closed transmission eigenchannels, consider that the noise spectral density $S$ for
electrons is a very sensitive measure of the distribution of transmission eigenvalues 
$P(\tau)$ studied in sections \ref{subsubsection2.1.5} and \ref{subsubsection2.1.5a}. 
To understand this point, consider that, for many scattering channels,
$N\gg 1$, the expression for $S$ can be conveniently rewritten as follows, $S=2e(2e^2/h)V\int_0^1 P(\tau)\tau(1-\tau) d\tau$, 
from which we may conclude that the distribution function $P(\tau)$ enters the spectral density $S$ through its first and second moment 
(i.e., $\int_0^1 P(\tau)\tau^n d\tau$ with $n=1,2$). With the first moment just being the average transmission, the
shot noise power thus additionally provides access to the second moment of $P(\tau)$, a 
quantity which will be stongly influenced by 
the presence of open and closed channels. In particular, if we replace $P(\tau)$ with the bi-modal distributions 
obtained from RMT and from the DMPK equation, one finds
very specific values for the Fano factor. 

Consider first the case where we assumed the scattering matrix to be distributed according to Dyson's 
circular ensemble with $P(\tau)=1/[\,\pi\sqrt{\tau(1-\tau)}\,]$ [see Eq.~(\ref{eq:onepoint})]. 
In this case the shot
noise Fano factor can be calculated by hand to take on the universal value
 $F=1/4$ \cite{jalabert_universal_1994,baranger_mesoscopic_1994}, corresponding to a shot noise
spectral density $S$ which is reduced to one fourth of the Poissonian value $S_P$ of Schottky \cite{schottky_uber_1918}.   
When taking, instead, the transmission eigenvalue distribution which we found for the wire in the diffusive regime, $P(\tau)\propto
1/(\,\tau\sqrt{1-\tau}\,)$ [see Eq.~(\ref{eq:bimodal})] one finds the shot noise to be suppressed to one third with a corresponding Fano factor of  $F=1/3$ \cite{beenakker_suppression_1992,nagaev_shot_1992},
 which is entirely independent of the mean free path $\ell^\star$ and of the system length $L$. 
In the transition from the diffusive to the ballistic limit 
(where all eigenchannels open up) the Fano factor vanishes, $F\to 0$ 
\cite{de_jong_mesoscopic_1992}, and in the transition to 
the localized limit (where all eigenchannels get closed) the Fano factor approaches one, $F\to 1$ 
\cite{frahm_equivalence_1995}.  
Note that in both of the non-trivial limits where the Fano factor takes on fractional values, low-temperature experiments with 
coherent electrons \cite{steinbach_observation_1996,oberholzer_crossover_2002}
 have meanwhile confirmed the theoretical predictions (see Fig.~\ref{fig:onethird}), 
thereby providing a convincing proof for the existence of open transmission channels in 
transport through chaotic and disordered media, respectively. As we will see below, going beyond this 
statistical evidence by accessing transmission eigenchannels in optics directly will only become possible through the techniques of wave-front shaping, see section \ref{section4}.

The above universal values for the Fano factor rely on the assumption that waves entering in a scattering
region get perfectly randomized before exiting this region. In fact, this assumption is the starting point for 
RMT and in weaker form, also enters the DMPK equation through the approximation of transverse isotropy. There are
of course many ways in which a specific scattering system can fail to fulfill these assumptions: First of all, a scattering region might neither be fully chaotic nor disordered 
\cite{agam_shot_2000,aigner_shot_2005,oberholzer_crossover_2002}, or its disorder
might feature spatial correlations which lead to very specific transmission statistics
\cite{izrailev_anomalous_2005}. Also any effects like absorption 
\cite{brouwer_transmission_1998,mendez-sanchez_distribution_2003} and
dephasing \cite{huibers_distributions_1998,baranger_effect_1995,brouwer_effect_1995} 
have a significant influence (see Fig.~\ref{fig:cond_distribution}).
Consider also that the way in which one couples to a disordered region (as, e.g., by barriers or point contacts) can
lead to the situation that part of the incoming flux is immediately backreflected, rather than being randomized. Such non-universal
contributions to the transport statistics can, however, be suitably described with tools like the Poisson kernel \cite{brouwer_generalized_1995}. 
Alternatively, one might also be confronted with systems like thin disordered interfaces 
which, on the one hand, scatter incoming waves strongly but which are shorter than the wavelength, on the other hand, such
that they fall outside of the predictions for ballistic, diffusive or localized samples; see 
\cite{schep_transport_1997} for a successful treatment of such cases.
 
A particularly interesting challenge to conventional theories arises for the case when the randomization in a given scattering
system affects only a sub-part of the scattered flux. This situation occurs, e.g., when 
``direct'' scattering processes 
are able to penetrate the random medium in a time that is below the time scale necessary for 
randomization to set in \cite{agam_shot_2000,gopar_problem_1998}.
For conventional strongly scattering media the fraction of such ``ballistic'' scattering states decreases
exponentially with the system size. In imaging, this strong suppression of ``ballistic light'' 
in turbid media is in fact one of the key challenges for techniques based on light in the visible
part of the spectrum, which is scattered significantly, e.g., in biological tissue \cite{ntziachristos_going_2010}. 
Also in the field of quantum shot noise a whole body of work exists
in which the influence of such non-universal contributions is investigated in detail
\cite{marconcini_analysis_2006,rotter_statistics_2007,schomerus_quantum--classical_2005,oberholzer_crossover_2002,sukhorukov_quantum--classical_2005,jacquod_breakdown_2004,aigner_shot_2005,agam_shot_2000,nazmitdinov_shot_2002,
silvestrov_noiseless_2003}. Generally speaking, one finds that ballistic scattering contributions reduce the Fano factor below the universal values found above. This is because the fully closed or fully open transmission eigenchannels (with $\tau=0,1$) associated with ballistic scattering are ``noiseless'' in terms of their contribution to shot noise \cite{silvestrov_noiseless_2003}. We will see in the next section and in section \ref{subsubsection5b.1a} that such ballistic noiseless states in electronic quantum transport correspond to  geometric optics states in light scattering, i.e., light rays to which the eikonal 
approximation applies and which have a well-defined time-delay  \cite{rotter_generating_2011}.

\subsection{Time-delay}\label{subsection2.4}

When speaking of dynamical aspects of scattering problems, well-defined time scales are required to provide
an estimate for the duration of a scattering process. Whereas many different definitions of such time scales are
available in the literature, the most rigorously defined and most commonly used ones are the {\it time delay} (also
called {\it delay time} or {\it group delay}) and the {\it dwell time}, which quantities measure the duration of a scattering process and the time
spent inside a designated region, respectively. As can be expected, these two time scales will turn out to be
quantitiatively similar for many practical purposes, but also the subtle differences between them provide instructive insights.

\begin{figure}
  \centering
  \includegraphics[angle=0,width=0.9\columnwidth]{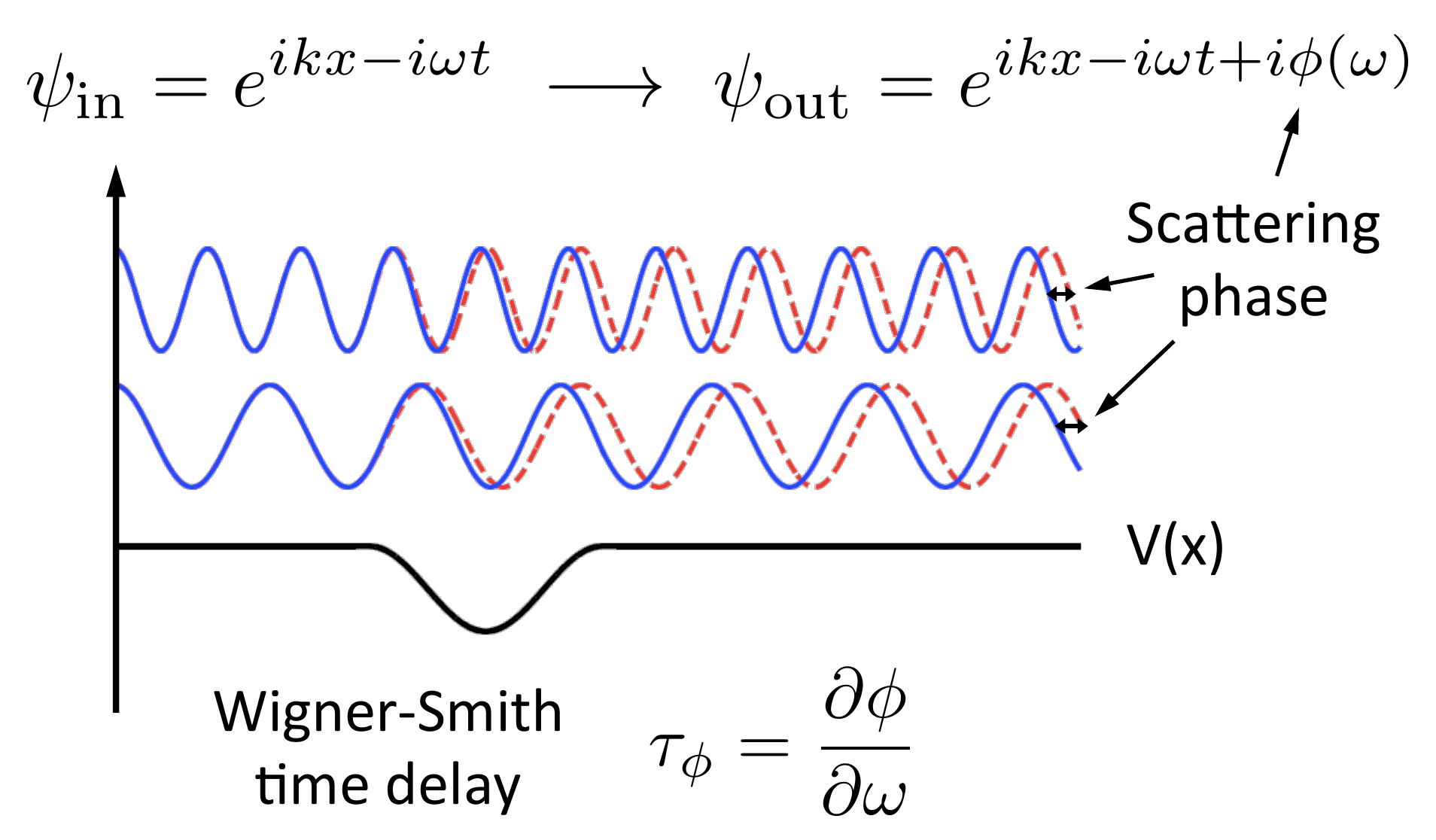}
  \caption{(color online). The Wigner-Smith time delay is calculated as the derivative of the scattering phase accumulated due to the presence of a scattering potential, here given as $V(x)$. The blue (solid) and the red (dashed) lines display schematically the scattering wave function in the presence and absence of this potential, respectively (two different scattering energies are shown and reflections by the potential are ignored).  Figure courtesy of R.~Pazourek \cite{pazourek_attosecond_2015}.}
  \label{fig:ews_delay}
\end{figure}

The foundations for work on time-delay were laid by Eugene Wigner and his student Leonard Eisenbud
who studied the single-channel scattering phase shifts in resonant quantum scattering 
\cite{eisenbud_ph.d._1948,wigner_lower_1955}. 
Their fundamental insight was that the time-delay $\tau_\phi(E)$ 
that an incident wave accumulates during a resonant scattering event (in one specific channel) as compared to non-resonant free propagation 
can be estimated by taking the energy derivative of the scattering phase shift $\phi(E)$ (see Fig.~\ref{fig:ews_delay}). 
The corresponding Wigner (or Wigner-Eisenbud) phase delay time is then given as $\tau_\phi(E)=\hbar\,\partial \phi(E)/\partial E=\partial \phi(\omega)/\partial \omega$ where $E$ is the 
scattering energy of a quantum particle and $\omega$ its angular frequency. Since the energy 
derivative at a sharp scattering resonance can be very large, also the corresponding time delay will, correspondingly, take on 
very large positive values at such resonant energy values. Note, however,
that the value of the time-delay can, in principle, also be negative as, 
e.g., for the case of scattering through a repulsive potential. In this case one speaks of a ``time-advance'' 
the value of which is limited by causality constraints. 

Subsequent work \cite{martin_time-delay_1976,jauch_time-delay_1972,ilic_modeling_2009} 
showed that the above definition of the
time-delay can be reformulated as follows, 
\begin{equation}
\tau_\phi(E)=\frac{1}{\sigma_{\rm scat} v
}\int d^3 r \left[|\psi_E({\bf r})|^2-1\right]\,.
\label{eq:dwelltime}
\end{equation}
This integral contains the single-channel 
scattering states $\psi_E({\bf r})$ and
extends over all of space. 
$\sigma_{\rm scat}$ is the scattering cross-section and $v$ is the velocity of the incident flux. 
Multiplied together, $\sigma_{\rm scat}v$ is 
equal to the incident flux on the scatterer, $J_{\rm in}$ (see section \ref{subsubsection2.1.3}).  
Neglecting a self-interference term outside the scattering region $\Xi$ (which usually averages out \cite{winful_delay_2003,smith_lifetime_1960}), 
the time-delay $\tau_\phi$ thus measures an excess in the ``dwell time'' 
\begin{equation}
\tau_d=\frac{1}{J_{\rm in}}\int_\Xi d^3 r |\psi_E({\bf r})|^2\,.
\label{eq:dwelltime2}
\end{equation}
inside the scattering region $\Xi$ as compared to propagation in free space. Both the scattering states $\psi_E({\bf r})$ and the incoming flux associated 
with them are normalized here such that the integral $\int_\Xi d^3 r |\psi_E({\bf r})|^2=1$ when the scattering region $\Xi$ is replaced by free space (for which case the excess dwell time is zero).

To understand intuitively why Eq.~(\ref{eq:dwelltime2}) represents a dwell time, consider that 
the integral which appears there measures the intensity
stored inside the scattering region $\Xi$. To obtain the time, which this intensity stays
inside the scattering region, one has to divide it by the outgoing flux $J_{\rm out}$, which, for
a stationary scattering state like $\psi_E({\bf r})$, is equal
to the incoming flux $J_{\rm in}$ (the absence of gain and loss in the medium is assumed here). 
The identity in Eq.~(\ref{eq:dwelltime2}) thus corresponds to
what one would expect from a simple classical picture. To emphasize this analogy we mention,
parenthetically, that also the dwell 
time of water molecules in a bath tub can be estimated equivalently based on the knowledge of the water volume
contained in the tub and the incoming water flux (provided that the latter is equal to the outgoing flux).  

The above expressions have originally been derived for the scattering of matter waves as described by 
the Schr\"odinger equation.
Using the analogy with the Helmholtz equation (see section \ref{subsubsection2.1.1}) these results can now be carried over to light scattering. 
Note, however, that in this case the electromagnetic energy density $u({\bf r},\omega)$ defined in 
 section \ref{subsubsection2.1.3} enters the definition of the dwell time $\tau_d$,  which gives rise to
 additional terms related to a potential energy density stored in the dielectric medium. This contribution
is particularly large when the dielectric constant of the medium is much larger than the 
 vacuum value. Details on these terms as well as on their relation to the  
 energy-dependence of the optical potential $V_{\rm light}$ introduced in 
section \ref{subsubsection2.1.1} are provided in \cite{lagendijk_resonant_1996}, see section 3.2.3 there.

The reason why the concept of time-delay has been so successful and widely used in a variety of different contexts is
because it allows one to extract temporal information out of spectrally resolved 
scattering quantities like the scattering phase shift.
A second major asset of the time-delay concept is its close relation to other physically relevant quantities, of which the stored
intensity inside a scattering region is just one. Another one is the absorption time 
$\tau_{\rm a}$, which measures the exponential decay of the light
intensity in an absorbing medium (see \cite{lagendijk_resonant_1996} for a review). 
Specifically, if we consider a light ray in a uniformly absorbing medium of constant refractive index $n=n_r+i n_i$
then the corresponding wave amplitude along the ray path can be written as follows, 
$\psi_\omega(x,t)=A \exp(iknx-i\omega t)$, with $A$ being an
overall amplitude and $x$ the spatial coordinate along the ray trajectory. 
The incoming wave intensity $|\psi_{\rm in}|^2=|A|^2$ will have decreased
exponentially due to absorption, $|\psi_{\rm out}|^2=|A|^2 \exp(-2 n_i k L)$ when leaving the medium.
The trajectory length $L$ inside the medium  is now easily related to a corresponding time, $\tau=L n_r /c$, which is now both the dwell time inside the medium (due
to its relation with $L$) and the absorption time (due to its relation with the decreased intensity $|\psi_{\rm out}|^2$). The simple physical reasoning behind this correspondence is that 
a wave suffers the more from absorbtion the longer it stays inside an absorbing medium. 
In the limit of small absorption, where $n_i k L\ll 1$, this time can be estimated as follows 
\cite{lagendijk_resonant_1996},
\begin{equation}
\tau=\lim_{n_i\to 0} \frac{n(1-|\psi_{\rm out}|^2/|\psi_{\rm in}|^2)}{2 n_i \omega}\,.
\label{eq:absorptiontime}
\end{equation}
Note that in this relation the ratio of outgoing
to incoming wave intensity emerges, which is known as the albedo of a scatterer, $a=\langle|\psi_{\rm out}|^2\rangle/\langle|\psi_{\rm in}|^2\rangle$.
For the visible part of the light spectrum the albedo (measured in reflection) ranges from values below 10\% for very dark 
substances (like coal) to almost 90\% for very bright substances (like snow). According to the above simple derivation, such albedo 
measurements provide very accurate information about the time light stays inside a given medium \cite{tiggelen_dwell_1993,feshbach_unified_1962}.
Note that for very inhomogeneously absorbing media the dwell time and the absorption time may
also be quite different from each other as the absorption time then depends significantly on whether 
regions of high absorption are visited by a scattering wave or not. 

 In a seminal paper  \cite{smith_lifetime_1960} it was shown that the time delay concepts 
can be straightforwardly extended to multiple channels. In particular, for flux-conserving systems without loss or gain a corresponding multi-channel
time-delay matrix ${\bf Q}$ can be defined based on the unitary scattering matrix 
${\bf S}$ in the following way:
\begin{equation}
{\bf Q}=-i\hbar\, {\bf S}^\dagger \frac{\partial {\bf S}}{\partial E}\,.
\label{eq:timedelayop}
\end{equation}
This Wigner-Smith (or Eisenbud-Wigner-Smith) time-delay matrix ${\bf Q}$ generalizes the concept of the 
``phase delay time'' from above
 to multiple channels. Care must, however, be taken with respect to the definition of the 
asymptotic states that are related by the scattering matrix: Depending on whether the asymptotic states incorporate the free space propagation between incoming and outgoing asymptotic region, the Wigner-Smith matrix either measures the times associated with the phase delays or with the phases themselves. 
 The matrix ${\bf Q}$ has 
the same number of $2N\times 2N$ complex elements as the scattering matrix itself (see section
\ref{subsubsection2.1.4}) and it is Hermitian by construction. Its real eigenvalues $q_n$ are referred to as the ``proper delay times'' 
which, when assuming an RMT distribution for the matrix elements of ${\bf S}$ can be shown to follow very specific distribution functions in 
the chaotic \cite{brouwer_quantum_1997} as well as in the diffusive limit \cite{ossipov_fingerprints_2003}
(see also \cite{mendez-bermudez_probing_2005} and the following review on this topic
\cite{kottos_statistics_2005}).

The Wigner-Smith time-delay matrix, in turn, shares a very deep connection to the density of states (DOS) $\rho(\omega)$ of an open scattering system. Specifically, for a finite open medium the DOS
can be defined as the sum over all quasi-bound states or ``resonances'', evaluated at the frequency $\omega$ \cite{breit_capture_1936},
\begin{equation}
\rho(\omega)=\frac{1}{\pi}\sum_m \frac{\Gamma_m/2}{(\Gamma_m/2)^2+(\omega-\omega_m)^2}\,,
\label{eq:definitiondos}
\end{equation}
where each of the Lorentzian mode profiles in this sum is spectrally normalized.
Following the work of Gamow \cite{gamow_zur_1928}
the resonance energies $\omega_m$ and their widths $\Gamma_m$ are the real and imaginary parts of complex resonance eigenvalues at which the scattering matrix ${\bf S}(\omega)$ has 
its poles (see last paragraph in 
section \ref{subsubsection2.1.4}). Based on this connection Krein, Birman, Lyuboshitz
and Schwinger 
\cite{lyuboshitz_collision_1977,schwinger_gauge_1951,birman_spectral_1992,krein_theory_1962} 
showed that the
DOS is directly expressible through the scattering matrix, $\rho(\omega)=[-ic/(2\pi)]{\rm Tr}\,{\bf S}^\dagger\, \partial {\bf S}/\partial \omega$, 
and thus through the trace of the Wigner-Smith time-delay matrix $\rho(\omega)=c/(2\pi){\rm Tr}\,{\bf Q}$, a connection which has meanwhile been verified also numerically 
\cite{yamilov_density_2003} and experimentally 
\cite{davy_transmission_2015}. From this relation we conclude that the DOS is directly proportional to
the sum of the time delays associated with all the $2N$ channels described by the scattering 
matrix \cite{smith_lifetime_1960,wigner_lower_1955}.
Since, in addition, the local DOS $\rho({\bf r}, \omega)$ 
(where the DOS $\rho(\omega)=\int dr^3 \rho({\bf r}, \omega)$) 
is also connected to the imaginary part of the Green's function, 
$\rho(\omega,{\bf r})=-2\omega{\rm Im}[G({\bf r},{\bf r}',\omega)]/\pi$ (assuming a scalar field, see 
\cite{wijnands_greens_1997}, chap. 4), the above relations also uncover a direct connection between 
the time-delay and  the Green's function (see also Krein-Friedel-Lloyd formula as discussed, e.g., in \cite{faulkner_scattering_1977}).

The close connection between the time-delay and the DOS also has a very fundamental insight in store that can be obtained through a result derived by Hermann Weyl 
\cite{weyl_uber_1911,arendt_mathematical_2009}. 
This so-called ``Weyl law'' states that the average DOS in a finite domain asymptotically (for increasing eigenfrequencies) 
follows a universal function (in frequency) that just depends on the volume and the surface area of the system, but not on the specific geometric details of the scattering potential. 
Through the  equivalence between the DOS and the time-delay, the latter is thus also just determined 
by the volume and the surface of the scattering domain -- a universal result that holds independent of whether the underlying medium leads to ballistic, diffusive scattering or even
Anderson localization \cite{pierrat_invariance_2014}. One very interesting consequence of this result is that, through the connection between the time-delay and the dwell-time 
(see Eqs.~(\ref{eq:dwelltime},\ref{eq:dwelltime2})) this ``universal'' time-delay can also be 
directly linked with the energy stored in the medium for unit incident flux in each 
of the scattering channels.

To show this explicitly we, however, first need to generalize the definition of the dwell-time 
in Eq.~(\ref{eq:dwelltime2})
from a single to  multiple scattering
channels -- in a similar way as we did earlier for the time-delay.  
For the corresponding definition of a ``dwell time operator'' to be meaningful, 
we demand that the expectation value of this operator, 
for a given multi-channel incoming state, yields the corresponding
dwell time $\tau_d$ of this state. Since, according to Eq.(\ref{eq:dwelltime2}), 
the dwell time involves the integral of the corresponding 
scattering state over the scattering volume, the definition of the 
dwell time operator needs to incorporate the knowledge on the scattering states. Following section 
\ref{subsubsection2.1.2} we know that any scattering state can be connected to its incoming waves by way of the Green's 
function ${\bf G}$, with the result that the dwell time operator ${\bf Q}_d$ is given as follows 
\cite{sokolov_simple_1997,ambichl_delay_2012},
\begin{equation}
{\bf Q}_d=\hbar^2\, {\bf W}^\dagger ({\bf G})^\dagger {\bf G} {\bf W}\,,
\label{eq:dwelltimeop}
\end{equation} 
where ${\bf W}$ is the energy-dependent coupling matrix from the scattering to the exterior region introduced in Eq.~(\ref{eq:sgreenmat}). We emphasize here that the above expression for ${\bf Q}_d$ 
involves the knowledge of the
Green's functions on all points inside the scattering medium such that an evaluation of ${\bf Q}_d$
based on Eq.~(\ref{eq:dwelltimeop}) is very complex (i.e., numerically very costly and experimentally 
close to impossible). Alternatively, one can connect this dwell-time
matrix with the definition of the Wigner-Smith time-delay matrix (which only involves the knowledge
of the scattering matrix): 
When restricting the action of the energy (frequency) derivative of the scattering matrix
in Eq.~(\ref{eq:timedelayop}) to only the explicit energy-dependence in Eq.~(\ref{eq:sgreenmat}) and neglecting the energy-dependendence
of the coupling matrix ${\bf W}$, the time-delay and the dwell-time operators for unitary
scattering systems are the same \cite{sokolov_simple_1997}. In this sense, the time-delay and the dwell-time differ only by the above mentioned ``self-interference''
term which involves also the evanescent modes in the near-field of the scatterer \cite{ambichl_delay_2012}.  

In a similar way, the connection between the dwell time $\tau_d$ and the absorption time 
$\tau_{\rm a}$ for the single-channel 
case suggests that such a relation might also 
exist on the more formal operator level. Following this idea, it was shown in 
\cite{savin_delay_2003} that for any uniformly absorbing medium
($n_i({\bf r})$ is uniform in space) with arbitrary spatial complexity ($n_r({\bf r})$ varying in space) 
the following relation holds between the scattering matrix ${\bf S}$ and the dwell-time operator 
${\bf Q}_d$,
\begin{equation}
{\bf 1}-{\bf S}^\dagger {\bf S}=\Gamma_a {\bf Q}_d\,,
\label{eq:unitarydeficit}
\end{equation}
where the parameter $\Gamma_a$ is a phenomenological absorption rate and both ${\bf S}$ 
as well as ${\bf Q}_d$ are evaluated in the presence of
absorption. The above relation suggests that in a uniformly absorbing medium the time-delay operator is nothing else but the
operator which measures the unitarity deficit or the ``sub-unitarity'' of the scattering matrix. 
Since, in addition, the scattering matrix connects the incoming with the outgoing states in a scattering problem, which, in turn, are related to each other through the albedo $a$, Eq.~(\ref{eq:unitarydeficit}) is in fact nothing else but
the multi-channel generalization of Eq.~(\ref{eq:absorptiontime}).

As a last point in this section, we mention that the above concepts on time-delay may also be used to work out appropriately defined velocities. This is particularly relevant for the case of resonant wave scattering in a disordered medium, for which the conventionally used group and phase velocities fail to satisfy the causality relations required from special relativity \cite{lagendijk_resonant_1996}. 
A viable alternative is here the transport or energy velocity $v_E$, which determines the speed of energy transport \cite{brillouin_wave_1960} and is thus strictly causal also near the resonances of scatterers in the medium. Characteristic differences between $v_E$  for light and for electrons have been discussed in \cite{lagendijk_resonant_1996}.

\section{Mesoscopic effects in optical systems: theoretical and experimental analogies}\label{section3}
The theoretical framework presented in section \ref{section2} has been and continues to be applied to a whole host of different questions 
arising in the context of mesoscopic scattering. Many experiments, in particular for coherent electronic transport through 
mesoscopic conductors like quantum point contacts, ``quantum billiards'', nanowires, etc.~have
meanwhile been
carried out in which many of the above predictions could be studied in detail.  Our emphasis here will be on predictions 
from mesoscopic scattering theory, which
could be realized both in {\it electronic transport} as well as in {\it optical} experiments that will here be compared with each other.
We will review the first generation of such experiments, where also in the optical context ``mesoscopic physics'' effects have been revealed without resorting to wavefront shaping techniques.
Let it be clear that when we speak of ``mesoscopic effects'' in optics, 
we do not refer to the signatures of the
quantum (optical) nature of the electromagnetic field; rather, we refer here to signatures of light
scattering that are intrinsically related to the finite mode number of a medium as well as
to correlations between these modes.  

\subsection{Conductance quantization}\label{subsubsection3.1.1}
\begin{figure}
  \centering
  \includegraphics[width=1\linewidth]{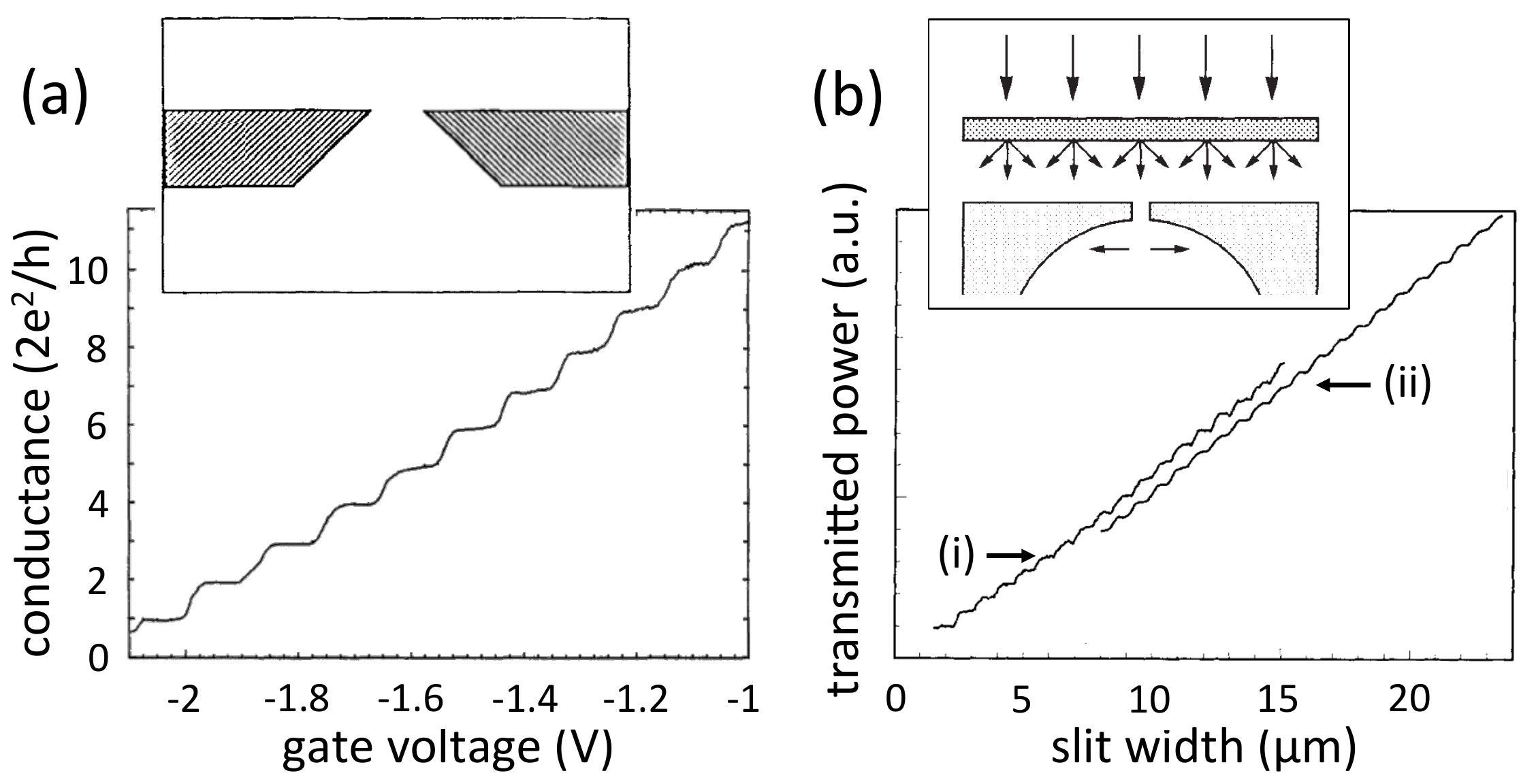}
  \caption{(a) Conductance quantization in coherent electron transport through a quantum point contact (see upper inset and similar point contacts at the openings of the quantum billiards in Fig.~\ref{fig:quantum_dot_figure}). The conductance shows steps at integer values of the conductance quantum $2e^2/h$, adapted from \cite{van_wees_quantized_1988}. (b) Power of light transmitted through a slit with variable width (see upper inset). Similar steps as in (a) occur here when the slit width corresponds to integer multiples of $\lambda/2$ with $\lambda$ being the light wavelength. Before propagating through the slit the light is sent through a diffuser realized (i) with a piece of paper and (ii) with an array of parallel glass fibers. Figure adapted from \cite{montie_observation_1991}.}
  \label{fig:cond_quant}
\end{figure}

One of the foundational experiments in  mesoscopic transport  was the demonstration of conductance quantization. By varying
the opening width of a so-called quantum point contact (through electronic gates on top of a hetero-junction) the conductance was observed to change in quantized steps of height $2e^2/h$, see
Fig.~\ref{fig:cond_quant}a 
\cite{houten_quantum_2008,van_wees_quantized_1988}. The 
origin of this effect is the quantization of the transverse momentum
in the quantum point contact; in other words, the electrons do get transmitted through individual transverse ``modes'', which can 
be labeled with a discrete quantum number $m$ or $n$ (see 
section \ref{subsubsection2.1.4} where we introduced this concept already).
Since each of these modes has a specific threshold that depends
on the width of the quantum point contact, the conductance 
increases in a step-wise fashion whenever such a threshold is crossed. The experiment by van Wees {\it et al.} was thus crucial to lend credibility to the Landauer formula (see Eq.~(\ref{eq:landauer}) in section \ref{subsubsection2.1.4})
that describes the conductance $G$ as a problem of coherent transmission $T$ through multiple modes. 
As this description is, however, solely due 
to the wave nature of electrons, one should observe it also with other types of waves, including electromagnetic
radiation. This idea was picked up in \cite{montie_observation_1991}, where a 
corresponding experiment was realized with light waves that were sent through a slit of 
tunable width, see Fig.~\ref{fig:cond_quant}b. To mimic the way in which electrons impinge
on the quantum point contact, the optical experiment featured a diffuse light source to distribute
the incoming flux equally over all available transverse modes. 
With this type of illumination the light intensity transmitted through the slit was observed to follow the same
step-like pattern as the electrons do in the mesoscopic analogue, see Fig.~\ref{fig:cond_quant}b. 

Surprisingly, this optical experiment, which is much easier to carry out 
than its preceding electronics counter-part, was performed only after the corresponding physics was first understood in a mesoscopic context. This point nicely underlines the main message of this review
and indicates that a strategy along these lines may have many more
interesting insights and surprises in store. Note here, in particular, that in optics, the transmission through and the reflection from
a scattering object are typically accessible in a mode-resolved way, which is not the case for electrons. This is also the case when
studying the scattering through an extended disordered region where both the incoming as well as the outgoing modes are quantized. To probe the total optical 
transmission $T=\sum_m T_m=\sum_{mn}T_{nm}$
as inherent in electronic conductance, one can use a diffuser (as in Fig.~\ref{fig:cond_quant}b)
to ensure a nearly isotropic spatial illumination, that excites all modes equally.
In contrast,  an 
illumination through a collimated laser beam with a well-defined incoming angle would probe the transmission through a suitable defined incoming mode 
$T_m=\sum_{n}T_{nm}$.	
The information on the transmission $T_{nm}=|t_{nm}|^2$ through the outcoupling modes $n$ is contained in 
the speckle-pattern appearing behind the scattering region, which contains the spatially resolved transmission pattern. 

\subsection{Conductance fluctuations}\label{subsubsection3.1.2} 
\begin{figure}
  \centering
  \includegraphics[width=1\linewidth]{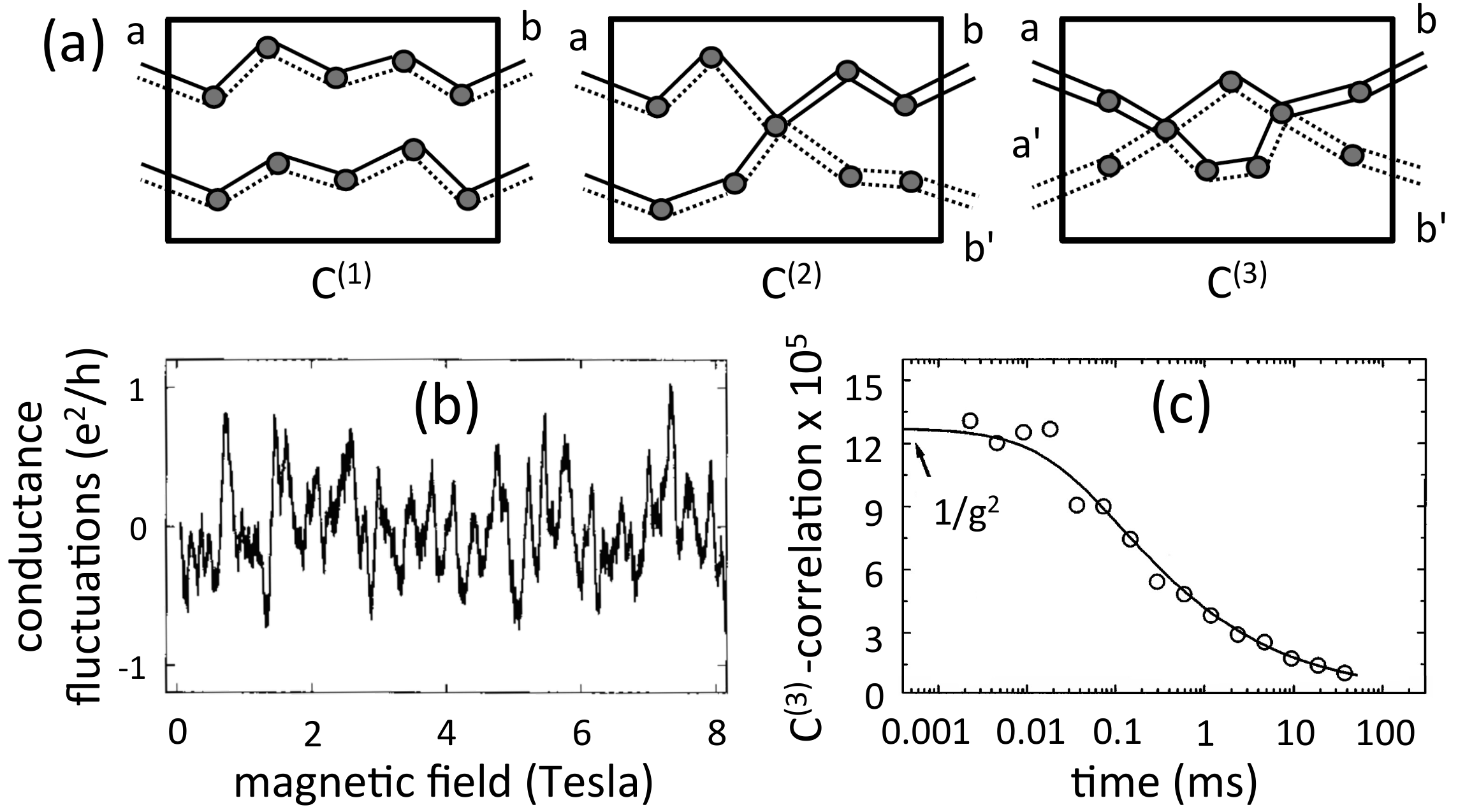}
  \caption{(a) Schematic of the different scattering paths in a disordered medium and their correlations due to crossings, adapted from \cite{scheffold_universal_1998}. Non-intersecting paths ($C^{(1)}$) as in the left panel give rise to short-range speckle fluctuations. Those paths with a single crossing ($C^{(2)}$) as in the middle panel are already correlated with each other. The universal conductance are, however, induced by paths with two crossing ($C^{(3)}$) as in the right panel. (b) Fluctuations of the electronic conductance $G$ with respect to its mean value $\langle G\rangle$, measured in a 310 nm long and 25 nm wide gold wire at 10 mK  as a function of a perpendicular magnetic field $B$, adapted from \cite{washburn_aharonov-bohm_1986}. 
  The variance of the conductance $\Delta G\approx 0.09\, e^2/h$ corresponds well to the theoretical prediction $(1/15) e^2/h$ spelled out in Eq.~(\ref{eq:gandvarg}). (c) Universal conductance fluctuations
  of light $C^{(3)}(t)$, measured dynamically in a turbid colloidal suspension, adapted from \cite{scheffold_universal_1998}. The experimental data is compared with a theoretical prediction using a dimensionless conductance of $g=89$ and a sample thickness 
 of $L=13.1\mu m$.}
  \label{fig:ucf_fig}
\end{figure}

Experiments that were crucial for uncovering the non-trivial correlations in  these different types of coherent transmission amplitudes 
were those 
reporting on conductance fluctuations in small metallic wires and rings 
\cite{washburn_temperature_1985,webb_observation_1985,umbach_magnetoresistance_1984}. 
These fluctuations  
observed in low-temperature measurements as a function of an applied magnetic field (see Fig.~\ref{fig:ucf_fig}b) were actually unexpected. Their origin was first 
believed to be ``finite size effects'' of the conductor; it was, however, soon revealed 
\cite{lee_universal_1985} that these fluctuations are due to the 
multiple disorder-scattering and the corresponding multi-path interference, which sensitively depends on all of the system parameters (like
the disorder configuration, the Fermi-energy, the magnetic field etc.). In this sense the conductance fluctuations are like a fingerprint of the
medium, which is highly complex but fully reproducible when measuring the conductance again a second time. An intriguing aspect of these 
conductance fluctuations is that their variance has a universal value of the order of $e^2/h$ (at zero temperature), which is
independent of the degree of disorder (in the diffusive regime) as well as of the sample size, 
hence the name ``universal
conductance fluctuations (UCF)''. As we demonstrated in sections \ref{subsubsection2.1.5} and \ref{subsubsection2.1.5a}, this surprising result as well
as the exact value of the universal fluctuations can be understood
based on the spectral rigidity of the transmission eigenvalues. 
Alternatively, one can also obtain very instructive insights into this phenomenon based on
so-called diagrammatic techniques \cite{berkovits_correlations_1994,feng_correlations_1988,lee_universal_1985} (see  \cite{dragoman_quantum-classical_2004,akkermans_mesoscopic_2007,montambaux_coherence_2006}
for reviews of these techniques).

Conceptually speaking, the universal value of the electronic conductance fluctuations is
a clear signature of the quantum coherence in the scattering process. We should thus expect to observe similar effects also with 
coherent disorder scattering of light. Since the optical speckle patterns contain sizable fluctuations
as well, one may be tempted to think that UCF are just a 
different aspect of speckle fluctuations. As it turns out, this is, however, not the case. 
To understand this in more detail consider the relation for the
variance of the fluctuations, which for the transmission of light is given 
by $\sigma=\langle T^2\rangle -\langle T\rangle^2$.
Writing $T=\sum_{mn}T_{mn}$ and $\delta T_{mn}=T_{mn}-\langle T_{mn}\rangle$, we obtain $\sigma=\sum_{mnm'n'}C_{mnm'n'}$, where
$C_{mnm'n'}=\langle\delta T_{mn}\delta T_{m'n'}\rangle$. For evaluating this expression for coherent scattering processes, a classical diffusion equation is clearly insufficient; instead, one can employ a perturbative approach in the limit of weak but multiple scattering, 
where the perturbation parameter is $1/(k\ell^\star)\ll 1$  with  $\ell^\star\ll L$
($\ell^\star$ is the transport mean free path as discussed
at the beginning of section \ref{section2}
and $L$ the medium thickness). In the corresponding expansion \cite{feng_correlations_1988}
the following contributions to the correlation function 
$C_{mnm'n'}=C_{mnm'n'}^{(1)}+C_{mnm'n'}^{(2)}+C_{mnm'n'}^{(3)}+\ldots$
can be distinguished based on their different contributing scattering diagrams, see Fig.~\ref{fig:ucf_fig}a
(only the first three terms in this expansion will be considered in the following). 

The first term $C_{mnm'n'}^{(1)}$ is always present (also in the absence of phase coherence) and of order zero in the expansion parameter $1/(k\ell^\star)\ll 1$. It corresponds to contributions from scattering paths that do not intersect while 
transmitting through the medium, see Fig.~\ref{fig:ucf_fig}a left panel. 
In the absence of such intersections, also correlations between 
modes are largely absent (the only correlations remaining in rather thin media give rise to the so-called memory effect discussed in sections \ref{subsubsection3.1.4} and \ref{subsection5.1}). 
In $C_{mnm'n'}^{(1)}$ the most dominant contributions to the transmission arise when 
the difference between both the incoming and the outgoing transverse momenta is zero, 
$\Delta q_n=q_n-q_{n'}=\Delta q_m=q_m-q_{m'}=0$. (Note that we have implicitly used here that our modes $m,n$ have a well-defined transverse momentum $q_n$.) 
As a result one finds that the fluctuations in the speckle pattern are of 
order of the average, $\langle\delta T_{mn}^2 \rangle=\langle T_{mn}\rangle^2$. This fact, which is also known as the Rayleigh law, is 
directly reflected in the granularity of a speckle pattern which features strong fluctuations between dark and bright spots. Why do these 
large fluctuations not translate to correspondingly large fluctuations of the total transmission/conductance? The answer is that the 
correlation term $C_{mnm'n'}^{(1)}$ only has contributions for the above very specific mode combinations and thus, although being formally of the largest scale, their relative contribution to fluctuations diminishes with the number of modes considered. Overall, $C^{(1)}$ correlations yield only a sub-dominant contribution to the total transmission fluctuations.

This is where the additional correlation functions $C_{mnm'n'}^{(2)}$ and $C_{mnm'n'}^{(3)}$ come into play, see Fig.~\ref{fig:ucf_fig}a middle and right panel, respectively. 
In the diagrammatic expansion those two contributions come from scattering paths with one and two quantum crossings in the transmission
process. Corresponding to the reduced likelihood for such
crossings to occur, the scale of these contributions is reduced. 
In the sum for the total correlation, this reduction is, however, compensated by a less restrictive angular selection in terms of the differences $\Delta q_n$, $\Delta q_m$: Whereas the $C_{mnm'n'}^{(1)}$
term features only short-range correlations, the 
 $C_{mnm'n'}^{(2)}$ and $C_{mnm'n'}^{(3)}$ terms feature long and infinite range correlations,
 respectively. It turns out, however, that not the long-range angular correlations
inherent in $C_{mnm'n'}^{(2)}$, but only the infinite-range correlations in $C_{mnm'n'}^{(3)}$ yield the desired universal 
contribution to UCF, $\sigma=\sum_{mnm'n'}C_{mnm'n'}\approx \sum_{mnm'n'}C_{mnm'n'}^{(3)}\approx 1$.

Optical experiments can go much further than simply re-measuring the universal value of UCF found already earlier in mesoscopic transport. In \cite{scheffold_universal_1998}   the
time-dependent correlation functions $C^{(2)}(t)$ and $C^{(3)}(t)$ 
were recorded in transmission through a small pinhole filled with a turbid colloidal 
suspension. The temporal fluctuations of the transmitted light were  due to the Brownian motion of scattering particles. 
By finding quantitative agreement with the theoretical predictions based on the 
above diagrammatic terms, 
it thus became possible to not only establish UCF in light scattering, but also to verify their microscopic
origins; see also corresponding experiments with microwaves \cite{gehler_channel_2016,shi_universal_2012}. 
As laid out in a review by Berkovits and Feng \cite{berkovits_correlations_1994} 
the above concepts on the correlation 
functions can be conveniently extended to describe also correlations in frequency as well 
as in the angle or in the spatial position of the 
emission from the disordered medium.  Note also the possibility offered in optics to probe other quantities, such as the $C^{(0)}$ correlations \cite{shapiro_new_1999,caze_near-field_2010}, that arise when the source is embedded inside the scattering medium, a setup that could be used for imaging purposes \cite{Skipetrov_nonuniversal_2000,carminati_speckle_2015}.

\begin{figure*}
  \centering
  \includegraphics[width=0.9\textwidth]{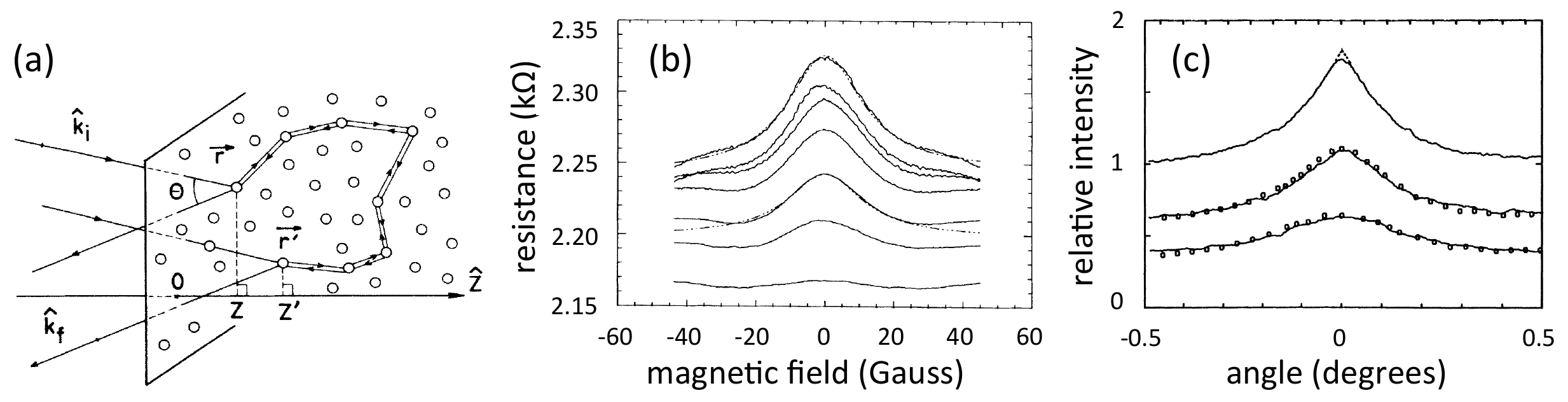}
  \caption{(a) Electron or light paths associated with the weak-localization effect, from \cite{akkermans_coherent_1986}. Due to disorder scattering, the reflected wave vector $\hat{\bf k}_f$ is rotated by an angle $\theta$ with respect to the incoming wave vector $\hat{\bf k}_i$.  (b) Resistance of an array of stadium-shaped quantum dots as a function of a perpendicular magnetic field, adapted from \cite{chang_weak_1994}.
 Different solid curves show experimental data at different temperatures (from top to bottom: $T=50$mK, $T=200$mK, $T=400$mK, $T=800$mK, $T=1.6$K, $T=2.4$K, $T=4.2$K). The dashed-dotted
 lines are Lorentzian fits. (c) Angular dependence of the light intensity backreflected from a disordered medium, adapted from \cite{wolf_optical_1988}. Different curves show experimental results for samples with different degree of absorption, as measured by the absorption mean free path $\ell_a$. From top to bottom: $\ell_a=\infty$, $\ell_a=810\,\mu$m, $\ell_a=190\,\mu$m. 
 The intensity is normalized with respect to the value at one degree, measured in the non-absorbing sample ($\ell_a=\infty$). The dots are predictions from diffusion theory. }
  \label{fig:weak_loc}
\end{figure*}

\subsection{Weak localization}\label{subsubsection3.1.3} 
Another fundamental phenomenon for which the angular correlations 
play an important role and which also relies on 
specific quantum crossings of scattering paths is 
the so-called ``weak localization effect' \cite{wolf_weak_1985,akkermans_weak_1985,bergmann_physical_1983,Khmelnitskii_localization_1984,abrahams_scaling_1979}
(see Bergmann in \cite{abrahams_50_2010} and \cite{montambaux_coherence_2006,dragoman_quantum-classical_2004,akkermans_mesoscopic_2007} for reviews).
The term ``weak'' refers to the overall weakness of this effect  
as compared to the ``strong'' (Anderson) localization for which weak localization is a precursor.
Our starting point here is the observation from mesoscopic transport theory (see  sections \ref{subsubsection2.1.5} and \ref{subsubsection2.1.5a}) that the transmission through a chaotic
scattering system or a disordered wire is reduced by a
small amount (again of the order of $e^2/h$) as compared to the value expected from a classical (i.e., 
incoherent) estimate. This suggests already that
interference effects which are based on the coherence of the scattering process, are at the heart of this phenomenon. 

To properly capture this effect, we employ a similar diagrammatic picture as in the previous section
in which the total transmission $T=|\sum_\alpha s_\alpha|^2$ 
can be written as a sum over all paths $\alpha$ with a corresponding complex amplitude $s_\alpha$ (the phase of which is given by the 
classical action). This expression does not only contain the incoherent summation 
over all individual probabilities $\sum_\alpha |s_\alpha|^2$ 
for scattering paths to go from one side of the disordered medium to the other (as inherent in the classical Drude formula), but also features 
interference terms $\sum_{\alpha\neq\beta} s_\alpha s^*_\beta$
(corresponding to products of scattering amplitudes for different paths). One might argue that these interference
terms average out to zero, since their random phases will lead to constructive and destructive interference with equal measure.
On close inspection, this argument turns out to be incorrect, however; this is because certain path-pairs have the same or very similar phase 
due to reciprocity (see section \ref{subsubsection2.1.4}) and may thus lead to a certain bias away from the classical result, as we will see in the following. 
Consider here, in particular, those paths that emanate from a source (outside of the medium) and return to it after scattering in the disordered medium (see illustration in Fig.~\ref{fig:weak_loc}a).
Since these loops can be traversed in two possible directions, we end up with two paths in the loop 
which have exactly the same phase as well as amplitude and thus always interfere constructively 
$|s_\alpha+s_\alpha|^2=4\,|s_\alpha|^2$, independently of the disorder configuration.
Since this contribution of such time-reversed path pairs to the reflection is larger as compared
to the classical result, where we would have $|s_\alpha|^2+|s_\alpha|^2=2\,|s_\alpha|^2$, they 
enhance the portion of the waves that are ``coherently back-scattered''
to the source by a factor of 2 and thus increase the reflection $R$.
In order to conserve the unitarity of the entire scattering process this increased reflection must be compensated by 
a corresponding decrease of the transmitted waves -- in perfect correspondence with our earlier observation (see  sections \ref{subsubsection2.1.5} and \ref{subsubsection2.1.5a}). 
The transmitted scattering paths that are responsible for this reduction can be shown to be self-crossing paths 
which feature loops in their scattering patterns which can be traversed both in a clock-wise and a counter-clock-wise direction \cite{akkermans_mesoscopic_2007}.

Although it turns out that the coherent backscattering contribution (resulting from
time-reversed path pairs) explains the weak-localization effect only partially \cite{hastings_inequivalence_1994}, a controlled breaking of reciprocity 
will eventually destroy weak-localization entirely and restore the classical (i.e., incoherent)
transport result. For electronic scattering problems this can easily be done by applying an
external magnetic field perpendicular to the scattering region. 
As shown in Fig.~\ref{fig:weak_loc}b the resistivity of an array of chaotic scatterers 
is, indeed, reduced
when the magnetic field is applied (note that the reduction is independent on the sign of the field, due to the Onsager relations, see section \ref{subsubsection2.1.4}).

For light waves, implementing a reciprocity-breaking mechanism is not so straightforward (adding absorption to the medium is, e.g., not sufficient as it only breaks time-reversal symmetry, but not reciprocity). A viable alternative is provided here by the access to the spatial degrees of freedom of light. Specifically, 
to measure the weak-localization of light its dependence on the angle of the light backscattered from the medium has been used as a key signature (see the angle $\theta$ in Fig.~\ref{fig:weak_loc}a
and its influence on the backscattered intensity in Fig.~\ref{fig:weak_loc}c). Since only those paths
that return directly to the source find a partner with the same phase, the enhanced reflection is concentrated in a very narrow 
back-reflection cone which is rather difficult to measure. Employing a Fraunhofer diffraction analysis, Akkermans \cite{akkermans_coherent_1986}
could show that the phase difference between two time-reversed reflected scattering path is given as $({\bf k}_i+{\bf k}_f)({\bf r}-{\bf r}')$,
where ${\bf k}_i$ and ${\bf k}_f$ are the initial and the final wavenumber, respectively, that impinge on the disordered medium at the first scatterer 
position ${\bf r}$ and leave it again at the last encountered scatterer position ${\bf r}'$
(see a corresponding illustration in Fig.~\ref{fig:weak_loc}a). When the backscattering is perfect, 
${\bf k}_i=-{\bf k}_f$, we obtain the enhancement by a factor of exactly two (as found above). This number can be reduced slightly when recurrent scattering events occur in the strong scattering limit \cite{wiersma_experimental_1995}. This maximum, however, degrades when the angle $\theta$ between
${\bf k}_i$ and ${\bf k}_f$ satisfies $|\theta -\pi|>\lambda/|{\bf r}-{\bf r}'|$, where $\lambda$ is the wavelength. 
For the shortest possible loop involving just two scattering events, the
typical value of $|{\bf r}-{\bf r}'|$ is given by the mean free path $\ell^\star$,
 resulting in a typical angular width of the 
backscattering cone $\theta\approx \lambda/\ell^\star$. Larger excursions of light paths in the medium with increased differences 
$|{\bf r}-{\bf r}'|$ are responsible for the peak of the cone, which was predicted to take on an 
approximately triangular shape \cite{akkermans_coherent_1986}. In the presence of absorption this shape gets rounded as dissipative mechanisms affect primarily  longer paths (see Fig.~\ref{fig:weak_loc}c). 

Sophisticated optical experiments could meanwhile measure the details of this cone lineshape, from which not only
the value of $\ell^\star$ but essentially the entire path length distributions in the scattering medium can be extracted \cite{akkermans_coherent_1986}
(compare also with similar studies for electronic scattering \cite{chang_weak_1994}). 
Note that in the optical regime the weak-localization effect also depends on the polarization of light, as has been
pointed out both theoretically \cite{akkermans_coherent_1986} 
and experimentally \cite{dragoman_quantum-classical_2004,albada_observation_1985}.
Recent experiments have even been able to go a step further in that they could actively suppress the 
coherent backscattering of light by exposing the medium to an ultra-fast pump pulse \cite{muskens_partial_2012},
which opens up interesting opportunities for actively controlling mesoscopic interference phenomena.

\subsection{Memory effect}\label{subsubsection3.1.4}
In section \ref{subsubsection3.1.2} on the universal conductance fluctuations we have already analyzed 
the different correlations that exist between modes involved in the scattering process across a disordered medium. Using diagrammatic scattering techniques, it was discovered in the same context that the correlation function $C_{mnm'n'}^{(1)}$ points to the existence of
correlations between incoming modes that have a similar transverse momentum, i.e., for which   
$\Delta q < 1/L$ \cite{feng_correlations_1988}. The inverse proportionality with respect to the 
thickness of the medium $L$ means that the angular range over which such correlations exist become smaller for increasing medium thickness -- a result that is notably independent of the value of the transport mean free path $\ell^\star$, and on the exact realization of disorder. Quite interestingly, this so-called ``memory effect'' 
was discovered based on 
mesoscopic transport theory, although in mesoscopic electron transport an experimental study of this
effect is not possible, since no angular resolution is available in electron scattering. It was realized very
quickly, however, that the memory effect can be directly mapped to optical scattering setups where
a laser beam with well-defined transverse momentum is sent onto a disordered medium \cite{freund_memory_1988}.  We will also see in section \ref{subsection5.1} how  this phenomenon has developed into a very useful and  practical effect with consequences  for imaging through or inside a disordered medium (with and without wavefront shaping). 

\begin{figure}
  \centering
  \includegraphics[width=1\linewidth]{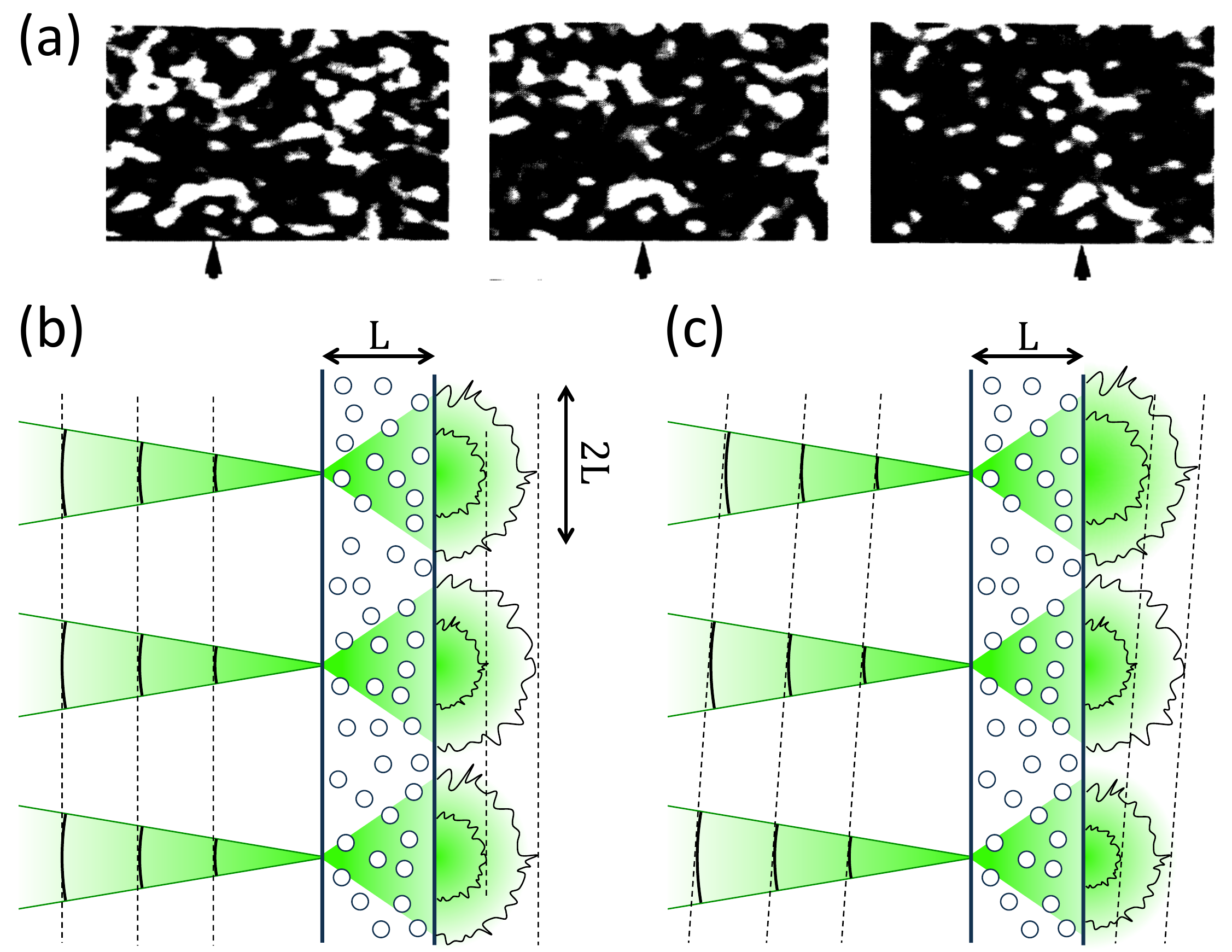}
  \caption{(color online). (a) Speckle pattern as recorded behind a thin disordered medium for different incident angles (0, 10 and 20 mdeg from left to right), adapted from \cite{freund_memory_1988}. The arrows below the three panels point to a specific pattern (arc above a bright spot) that serves as a convenient visual reference for seeing the rightward move of the speckle pattern with increasing tilt of the input laser.   (b),(c) Schematic illustration to explain this ``memory effect'': (b) A plane wave, represented here by three focal spots with the same phase, impinges on a disordered slab of thickness $L$ and creates a speckle pattern in transmission. The distance of $2L$ between the input spots is the minimal one for which the output speckle do not yet significantly overlap. (c) When tilting the incoming laser a phase gradient is imposed on the incoming wave. Provided that the tilt is smaller than a critical angle, $\theta\lesssim \lambda/(4L)$, this gradient is faithfully mapped onto  the transmitted wave, resulting eventually in a  shift of the speckle image recorded at a screen in the far field as in (a).}
  \label{fig:memory_fig}
\end{figure}

Within the angular range discussed above, a  
small angular rotation in the input beam then leads to a rotation  of the output speckle pattern by the same angle
(see Fig.~\ref{fig:memory_fig}), and correspondingly to  a shift at a distance from the medium.  
For larger angles,  the transmitted speckle pattern will rapidly become totally uncorrelated with  the original (un-shifted) speckle image. 
 This behavior can be intuitively understood by considering that shifting the angle of 
incident light corresponds to a linear phase shift between the Huygens-spots at which the incoming beam hits the disordered medium (see Fig.~\ref{fig:memory_fig}b).
In the diffusive regime, each of these spots will produce a cone of scattering pathways which reach the back side of the medium 
as a circular speckle halo, the diameter of which is about $2L$, where $L$ is the medium thickness. To successfully transfer the phase ramp from the input 
beam onto these output speckle patterns, the speckle disks at the backside of the medium may not acquire a phase that is larger than about $\pi$ from one disk to its nearest non-overlapping neighbor (see Fig.~\ref{fig:memory_fig}c). Using simple trigonometry, this condition, which is intuitively necessary to prevent phase mixing, can be translated to the requirement $\Delta q < 1/L$ found already earlier using diagrammatic techniques \cite{feng_correlations_1988}. Practically, this restriction limits the thickness of the disordered optical medium to a few tens of micrometers in order to have a measurable effect.
First experiments \cite{freund_memory_1988} 
on the optical memory effect have been performed soon after the theoretical proposal
\cite{feng_correlations_1988}, in which the predicted shift 
of the speckle pattern could be unambiguously identified (see Fig.~\ref{fig:memory_fig}a). 

When measuring correlations in reflection from a disordered medium, $C_{mnm'n'}=\langle\delta R_{mn}\delta R_{m'n'}\rangle$
rather than in transmission, one ends up in the interesting situation that both the memory effect as well as weak-localization 
corrections come into play. 
It turns out that, to first order, the angular width of the coherent backscattering cone and the memory effect angle are the same in reflection. This is because both effects rely on the diffuse spot size,  i.e., 
the width of both peaks is 
now related to the mean free path, $\Delta q < 1/\ell^\star$, which may be quite different from the 
memory effect angle in transmission,
that is related to the thickness of the disordered sample, $\Delta q < 1/L$.
The interplay between both effects has been studied in 
\cite{berkovits_time-reversed_1990}, again through the different 
contributions $C_{mnm'n'}^{(1)},C_{mnm'n'}^{(2)},C_{mnm'n'}^{(3)}$. It turns out that due to reciprocity these correlation functions get additional peaks as compared to the corresponding expressions for transmission. Consider, e.g., the
first contribution without quantum crossings, $C_{mnm'n'}^{(1)}$, which, in reflection, not only has a single peak at $\Delta q=0$ 
(corresponding to $q_n=q_{n'}$ and $q_m=q_{m'}$), but also a second one at $\Delta q=q_n+q_m$ (corresponding to
$q_{m}=-q_{n'}$ and $q_{n}=-q_{m'}$) which is due to the time-reversed contributions. Similar arguments can also be made for the
next contribution $C_{mnm'n'}^{(2)}$ (see \cite{berkovits_time-reversed_1990} for details and \cite{dragoman_quantum-classical_2004} for a review). 
Corrections to these results may be necessary due to internal surface reflections, as pointed out in \cite{freund_surface_1990}. 
Very recent acoustical measurements use the memory effect in reflection to obtain information on the path distribution in a random medium
close to the transition to Anderson localization, finding a strong recurrence of scattering paths at the point where they enter the medium \cite{aubry_recurrent_2014}.

\subsection{Distribution of transmission eigenvalues}\label{subsubsection3.1.5}
As we saw in the theoretical calculations presented in  section \ref{section2}
of this review, many interesting transport effects have their 
origin in the statistical distribution of the so-called ``transmission eigenvalues'' $\tau_n$, which are the eigenvalues of the 
Hermitian matrix ${\bf t}^\dagger {\bf t}$ (or the squared singular values of ${\bf t}$ itself). In principle, also the weak-localization 
correction to the conductance and the UCF can be expressed through the distribution of the $\tau_n$ and their correlations, 
respectively. Mesoscopic experiments, however, went much further in addressing also the interesting consequences of the
bi-modal distribution of the transmission eigenvalues, $P(\tau)$, which we derived earlier both for chaotic cavities, see Eq.~(\ref{eq:onepoint}), as well
as for diffusively scattering waveguides, see Eq.~(\ref{eq:bimodal}). 
Since the first moment of this distribution, 
$\langle \tau\rangle=\int_0^1 d\tau P(\tau)\tau$, corresponding to the average transmission per incoming channel, is basically unaffected by 
the bi-modal shape of $P(\tau)$, measurements of the electronic conductance alone (which is just proportional to $\langle \tau\rangle$)  
do not reveal any signatures of the bi-modality. This is different for other experimental observables, which depend on 
the higher moments of this distribution, $\langle \tau^n\rangle=\int_0^1 d\tau P(\tau)\tau^n$, as, e.g., the quantum shot-noise power
of electrons which probes the second moment $\langle \tau^2\rangle$ in addition to the first (see section \ref{subsection2.2} for more details). The corresponding mesoscopic transport measurements with cavities  and nano-wires
\cite{steinbach_observation_1996} (see \cite{blanter_shot_2000} for a review)  
could not only observe the shot-noise suppression below 
the Poissonian value, corresponding to different predictions for the values of the Fano factor $F$ (see Fig.~\ref{fig:onethird});
in fact, even the deviations from such universal behavior could be measured \cite{oberholzer_crossover_2002} and understood in detail \cite{marconcini_analysis_2006,jacquod_breakdown_2004,sukhorukov_quantum--classical_2005,agam_shot_2000,rotter_statistics_2007,aigner_shot_2005}. 

\begin{figure}
  \centering
  \includegraphics[width=0.8\linewidth]{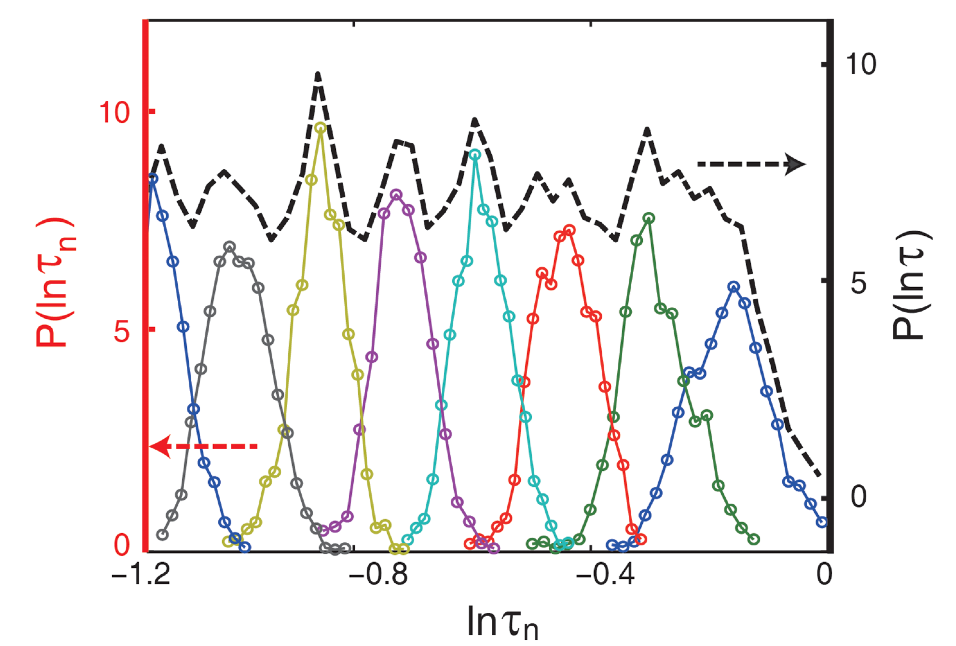}
  \caption{(color online). Experimental results from \cite{shi_transmission_2012} on the distribution of individual transmission eigenvalues $\tau_n$ (labeled by different colors) in microwave scattering through a strongly disordered quasi-1D geometry. As predicted theoretically, Anderson localization leads to a ``crystallization'' of transmission eigenvalues, corresponding to an equidistant spacing (on a logarithmic scale) between neighboring peaks of the corresponding distribution functions $P(\ln \tau_n)$. The top curve in black shows the distribution of the overall transmission $P(\ln \tau)$, where $\tau=\sum_n \tau_n$. }
  \label{fig:crystallization}
\end{figure}

Whereas a fair amount of convincing evidence for the bi-modal law has thus been put forward, no {\it direct} measurement of the transmission
eigenvalues or of their distribution could be achieved in the mesoscopic context. Also for electromagnetic waves such measurements
are also challenging, as the knowledge of the {\it complete} transmission matrix ${\bf t}$ is required to have access to its eigenvalues and eigenvectors, which
determine the corresponding transmission channels. In particular in optics, where the number of such channels is huge (as being
proportional to the sample cross-section and to the inverse squared of the wavelength, $N\propto A/\lambda^2$) measuring the entire transmission
matrix of a disordered sample is currently  still out of reach \cite{yu_measuring_2013,popoff_measuring_2010}. This condition is, however, much more relaxed for 
waves with a longer wave-length, as for micro-waves or also for acoustic waves. First micro-wave measurements \cite{shi_transmission_2012}
on the full transmission matrix through metallic tubes, filled with randomly placed and strongly scattering aluminium spheres, confirmed already
many of the interesting predictions which we discussed in the theory section \ref{section2}: 
In the diffusive regime, theory predicts 
that the largest transmission eigenvalue $\tau_1$ is close to unity and thus 
corresponds to an open channel. In the localized regime, in turn, this largest transmission eigenvalue
dominates the total 
transmission $T=\sum_n\tau_{n}$, such that $\tau_1\gg \tau_{n>1}$. The cross-over between these two regimes involving a ``crystallization of transmission eigenvalues'' (see section \ref{subsubsection2.1.5a}) with an equidistant spacing between the ``crystal sites'' ($\ln \tau_n$) was, indeed, observed in a microwave experiment (see Fig.~\ref{fig:crystallization}). 

In spite of this very good theory-experiment correspondence, the elusive bi-modality of the transmission eigenvalue distribution could so far 
not be verified with microwaves. An alternative strategy has recently been put forward based on the propagation of elastic Lamb waves
in a two-dimensional macroscopic metal stripe into which holes were drilled to emulate disorder 
\cite{gerardin_full_2014}. Although the scattering matrix
recorded here with laser interferometry was also not fully unitary, the bi-modality of $P(\tau)$ could be verified with this setup.
An interesting advantage of this experimental setup is that it not only allows one to measure all transmission and reflection amplitudes, 
but, in fact, also the scattering wave functions inside the disordered medium in analogy to similar scanning techniques
used for electrons \cite{topinka_coherent_2001}, micro-waves \cite{hohmann_freak_2010}
 and optical fields \cite{fallert_co-existence_2009}. 

We will discuss in section \ref{section5} the optical experiments dedicated to unraveling or exploiting open and closed channels.

\section{Optical wave front shaping in complex media}\label{section4}

\begin{figure}
  \centering
    \includegraphics[width=0.90\linewidth]{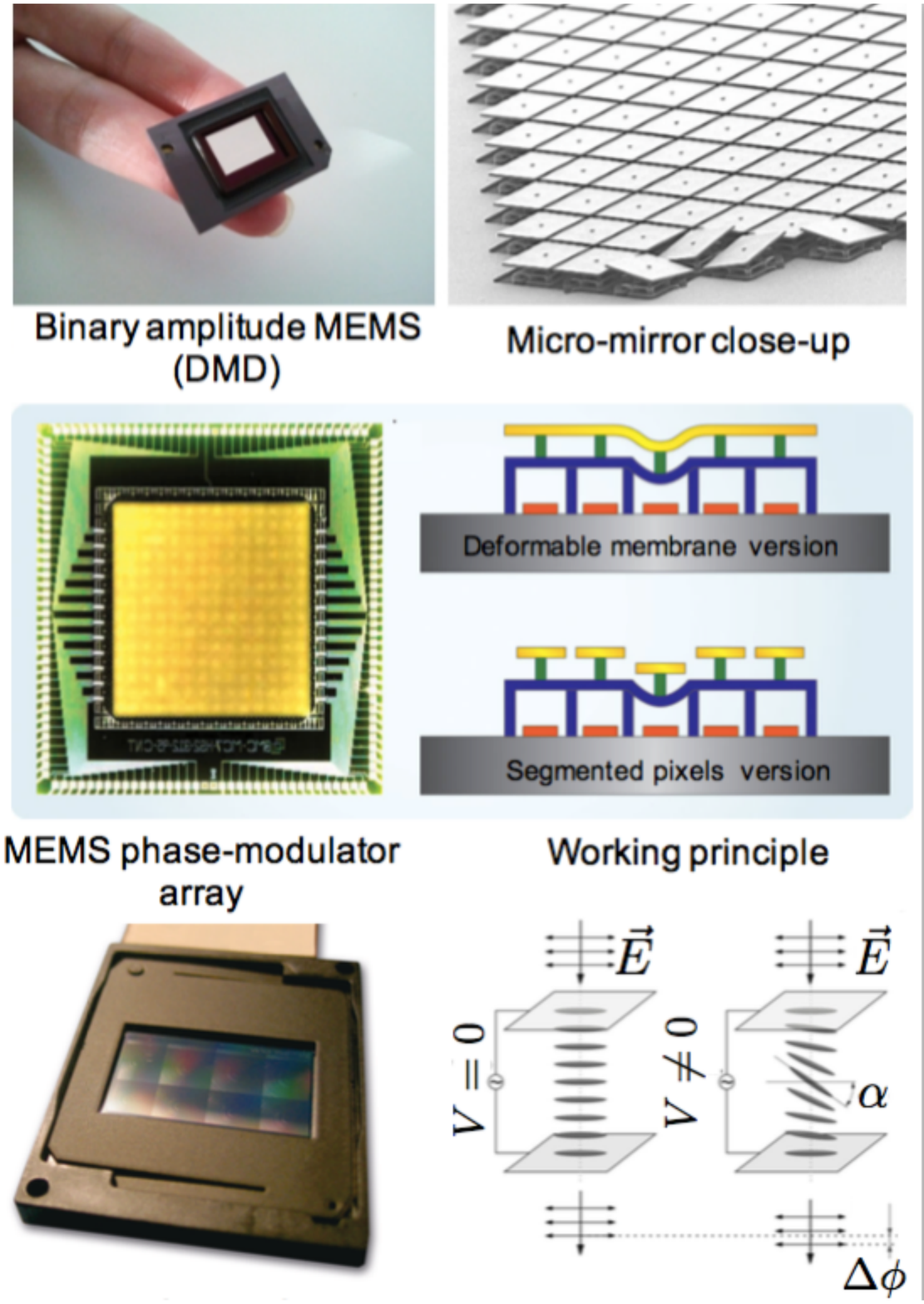}
  \caption{ (color online). Widely used types of digital spatial light modulators (SLMs). (top) MEMS-based binary amplitude modulators from Texas Instruments (adapted from \cite{rabinovitz_digital_2011}). (middle) MEMS-based phase-only SLM, available both with isolated pixels or with a deformable membrane (adapted from \cite{bifano_adaptive_2011}). (bottom) Liquid crystal based phase-only SLM. left: photo (courtesy of Holoeye Photonics AG), right: working principle in phase-only mode (courtesy of Monika Ritsch-Marte).}
  \label{fig:SLMtechnos}
\end{figure}

Progress in semi-conductor and electronic engineering has led to the emergence of a now vast range of techniques and devices to actively manipulate light, in particular spatial light modulators (SLMs). SLMs are mostly based on either liquid crystal technology \cite{lueder_liquid_2010} or on microelectromechanical systems (MEMS) \cite{gad-el-hak_mems_2010, gehner_MEMS_2006, cornelissen_MEMS_2012,hornbeck_DMDTM_2001}, see Fig.~\ref{fig:SLMtechnos}. They are nowadays offering control of up to a few millions of spatial degrees of freedom (pixels) of light, in phase or amplitude \cite{conkey_high-speed_2012, van_putten_spatial_2008, goorden_superpixel-based_2014} and are meanwhile widely used in imaging and microscopy \cite{maurer_what_2011}. The advent of digital image sensors (mainly CCD and CMOS) also allows to detect a correspondingly large number of degrees of freedom in intensity or in amplitude with the help of digital holography \cite{yamaguchi_phase-shifting_1997,leith_microscopy_1965, cuche_spatial_2000}.In the last 50 years, deformable mirror technology \cite{babcock_possibility_1953} and adaptive optics concepts have revolutionized imaging through the atmosphere \cite{lee_weak_1969} and thereby also earth-based astronomy \cite{roddier_adaptive_1999}.

The aim of  the first two parts \ref{subsection4.1} and \ref{subsection4.2} of this section  is  to review these experimental techniques. In  part \ref{subsubsection4.3} and \ref{subsubsection4.4}, we will show how these methods and concepts have been applied successfully to complex media. Starting point will be the paradigmatic case of the so-called ``opaque lens" concept \cite{cartwright_opaque_2007},  which has a large range of applications, in particular in imaging. We will detail a few more specific systems of particular interest in part \ref{subsection4.5}.

While many techniques described here are nowadays standard in optics, the reader from the mesoscopic physics community might not be familiar with them. We will therefore take a  pedagogical and historical approach,  and show how optical elements have evolved from passive to  active and finally to digital elements, which can meanwhile not only  compensate weak  perturbations of the wave front, but also  strong perturbations as in the multiple scattering regime.

\subsection{Wavefront shaping concepts and tools}\label{subsection4.1}

\subsubsection{Classical optical elements}\label{subsubsection4.1.1}

Classical optics relies on a variety of linear optical elements such as mirrors, lenses, prisms, to conveniently manipulate light for bending, deflecting or diffracting to achieve a particular goal. In the paraxial approximation fields propagate at low incidence in a given direction (with $z$ being the propagation axis) and most  optical elements can be modeled as transforming the optical field by modulating it spatially in phase and amplitude in a thin layer in the $(x,y)$ plane. For instance,  a lens adds a quadratic phase to the beam, thus focusing a plane wave to a diffraction limited spot.  A grating will periodically (in space) modulate  the phase or the amplitude of the transmitted or reflected wave, thus diffracting light in well-defined directions. An amplitude mask, be it a slit, a hole, or a complex image, will transmit and diffract a partial field. In all these cases, we can write the effect of the optical element formally as a well-defined and fixed spatial mask in  phase and amplitude. In between the thin optical elements, one usually considers homogeneous media (glass, air, etc.) where the propagation can  be  described using a whole variety of approximations ranging from paraxial rays to a full electromagnetic description. In the propagation through thin optical elements and homogeneous media spectral dispersion can be easily taken into account.

Formally, if we consider a monochromatic paraxial  field ${\bf E}_\omega (x,y,z)$ at frequency $\omega$, its transmission through an optical element will be described by a transmission function ${ t}_\omega (x,y)$, which can be a phase-only (complex of norm unity), amplitude-only (real), or a phase and amplitude function simultaneously.  If ${t}_\omega(x,y)$ depends explicitly on $\omega$ (such as a prism or a grating), then the optical element is dispersive. For conventional passive optical elements, this transmission is less or equal to one in absolute magnitude, and does not change over time (in contrast to temporal modulators discussed in subsection \ref{subsubsection4.2.4}). Obviously, the polarization of light and polarizing elements can also be taken into account, by writing the  transmission as a tensor.

All imaging systems, such as microscopes, telescopes, etc.~can be modeled by such a formalism. Getting a sharp image on a detector means carefully designing the optical system to ensure  low chromatic or spatial aberrations. This requires high quality optical components (planeity of the mirrors, curvatures of the optics, dispersion) to ensure stigmatism, i.e., diffraction limited images. Carefully designed optics can compensate for natural dispersion and simple aberrations. For instance, high numerical microscope objectives  can be pre-compensated for the spherical aberrations and the chromatic dispersion introduced by a glass cover slide. We will see how this imaging paradigm is modified when the propagation occurs in a complex medium. 

\subsubsection{Active and adaptive optics}\label{subsubsection4.1.2}

While well-determined transformations, such as those introduced by a homogeneous medium or by imperfect optical elements, can be readily compensated in an optical system, the problem of imaging through a medium with unpredictable inhomogeneities of the refractive index has remained a major challenge. Even weak spatial variations of the refractive index  can introduce significant  local phase variations when adding up during propagation \cite{lee_weak_1969}.  In particular, these local phase variations  will  degrade the so-called point spread function, i.e., the resolution of the optical system. One obtains  a blurred image, and we say the system is aberrating. An example of high practical importance is atmospheric turbulence that appears on many different temporal and spatial scales \cite{kolmogorov_dissipation_1941}. 

\begin{figure}
  \centering
  \includegraphics[width=0.8\linewidth]{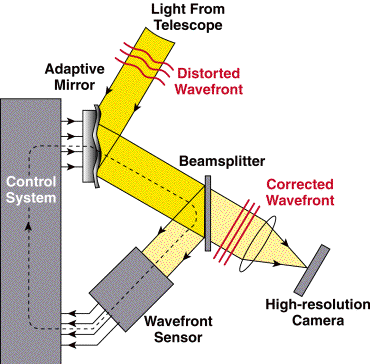}
  \caption{ (color online). Principle of astronomical adaptive optics. The aberrations from a star can be measured with a wavefront sensor, and corrected in real time on a deformable mirror, restoring sharpness of the image on a camera. (Image: Center for Adaptive Optics, University of Santa Cruz.) }
  \label{fig:AOprinciple}
\end{figure}

Most importantly, these fluctuations cannot be corrected by a passive element since they are by nature unpredictable and often time-dependent. Accordingly, they have posed a limit for the spatial resolution of conventional passive optical systems in imaging, as well as for the bandwidth in free-space telecommunications. A solution to this problem was proposed by  Babcock in the 1950s \cite{babcock_possibility_1953}, who suggested the use of an active deformable multi-element that, when coupled with a so-called wavefront sensor,  can correct in real time for these spatial inhomogeneities. This was the base for the field of adaptive optics, that is now routinely used in terrestrial astronomical telescopes \cite{roddier_adaptive_1999}. The technique requires fast measurement of the wavefront, then  analog feedback over multiple elements to compensate for the aberrations faster than the medium evolves (see Fig.~\ref{fig:AOprinciple}).

In essence, if the aberrating medium can be modeled as a single thin layer of thickness $L$ with local change in refractive index $n(x,y)$, then its effect on light is to add an additional length $\Delta z(x,y)=L n(x,y)$. This perturbation can be canceled by measuring the spatial phase of the light (the wavefront, see subsection \ref{subsubsection4.2.1} for a description of techniques to measure it) at the output of the layer, then correcting this wavefront to re-obtain the original unperturbated wave by inserting  a deformable mirror of shape $-\Delta z(x,y)$ at a later stage in the optical system, in a plane that images the original plane (called ``conjugate plane"). This correction method necessitates the following two remarks: (i) As long as the index is independent of the wavelength, the correction is by nature \textit{broadband} because the path difference  is corrected, rather than the phase:  all wavelengths are corrected simultaneously. (ii) Since the mirror completely corrects the aberrations of the layer, any incoming wavefront will be corrected: the correction is said to be \textit{widefield}. Note that the quality of the correction depends on the ability to perfectly map the path length distribution $\Delta z(x,y)$ on the deformable mirror, which is limited by the number of actuators and by their achievable  displacement. Deformable mirrors for astronomy  typically have 100-1000 actuators with relatively long displacement range (10-100 microns or more), that are optimized for atmospheric turbulence of low transverse spatial frequency.
Also, since propagation is invariant by time-reversal, it was realized early that the correction not only compensates the incoming light in order to form a sharp image on a detector (for astronomy for instance), it also corrects a backward propagating wave, that would traverse the layer in the opposite direction. This feature can be used in the context of long distance bidirectional free-space communications \cite{zhu_free-space_2002}. Adaptive optics for  turbulence mitigation was the first instance in a whole series of digital tools for optical wavefront shaping that we will review in the following parts. 

\subsubsection{From aberrating layers to aberrating volumes}\label{subsubsection4.1.3}

One major limitation of the above approach stems from the fact that  aberrating layers like the atmosphere  are in reality aberrating \textit{volumes}, meaning that they have to be described by a three-dimensional  distribution of the  refractive index $n(x,y,z)$. The  transmission thus has to be modeled by a more complex transmission matrix $\bf{t}$ than the one defined in section \ref{section2} and it cannot be fully corrected with a deformable mirror.  If one  measures the wavefront issued from a point source (a star for instance), then the wavefront corrections applied on the mirror will perfectly correct the wavefront originating from this source, as if the aberrations were very thin, but will only  correct a  small angle (or a small field of view) around it. This so called ``isoplanetic patch", depends on the thickness and on the distance of the aberrating volume to the detector, and bears some strong similarities with the memory effect in the multiple scattering regime (discussed in sections \ref{subsubsection3.1.4} and \ref{subsection5.1}). In order to correct more accurately for such a complex propagation,  the concept of multiconjugation has been introduced \cite{beckers_increasing_1988}. There, multiple deformable mirrors in series are placed in conjugated planes  of  multiple depths inside the scattering volume, leading to a correction over of a larger field of view.

Another limit of the adaptive optics  approach above is that it works only in the weak aberration regime. This means that locally the variations of the refractive index must be small in amplitude, and of low spatial frequency, such that the perturbation of the wavefront is small locally   and backscattering is limited. In addition, also  the overall perturbation must  be of low spatial frequency  in order to be corrected with a few actuators on the deformable mirror. When the index of refraction varies with larger amplitude and over smaller scales, then scattering becomes stronger and  the spatial frequencies of the fluctuations to be compensated (as well as their amplitude) exceed the capabilities of deformable mirrors: the adaptive optics approach breaks down. 

\subsubsection{Optical phase conjugation}\label{subsubsection4.1.4}

\begin{figure}
  \centering
    \includegraphics[width=0.850\linewidth]{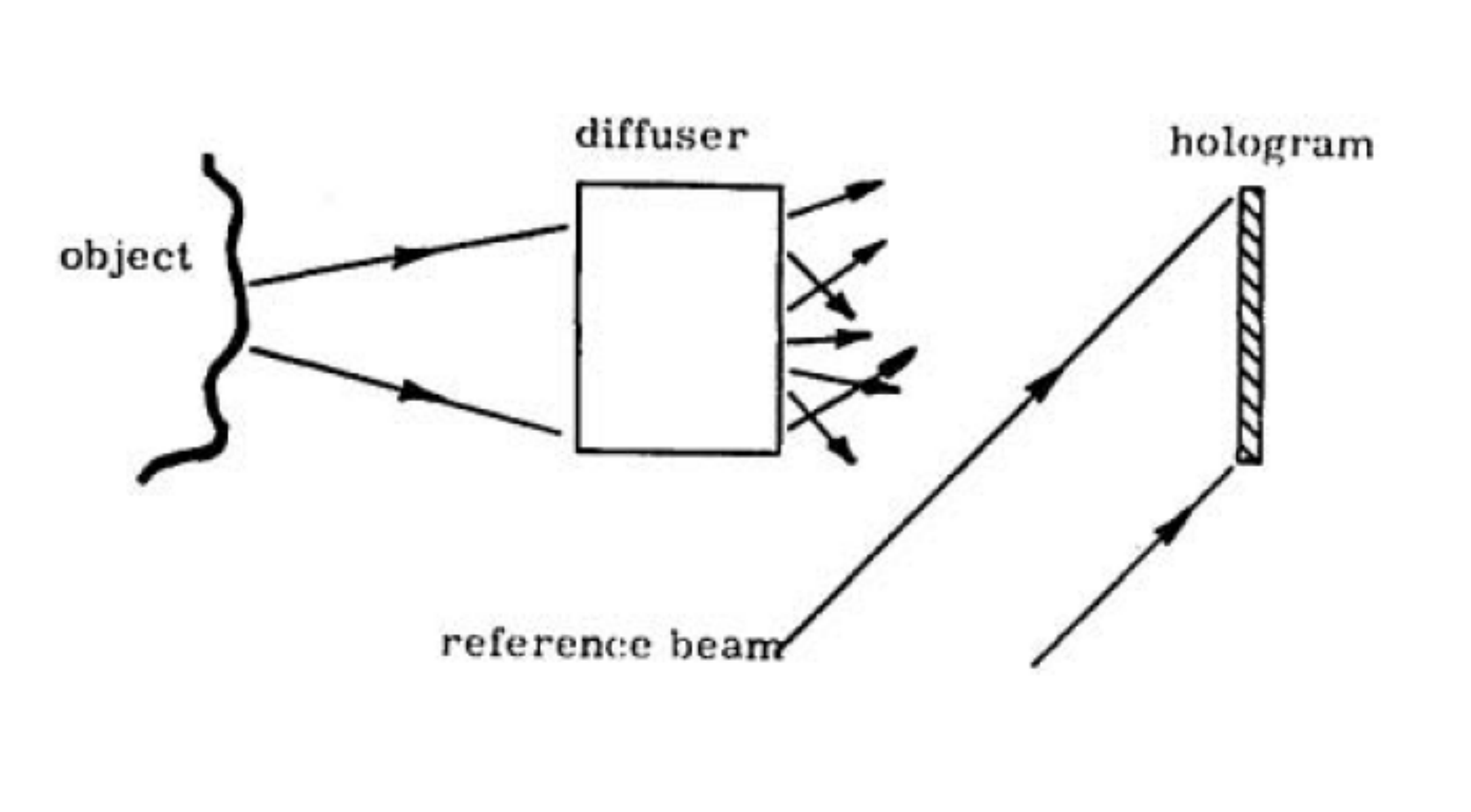}
  \includegraphics[width=0.85\linewidth]{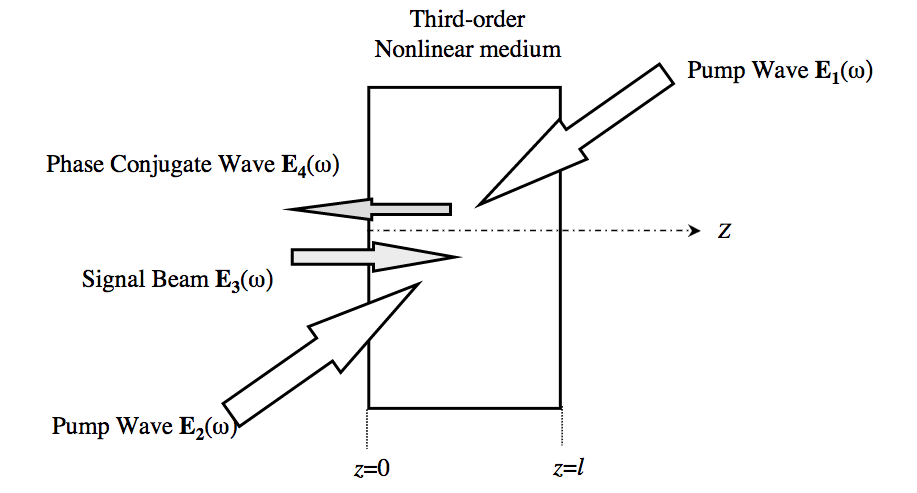}
  \caption{Two implementations of optical phase conjugation (OPC). (top) OPC via holographic recording (from \cite{leith_holographic_1966}). The hologram is the interference of the signal with a reference beam and is stored on a photographic plate.  The phase conjugate beam can be emitted at a later time by shining  the reference beam on the plate.  (bottom) OPC by four wave mixing. The signal beam is incident on a non-linear crystal. Two counter-propagating waves (the pump) interfere with the signal beam and  generate automatically and instantly the phase-conjugate wave via a non-linear process \cite{he_optical_2002}.}
  \label{fig:OPC}
\end{figure}

The arsenal of techniques for compensating a wavefront distortion found an interesting extension in the 1960s when it was realized that one can generate the phase-conjugate  of a wave. Such a transformation, that corresponds to changing the sign of the complex phase of the field,  makes the wave propagate backward and ``undo" the distortion, thanks to the  time reversal  invariance and reciprocity of the wave equation.  This concept was first realized with a holographic method \cite{leith_holographic_1966} (see Fig.~\ref{fig:OPC}). A hologram can be recorded by interfering  both the distorted wave and a reference plane wave on a photosensitive plate. By illuminating the plate with the same reference plane wave, the  scattering of this reference beam by the hologram carries the phase-conjugate of the incident beam, and can repropagate through the medium (however complex it may be) and reform the initial object or focus \cite{leith_holographic_1966}.  Based on the emergence of non-linear optics in the 1960s and 1970s, it was suggested by Yariv \cite{yariv_transmission_1976} that this holographic  optical phase conjugation could be performed in real time, using various non-linear processes. An implementation of this concept  was first realized   via four-wave mixing \cite{bloom_conjugate_1977, yariv_amplified_1977, yariv_four_1978}, and  via stimulated Brillouin scattering \cite{kralikova_image_1997}, in liquid crystals \cite{karaguleff_optical_1990}, or using three wave mixing \cite{voronin_compensation_1979, Ivakhnik_compensation_1980}. Many more details on optical phase conjugation (OPC) can be found in  \cite{fisher_optical_2012}.

In the presence of gain or loss without saturation, the medium is still reciprocal (as discussed in section \ref{subsubsection2.1.4}), which is a sufficient condition for a phase conjugated  wave to effectively retrace its path. We will discuss  wavefront shaping in presence of gain or loss in section \ref{section5}.  Also, even \textit{partial} phase conjugation will refocus, albeit with a lower efficiency.
 In the monochromatic domain, optical phase conjugation has led to numerous applications, in particular for lasers \cite{brignon_phase_2004}. In the time-domain, this has also led to spectacular demonstrations in acoustics \cite{fink_time-reversed_2000}, where transducers naturally provide amplitude detectors and arbitrary wave generators. 

In a nutshell, all these techniques work independently of the complexity of a medium, provided it is reciprocal. Although it had been developed primarily for aberrating layers, OPC even works for a strongly scattering medium.  An important limitation of OPC is, however, that it cannot by itself generate a given wavefront; it requires instead an input field or, in other terms, a physical source that will emit a wave. We will see in section \ref{subsubsection4.4.1} how digital tools nowadays allow not only for digital-OPC, but also  more advanced  OPC-like operations, in particular focusing behind a complex medium without a source.

\subsection{Digital tools for wave manipulation in optics}\label{subsection4.2}

In the optical realm, there is no generic or universal spatial and temporal modulator for light that is able to generate an arbitrary spatial and temporal waveform, quite  in contrast to what is possible for acoustics using  a transducer array, or in the radio-frequency domain using an antenna array. Whereas for these latter types of waves the modulators, sources, and detectors  can also all be the same device, this is unfortunately not the case in the optical domain. As we are mostly interested  in coherent manipulation in this review, we will  focus our discussion on coherent sources such as lasers, that can be either broadband (for pulsed operation) or  narrowband (for monochromatic emission). Other sources will be relevant as well, such as superluminescent diodes (SLED) or supercontinuum sources, which are spatially coherent but temporally incoherent. These light sources can then be controlled spatially and temporally by a wide range of modulators, that can either be analog or digital. The aim of this section is to give a flavor to the non-optician reader of what can and what cannot be done in optics.

The emergence of microelectronics has pushed  the generation and detection of signals into the digital age. Instead of recording an image on a photographic film, and transmitting an image through a fixed optical element, multi-element arrays are now routinely used for spatial modulation or spatial detection in cameras (charged coupled device (CCD), or complementary metal–oxide–semiconductor (CMOS) technologies in particular). We will focus our description on the modulation part, which is less well-known and more critical for the purpose of controlling light in complex media. For spatial modulation we can distinguish between amplitude modulators, i.e., display devices (such as  the liquid crystal dislay (LCD) of modern televisions) and  phase-modulators (that are also based on liquid crystal technologies). These modulators are usually referred under the generic term of spatial light modulators (SLMs). 
While it is by no means the purpose of this review to enter into the technical details of these devices, we will try to review their design principle, as well as the current state of the art of their performance. We will leave out of this review the deformable mirrors that have been mentioned in the previous section for adaptive optics, but are in little use in complex media studies. Two aspects of SLM devices and digital cameras make them invaluable tools for mesoscopic physics investigation. The first one is that the size of individual actuators or detectors can be of the order of a few microns, which means that in a microscopy system the diffraction limit can easily be reached, both for detection and for modulation. The second advantage is the massively parallel nature of these arrays, which can in principle give easy access simultaneously to up to a few millions of degrees of freedom, at a very small cost. This feature provides an unprecedented technological toolset for manipulating or detecting optical waves in complex media.

\subsubsection{Matricial detector arrays}\label{subsubsection4.2.1}

CCD and CMOS cameras are now in everyday use for imaging in the visible range. Other types of sensor arrays for different wavelength domains  also exist, such as InGaAs detector arrays for  the near infrared range, for instance. Typically all such electronic devices  are too slow to access the waveform of an optical field in real-time, in contrast to what is possible with acoustics and with radiofrequency signals. Whatever the conversion process, photodetectors are sensitive only to the intensity of the field, i.e., they integrate the local energy over a certain duration that is much longer than the period. As a result,  the duration of an optical pulse cannot be measured when it is shorter than the integration time of the photodetector. Another essential limitation in optical detection is that the phase of the field is not directly accessible.

 Fortunately, for monochromatic fields, several techniques allow to retrieve the complex amplitude and phase of an image. The most widely used method is digital holography, that relies on recording a stationary interference pattern (or ``hologram")  on the camera,  between the monochromatic image  to record (called the signal) and a plane wave at  the same frequency (called the reference). The complex field of the image can be inferred either by taking several holograms (at least three) while adding a global phase to the reference, resulting in what is called phase-shifting digital holography \cite{yamaguchi_phase-shifting_1997}. The complex field can also be retrieved in a single image, using a tilted plane wave as a reference followed by a digital filtering, a technique called off-axis holography \cite{leith_microscopy_1965, cuche_spatial_2000}.  Finally, some techniques allow complex field recovery by taking images in different planes and solving an inverse problem, e.g., based on the transport of intensity equations \cite{teague_deterministic_1983}. Alternatively, computational tools, such as phase-retrieval of phase-diversity algorithms can, in some instances, allow to retrieve the phase of an unknown field from intensity images  (see  \cite{fienup_phase_1982} for a review).
 
When the signal is smooth and slowly varying in space, and when one in interested only in measuring the wavefront, other  sensing techniques provide easy access to the local phase gradient without a reference, even when the illmumination is incoherent. This is the domain of wavefrond sensing, dominated by the  so-called Shack-Hartmann wavefront sensor \cite{Platt_history_2000}, which is widely used in adaptive optics experiments, but  other types of wavefront sensors exist \cite{ragazzoni_pupil_1996, bon_quadriwave_2009}. Note that all wavefront sensors combine a diffractive element (a lenslet array for the Shack-Hartmann sensor for instance) followed by a CCD or CMOS sensor.

\subsubsection{Matricial spatial light modulators}\label{subsubsection4.2.2}

In section \ref{subsubsection4.1.2}, we have already described deformable mirrors, that  could be used for a digital and smooth deformation of a wavefront as needed for adaptive optics. Thanks to the speed and the long course of the actuators, deformable mirrors are well adapted to fast and broadband compensation of atmospheric perturbations, but they have a limited number of actuators, and an inherent cross-talk between the pixels due to the fact that the actuators deform a common membrane. While this is an advantage for smooth compensation of aberrations, these features are  impractical when  highly complex wavefronts are needed; in this case SLMs with very high spatial resolution and independence of the spatial control  are crucial. 

 Matricial SLMs (see Fig.~\ref{fig:SLMtechnos}) satisfy these stringent conditions. They comprise an array of pixels that can locally modulate the light. In practice, a  wave incident on the SLM is transmitted or reflected off each pixel with a locally determined  partial transmission or reflection (for amplitude modulation) or with a path retardation or advance (i.e., a phase delay).  The resulting spatially modulated wave can be designed at will, in general by setting the voltage applied to each pixel. There is unfortunately no ``ideal" SLM that can generate arbitrary phase and amplitude modulations. Most SLMs are usually either phase-only or amplitude-only, depending on the technology used. Note also that SLMs are typically relatively broadband: the path retardation or attenuation affects all wavelengths the same way. Still, for phase-modulation, a given path retardation means that the phase-retardation is  wavelength-dependent, which can be problematic for broadband light. Another feature of  these light modulators is that they are typically slow (a few kHz at most) such that they cannot perform any fast temporal modulation. This  limitation usually comes from the physical mechanism at the heart of the modulation process such as a mechanical displacement or a mirror or the slow rotation of a liquid crystal. In this sense, the control is truly spatial  only  (with a slow evolution in time). We will detail in the next section how the temporal or spectral degrees of freedom of the light can be controlled, when true temporal control is required. 

 The most straightforward way to spatially modulate light is to  reflect it off a  reflective array of micro-actuators  (i.e., a  set of movable mirrors) that are electrically or magnetically actuated \cite{gad-el-hak_mems_2010}. These are the so-called digital micromirror devices, and they belong to the very general class of MOEMS (micro-opto-electro-mechanical systems). A linear (piston-like) movement of each mirror will result in a pure path length change $\Delta z$, resulting in  a modulation of the phase $\phi=2 \pi \Delta z/{\lambda}$.  Pure  phase modulators are not very widespread but exist, up to $64 \times 64$ pixel arrays \cite{gehner_MEMS_2006, cornelissen_MEMS_2012}, and are used for adaptive optics. Spatially modulating the intensity of light, on the other hand,  would require a continuous change of reflectivity, which is actually very difficult to achieve in practice with MOEMS. Still, MOEMS amplitude modulators are now mass-produced for display applications, but they are  dynamic binary modulators based on the technology  developed by Texas instruments Corp., where the mirror can  switch very rapidly between two angular positions (one where light is reflected,  and one where light is deflected towards an absorber). Intermediate intensity levels can be achieved, but only on average, thanks to  a very fast mechanical switching time. The level of reflected intensity is then obtained by controlling the ratio of on- and off-time \cite{hornbeck_DMDTM_2001}. 
To be more concrete, we provide some typical numbers in the following: The DLP 0.95  1080p chipset from Texas Instruments has a $1920 \times 1080$  matrix of micromirrors, each a square of a lateral size of 10.8 $\mu$m with a tilt angle of $12^\circ$ between the two switching positions, and a modulation speed of  23 kHz. 


Another widespread technology for spatial light modulators relies on  liquid crystals. In essence, a modulation of the local polarization, and pure-phase control of the light can be locally achieved by modifying the orientation of a liquid crystal based on its birefringence. A local orientation of the liquid crystal  can be achieved electronically or optically \cite{lueder_liquid_2010}. With additional polarization optics, amplitude modulation can also be realized, a feature which led to the development of liquid crystal displays (LCD).  Some devices work in transmission, but most devices work in reflection.  We will focus our discussion on phase modulators, that are used in the vast majority of the experiments covered in this review. Since phase modulation is achieved by a small change of the refractive index, only limited retardation can be achieved. With  at least $2\pi$  phase retardation,  it allows for arbitrary phase form generation at a given  wavelength. However, unlike  deformable mirrors, it does not permit broadband retardation patterns because of the problem of ``wrapping" of the phase for path delays of more than a wavelength. In applications where speed is  required, liquid crystal-based SLMs are sometimes too slow: the limiting speed factor is the unavoidable  time required for the liquid crystals to orientate. Typical twisted nematic liquid crystal SLMs have around $10^6$ pixels, like digital micromirror devices (DMDs).  They allow pure phase modulation over $2\pi$ in the visible to near-infrared range, at the cost of a refresh rate of  a few tens of Hz typically. Other types of liquid crystals  are also used as SLMs, e.g., based on the ferroelectric effect.  These devices are usually limited to binary phase  modulation and are to date much less used for wavefront shaping in complex media, despite having a very fast switching time of only 100 microseconds, and a refresh rate of up to several kHz.

 There are  a number of techniques that allow the use of these SLMs to generate  any wavefront shape in both phase and amplitude. One possibility is to combine a phase and an amplitude modulator in series, but this approach is expensive, complex to implement, and therefore rarely used. Thanks to the propagation law of Fourier optics, it is also possible to compute the wavefront from a single  modulator (amplitude or phase, binary or not)  that will  generate  an arbitrary object in a Fourier plane \cite{lee_iii_1978, conkey_high-speed_2012} (for instance in the focal plane of a lens). It is also possible, by appropriate filtering in the Fourier space, to generate an arbitrary phase and amplitude object from a phase-only modulator \cite{van_putten_spatial_2008}, and even from a binary modulator \cite{goorden_superpixel-based_2014}. These techniques can be useful when one wants to generate a specific input state with high accuracy, as we will see in the next section.  Note also that  SLMs  are two-dimensional masks: while they can generate complex spatial fields, they cannot perform an arbitrary  transform that would be necessary to perfectly compensate for a volume scattering or an arbitrary transmission matrix $\bf{t}$. Some works, however, have suggested theoretically and experimentally that it is possible to use several SLMs in series, or multiple reflections off a single SLM, to generate an arbitrary spatial unitary transform \cite{morizur_programmable_2010}. While promising, these implementations of unitary transforms are not yet mature enough to have found applications in complex media, since the number of degrees of freedom that can be programmed is still limited to a few tens of modes maximum. First successful applications to  few-modes fibers have emerged recently \cite{labroille_efficient_2014}. 

In terms of applications, SLMs are now widely used in optical laboratories as reconfigurable diffractive optical elements. They are extremely versatile tools to generate complex spatial states of light. In particular,  the generation of high orbital angular momentum states \cite{gibson_free-space_2004}, of states with multiple focii for optical manipulation \cite{grier_revolution_2003}, and of complex illuminations patterns for microscopy \cite{maurer_what_2011} should here be mentioned.

\subsubsection{Other types of spatial light modulators}\label{subsubsection4.2.3}

In addition to SLMs, several other devices allow spatial control of an optical beam. The most relevant ones in the context of this review are deflectors, i.e., devices that are able to deflect and rapidly scan angularly a laser beam, thus producing a translation of the beam in the far field. Two orthogonal  deflectors can be used in series for 2D deflection. They are commonly used in such a configuration for scanning microscopes \cite{mertz_nonlinear_2004} to raster scan a 2D region with a focused laser beam in order to recover an image. There are two main types of such deflectors: The first one are acousto-optic deflectors (AODs), that use diffraction of a laser beam on an acoustic wave. By varying the acoustic frequency, the deflection angle can be varied \cite{gottlieb_electro-optic_1983}. The second type are  scanning mirrors,  or galvanometers: A lightweight mirror is placed on a fast rotating stage, that produces angular rotation of the mirror. While deflectors cannot generate arbitrary wavefronts, they are well adapted to raster scan the different input modes of a medium.  Both device types work similarly, but each has specific advantages. Mirrors can be large (several millimeters) and therefore may reflect and deflect complex wavefronts, which can be useful when one wants to exploit the memory effect (see discussion in \ref{subsection5.1}).  We will see examples of such devices used in conjunction with matricial SLMs. AODs on the other hand,  require diffraction limited beams to pass through a small aperture, but can be orders of magnitude faster (in the MHz regime versus kHz for mechanical scanners). 

\subsubsection{Temporal modulators}\label{subsubsection4.2.4}

We have seen in the previous section how to spatially shape an arbitrary wave. We will now review the different ways in which one can  achieve temporal control in optics. Note that this control is normally achieved on a single spatial mode only. 

For a laser source (e.g., a monochromatic laser at frequency $\omega_L$), it is possible to use an electro-optic modulator (EOM) or an acousto-optic modulator (AOM) to  temporally modulate a beam (in amplitude or in phase) by modifying the index of refraction of the propagation medium. The effect of this modulation is best understood as a modulation of the transmission in the frequency domain, $T(\omega)$. Since the frequency domain that is available for this kind of modulation ranges from the MHz range (AOM) to a few GHz (EOM), the effect of this modulation on an optical signal can be understood as a slow envelope modulation. In practice, a modulation at $\omega_M$ will create new sideband frequencies in the spectrum at $\omega_L \pm \omega_M$ \cite{wooten_review_2000}. In the context of scattering through complex media,  temporal control on this spectral range would primarily be interesting if the medium exhibits a temporal response on a similar time-scale (nanoseconds, to microseconds), which is important for either very large scattering systems or for long optical fibers.

The second kind of temporal modulation is spectral shaping, where one tries to directly control the optical waveform. Since the optical frequencies involved prevent the direct use of electronics, one needs to  manipulate instead the spectral components of the light to control its temporal profile. For this technique, one requires  a broadband coherent source such as a mode-locked laser. This type of source consists of  a laser  with broad inhomogeneous gain where several frequencies can lase simultaneously.  A mode locking mechanism (usually a Kerr effect, or a spatial filter) favors emission from all theses modes with a fixed  relative phase. In such a situation, the laser emits  ultrashort pulses with a Fourier limited duration, and a repetition rate given by the round-trip time in the cavity. A very typical system, the Titane-Sapphire femtosecond laser, shows  gain in the 700-1100 nm band. Most sources emit with a central frequency in the near-infrared around 800nm, with a bandwdith of 10-100 nm, thus providing pulses in the 10-100 femtoseconds range, typically. The repetition rate is usually around 80 MHz for these lasers, meaning that around 12 ns separates two successive pulses. 

\begin{figure}
  \centering
  \includegraphics[width=1\linewidth]{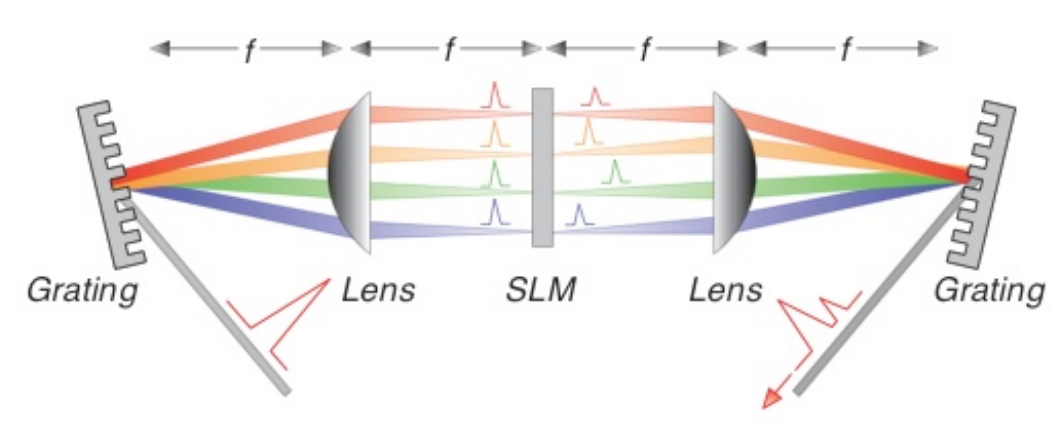}
  \caption{(color online). Principle of Fourier transform pulse shaping. An ultrashort pulse is incident on a dispersive element (here grating), that separates spatially the different spectral components, which are then individually addressed by means of an SLM, thus generating a controllable spectral phase function. The different spectral components are later recombined into a single spatial mode to form a   pulse with a well-defined temporal shape. (Figure adapted from \cite{weiner_femtosecond_2000}.)}
  \label{fig:4fline}
\end{figure}

Starting from such Fourier-limited pulses, it is  possible to modify at will the temporal shape of the pulses by controlling their spectral phase. For this domain of pulse shaping we refer the reader to several detailed reviews \cite{weiner_femtosecond_2000, monmayrant_newcomers_2010, cundiff_optical_2010}.  While there are different ways to shape temporally an ultrashort pulse, we will focus in our discussion on the most common implementation, i.e., the  Fourier-transform pulse shaping (see Fig.~\ref{fig:4fline}). In essence, a dispersive element (prism or grating) disperses the different spectral components. In a Fourier plane of the dispersive element (e.g., in the focal plane of a lens), each spectral component is focused at a different position. A linear spatial light modulator in this plane then adds an arbitrary phase $\phi(\omega)$. When recombining all spectral components,  a pulse  is formed with the chosen temporal shape governed by the spectral phase imposed by the SLM. The pulse shape is arbitrary, but the maximum arbitrary pulse duration is determined by the resolution of the apparatus (i.e., limited by the pixel size and the spectral resolution of the dispersion line), and the refresh rate is limited by the SLM speed.

\subsection{The thin disordered slab: an opaque lens}\label{subsubsection4.3}
In section \ref{section2}, we have treated the important case of a disordered wire, and introduced in this context the concepts of ballistic wave scattering, transport mean free path, modes, the scattering matrix, etc. These concepts from the mesoscopic formalism will now be mapped to the optical domain. However, there is no exact optical equivalent of the disordered wire. In particular, the impenetrable and lossless boundaries of a wire cannot be easily reproduced in optics. Multimode optical fibers could be considered a stricly bounded complex system with a limited number of modes (and will be described in section \ref{subsubsection4.5.1}), but they  behave very differently from the disordered wire since there is no significant bulk disorder (scattering mostly comes from the boundaries), and almost no backscattering. 
Rather than a multimode fiber,  the  paradigmatic system in optics is the slab geometry: a disordered slab of finite thickness $L$ and of infinite lateral extension, with a transport mean free path $\ell^\star$ that is short enough to push the system into the multiple scattering regime, $\ell^\star\ll L$.  From now on we will refer to this system as the ``opaque lens"  \cite{cartwright_opaque_2007}.

\begin{figure}
  \centering
  \includegraphics[width=1\linewidth]{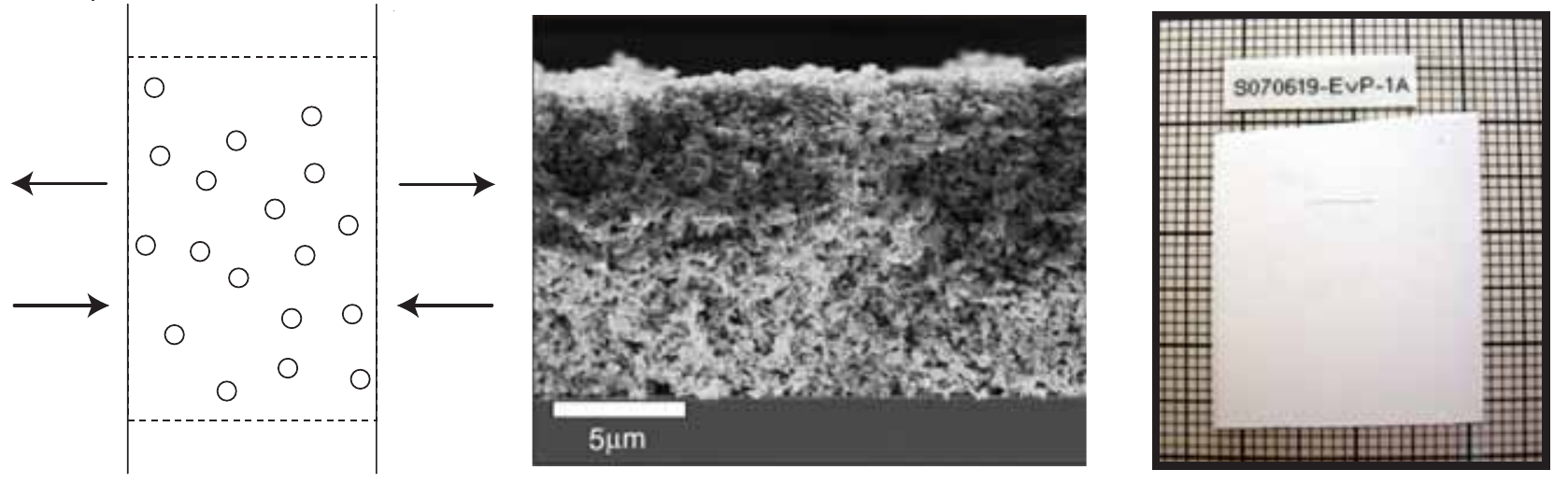}
  \caption{(color online). The opaque lens in optics: a thin disordered slab of scattering material. (left) Schematic model (adapted from \cite{yaqoob_optical_2008}). (center)  Scanning electron microscope (SEM) side view.  (right)  Photograph of a ZnO opaque lens. (Middle and right panels adapted from \cite{vellekoop_controlling_2008}.)}
  \label{fig:opaquelens}
\end{figure}

A common experimental realization of the opaque lens is typically a layer of dielectric scatterers of micrometer or sub-micrometer size, randomly packed, deposited on a transparent holder, such as a glass slide. To ensure sufficient scattering, the layer should be thick enough (a few micrometers to a few tens of micrometers), producing a  white and opaque appearance, provided that the material is non-absorptive (see Fig.~\ref{fig:opaquelens}). This realization  has the specific advantage of being easy to fabricate and extremely stable. Wavefront shaping techniques are by nature relatively slow, making the stability of a particular realization of disorder an essential requirement. Of course, this simple system can be mapped to several practical situations and materials. Snow, biological tissues, white paper, egg shell, bones, are just a few examples  of thin or thick materials that can be understood with the same formalism. An important difference between these systems and electronic mesoscopic systems is the very large number of modes supported: Since objects in optics are usually macroscopic, often in the millimeter scale or larger, the number of optical modes, which  scales as $A/\lambda^2$ (with $A$ being the transverse area and $\lambda$ being the wavelength), can easily be in the $10^6$ range or higher. 
Most of the results extend also to dynamical systems (such as milk, fog, clouds etc.) although the geometry and the fast dynamics makes them very challenging for wavefront shaping.  

To describe an opaque lens more formally, let us now consider a three-dimensional slab of finite thickness $L$, that features  complex inhomogeneities of the refractive index, such as to be fully disordered. Such a case is realized for a slab composed of a random packing of particles of high refractive index of different sizes, in a matrix of low refractive index (as, e.g., in air). These inhomogeneities scatter light in a very complex way, but as we have seen in section \ref{section2}, this does not prevent us from describing such a system  with the formalism of the scattering or transmission matrix, as long as the system is linear (see Fig.~\ref{fig:opaquelens}).
 
  We will consider the case  when the light that is transmitted through the sample has been multiply scattered (i.e., $\ell^\star \ll L$) and absorption can be neglected (although absorption or gain do not necessarily break the linearity assumption). We also need to make sure that the scattering strength  is not too strong ($k \ell^\star \gg 1$), to avoid effects related to strong localization.  In essence, we suppose that light is subject to a purely diffusive process when going through the slab. While we have described in section \ref{section3} that different mesoscopic effects can be present in such a situation, these effects are typically elusive. Especially in the case of only partial control and detection, we can often consider to first order that no mesoscopic correlations are present.
  
 In this limit, we can assume that any coherent monochromatic wavefront incident on one side of the slab will be  multiply scattered such as to produce a fully developed speckle  on the other side \cite{goodman_fundamental_1976}.  Two speckle patterns, resulting from  two different illuminations, will be completely uncorrelated, provided the inputs are sufficiently uncorrelated. Still, each speckle is the deterministic result of the multiple scattering process and is specific to the medium and to the chosen input wavefront.  Because of the high complexity of the multiple scattering propagation, it is  in practice impossible  to calculate a priori the speckle pattern. However, as discussed in \ref{subsubsection3.1.2}, speckle patterns have some well defined statistical properties. While complete books are devoted to these phenomena (see, e.g.,  \cite{goodman_speckle_2007}, we will only recall here some fundamental aspects for polarized, monochromatic and spatially coherent incident light: (i)  Due to the multiple scattering process, the incident polarization is  mixed during the propagation. After the medium, the light has a well defined but unpredictable polarization state at any point: The vectorial field  distribution  is the sum of intensities of two uncorrelated fully developed speckles for two orthogonal polarization states (e.g., vertical and horizontal polarization). (ii) For a given polarization, the distribution of intensities  follows the Rayleigh law  (for each polarization). (iii) The characteristic grain size depends on the geometry of the system, it corresponds to  diffraction-limited spots, and is related  to the spatial correlation $C^{(1)}$ (defined in section \ref{subsubsection3.1.2}). 
  
\subsubsection{Transmission matrix in the spatial domain}\label{subsubsection4.3.1}  
  
In this section, we consider the transmission matrix of a disordered slab, i.e., of an  opaque lens. This transmission matrix will be labeled as $\tilde{\bf{t}}$, and links the fields of the $N$ input to the $M$ output pixels of the SLM and of the detector, respectively. 

 We have seen in Eq.~(\ref{eq:smatrix}) of section \ref{section2} that the transmission matrix $\bf{{t}}$ connects the incoming field modes from the left to the outgoing field modes on the right, written  $\bf{c}^+_r=\bf{{t}}\,\bf{c}_l^+$.  The difference between the experimentally measured $\tilde{\bf{t}}$ and the full transmission matrix $\bf{{t}}$ of the disordered medium, will be that $\tilde{\bf{t}}$  also comprises the propagation of the field from the SLM to the medium and from the medium to the detector. In addition $\tilde{\bf{t}}$ only contains a small part of the full transmission matrix $\bf{t}$, usually decomposed in pixels, which do not constitute a complete basis.  Since additionally the pixels, both in modulation and detection, are typically illuminated at close to normal and constant incidence, we can usually neglect in practice the problem of flux normalization discussed in \ref{subsubsection2.1.4}. For simplicity, we will note these outgoing (incoming) modes as $\tilde{\bf{E}}^{\rm out}$ ($\tilde{\bf{E}}^{\rm in}$), in which notation the field on the $m$-th output pixel is   ${\tilde{E}^{\rm out}_m}= \sum_n {\tilde{t}}_{mn}{\tilde{E}^{\rm in}_n}$, or equivalently $\tilde{\bf{E}}^{\rm out}= \tilde{\bf{t}}\,\tilde{\bf{E}}^{\rm in}\,$.   We will explain in section \ref{subsubsection4.3.3} how to retrieve $\tilde{\bf{t}}$.
 
 In the same way as the singular values $\sigma_i$ of ${\bf t}$ and the corresponding transmission eigenvalues $\tau_i=\sigma_i^2$   give access to the physics of wave propagation through the disordered wire (section \ref{section2}), we will now work  with the singular values and singular vectors of the transmission matrix $\tilde{\bf{t}}$ for the opaque lens, first in the monochromatic picture, and then including  the temporal and spectral aspects. 

 In the limit defined above, where one has access only to a small number of well separated modes of the open system, i.e., where each input mode $n$ gives rise to an independent speckle uncorrelated with the others, the singular values of the transmission matrix $\tilde{\bf{t}}$ are expected to follow the so-called Mar\v{c}enko-Pastur law \cite{marcenko_distribution_1967} that describes the singular value distribution (SVD) of rectangular random matrices without correlations. In essence, this law states that, for a $N\times M$ random matrix ($M \geq N$) of uncorrelated identically distributed elements, the distribution of  singular values $\widetilde{\sigma}$ (normalized to the average transmission)  depends only on the ratio $\gamma=M/N\geq 1$ and converges to,
\begin{eqnarray}
P_\gamma(\widetilde{\sigma}) &\approx &\! \frac{\gamma}{2 \pi \widetilde{\sigma}} \sqrt{(\widetilde{\sigma}^2\!-\!\widetilde{\sigma}^2_{min})(\widetilde{\sigma}^2\!-\!\widetilde{\sigma}^2_{max})} \,\,\,  \forall \widetilde{\sigma} \in [\widetilde{\sigma}_{\mathrm{min}}, \widetilde{\sigma}_{\mathrm{max}}]    \nonumber \\
&\approx & 0 \quad \text{otherwise, } 
\label{eq:Marcenkopastur}	
\end{eqnarray}
where $\widetilde{\sigma}_{min}=1-\sqrt{\gamma^{-1}}$ and $\widetilde{\sigma}_{max}=1+\sqrt{\gamma^{-1}}$. Correspondingly, the normalized SVD is bounded in the domain $[1-\gamma^{-1}, 1+\gamma^{-1}]$. We have supposed $M\geq N$, but the SVD remains the same if we reverse the role of $N$ and $M$. The interesting case of the square matrix $N=M$ gives rise to a circular distribution of the singular values in the interval $[0,2]$ and is usually referred to as the ``quarter-circle law". Still, for $N,M$ finite, Eq.~(\ref{eq:Marcenkopastur}) is an approximation and the eigenvalue density outside the indicted interval is exponentially small, but not zero.

\begin{figure}
  \centering
  \includegraphics[width=0.90\linewidth]{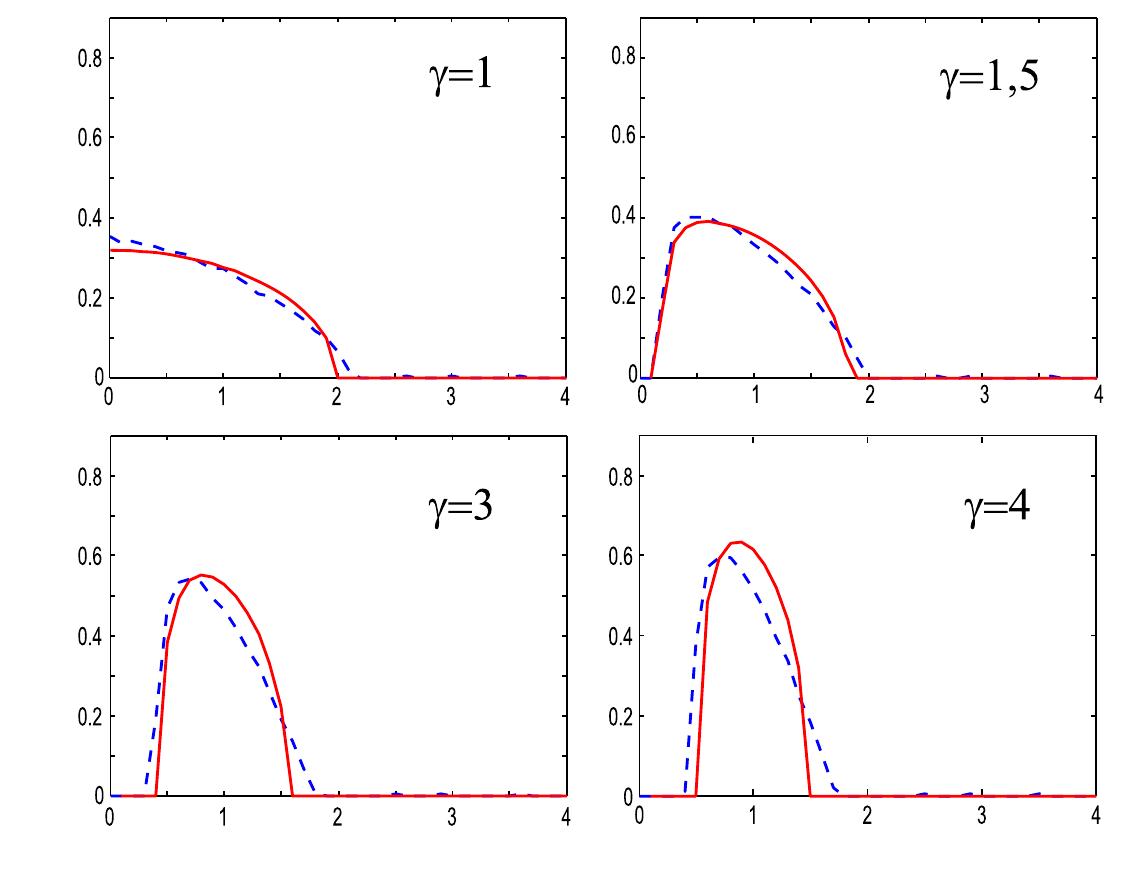}
  \caption{(color online). Normalized distribution of singular values $\tilde{\sigma}$ of transmission  matrices  of opaque lenses  as a function of the ratio $\gamma$ between the number of output to input modes $M/N$.  (Values of $\tilde{\sigma}$ are normalized to the average transmission and can thus be larger than 1.)  Dashed blue line: experimental data,; red line: independent identically distributed random matrices. The discrepancy can be attributed to residual correlations in the measured matrices.  For the value $\gamma=1$ one obtains the so-called ``quarter circle law". Adapted from \cite{popoff_controlling_2011}.}
  \label{fig:marcenkopastur}
\end{figure}

This result is routinely used in telecommunications to assess the bit rate and the error rate of data transmission \cite{chizhik_multiple-input-multiple-output_2003}. The Mar\v{c}enko-Pastur distribution has also been measured experimentally through a multiply scattering medium in acoustics  \cite{sprik_eigenvalue_2008, aubry_random_2009}.  In the next section, we will show how the transmission matrix $\tilde{\bf{t}}$ can be measured and the same distribution can be experimentally recovered from it also in optics \cite{popoff_controlling_2011} (see Fig.~\ref{fig:marcenkopastur}). The question that immediately arises at this point is how this result relates to the bimodal distribution of transmission eigenvalues derived in section \ref{section2}. The crucial point to observe here is that in the experimental conditions we have considered,  we access only a small fraction of the modes. Under these conditions,  the bimodal distribution reduces to the  Mar\v{c}enko-Pastur law when decreasing the number of input and output channel considered, whereby correlations get increasingly lost \cite{goetschy_filtering_2013}. To approach the regime where mesoscopic correlations start playing a role, particular care has to be taken to control and measure a large fraction of the modes,  also for large solid angles (e.g., by using high numerical aperture optics), and for both polarizations. This particular topic will be covered in section \ref{subsection5.2}. For the time being, we conclude that in most practical cases,  we can consider the disordered slab as a ``perfect" mixer for light, obeying Mar\v{c}enko-Pastur's law. 

So far we have not discussed what happens in reflection from the disordered slab. In principle, the reflection matrix can also be defined in the same way as for transmission. However, reflected light comprises not only multiply scattered light, but also singly scattered components, as well as all components in between. The reflection will therefore not be as perfectly ``mixed" as the transmission, in particular in terms of the polarization of light that is conserved for single scattering and can be partially conserved for few scattering events \cite{mackintosh_polarization_1989}. Weak localization effects such as the coherent backscattering cone described in section \ref{section3} are also  present, although their signature has  not been observed directly in the optical reflection matrix (see \cite{aubry_random_2009, aubry_recurrent_2014} for realizations in acoustics). Still, most of the results and experiments described below translate almost perfectly  from transmission to reflection. We will see how these deviations from perfect mixing can be retrieved and exploited for imaging in section \ref{subsubsection4.5.2}.
	
\subsubsection{Temporal and spectral aspects}\label{subsubsection4.3.2}

\begin{figure}
  \centering
  \includegraphics[width=1\linewidth]{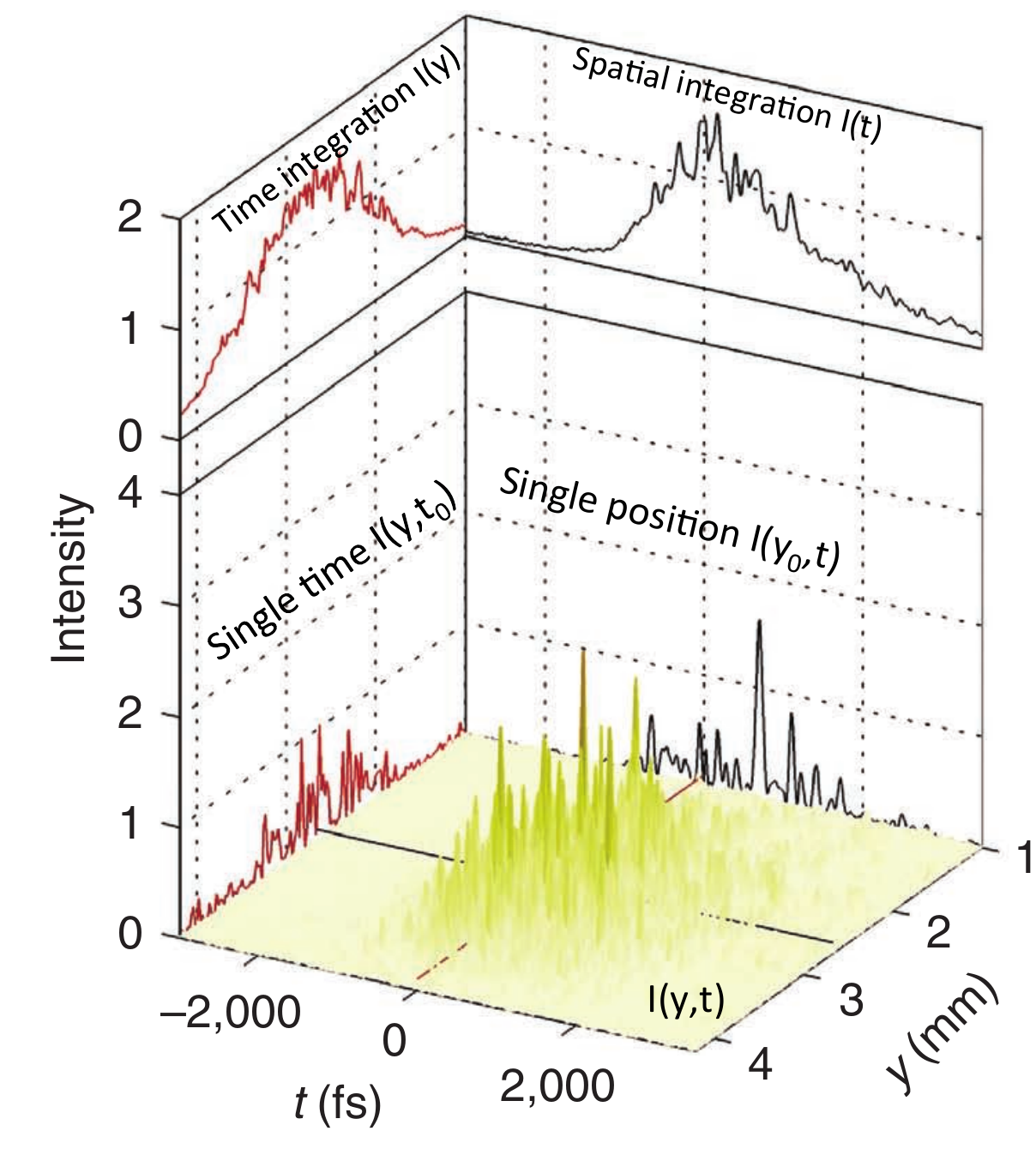}
  \caption{(color online). Representation of a spatiotemporal speckle, resulting from the propagation of a focussed ultrashort pulse through a thin ZnO Sample. The speckle is measured along one spatial dimension and as a function of time. One observe a complex spatiospectral structure $I(y,t)$ with speckle statistics, that are apparent when looking at a temporal or a spatial section for a given time or position, i.e., $I(y_0,t)$ and $I(y,t_0)$. When looking at a projection (integration) along the temporal or spatial coordinate, i.e., $I(y)$ or $I(t)$, one retrieves respectively the diffuse halo, and the average temporal broadening of the pulse. (Figure adapted from \cite{mccabe_spatio-temporal_2011}.)}
  \label{fig:temporalbehavior}
\end{figure}

In the temporal domain, light enters a disordered slab of thickness $L$, diffusely propagates in it, and exits on either side, or is absorbed. Knowing the diffusion parameters for light, such as the diffusivity $D=v_E\,\ell^\star/3$ (here $\ell^\star$ is the transport mean-free-path and $v_E$ the energy velocity defined at the end of section \ref{subsection2.4}), it is possible to recover the so-called Thouless time $\tau_D$ of the medium \cite{thouless_maximum_1977}. The parameter $\tau_D$ corresponds to the average time that a photon, already in the medium, takes to reach the medium boundaries, and is related to the Thouless number defined in section \ref{subsubsection2.1.5a}. The time $\tau_D$ scales with $L^2/(\pi^2 D)$, and corresponds to  a spectral bandwidth $\Delta \omega_D=1/\tau_D$.   While this time intrinsically describes the photon lifetime in the medium, it is not exactly the relevant quantity for transmission and reflection, where a photon first needs to enter the medium, before exiting on either side. For the thin but multiply scattering slab geometry, where $\ell^\star\ll L$, the reflection time $\tau_R$ is typically much shorter than the transmission time and than the Thouless time $\tau_D$, since a photon typically only explores a small volume of depth $\ell^\star$ before exiting on the same side; as such, $\tau_R$ is of the order of $\ell^\star/v_E$. Meanwhile, the transmission time $\tau_T$ will be on average slightly longer than $\tau_D$ since the photon must first enter the medium before exiting \cite{landauer_diffusive_1987, vellekoop_determination_2005}, but the distribution of transmission times will have an exponential tail of exponent $-t/\tau_D$. A rigorous way to define and to assess these scattering times is through the concepts of ``time delay'' and ``dwell time'', discussed in section \ref{subsection2.4}. For practical purposes it is often convenient to use Monte Carlo simulations \cite{patterson_time_1989} or to measure the times experimentally \cite{vellekoop_determination_2005, curry_direct_2011, mccabe_spatio-temporal_2011} (see Fig.~\ref{fig:temporalbehavior} for a spatiotemporal speckle and its spatial and temporal average). To these transmission and reflection times spectral bandwidths $\Delta \omega_T$ and $\Delta \omega_R$ are associated, which will in turn correspond to a spectral correlation of the respective transmission and reflection matrices of the slab. 

\subsubsection{Accessing the monochromatic transmission matrix of an opaque lens}\label{subsubsection4.3.3}

The monochromatic transmission matrix of a complex medium can indeed be measured \cite{popoff_measuring_2010}. In essence, it is possible to send a set of input spatial modes,  to record for each of these modes the transmitted amplitude, and to determine directly from these input-output measurements the transmission matrix elements $\tilde{t}_{mn}$ linking the input mode pixels of the SLM  to the output pixels on a CCD camera.

\begin{figure}
  \centering
  \includegraphics[width=1\linewidth]{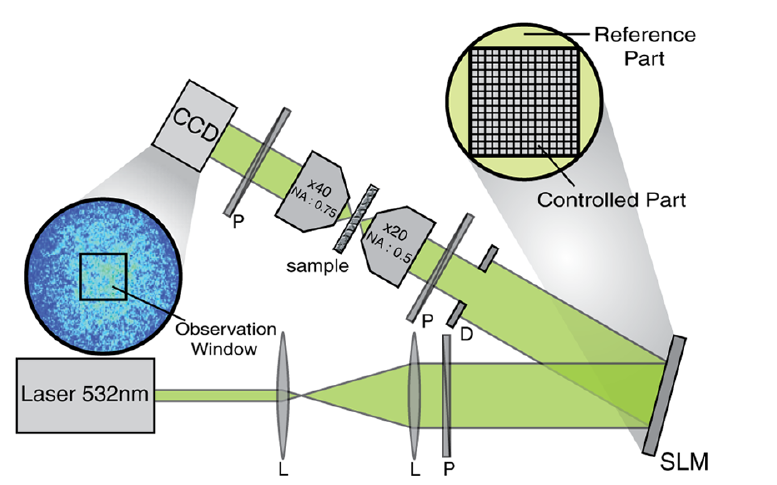}
  \caption{(color online). Setup of the first measurement of the transmission matrix. The laser is expanded and reflected off an SLM. Part of the SLM is unmodulated and serves as a reference for the interferometric measurement of the output field. The phase-modulated beam is focused on the multiple-scattering sample and the output intensity speckle pattern is imaged by a CCD camera. Additional elements: lens (L), polarizer (P), diaphragm (D). (Figure adapted from \cite{popoff_measuring_2010}.) }
  \label{fig:firstTMmeasurement}
\end{figure}

In an initial implementation (see Fig.~\ref{fig:firstTMmeasurement}), the  input spatial modes were generated using a liquid crystal SLM, and the output amplitudes were obtained using phase-shifting holography \cite{yamaguchi_phase-shifting_1997}, i.e., by recording several images on the CCD, resulting from the interference of the output wave to be measured with a reference wave with different phase shifts.  Later on, several variants  were used to either modulate or detect the amplitude of the field and recover the transmission matrix. In \cite{choi_overcoming_2011}, the medium was illuminated with a plane wave, with an angle of illumination  that was varied  using a galvanometer-mounted tilting mirror \cite{choi_tomographic_2007}. While this method allows to measure every input angle directly, thanks to the movable mirrors, it does not permit to generate a given arbitrary wavefront directly. On the detection side,  in order to record the amplitude hologram of the output speckle in a single image rather than in a sequence, off-axis holography was implemented \cite{kim_maximal_2012, akbulut_measurements_2013}.  Using a set of polarization beamsplitters and polarization optics, it is also possible to control or to detect both polarizations at the same time, thus accessing a polarization-resolved transmission matrix \cite{tripathi_vector_2012}. Finally, using phase-retrieval algorithms, it is even possible to infer the phase and amplitude of the field from intensity measurements, without the need of the reference \cite{Dremeau_reference-less_2015}. Once the transmission matrix has been measured, one can either use it to study the medium, e.g., by looking at the modes (see Fig.~\ref{fig:marcenkopastur} and the discussion in section \ref{subsection5.2}), or to control the light transmitted through the medium, as we will see in section \ref{subsubsection4.4}.

\subsubsection{Accessing the temporally  or spectrally-resolved transmission matrix}\label{subsubsection4.3.4}

In the monochromatic approach, one characterizes the behavior of the medium at a specific wavelength, at the expense of ignoring the richness of the spectral and temporal behavior of light in the medium. This additional information can, in turn, be extremely useful when trying to either control spectrally or temporally the transmitted light. It also provides additional insights into the modes of the medium. 

Two approaches have been introduced in order to explore this additional dimension. The first one is based on accessing a spectrally-resolved transmission matrix, which amounts to measuring  a monochromatic  transmission matrix at many frequencies. In this way the spectral  behavior of the medium can be fully determined, provided the measurement is done with a spectral resolution comparable to, or better than the spectral correlation  of the medium. This was achieved, e.g., by using a tunable continuous-wave laser, and measuring several monochromatic transmission matrices for a set of closely-spaced wavelengths \cite{andreoli_deterministic_2015, mounaix_spatiotemporal_2016}.

Another possibility, complementary to the first one, consists in measuring a time-resolved matrix, which can be conveniently achieved when using broadband light via low-coherence interferometry. Since the interference between the transmitted light and the reference beam only takes place when their path length difference lies within the (short) coherence length of the source, it means that the recorded interferogram  only contains information about a given fraction of the light, which had a time of flight defined by the path delay of the reference beam. By varying the length of the reference arm, it is therefore possible to achieve a time-resolved measurement. This technique was implemented in reflection \cite{choi_measurement_2013, Kang_imaging_2015} as well as in transmission \cite{mounaix_deterministic_2016}.

\subsection{Light manipulation through the opaque lens}\label{subsubsection4.4}

Digital tools have provided a way to change the configuration of the  light incident on an opaque lens in a controlled way. We will see that, even prior to the measurement of the transmission matrix of a complex medium, wavefront shaping tools have allowed some light control through the opaque lens. We take here a didactic rather than a historic  approach to introduce the different techniques and concepts that have been applied to this problem.

\subsubsection{Time reversal, analog and digital phase conjugation through the opaque lens}\label{subsubsection4.4.1}

The concepts of phase conjugation and time reversal tell us that, thanks to the reversibility and reciprocity of the wave equation, an initial input wave can be recovered, when the wave resulting from the scattering of the incident wave by the slab is phase-conjugated and sent back through the medium. In practice, such a procedure requires perfect phase conjugation, and thus a collection of all the scattered light on both sides of the slab. In real experiments with the slab geometry, however, we generally  have  access only to  one side of the medium and to a limited fraction of the incident light as well as of the scattered light. As was shown both in optics and in acoustics,  in the case of multiply scattering materials (and also in the case of chaotic cavities) even limited phase conjugation or incomplete time reversal can partially reconstruct the initial wave \cite{draeger_one-channel_1997, derode_robust_1995, calvo_exact_2010}. In essence,  the different modes that are phase-conjugated contribute in a constructive way to re-inject energy into the initial mode. If the initial mode originates from a particular position, the wave will refocus to this position, with an efficiency that depends on the fraction of the energy that is phase-conjugated. 

Optical Phase Conjugation (OPC) was first performed by recording a hologram on a photographic plate \cite{leith_holographic_1966}.  However, based on the emergence of non-linear optics in the 1960s and 1970s, it was suggested by Yariv \cite{yariv_transmission_1976} that this holographic  optical phase conjugation could be performed in real time, using various non-linear processes. An implementation of this concept  was first realized   via four-wave mixing \cite{bloom_conjugate_1977, yariv_amplified_1977, yariv_four_1978}, and  via stimulated Brillouin scattering \cite{kralikova_image_1997}, in liquid crystals \cite{karaguleff_optical_1990}, or using three wave mixing \cite{voronin_compensation_1979, Ivakhnik_compensation_1980}. For a review on OPC, see \cite{fisher_optical_2012}.

For complex media investigations, OPC suffers, however, from several shortcomings that are mainly due to the limitations  of the physical effects giving rise to the phase conjugate of an optical wave. Non-linear wave mixing \cite{gower_optical_1994} is usually complex to implement, as it requires  non-linear crystals, specific wavelengths and often intense laser sources. Nonetheless, phase conjugation has been employed since its early days to refocus through a complex medium \cite{yariv_compensation_1979}.   Photorefractive crystals are another alternative for OPC, which, albeit being slow, was successfully used to refocus through thick biological tissues \cite{yaqoob_optical_2008}. Recently, new photorefractive  materials  have provided very high conjugation speeds comparable to fast SLMs \cite{farahi_time_2012, liu_optical_2015}. Gain Media (such as laser crystals)  also typically provide very fast OPC but work only for a narrow spectral range. They allow  for amplification of the phase-conjugated wave \cite{Feinberg_phase-conjugating_1980} and have been used for imaging in turbid media \cite{jayet_optical_2013}. Three wave mixing is fast and broadband, but is only effective over a very small angular range, yet was used for imaging through turbid media \cite{devaux_image_1998}. Despite all its constraints, OPC has the advantage that it can conjugate a very large number of modes simultaneously over the  surface of the OPC material \cite{xu_time-reversed_2011}, typically one or two orders of magnitude larger than what is currently achievable by digital means. OPC therefore remains a very competitive technique, especially for biomedical applications. 

\begin{figure}
  \centering
  \includegraphics[width=0.7\linewidth]{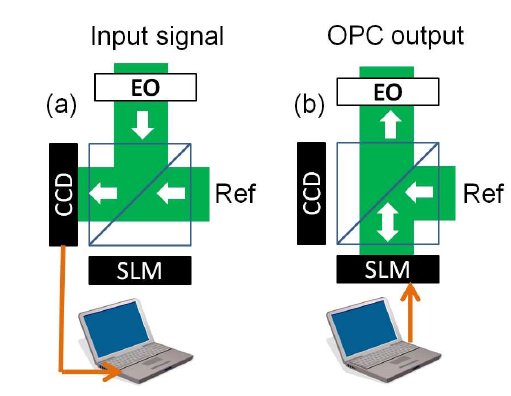}
  \includegraphics[width=1\linewidth]{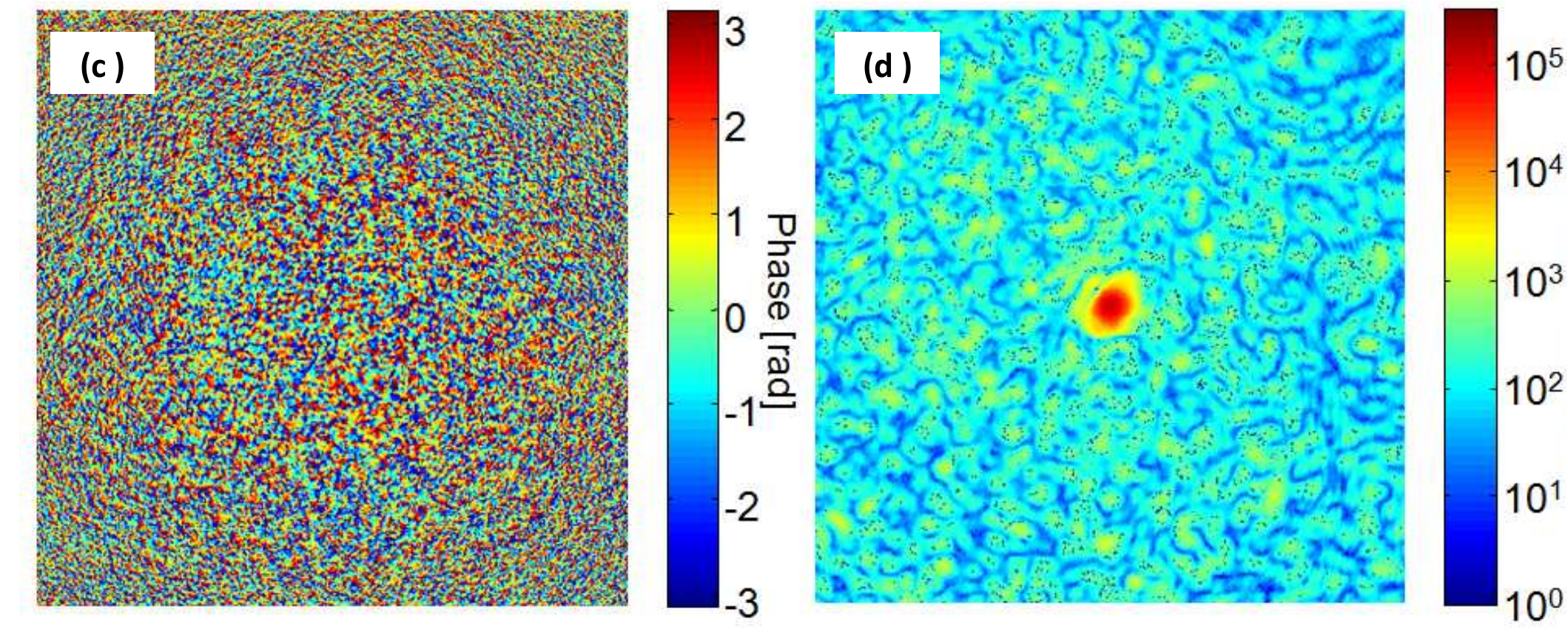}
  \caption{(color online). Principle of digital optical phase conjugation (DOPC):  (a) in a recording step, the interference pattern between a reference wave and the signal to be phase conjugated is recorded on a CCD, phase-shifting their relative phase difference with an electro-optic modulator (EO). In a digital playback step (b), the recorded pattern is displayed on the SLM, and the same reference beam, diffracting on the SLM, generates the phase-conjugate beam: (c) the DOPC measured phase profile and (d) the measured signal at the OPC output before the opaque lens, showing a strong focus in the center, on the original signal mode. (Figure adapted from  \cite{cui_implementation_2010}.)}
  \label{fig:DOPC}
\end{figure}

Thanks to the emergence of digital SLMs, it is now possible to envision a digital counterpart of optical phase conjugation (DOPC). Provided one can measure the complex amplitude of a field,  an SLM can in principe generate its phase conjugate. First experimental demonstrations of this concept were performed to co-phase several beams through a fiber bundle \cite{paurisse_phase_2009}, then applied to a thin scattering slab \cite{cui_implementation_2010} and  later to multimode fibers \cite{lhermite_coherent_2010, papadopoulos_focusing_2012}.  In all cases, an input  beam is incident on the  medium and the transmitted light is recorded on a camera, that has to be matched pixel to pixel to a spatial light modulator situated in a conjugated plane by means of a beamsplitter (see Fig.~\ref{fig:DOPC}). One then needs to recover the field on the camera, display the corresponding pattern on the SLM so that a laser beam reflected off this SLM carries the phase-conjugated wavefront. In its simplest implementation,  off-axis digital holography provides the necessary tool for both operations: A reference plane wave interferes at an angle with the unknown input wave, producing an interferogram on the camera that contains the phase and amplitude information of the  unknown input field. It can be shown that the same tilted reference plane wave diffracting off the same interferogram now displayed on the SLM will generate the phase conjugate of the unknown wavefield, thus producing a refocusing on the source.

Although it cannot match conventional OPC in terms of number of modes,   DOPC has several advantages compared to its analog counterpart, in particular to conjugate a speckle field. Specifically, since the complex wavefronts are actually recorded in a memory, it is possible to record the output for different waves and then replay them at a later time in any order, which is not possible in analog phase conjugation, where the hologram engraved in a crystal is transient by nature. Even more interestingly, it is possible to modify the interferogram before displaying it on the SLM, or combine several hologram together, which also brings an additional flexibility compared to its analog counterpart, a feature that will become particularly important as we will see in later examples related to imaging.  Still, rather  cumbersome alignments are required \cite{cui_implementation_2010}, although simplified implementations were proposed \cite{hillman_digital_2013}. A promising direction to simplify DOPC  is to design a unique monolithic device that would play the role of the detector and of the modulator simultaneously \cite{laforest_towards_2012}.

\subsubsection{Focusing and iterative optimization}\label{subsubsection4.4.2}

\begin{figure}
  \centering
  \includegraphics[width=1\linewidth]{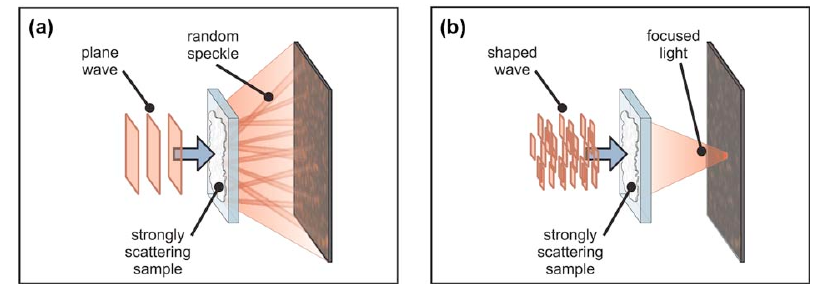}
    \includegraphics[width=1\linewidth]{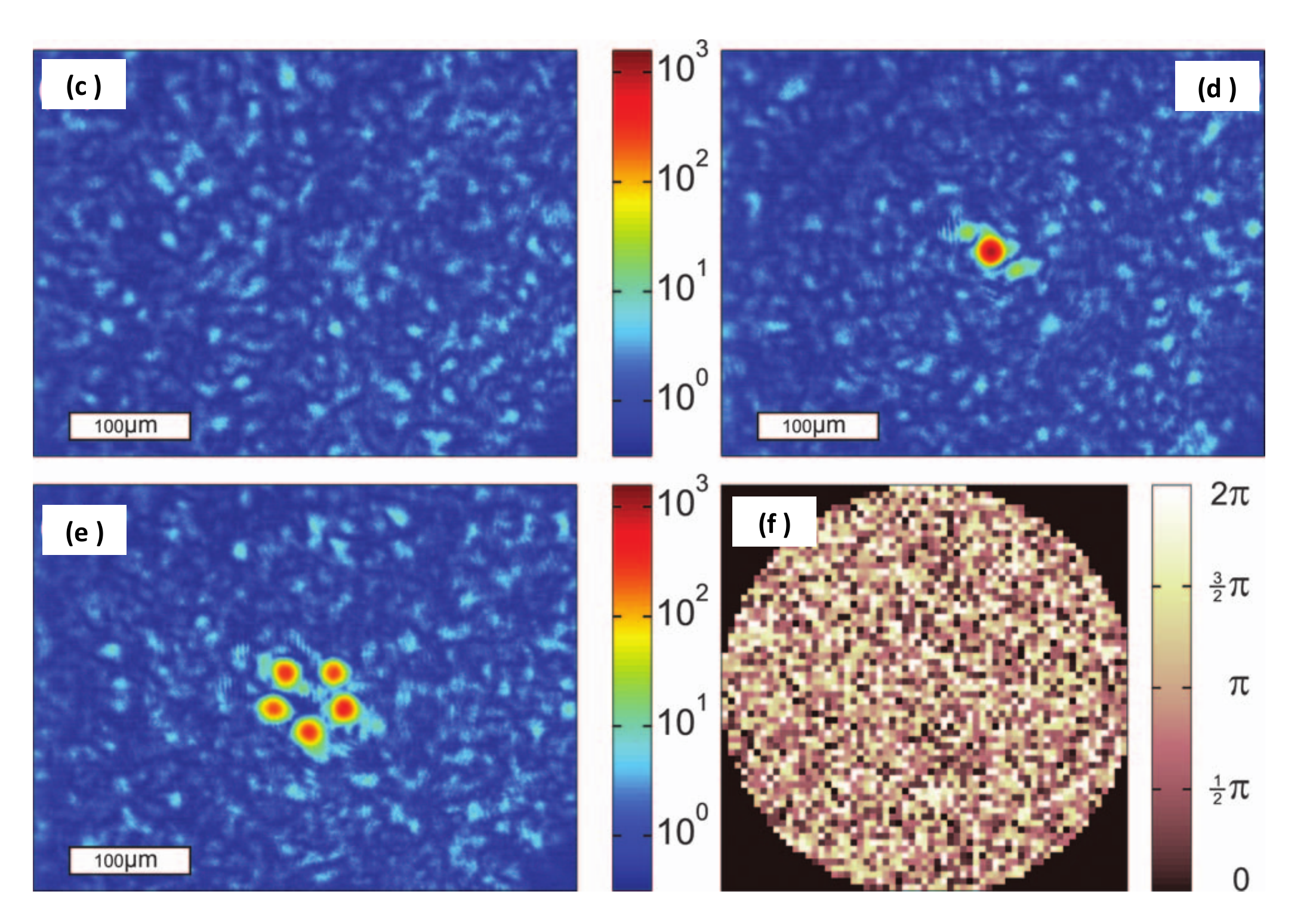}
  \caption{(color online). Principle of wavefront optimization through a complex medium: A plane  wave (a) incident on an opaque multiply scattering layer of white paint gives rises to a speckle field  on the far side (c). After optimization of the wavefront, an optimally shaped wave (b)  gives rise a speckle field that has a  strong focus at a chosen position (d). (e) A wave front can also be focused on several positions at the same time. (f)  Typical phase mask on the SLM after optimization, showing the apparent randomness and high  complexity of the obtained solution. (Figure adapted from \cite{vellekoop_focusing_2007}.)}
  \label{fig:optimmosk}
\end{figure}

The seminal experiment by the group of Allard Mosk in Twente  published in 2007 \cite{vellekoop_focusing_2007} expanded this concept of OPC  to a new level by removing the need for a source. In essence, instead of recording a wavefront from a source and then re-emitting its phase conjugate, thus achieving refocusing, the authors proposed an iterative optimization technique to find the optimal wavefront at the input of a disordered slab, that would maximize the intensity at a given position at the other side of the slab.  In this experiment, each pixel controlled on the SLM is assumed to generate an independent speckle on the far side of the disordered slab, on a camera. The resulting speckle on the camera therefore corresponds to the coherent sum of all the speckle contributions from all input pixels, which by itself is also a fully developed  intensity speckle pattern, since each speckle grain at the output is the result of a sum of different contributions with uncorrelated phases. In the formalism of the transmission matrix, it means that the field on pixel $m$ is given by $E_m=\sum^{N}_{n=1} A_n t_{mn} e^{i \phi_n}$, where $A_n$ is the field incident on input pixel $n$,  $\phi_n$ is the phase delay (or advance) imposed by the SLM at the pixel, and $t_{mn}$ is the transmission matrix of the complex medium between the pixels of the SLM and those of the CCD.  By optimizing the phase $\phi_n$ at each input pixel (i.e.,  by modifying the  spatial wave front) to maximize the intensity on a given output position, it is possible to converge to a constructive interference at this target position. A simple qualitative way to understand this process is to remember that the intensity distribution of a speckle is the consequence of the fact that each speckle grain is a sum of phasors (complex amplitudes) with uncorrelated and evenly distributed phases. When optimizing the phase of $N$ input pixels in the above way, the situation where the $N$ contributions add with uncorrelated phases is changed to a situation where $N$  contributions all add in phase. This corresponds to an increase of the final amplitude of the order of $\sqrt{N}$, and accordingly to an increase of the final intensity that scales with $N$ \cite{vellekoop_focusing_2007}, where $N$ is the number of pixels controlled. In the first realization \cite{vellekoop_focusing_2007}, a focus more than 2000 more intense than  the average of the unoptimized speckle background was observed (see illustration in Fig.~\ref{fig:optimmosk}).

This important result deserves extensive comments. Firstly, the above methodology assumes full independence between the different speckles generated by each pixel, i.e., no correlations must be present, which is one of the main assumptions for the opaque lens. This makes the optimization process very simple, since there is a unique optimum (up to a global phase), that  any algorithm can find. Only when noise or decorrelation comes into play will different algorithms perform differently \cite{vellekoop_phase_2008}. We will detail in section \ref{subsection5.2} what happens when correlations are present and how the results are modified. Secondly, it is interesting to link this approach with  phase conjugation. Indeed, it can be shown that the final wavefront is very close to the phase-conjugate solution or, more precisely, to the phase conjugate of the field emitted by a source placed at the target position, with the important  advantage, however, that no source is required. An important difference with respect to OPC is that the SLM is  phase-only, so while the spatial phase corresponds to the phase-conjugate solution, the amplitude cannot be controlled and depends on the illumination: it is constant for a plane wave incident on the SLM. Still, the wave  produces a focus, corresponding to the earlier insight in acoustics that the phase is the most important parameter when the aim is to put a maximum energy at a given point:   In terms of signal-to-noise ratio, it was even shown that the focusing efficiency is nearly equivalent to the one expected for perfect phase conjugation \cite{derode_random_2001, derode_random_2001-1,derode_ultrasonic_1999}. Additionally, with the assumption of independence of the input modes and of uncorrelated elements of the transmission matrix,  the background speckle is not statistically modified when the wavefront is optimized, nor is the energy of the total transmission. We will study deviations from this behavior  in more details in section \ref{subsection5.2}.

Another important feature of the focusing effect is the spatial size of the focus. The output speckle field has a well-defined grain size  which corresponds to its $C^{(1)}$ spatial intensity correlation (see section \ref{subsubsection3.1.2}).  The optimization procedure can create locally a constructive interference, and the spatial extent of this focus is given by the correlation distance, i.e., of the size of a speckle grain. This has two important consequences that have led to the concept of  the ``opaque lens" \cite{vellekoop_exploiting_2010}: (i) The optimized focus is perfect, in the sense that it is diffraction-limited, without aberrations, and it sits on a speckle background that can be orders of magnitude lower in intensity.  (ii) The size of the focus is only given  by the $C^{(1)}$ correlation of the medium in this plane, which is independent of the entire optical system located in front of the slab and its possible imperfections. Hence one can overcome the diffraction limit imposed by the limited angular apertures and the imperfections of the optical system. It is interesting to note that the same concept was previously proposed in adaptive optics to maximize a focus intensity by dithering the phase of multiple elements \cite{bridges_coherent_1974}, albeit with only a few degrees of control. In contrast to this last work, the optimization through a multiply scattering medium requires a very large number of degrees of freedom to be effective, but takes advantage of the statistical properties of the speckle to have a well-defined focusing efficiency and focus size. We also refer the reader to a review on optimization methods \cite{vellekoop_feedback-based_2015}. 

All the techniques described  above can be used to focus light to a single speckle grain. Digital phase conjugation and optimization techniques readily provide the wavefront that focuses light to one or multiple targets, with the difference that optimization provides a phase-only approximation of the exact phase-conjugated field. In the case that the transmission matrix is recorded, the corresponding wavefront can also be directly computed, and displayed using an SLM. As described in \cite{popoff_measuring_2010}, the input field $\tilde{\bf{E}}^{\rm in}$ that approximates the desired target $\tilde{\bf{E}}^{\rm target}$ can be deduced from the matrix as follows,
\begin{equation}\label{eq:phaseconjugation}
\tilde{\bf{E}}^{\rm in}= \bf{\tilde{t}}^\dagger \,\tilde{\bf{E}}^{\rm target}\,,
\end{equation}
where $\tilde{\bf{E}}^{\rm target}$ is set to 1 at the desired focus (or focii)  position, and 0 elsewhere. To understand this, let us go back to the fundamental relation $\tilde{\bf{E}}^{\rm out}= \tilde{\bf{t}}\,\tilde{\bf{E}}^{\rm in}\,$, which would suggest that to get a desired output field, one needs to invert the transmission matrix. Inversion is, however, rather unstable; an inversion is also sub-optimal for focusing since it would try to match the output as closely as possible, including minimizing the field outside of the focus. Taking instead the Hermitian conjugate of $\tilde{\bf{t}}$ as in Eq.~(\ref{eq:phaseconjugation}) amounts to a time-reversal -- or phase conjugation -- of the transmitted field \cite{prada_eigenmodes_1994}. Since, however, the reflected field and the unmeasured modes are here not part of the time-reversal (for a unitary $\tilde{\bf{t}}$ inversion and Hermitian conjugation would be equivalent), this reconstruction is not perfect, but turns out to be stable to measurement noise. Depending on the modulation scheme, a phase-only approximation \cite{popoff_measuring_2010} or a more exact phase and amplitude input (see, e.g., \cite{kim_maximal_2012}) can be generated. All theses techniques basically provide the same phase-conjugated field as a solution and share a comparable efficiency (i.e., proportional to the number of controlled input pixels). For imaging purposes also more advanced operators can be useful, as discussed in section \ref{subsubsection4.4.4}.

Several general remarks can be made at this point: Phase-only and full modulation both provide comparable focusing efficiencies, up to a factor 2 in intensity.  Phase-only modulation does not diminish the overall speckle intensity, and is optimal for delivering the maximal amount of energy to a given point \cite{vellekoop_focusing_2007}. Amplitude-only modulation has also been shown to permit focusing \cite{akbulut_focusing_2011, Dremeau_reference-less_2015}, by essentially turning off  a fraction of the input pixels to leave only pixels that contribute constructively  at the focus, thus simultaneously reducing the background. Joint phase and amplitude modulation provides a compromise between signal-to-noise ratio and focusing efficiency, by diminishing the contribution of pixels that contribute little to the focus, but significantly to the background. A general comparative discussion of the focusing efficiency in the context of acoustics can be found in \cite{tanter_time_2000} and is fully applicable in optics. In transmission, an additional control of the polarization state does not  change the overall performance, except by doubling the number of modes effectively controlled.

Extensive studies in  acoustics and in the radiofrequency domain are dedicated to the minimum attainable size of the focus \cite{lerosey_focusing_2007}. As we recalled already before, far away from the scattering region the focus is limited by the speckle grain size (i.e., by the  $C^{(1)}$ correlation function), and as such it can surpass the diffraction limit of conventional optics \cite{vellekoop_exploiting_2010, choi_overcoming_2011} and can get very close of the  limit of $\lambda/(2 n)$ allowed by diffraction. For instance in \cite{van_putten_scattering_2011} it was shown that the focus could be made smaller than $100nm$ for a monochromatic laser at $561nm$ and a refractive index of $n=3.41$. The possibility to modify the size of the focus by optimizing on a sub-part of the momentum space forming the speckle (using a spatial mask) was proposed and realized in \cite{di_battista_momentum_2015}: By exploiting a fraction of the speckle with a smaller $C^{(1)}$ correlation, a focus notably smaller than the average speckle grain was demonstrated, albeit at the cost of a much lower efficiency. This was extended to arbitrary point spread functions with a transmission matrix approach in \cite{boniface_transmission-matrix-based_2017}.
However, to break the diffraction limit and generate a focus smaller than $\lambda/(2 n)$, it is necessary to  have access to evanescent waves \cite{lerosey_focusing_2007, carminati_theory_2007,pierrat_subwavelength_2013}, such as just above the surface or inside the medium itself. Another approach is to use an active sink, as proposed in  \cite{carminati_reciprocity_2000} and realized experimentally in acoustics \cite{de_rosny_overcoming_2002}. The possibility of using a passive sink in optics has also been put forward in \cite{noh_broadband_2013}.  Different ways of focusing {\it inside} the medium have been considered. One of them proceeds with a source for digital phase conjugation \cite{hsieh_digital_2010}, another one works with a probe for iterative optimization \cite{vellekoop_demixing_2008}. In both cases, an unambiguous proof that the focus is sub-Rayleigh is difficult to obtain. In contrast,  focusing at the surface of the scattering medium can be achieved using a scanning near field optical technique \cite{park_subwavelength_2013}. Note that when using resonant systems, such as structured metallic layers that exhibit plasmon resonances, a sub-wavelength focus has also been demonstrated \cite{gjonaj_active_2011}. 

 The ideal signal-to-background ratio $\eta$ in focusing is easy to calculate and  depends essentially on the number of modes controlled, e.g., for phase-only optimization it was calculated to be $\eta=\frac{\pi}{4}(N-1)+1$  \cite{vellekoop_focusing_2007}.  However, the effective enhancement depends on the experimental conditions as well as on the algorithm used and on the number of iterations steps for the optimization. It was also shown that most methods are affected by the signal-to-noise ratio of the detection and are ultimately limited by shot noise fluctuations \cite{yilmaz_optimal_2013}. Another limiting factor is the stability of the medium, which is affected by small changes over time that decrease the efficiency of the process. In practice, most reported enhancement factors range between $20\%$ and $80\%$  with respect to the ideal case. Genetic algorithms have been shown to be particularly efficient in low signal-to-noise situations \cite{conkey_genetic_2012}.

 Focusing to multiple points or to areas larger than a speckle grain is possible with all the techniques described above. The total energy deposited via phase conjugation only depends on the number of degrees of control. As a consequence, the energy distributed over one or many targets is the same, but the signal-to-noise ratio is reduced by the number of modes that one seeks to control \cite{tanter_time_2000}. This insight was evident already from the first optimization experiment \cite{vellekoop_focusing_2007}, where focusing to five spots was realized (see Fig.~\ref{fig:optimmosk}e), and in \cite{popoff_measuring_2010}, where focusing on three spots yielded  a three-fold reduction in the focus to background intensity ratio. Interestingly, if the transmission matrix of a medium is known,  it is possible not only to perform digital phase conjugation by using the conjugate transpose operator as in Eq.~(\ref{eq:phaseconjugation}), but also to go beyond the theoretical limits of phase conjugation, or to optimize a different metric by using a more advanced operator, such as inversion for instance \cite{popoff_image_2010} (see next section). Depending on the desired goal, the figure of merit can not only be the intensity, but also the contrast, or any measurable quantity to be optimized.  Note also that different algorithms, when optimizing multiple points, might perform differently in achieving equal intensity on every target.

	One interesting application of light focusing is optical trapping \cite{grier_revolution_2003, dholakia_shaping_2011}, where optical gradient forces are exploited to ``trap" a nano- or micro-particle at the focus of a tightly focused beam. In this domain, SLMs have been largely exploited to generate multiple foci for trapping and moving around multiple particles \cite{curtis_dynamic_2002}. While adaptive optics has long been proposed as a means to correct the focus quality \cite{wulff_aberration_2006}, it was only recently realized that such techniques could be used for optical trapping in complex media \cite{cizmar_situ_2010}.
	    
\subsubsection{Imaging}\label{subsubsection4.4.4}

Focusing or re-focusing a wave behind a scattering medium  is indeed an important milestone also for imaging. In particular, the ability to scan a focus is at the basis of many imaging techniques (e.g.,  multi-photon microscopy), since focusing at different points would allow in principle to form an image, e.g., by fluorescence measurement, as described in \cite{vellekoop_scattered_2010}. We will see in  section \ref{subsection5.1} that in some occurrences, the so-called ``memory-effect"  allows one to scan an optimized focus over a narrow angular range, without the need to run an optimization algorithm for every point or to measure a transmission matrix.

 Directly recovering a  spatial distribution of intensity or phase from an object  (i.e., direct imaging)  from its transmitted speckle pattern is more challenging. The phase conjugation operation is by nature limited in the signal-to-noise ratio when trying to form a complex shape (as determined by the ratio of input to output degrees of freedom). Optical phase-conjugation, such as in \cite{yaqoob_optical_2008}, benefits from a very large optical etendue: Phase-conjugation is effective for a very large number of speckle grains within the non-linear crystal, thus allowing to reform a complicated image. In a similar way and in perfect analogy to Eq.~(\ref{eq:phaseconjugation}), the transmission matrix allows to reconstruct an object field $\tilde{\bf{E}}^{\rm obj}$  from the output field $\tilde{\bf{E}}^{\rm out}$ through the phase conjugation operation,
\begin{equation}
\tilde{\bf{E}}^{\rm obj}= \tilde{\bf{t}}^\dagger \,\tilde{\bf{E}}^{\rm out}.
\end{equation}\label{eq:imagereconstruction}
 First reconstructions were limited to very simple objects (one or two pixels turned on), using a square transmission matrix \cite{popoff_measuring_2010}. Later, a more complex object (a resolution target) of $N=20.000$  pixels could be  retrieved in \cite{choi_overcoming_2011}, by exploiting a transmission matrix measured over a very large number $M$ of output pixels (the full camera).  An additional advantage of the transmission matrix is that it is not restricted to phase conjugation. Other operators than phase-conjugation were successfully implemented  in \cite{popoff_image_2010}, to demonstrate  image recovery using, e.g.,  the so-called Tikhonov regularization \cite{tikhonov_solution_1963}. This regularized inversion operation,  has a much better performance than phase-conjugation and is robust to experimental noise.

\subsubsection{Deterministic mixing}\label{subsubsection4.4.5}

Another feature that can be exploited is the strong and optimal mixing produced by the opaque lens, linked to the fact that its transmission singular values follow the Mar\v{c}enko-Pastur distribution. Specifically, the complex mixing of light by a multiply scattering material, that is too complex to be copied or mimicked, has been considered for cryptography and security \cite{pappu_physical_2002}. Important implications for the  information capacity of such a medium have also been discussed for communication \cite{skipetrov_information_2003}. In the context of wavefront shaping, this natural optimal mixedness can be exploited in numerous ways and has probably many more potential applications. Among the emerging ideas that directly exploit this deterministic and efficient mixing, one can cite the generation of quantum-secure classical keys \cite{goorden_quantum-secure_2014} relying on the one-to-one association between an optimized wavefront and a medium for few-photon states. 
Another interesting application is compressive imaging that can provide the reconstruction of a sparse object with only a few measurements, provided each local measurement carries global information about the object \cite{Donoho_compressed_2006, candes_near-optimal_2006}, which is the case for a detector behind a disordered medium, as demonstrated in  \cite{liutkus_imaging_2014}.

\subsubsection{Polarization control}\label{subsubsection4.4.6}

When considering the vectorial nature of light, a question that naturally arises is that of  polarization control. We have seen earlier that at each elastic scattering event, the polarization of the scattered wave is modified in a deterministic way. In a classical picture, during a scattering event from incident wavevector $\bf{k}$ to a scattered wavevector $\bf{k'}$, the input and output polarization vectors $\bf{n}$ and $\bf{n'}$ are related by $\bf{n'} \propto \bf{n} - (\bf{k'}. \bf{n }) \bf{k'}$  \cite{mackintosh_polarization_1989}. For this reason, forward and backward scattering events tend to maintain polarization, while large angle scattering tends to modify polarization more strongly. In the opaque lens case, all  transmitted light has been scattered a sufficient number of times to ensure a fully mixed polarization. As a consequence, the speckle resulting from the propagation through the opaque lens  also has a complex polarization state at any point, which is in general elliptic. In essence, it is the sum of two orthogonally polarized speckle, uncorrelated to each other. This feature has been proposed as a way to exploit the opaque lens for polarimetric measurements \cite{kohlgraf-owens_finding_2008, kohlgraf-owens_transmission_2010}. 

It is important to point out that the polarization behavior of the opaque lens, which is universal in transmission, is clearly  system-dependent in reflection. Firstly, a single-scattering contribution is always present. Just like for a reflection off a mirror at normal incidence, it has the same polarization as the input in backreflection direction for a linear input polarization, and a reversed helicity for an elliptical input polarization. Secondly, the reflected light also has contributions from multiply scattered light with few scattering events, that retain partial polarization memory. This second contribution depends strongly on the scattering properties of the medium \cite{mackintosh_polarization_1989}. When working in reflection, it is possible to eliminate the single scattering contribution and a fraction of the light that endured few scattering events by selecting a specific polarization of the reflected light (e.g., the orthogonal polarization for linearly polarized input light), a trick used for measuring the coherent backscattering cone \cite{van_albada_observation_1990}.

\begin{figure}
  \centering
  \includegraphics[width=1\linewidth]{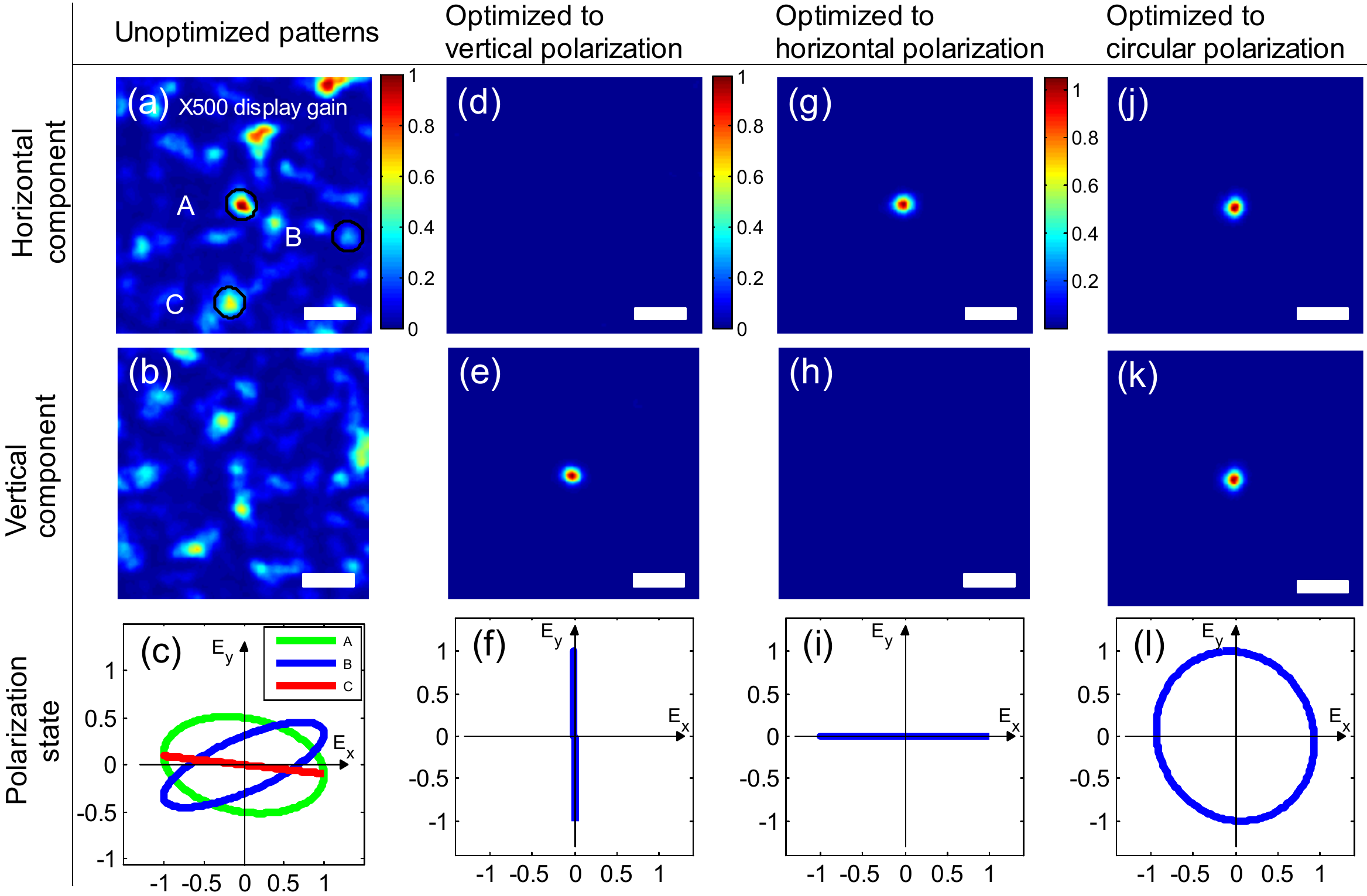}
  \caption{(color online).  Illustration of  the full control of the output polarization state.  Since the output polarized specke is composed of two uncorrelated speckle patterns of orthogonal polarizations (left column), it is possible to focus independently on one or the other polarization (second and third column). Combining the two wavefronts that focus at the same position for either polarization state, one can  generate at this point a focus with arbitrary polarization (here, circular, last column). (Figure adapted  from \cite{guan_polarization_2012}.)}
  \label{fig:polarizationcontrol}
\end{figure}

In the context of wavefront shaping, early experiments \cite{vellekoop_focusing_2007, popoff_measuring_2010} were  realized by placing a polarizer between the opaque lens and the detector, so as  to obtain on the camera an intensity pattern corresponding to a single scalar speckle rather than the sum of two uncorrelated speckles \cite{goodman_fundamental_1976}. Only later it was demonstrated experimentally that this additional degree of freedom can be turned to an advantage and allows to generate an arbitrary polarization state at  the focus \cite{guan_polarization_2012, tripathi_vector_2012} (see Fig.~\ref{fig:polarizationcontrol}). Since the total speckle is the sum of two orthogonally polarized speckles, one can generate any polarization state (linear, circular, elliptic) at will. The final quality of the polarization state follows the usual signal-to-noise restrictions common to all phase-conjugation techniques discussed above.

\subsubsection{Temporal and spectral control}\label{subsubsection4.4.7}

\begin{figure}
  \centering
  \includegraphics[width=1\linewidth]{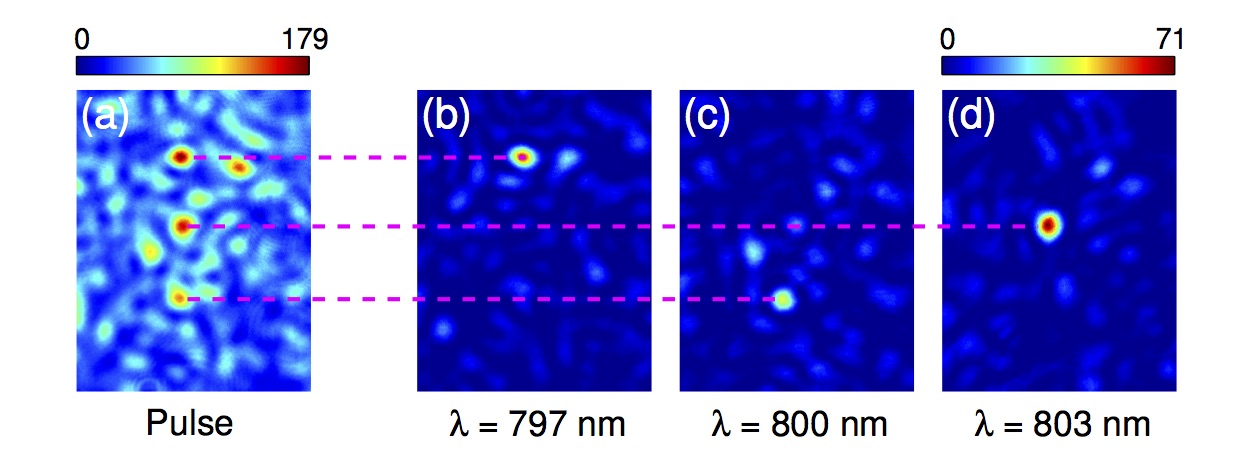}
  \caption{(color online). Spatiospectral control of broadband light, adapted from \cite{andreoli_deterministic_2015}. Using the information gathered from the multispectral transmission matrix of an opaque lens, it is possible to (a) spectrally focus different spectral components of a broadband pulse at arbitrary positions. (b-d) Scanning in continuous mode the same laser demonstrates that each focus corresponds to a different wavelength. In this way the opaque lens is turned into a generalized grating with a spectral resolution given by its spectral correlation bandwidth.}
  \label{fig:generalizedgrating}
\end{figure}

While all focusing and imaging experiments through the opaque lens discussed so far have considered a specific wavelength only, i.e., a monochromatic light source in conjunction with the spatial degrees of freedom of the medium at this wavelength, we have seen that the behavior of the opaque lens is strongly wavelength-dependent -- a feature has been exploited for a long time to retrieve diffusion properties of the medium \cite{vellekoop_determination_2005,curry_direct_2011}.  Correspondingly, in the context of focusing, it has been  shown that if a phase pattern  generates a focus for a given wavelength, then the focus will be resilient to a small wavelength variation. The corresponding bandwidth is exactly the frequency-bandwidth of the medium \cite{van_beijnum_frequency_2011}, a property that has been exploited to use the medium as a spectral filter \cite{small_spectral_2012}. Performing optimization with polychromatic light is possible, and  has been shown to result in a narrowing of the spectrum \cite{paudel_focusing_2013}. When measuring a multispectral transmission matrix \cite{andreoli_deterministic_2015}, it is also possible to focus several spectral components at a single or at multiple positions (see Fig.~\ref{fig:generalizedgrating}).

In the acoustic time-reversal community, it had been realized from the early days on that spatial and temporal degrees of freedom of a complex medium could be coupled \cite{fink_time_1997}. In particular, it was understood that time reversing the signal received at a given location would allow spatiotemporal focusing on the source at a specific time. Also, thanks to the reciprocity of the propagation, re-emitting at the source the time-reversed signal from the detector would yield spatiotemporal focusing on the detector. Nonetheless, when performing such a single-channel time reversal experiment in an open system, the observed spatiotemporal focusing is truly a temporal focusing only: When integrating  the energy at the source position over time a significant energy increase is not observed. To truly enhance the total energy at the source position either requires multiple detectors to be time-reversed simultaneously, or a closed system such as a chaotic cavity \cite{draeger_one-channel_1997}. 

\begin{figure}
  \centering
  \includegraphics[width=1\linewidth]{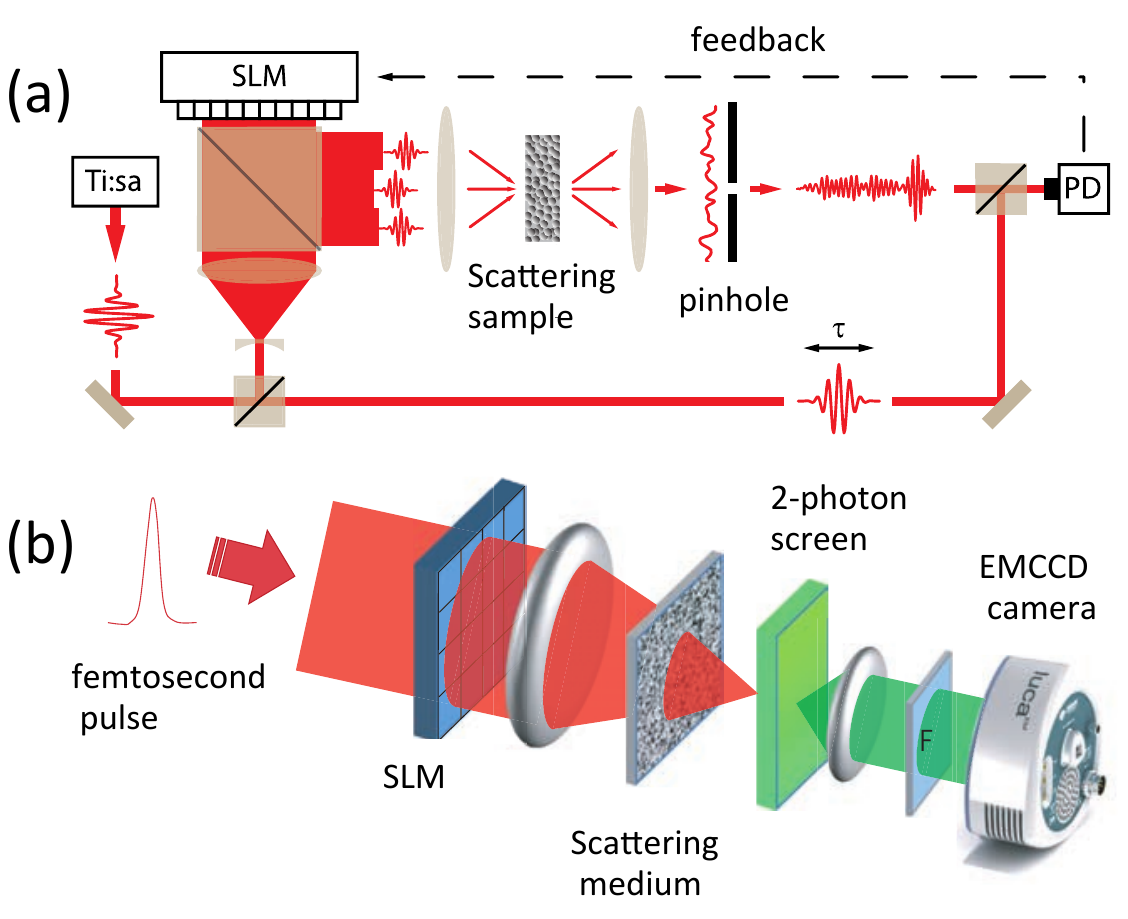}
  \caption{(color online). Schemes for temporal focusing via spatial-only shaping. (a) Spatial shaping and optimization on the intensity at a given time-delay \cite{aulbach_control_2011}. (b) Spatial shaping and optimization of a nonlinear signal \cite{katz_focusing_2011}. 
 }
  \label{fig:spatiotemporalfoc}
\end{figure}

An analogue of such a single-channel time reversal in optics was performed in \cite{mccabe_spatio-temporal_2011}. A short pulse from a femtosecond laser was sent through a layer of paint and the complex spatiotemporal speckle figure was recorded on the far side using an imaging spectrometer (see Figs. \ref{fig:temporalbehavior} and \ref{fig:spatiotemporalfoc}). Note that due to the difficulty of measuring and controlling an optical signal directly in the time domain,  both the temporal measurement and the temporal emission were performed in the spectral domain: The temporal speckle was measured using spatially and spectrally resolved Fourier-transform interferometry (SSI) \cite{tanabe_spatiotemporal_2002}, and the pulse was time-shaped using a spectral shaper \cite{monmayrant_newcomers_2010}. In this experiment, the time-reversed signal (measured via SSI)  was sent from the source (the pulse-shaper at the output of the femtosecond laser) to the detector. At the output position, it was shown that the pulse was compressed temporally close to the initial pulse duration, thanks to the reciprocity of the wave equation. Still, integrated over the pulse duration, the total intensity at the target spot was not increased. Just like spatial focusing can be seen as an extension of adaptive optics to multiply scattering material, this work can be seen as an extension of temporal pre-compensation  of dispersion of an ultrashort pulse \cite{delagnes_compensation_2007} to the opaque lens. 

While temporal control provides temporal focusing, and spatial control spatial focusing, spatial and temporal degrees of freedom are coupled in a complex medium. We will now describe this spatiotemporal coupling and related experiments exploiting this effect in the opaque lens, i.e., a diffusive slab in the multiply scattering regime. 
As shown in \cite{mccabe_spatio-temporal_2011}, a spatiotemporal speckle along a line, as measured behind an opaque lens by SSI, has a scalar field distribution $E(x,\omega)$, characterized by a short range spatial correlation function $\langle E(x,\omega) E(x',\omega)\rangle$  that has a well-defined width that is given by the speckle grain size, and by a spectral correlation function $\langle E(x,\omega) E(x,\omega')\rangle$, whose width is directly  related to  the traversal time  of the medium $\tau_T$ (defined in section \ref{subsubsection4.3.2}). These spectral correlation functions can  be retrieved from the wavelength correlation within the transmission matrix $\tilde{\bf{t}}(\omega)$. In the time domain, a given input time $t$ correspondingly couples to all  times $t' > t$ within a few $\tau_T$, and that coupling strongly depends on the input and output position. This in turn, means that there can be spatiotemporal couplings within this range. 

This behavior can  also be understood within the mode formalism: The existence of a well-defined traversal time and of this spatiotemporal coupling can be linked to the fact that in the diffusive regime, the mean spacing between the modes is much smaller that their average linewidth, as characterized by the Thouless number $\delta$ defined in section \ref{subsubsection2.1.5a}. If one sends an optical pulse of duration shorter than the average transmission time of the medium $\tau_T$, its spectrum is therefore broadband compared to the average distance between the modes. As a consequence, it  couples to many different transmission modes, at different frequencies, that recombine in a complex way after the medium, thus producing a complex spatiotemporal pattern \cite{wang_transport_2011}.  This behavior is  modified in the localization regime, where modes are spectrally isolated and where a short pulse might only couple to one of a few transmission channels only \cite{pena_single-channel_2014}.

A consequence of the spatiotemporal coupling is that spatial shaping can generate a temporal focus at a given position, by setting a constructive temporal interference at a given time between different frequency components, as pointed out in \cite{lemoult_manipulating_2009}. This was exploited in two seminal works in optics, where spatial-only phase control over a broadband pulse was shown to be able to induce temporal focusing as well as time-integrated spatial focusing. Both approaches were based on a 2D-SLM optimization algorithm, but using different signals as a feedback. In \cite{aulbach_control_2011}, the optimization was performed on the intensity at a given position and time using a heterodyne pulsed detection, while \cite{katz_focusing_2011} used a two-photon absorption signal at a given position. As this signal  depends on the square of the intensity, it is proportional not only to the total energy integrated in time, but also to the average pulse duration. Both approaches are summarized in Fig.~\ref{fig:spatiotemporalfoc}. As an alternative to optimization techniques, also a matricial approach can be taken. Measuring the multispectral transmission matrix (or MSTM)  \cite{andreoli_deterministic_2015}, it is possible to demonstrate arbitrary pulse shaping, provided the spectral phase can be addressed. This was demonstrated in \cite{mounaix_spatiotemporal_2016}, where not only pulse recompression was shown, but also more advanced temporal functions were realized such as two pulses with a controllable delay. 

Another possible approach for temporal control that has been investigated is a time-resolved matrix measurement \cite{choi_measurement_2013, Kang_imaging_2015}. These approaches have been  performed in reflection geometry, mainly to achieve depth-sectioning and light delivery at a certain depth, in analogy with optical coherence tomography, using ballistic light. However they also allow for temporal focusing at the detection plane and are an alternative to spectral or non-linear measurements.  

\subsection{Other complex scattering systems} \label{subsection4.5}

\subsubsection{Multimode optical fibers}\label{subsubsection4.5.1}

A very interesting system has recently emerged as a complex medium in optics: multimode optical fibers. Optical fibers are composed of a core material, where light is guided, and of a cladding, the function of which is to confine light within the core, so that the fiber behaves like a wave guide. The confinement can be achieved in various ways: total internal reflection on an index-step, index gradient, or photonic bandgap confinement. In all cases, the guiding occurs within a certain angular cone, that defines an effective numerical aperture. 
Depending on the diameter of the core and on the wavelength, one or many transverse modes can be supported. If only one tranverse mode propagates within the core, then the fiber is called a single-mode fiber (SMF). When many cores are  implemented in a single fiber, one speaks of a multicore fiber (MCF). If the diameter of the core is increased, then the fiber can support a few modes and is called a few mode fiber (FMF), while for  many modes one speaks of a multimode fibers (MMF).  The available number of modes scales with the diameter $D$ of the core and with the numerical aperture NA as $(D\times {\rm NA}/\lambda)^2$, resulting in up to  thousands of modes. FMFs and MMFs  are increasingly considered in the context of high bitrate fiber communications for space-division multiplexing (SDM), which means using several transverse modes as independent transmission channels  \cite{berdague_mode_1982} (see \cite{richardson_space-division_2013} for a review). Spatial degrees of freedom thus come as extra degrees of freedom for information transmission, complementary to wavelength and polarization, to carry more than one channel of information on a single optical fiber. While multicore fibers (i.e., fiber bundles) have long been used as independent spatial modes (despite some cross coupling), using a single core FMF or MMF remains a challenge because of the complex nature of the  modes.

\begin{figure}
  \centering
  \includegraphics[width=1\linewidth]{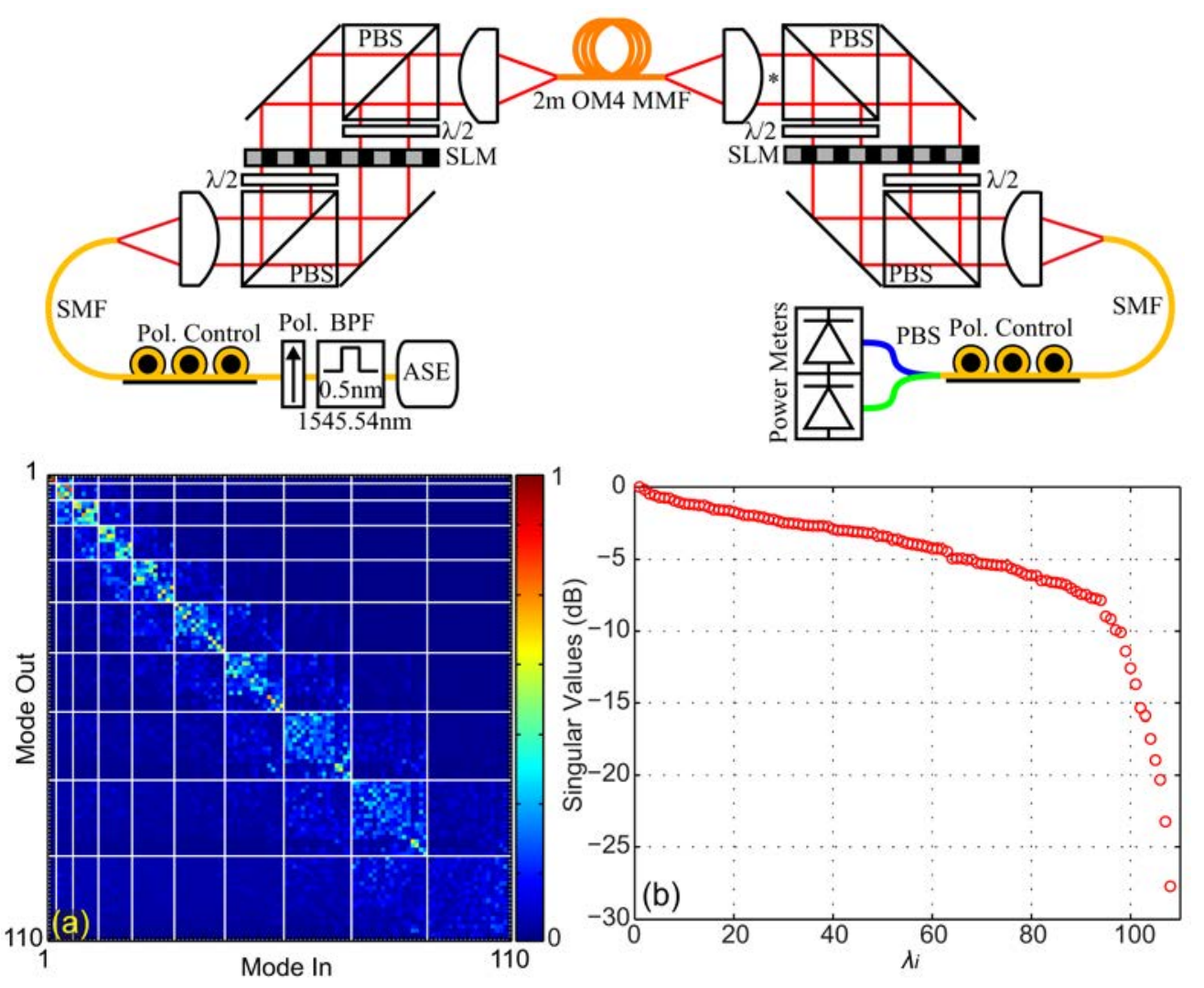}
  \caption{(color online). Measurement of a $110\times 110$ transmission matrix of a multimode fiber in the LP modes basis. (top) Experimental setup: two SLMs on each side to control two polarizations  are recombined on a polarizing beam splitter (PBS), allowing the near-perfect detection and injection of a well-defined mode at the input and output of the fiber to measure the corresponding coefficient on the matrix. (bottom left) Amplitudes of the transmission matrix in the LP modes basis.  (bottom right) Singular value decomposition, showing the mode-dependent losses. In the considered weak mode coupling limit, the transmission matrix is relatively block diagonal as subfamilies of LP modes are preferentially coupled. (Figure adapted from \cite{carpenter_110x110_2014}.)}
  \label{fig:carpenter}
\end{figure}

 An ideal MMF should support well defined modes, where the linearly polarized (LP) mode family is generally used \cite{snyder_optical_1983} to describe eigenmodes of the fiber. 
 However, due to fabrication imperfections and bendings, the ideal linearly polarized (LP) 
 modes are not in general the eigenmodes of the system: The eigenmodes of a large-core MMF tend to be different from the ideal 
 case. Nonetheless, it is possible to generate the so-called ``principal modes" of an MMF \cite{fan_principal_2005, shemirani_principal_2009}, which are unaffected by modal dispersion to first order of frequency variation. Their design principle bears very strong similarities with the eigenstates of the Wigner-Smith time-delay matrix described in section \ref{subsection2.4} (a detailed comparison between principal modes and so-called ``particle-like states'' will be provided in section \ref{subsection5b.1}). Only in the case of an ideal straight fiber  where all modes have different group velocities do the principal modes coincide with the LP fiber modes. In practice, the exact phase delay is very sensitive to  experimental conditions, and each mode has its own group delay, which means that a short (broadband)  pulse will be stretched. Interestingly, the distribution of delay times follows a semi-circle law as was shown based on a suitable random matrix model \cite{ho_statistics_2011}.  Even for bent or long fibers, principal modes tend to remain well isolated from each other. Measurements are challenging \cite{milione_determining_2015}, but have meanwhile been reported (see \cite{carpenter_observation_2015,xiong_spatiotemporal_2016,carpenter_110x110_2014} and Fig.~\ref{fig:carpenter}). Without these techniques, a monochromatic input is typically injected into more than one principal mode and will give rise at the output to a complex superposition of these modes, i.e., a speckle pattern. 

Many of the concepts developed to take advantage of opaque lenses for imaging and spatial and temporal control can therefore be translated to MMFs. 
A first set of experiments in this context is related to MCFs  and FMFs, that were studied in terms of their potential as high power fiber lasers and fiber amplifiers. For this class of problems, the difficulty is to exploit several transverse modes to achieve higher intensity. Unfortunately, due to the dispersion between the modes, the output laser mode is typically very multimode spatially, which is detrimental when high spatial quality is required. In the context of multicore fiber arrays, maintaining or retrieving a common phase between the different output modes is necessary in order to maintain a high transverse spatial quality (i.e., a beam quality factor close to unity). Corresponding cophasing methods, similar to phase-conjugation, can be either passive \cite{lhermite_passive_2007} or active using piezo-electric fiber stretchers \cite{yu_coherent_2006} or spatial light modulators \cite{lhermite_coherent_2010,bellanger_coherent_2008}. 

\begin{figure}
  \centering
  \includegraphics[width=1\linewidth]{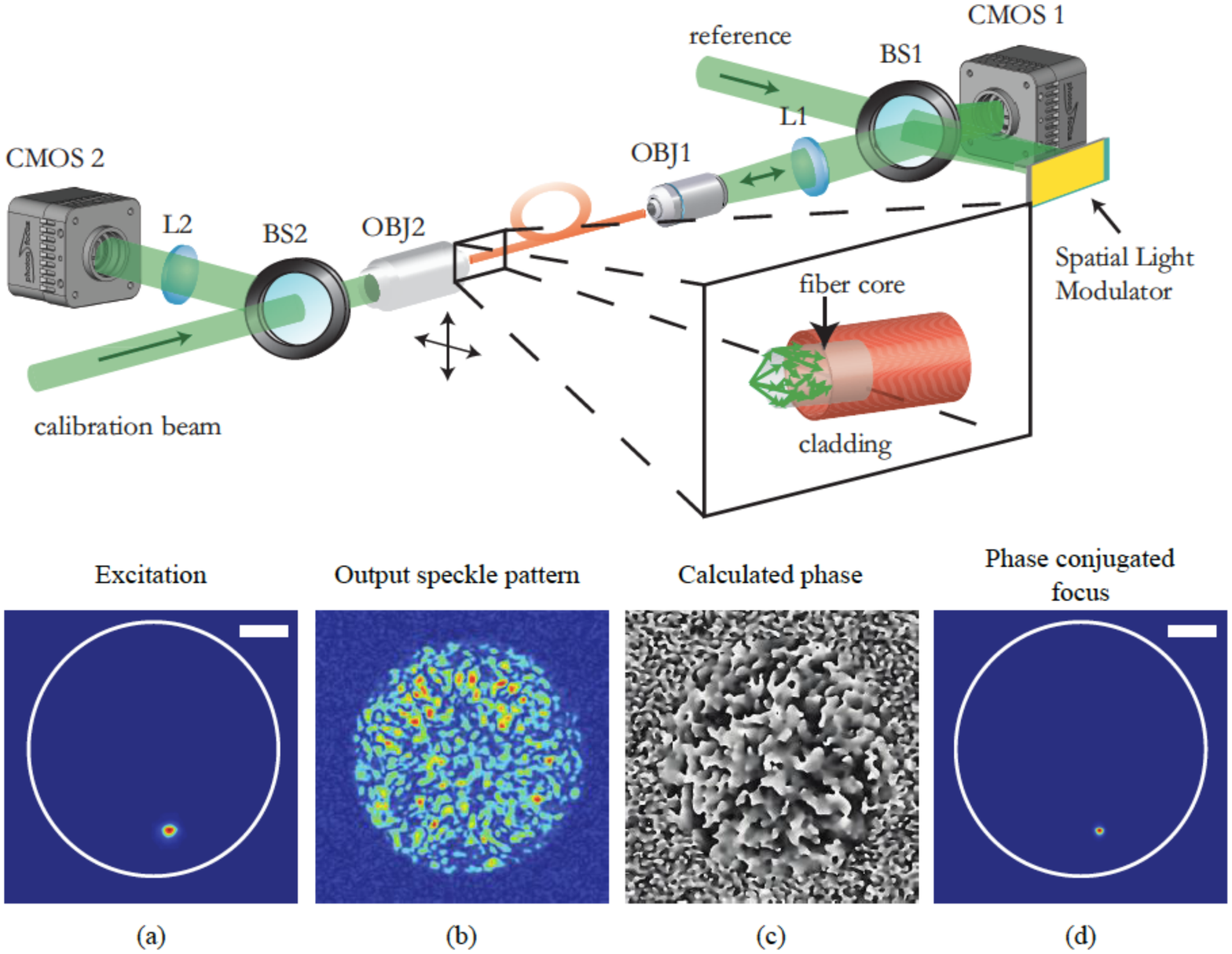}
  \caption{ (color online). Digital Optical Phase conjugation through a multimode fiber. (top) Experimental setup with DOPC (right-hand side of the fiber) and injection plus imaging (left hand side of the fiber). (bottom) A DOPC experiment: (a) Injection of a focused spot. (b) Output speckle. (c) Hologram used for digital phase conjugation. (d) Refocused spot. (Figure adapted from \cite{papadopoulos_focusing_2012} and \cite{farahi_dynamic_2013}.)  }
  \label{fig:farahi}
\end{figure}

Following the progress made in scattering media, this kind of monochromatic phase-conjugation has been extended to imaging. In particular, it was realized  that, just like in a scattering medium, wave front shaping could allow the formation of a sharp focus on the far side of a fiber, be it by optimization \cite{di_leonardo_hologram_2011, cizmar_shaping_2011} or via digital phase-conjugation \cite{morales-delgado_delivery_2015,cizmar_exploiting_2012,  papadopoulos_focusing_2012, caravaca-aguirre_real-time_2013}, and even by way of measuring the transmission matrix of the fiber \cite{bianchi_multi-mode_2012, choi_scanner-free_2012}. One example of such focusing techniques is shown in Fig.~\ref{fig:farahi}.

Thanks to the ability to focus to single or multiple points, or reconstruct an image, a wide variety of imaging modalities were proposed and realized,  in particular fluorescence microscopy \cite{cizmar_exploiting_2012, papadopoulos_high-resolution_2013} and photoacoustic microscopy \cite{papadopoulos_optical-resolution_2013}. In all these endoscopic applications images are retrieved in depth, with a diffraction limited resolution given by the numerical aperture of the fiber.  Beyond imaging, the possibility to create one or several foci at the tip of an MMF  was also shown to allow optical trapping of dielectric particles \cite{bianchi_multi-mode_2012}. 

Of course, the exact mode-mixing in the fiber strongly depends on its specific configuration.  The stability of the fiber  can be extremely good when left untouched, but movements, temperature drifts etc.~may degrade the stability of the focusing considerably. Overall,  this sensitivity  scales with the numerical aperture of the fiber, its length and its diameter. For example, the resilience of the transmission matrix to bending of the fiber was investigated in \cite{choi_scanner-free_2012}, where it was shown that it remained exploitable when moving the tip of the fiber by one centimeter, for a 1 m long,  200 $\mu$m diameter fiber of numerical aperture 0.48. Different methods were proposed to compensate for fiber movements, e.g.,  fast optimization to a point \cite{caravaca-aguirre_real-time_2013}. Another approach relies on  storing a set of phase-conjugate patterns for different fiber positions, and determining the fiber position at any time using a so-called ``virtual holographic beacon"\cite{farahi_dynamic_2013} and correlating the emission from this beacon with the set of measurements to recover the fiber position and use it for imaging. A more recent approach relies on predicting the TM by evaluating the effect of propagation and bending on the phase retardation of each principal mode \cite{ploschner_seeing_2015}.

In practice, many of the results of the opaque lens on imaging and focusing apply to MMFs as well, but  they also have several unique features that arise from their particular propagation properties. In the weak coupling limit  the correlations in the  spatial pattern on the far side of the fiber strongly depend on the injection mode, because the major contributions in the transmission matrix are centered around the diagonal (see Fig.~\ref{fig:carpenter}). The distribution of the input mode  in $k$-space (i.e., the angular range) matters: In particular, injecting a plane wave at low incidence will populate preferentially the low order fiber modes, while a strongly  focused wave will result in a decomposition over higher order modes, that will partially survive propagation (at least for short  distance)  and result at the output in variable speckle grain sizes. Symmetries are also important: A focused beam will produce qualitatively  very different speckle patterns depending on the input position.

Moreover,  since the number of modes is well-defined and the numerical aperture is limited, the TM can be completely measured, as in \cite{xiong_spatiotemporal_2016,carpenter_observation_2015,choi_scanner-free_2012}. In addition, since most of the light is transmitted in the forward direction, most singular values of the transmission value are of modulus close to unity (although, in practice, absorption and imperfect injection degrades the flatness of the distribution, see Fig.~\ref{fig:carpenter} (bottom right)). This means that, in contrast to scattering systems, imaging is much more robust to noise and image reconstruction can be straightforwardly achieved using phase-conjugation \cite{choi_scanner-free_2012, cizmar_exploiting_2012}.  Polarization mixing is present during propagation in  MMFs \cite{shemirani_principal_2009} and can be compensated via phase-conjugated techniques \cite{mcmichael_correction_1987}. In the endoscopic works with digital wavefront control, a control of both polarizations has been achieved  \cite{cizmar_shaping_2011}. Finally, in contrast to the opaque lens, the number of modes in an MMF is limited, therefore it is possible to control near-perfectly the ouput  pattern with an SLM  (with $95\%$ fidelity reported in  \cite{loterie_digital_2015}). A consequence for focusing is that the fraction of light intensity that can be brought to the focus can be close to unity, thereby strongly diminishing the background speckle. This in turn means that the speckle grains are not completely independent as in the opaque lens, but are correlated, due to energy conservation  and due to the fact that one can achieve almost complete modal control. (We will see how speckle correlations affect the opaque lens in the next section \ref{section5}.)

\begin{figure}
  \centering
  \includegraphics[width=0.8\linewidth]{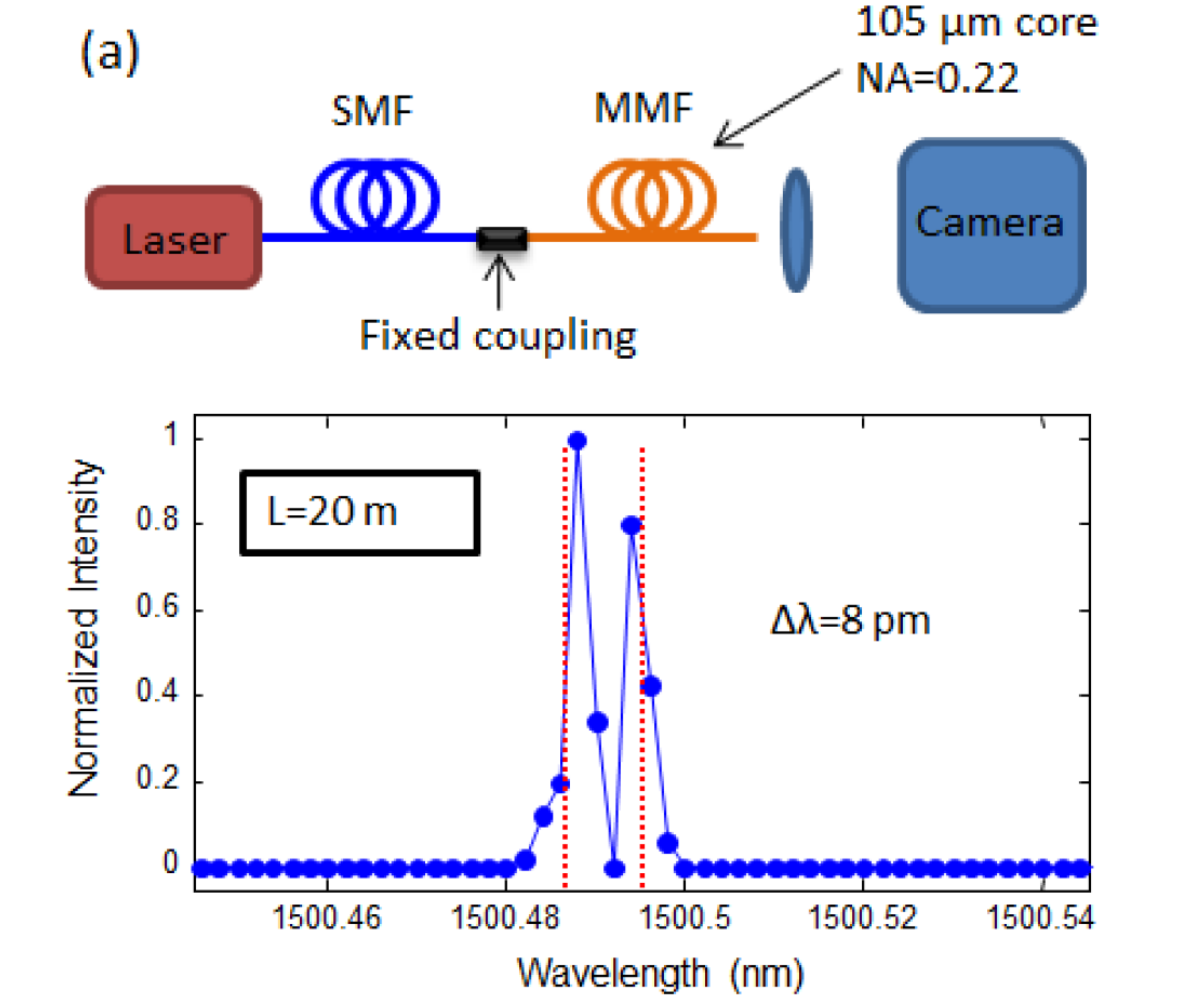}
  \caption{(color online). Multimode fiber based spectrometer.  (top) Experimental setup. (bottom) Example of laser line determination with 8pm accuracy. Two laser lines separated by 8pm  can be resolved through the reconstruction from the speckle pattern. (Figure adapted from \cite{redding_all-fiber_2013}.)} 
  \label{fig:redding}
\end{figure}

As discussed above, the temporal or spectral behavior of MMFs is highly complex and of immediate relevance for telecommunications. Like opaque lenses, spatiotemporal coupling is present and can in principle be exploited. For instance, temporal focusing of an ultrashort pulse by DOPC has  been demonstrated \cite{morales-delgado_delivery_2015}. An interesting application that has also been proposed is to use an MMF as a high resolution spectrometer \cite{redding_using_2012, redding_all-fiber_2013}. In essence, a fixed spatial input (a SMF) serves to inject a well-defined spatial mode, that generates on the far side a  complex speckle that depends on the wavelength, with a sensitivity proportional to the fiber length. The system must first be calibrated with a tunable monochromatic source. In a second step, a complex  spectrum is injected and it produces on the distal side a superposition of many different speckle patterns that add incoherently, i.e., in intensity rather than in amplitude. The spectrum responsible for this pattern can finally be retrieved by inversion. As shown in Fig.~\ref{fig:redding}, the resolution can be extreme for long fibers (the authors demonstrate 8pm resolution for a 20 m fiber of 105 $\mu$m core, with 0.22 NA). 

\subsubsection{Biological tissues}\label{subsubsection4.5.2}

A complex system of high interest for imaging and wave front shaping is obviously biological tissue. While optical imaging in biological tissues is a vast field, in particular with an endless variety of coherent or incoherent imaging techniques able to retrieve ballistic information from the multiple scattering background and depth resolution (see \cite{ntziachristos_going_2010, wang_photoacoustic_2012, wang_biomedical_2012,yu_recent_2015} for   reviews).
A commonly accepted order of magnitude for the scattering mean free path $\ell$ of tissues is of the order of 100 $\mu$m in the visible range, but due to the high forward anisotropy of the scattering, the transport mean free path $\ell^\star$ is usually of the order of a millimeter. Obviously, there is a strong variability from tissue to tissue, and a vast amount of literature exists on   measurements of scattering properties of tissues (see, e.g., \cite{cheong_review_1990}).

While most of the concepts developed in the framework of a multiple scattering slab can be adapted to biological tissues, such as focusing or imaging  with wavefront shaping, there are some specific questions to be addressed when dealing with imaging in biological tissues. The first one is the problem of decorrelation, that is inherent to soft media: The distribution of the refractive index changes relatively rapidly with time, similar to what is encountered in atmospheric adaptive optics. Measuring the transmission matrix  or running an optimization algorithm must thus be performed on a timescale comparable with the decorrelation of the medium, typically of the order of a millisecond for in-vivo tissues. 
Another important aspect is that  tissues are typically very thick (a few centimeters) and the distal side of it might not be accessible. At a more fundamental level, one usually wants to image or focus inside rather than through a scattering medium. For this reason, the concept of a thin slab is not directly relevant: although it is reasonable to consider that the medium up to the depth $D$ at which one wants to image is an opaque slab of thickness $D$ to be traversed, this is, in fact, only partially true. Consider here, e.g., that light can also propagate deeper than $D$ before diffusing back to the plane at depth $D$ that is of interest. 

Workarounds for this problem of accessing the region of interest have meanwhile been developed. One of them is to try to get access to the local light intensity deep inside the tissue using a complementary technique such as acoustics. Particularly  promising progress has been made in acousto-optics \cite{xu_time-reversed_2011, wang_deep-tissue_2012, si_fluorescence_2012}, and in photo-acoustics \cite{kong_photoacoustic-guided_2011, Chaigne_controlling_2014, Chaigne_light_2014} (and for a review \cite{bossy_photoacoustics_2016}. In all these techniques, the  resolution is governed by the acoustic wavelength, which is typically much larger than the optical wavelength. Nonetheless, it is possible to overcome the acoustic resolution and get closer to optical speckle scale resolution, either by careful spatial coding \cite{Judkewitz_speckle-scale_2013} or by  exploiting non-linearities \cite{conkey_super-resolution_2015,  lai_photoacoustically_2015}. Another option is to rely on reflection measurements only. For instance, optimizing the wavefront to maximize the total non-linear reflected signal can force light to focus at depth \cite{katz_noninvasive_2014, tang_superpenetration_2012}. It is also possible to measure the reflection matrix \cite{choi_measurement_2013} and exploit its statistical properties for imaging.  Another possible idea is to use differential measurements to focus on a moving target \cite{zhou_focusing_2014}.  Most of these techniques are detailed in a recent review \cite{horstmeyer_guidestar-assisted_2015}. 

\section{Mesoscopic physics and wave front shaping}\label{section5}

Wavefront shaping techniques have led to remarkable progress for imaging in or through complex media. We will see in this chapter how these techniques can be used to unravel and exploit mesoscopic effects. We will  explain in section \ref{subsection5.1} how the memory effect has emerged as a powerful tool for imaging, and describe in section \ref{subsection5.2}  recent optical experiments and theoretical works, where first evidence on the existence of open and closed channels have been discussed. In the next part \ref{subsection5b.1} we will describe the properties of time-delay eigenstates in different contexts and in \ref{subsection5b.3} new avenues for wave front shaping in media with gain and loss will be discussed.

\subsection{Memory effect}\label{subsection5.1}

As we have seen in section \ref{section3}, the information on the incident wavefront is not lost when traversing a multiply scattering medium. A special role in this context plays the memory effect, where spatial variations of the incident wavefront are partially mapped onto easily predictable changes in the transmitted speckle. In transmission, we have seen in section \ref{subsubsection3.1.4} that the thickness of the medium $L$ determines the typical transverse spatial features that are conserved, which implies that  this effect is independent of the strength of the scattering or of the exact scattering properties of the medium.

While the mechanism behing the memory effect is very general, we have seen that a particularly important case is the one of a linear phase ramp on the input wavefront (see Fig.~\ref{fig:memory_fig}). This angular rotation is transferred to the far side of the medium, provided that the transverse wavevector $q$ of the phase-ramp is changed only slightly, $\Delta q <1/L$, corresponding to a varation of
the angle of rotation below the so-called memory effect angle, $\theta < \lambda/(2 \pi L)$. More precisely, the transmitted speckle decorrelates over a characteristic angle determined by the  memory effect. The shape of the angular correlation function was predicted in \cite{feng_correlations_1988} to be a sinh, but the first experimental realization \cite{freund_memory_1988}, showed that this function is closer to an exponential decrease. The phase shift of the speckle pattern corresponds to a rotation by the same angle as the incident one, When the field propagates  away from the sample. Far away from the sample (as on a distant screen) the speckle field will thus be translated.  As  shown already in \cite{freund_memory_1988}, the memory effect is also present in reflection, but the corresponding angle is then not given by the thickness but by the transport mean free path $\ell^\star$, which measures the extent of the diffuse spot in reflection (as discussed in section \ref{subsubsection3.1.3}).

A very fundamental insight is that, based on the memory effect, not only linear phase ramps, but actually any arbitrary modification of the input wave front can be transferred through the medium (with a cut-off spatial frequency determined by the medium thickness $L$). A quadratic phase can, e.g., generate a longitudinal shift of the resulting speckle far away from the sample, etc. In a visionary paper as early as 1990 \cite{freund_looking_1990}, Isaac Freund realized that this means that an opaque layer could be used  for several functions, such as a lens, a grating, a mirror, as well as for imaging.  Unfortunately, the possibilities of shaping the speckle to a focus were not yet available at that time, and the proposal relied on image correlations between the speckle of interest and  a reference speckle. 

\subsubsection{Imaging using the memory effect}\label{subsection5.1.2}

\begin{figure}
  \centering
  \includegraphics[width=0.9\linewidth]{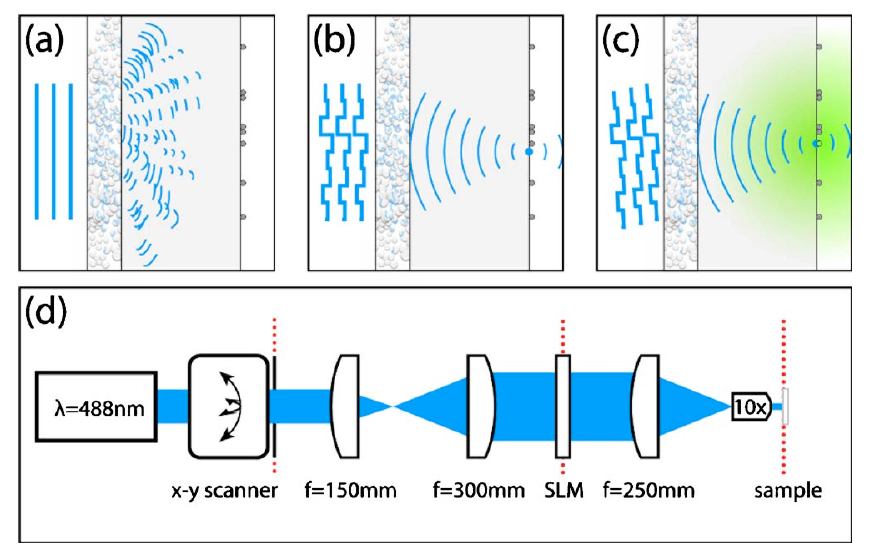}  
  \caption{(color online). Principle of imaging using wavefront shaping and the memory effect. (Figure and caption adapted from \cite{vellekoop_scattered_2010}.) 
(a) A thin scattering layer blocks a fluorescent structure from sight; all incident light is scattered. (b) By use of interferometric focusing (e.g., phase conjugation or wavefront shaping), scattered light is made to focus through the layer. (c) Imaging: the focus follows rotations of the incident beam over a short angular range. (d) Simplified schematic of the experimental setup for 2D imaging. A laser beam is raster scanned, and its wavefront is shaped with a spatial light modulator (SLM). Dotted lines are conjugate planes. }
  \label{fig:aegerterME}
\end{figure}

 While most early work considered a plane wave input, the memory effect is also effective for an arbitrary initial input wavefront. In particular, it also works when  the wavefront has been shaped to obtain a speckle that contains a bright focus. If the medium is thin and if the focus has been achieved at a distance, it means that the focus can be translated. This approach is particularly interesting for imaging since, using a single focus and raster scanning it around, it is possible to recover an image, for instance of a fluorescent object, as was first demonstrated in \cite{vellekoop_scattered_2010} (see Fig.~\ref{fig:aegerterME}). In such a setup, it does not matter how the focus has been initially obtained, as, e.g.,  by optimization of the intensity on a CCD camera \cite{vellekoop_scattered_2010,van_putten_scattering_2011}.   When using the transmission matrix method, one naturally has the ability to focus at any measurement position. If the memory effect is present, it can be retrieved from correlations between  lines of the transmission matrix corresponding to neighboring positions, as demonstrated in \cite{popoff_controlling_2011,Chaigne_controlling_2014}. Still, the ability to move a focus by adding an angular tilt is interesting, particularly because it means fast scanners can be used to rapidly raster scan a focus, possibly orders of magnitude faster than a pixellated SLM.

It is also possible to use the memory effect without having access to the focus region, as was demonstrated  by phase-conjugating the second harmonic signal generated by an implanted probe \cite{hsieh_imaging_2010}, using the photoacoustic effect to remotely monitor the light intensity \cite{Chaigne_controlling_2014}, or after optimizing a non-linear signal to a focus \cite{tang_superpenetration_2012, katz_noninvasive_2014}.  Also, by adding not only a linear phase shift but also a quadratic phase ramp, the technique can be extended to the third dimension both for scanning and imaging \cite{yang_three-dimensional_2012, ghielmetti_scattered_2012, ghielmetti_direct_2014}.

Once a wavefront has been shaped by an SLM to focus at a given position, it means that a  source placed at this position would be transformed by the same SLM into a plane wave, and therefore can be conjugated to a focal spot by a subsequent imaging system. Based on this concept, it was then realized in \cite{katz_looking_2012} that scanning the focus is not the only way to recover an image. As discussed by Isaac Freund \cite{freund_looking_1990}, the memory effect features an isoplanetic angle over which the speckle remains correlated. In other words, the correction of the wavefront compensates the scattering medium for a small angle and for a small range of frequency, irrespectively of the illumination. In \cite{katz_looking_2012}, 
a point source generates a speckle after a scattering medium, after which an SLM is placed and the wavefront is optimized to form a focus on a CCD camera, therefore performing the analog of a focusing experiment, except that the SLM is placed after the medium rather than before. The SLM is conjugated with the output plane of the scattering medium in order to maximize the memory angle range. If the point source is displaced, so is the focus image on the camera, provided the displacement is smaller than the one allowed by the memory effect. The point source is then replaced by an extended source (an object), and an image is directly obtained on the camera. Crucially, the correction even works if the object is illuminated by spatially and temporally incoherent light, but since the correction of the wavefront is only valid within a given angle around the focus and for a given spectral bandwidth  related to the spectral correlation of the medium around the calibration frequency, only this fraction of the light is well conjugated.

\begin{figure}
  \centering
  \includegraphics[width=1\linewidth]{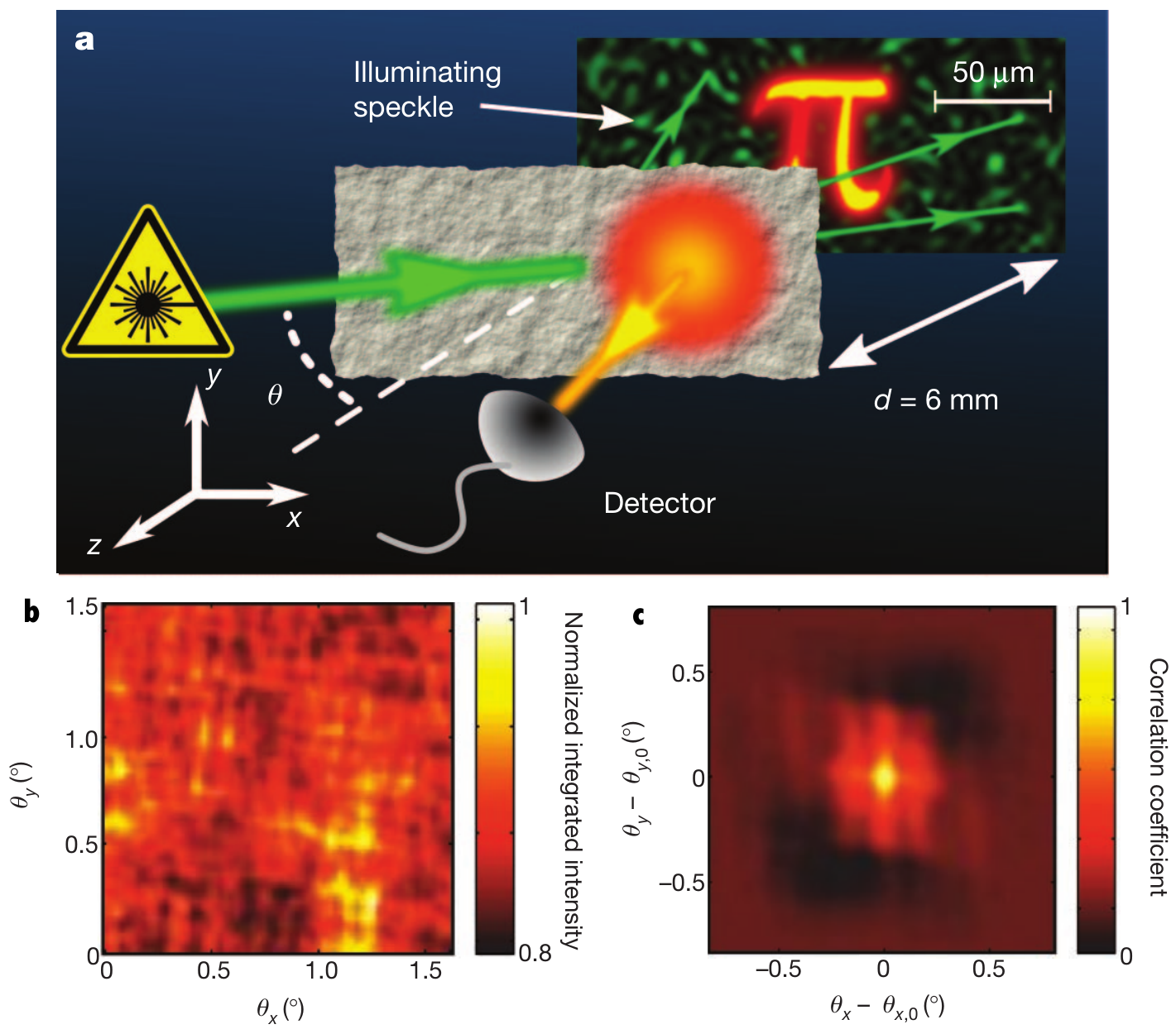}  
  \caption{(color online). (a) Schematic of the apparatus for non-invasive imaging through strongly scattering layers. A monochromatic laser beam illuminates an opaque layer of thickness $L$ at an angle $\theta$. A fluorescent object is hidden a distance of 6 mm behind the layer. The fluorescent light is detected from the front of the scattering layer by a camera. (b) Integrated fluorescent intensity on the camera, as a function of the incident angle,
$\theta=  (\theta_x, \theta_y)$. (c) Autocorrelation of the intensity, averaged over nine scans taken at different
values of the starting incidence angle, $\theta_0$, to average over the different realizations of the speckle. From the autocorrelation, the original image (here the letter $\pi$) can be retrieved by phase-retrieval. (Figure adapted from \cite{bertolotti_non-invasive_2012}.)
  } 
   \label{fig:noninvasive}
\end{figure}

Finally, several works reverted to the original idea of Freund \cite{freund_looking_1990} of using the memory effect without shaping or focusing to image behind a turbid layer. All these approaches exploit the fact that a speckle, despite being a very complex pattern, has a well-defined peaked autocorrelation function. In \cite{bertolotti_non-invasive_2012}, a fluorescent object placed at a distance behind a scattering layer is illuminated by a speckle that is translated (by scanning the illumination angle), and its fluorescence is collected as a function of the  shift of the speckle, thus forming an image (see Fig.~\ref{fig:noninvasive}). While the resulting image is very complicated and speckle-like, its autocorrelation is actually the product of the autocorrelation of the object and of the autocorrelation of the speckle, with the latter having a diffraction-limited peaked function. Therefore, one has access to the autocorrelation of the object with a resolution given by the speckle grain size. Using a reconstruction technique known as phase-retrieval \cite{fienup_phase_1982}, the image of the object can be retrieved from its autocorrelation function. In \cite{yang_imaging_2014}, the same technique was used to recover the image of blood cells behind a scattering layer of tissues. In \cite{katz_non-invasive_2014}, another case was studied, where a semi-transparent  object, illuminated by spatially incoherent light, could be retrieved from the autocorrelation of the speckle it produced behind a scattering layer. In all these approaches, no calibration is required, since the exact scattering properties of the scattering layer are not important. However, in most of these work, a scattering layer, rather than a scattering volume was used, to ensure light transmission and a very pronounced memory effect. 
 
 \subsubsection{Beyond the conventional memory effect}\label{subsection5.1.4}
 
While the memory effect is mostly considered in transmission, it is also present in reflection \cite{freund_memory_1988} and can be exploited.
 As mentioned in section \ref{subsubsection3.1.4} the limiting
memory effect angle in reflection 
depends on the transport mean free path $\ell^*$, which determines the size of the diffusive halo for  focused incident light. For a sufficiently strongly scattering material such as a paint layer, where the mean free path can be on the order of a  micrometer, the angular range of the memory effect can be of several degrees. The relatively large fields of view resulting from this estimate have been 
exploited for instance in \cite{katz_looking_2012} and in \cite{katz_noninvasive_2014} to image ``around corners''. The memory effect has also been studied in the context of time-reversal  \cite{freund_time_1991} and polarization \cite{freund_stokes-vector_1990}.

An important point in the quest towards exploiting the memory effect for biological imaging (see also section \ref{subsubsection4.5.2}) is to assess whether some memory effect can be present \textit{inside} a biological tissue. For some biological systems the scattering occurs mainly on a thin layer at the surface, and the rest of the sample is mostly transparent. This is, e.g., the case for the drosophilia puppa at its early development stages, where  the fly embryo is covered in a thin (8 micrometers) but very scattering layer of cells \cite{vellekoop_focusing_2010}. Inside a volumic scattering medium, which is the case of interest for deep biomedical imaging, one would expect from mesoscopic theory that the memory effect is not present, since in the derivation it is supposed to be only observable at a distance from the scattering layer. However, some works indicate that this view is conservative. Indeed, tissues typically have a scattering mean free path $\ell$ of 50-100 micrometers \cite{cheong_review_1990} and more importantly, they scatter mostly forward with g-parameters (the average of the cosine of the mean scattering angle) often larger than $0.9$. This means that, at small depth (millimeters), there are still forward scattered photons, and the diffuse halo is narrower than the one given by a fully diffusive model. 
Experimentally, in some instances, a thin scatterer (onion layer, or chicken breast slice, fixed brain slice) could be used for imaging within its memory effect range, which was characterized to be larger than predicted by a diffusive model \cite{katz_noninvasive_2014, schott_characterization_2015}, a strong indication that the memory effect should be present inside a medium. In \cite{tang_superpenetration_2012}, a focus was obtained at depth inside brain tissues (800 micrometers) and was scanned over a few micrometers. In forward scattering media such as biological tissues, a ``translational" memory effect was identified: A lateral shift of the input wavefront resulted in a lateral shift of the focus \cite{judkewitz_translation_2015}, an effect valid inside the medium rather than at a distance. 

 Analogues of the angular memory effect were also demonstrated in MMFs. Here the memory effect is not transverse, but longitudinal. This comes from the fact that  a  plane wave with a given $k$-vector that is injected,  is mixed angularly but not radially, provided the fiber is not too long or twisted, producing at the output a narrow cone of speckle with the same transverse angle of incidence. The width of this output cone corresponds to an azimutal correlation of the speckle. This means that any radial curvature  to the initial wavefront can be transferred to the output. In \cite{cizmar_exploiting_2012} light is brought to a focus by  wavefront-shaping at the distal end of an MMF. Using those azimutal correlations, the focus was shifted axially, but also elongated to produce a Bessel beam \cite{mcgloin_bessel_2005},  a doughnut-shaped focus for stimulated emission depletion (STED) \cite{willig_sted_2007} and more generally for engineering the point-spread-function. There is also a rotational memory effect, coming from the fact that a MMF conserves some rotational symmetry, which can be used to rotate a focus around its center of symmetry \cite{amitonova_rotational_2015, rosen_focusing_2015}.

\subsection{Bimodal distribution of eigenchannels}\label{subsection5.2}

We have seen in the previous section how the transmission matrix (TM) of a disordered slab, a multimode fiber, or of any linear optical system  can be measured. However, it is very important to stress the difference between the TM $\tilde{\bf t}$ as measured in experiments and the full TM $\bf{t}$ as described in sections \ref{section2} and \ref{section3}. The experimental TM $\tilde{\bf{ t}}$ takes as input modes the different modes that can be generated and detected by modulating the input beam, i.e., typically  pixels on an SLM and on a camera, respectively. In the opaque lens, the number of controlled modes is typically much smaller than the number of available input modes of the medium. In addition, the number of detected modes is much smaller than those being populated at the ouput of the slab. One of the most striking features of mesoscopic transport, the bimodal distribution of eigenchannels (see Fig.~\ref{fig:bimodaldist}), is however elusive when only measuring a sub-part of the full TM of the system. In all the different works on the subject, the ratio of modes that are controlled, detected or illuminated, is always the limiting parameter. There is no consensus on notation or even on the definition of this ratio, that is defined and labeled differently in every paper, and that depends on the specific situation. In this section, we chose to leave the different definitions (and notations) as they were used in the literature, and point out when they differ.

\subsubsection{Accessing the bimodal distribution}\label{subsection5.2.1}

\begin{figure}
  \centering
  \includegraphics[width=0.9\linewidth]{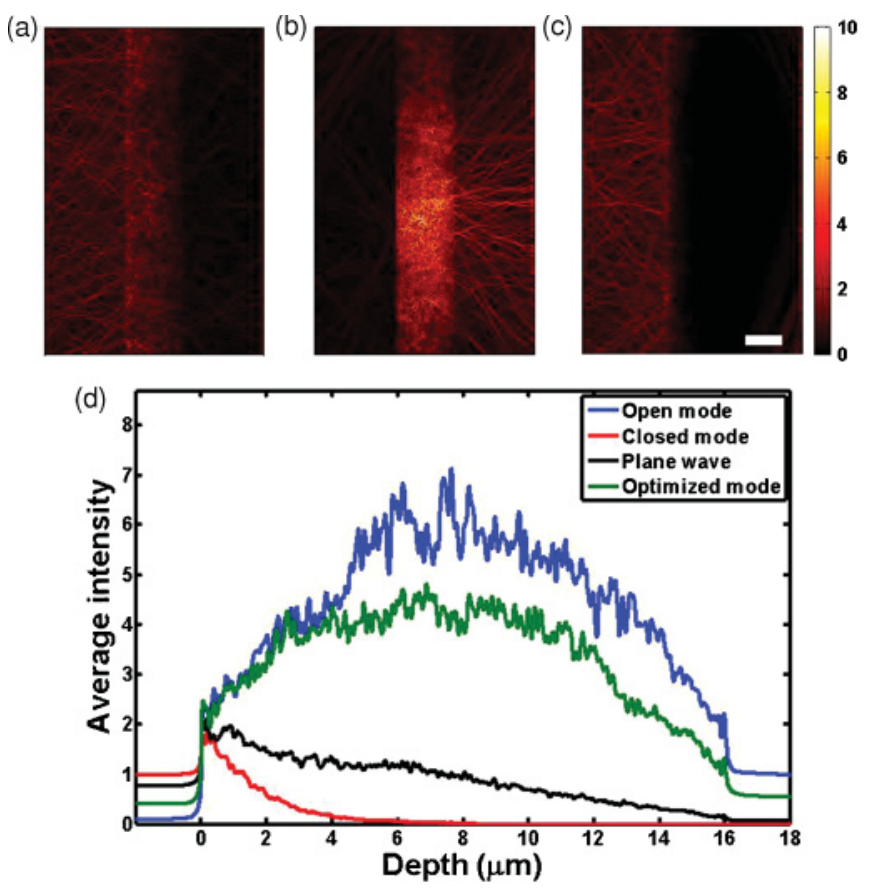}
  \caption{ (color online). Simulation of scalar field distributions of transmission eigenchannels inside a 2D disordered slab with 299 channels. The medium of height 130 $\mu$m and thickness 16 $\mu$m is shown in the middle of each of the three top panels (a)–(c). Field distributions (a) of a plane wave whose incident angle is $11.5^\circ$, (b) of an open transmission eigenchannel (transmission of 0.955) and (c) of a closed transmission eigenchannel (close to zero transmission), respectively. The incident field is subtracted on the left-hand side of the medium. Color bar: amplitude normalized to the input wave. Scale bar: 10 $\mu$m. (d) Averaged intensity along the $x$ direction as a function of the depth in the $z$ direction for the same three different input wavefronts, plus the wavefront corresponding to a focusing optimization, as in \cite{vellekoop_universal_2008}. The disordered medium fills the space between 0 and 16 $\mu$m in depth. The intensity is normalized to that of a normally incident plane wave. (Figure adapted  from \cite{choi_transmission_2011}.)}
  \label{fig:choibimodal}
\end{figure}

Since a complete channel control for accessing open and close channels is currently not available, it is very helpful to resort to simulations, as in \cite{choi_transmission_2011}, where the full monochromatic transmission matrix of a disordered slab is numerically evaluated.  The numerical data is in good agreement with RMT \cite{nazarov_limits_1994}, and  the resulting modes when injecting open and closed channels are evaluated. A very striking result is shown in Fig.~\ref{fig:choibimodal}, where the intensity distribution inside the medium and the scattered fields are computed for a plane wave input as well as when injecting the optimal wavefront for exciting an open or closed channel. One can see a very dramatic difference in the intensity distribution along the longitudinal direction when comparing these different cases. When injecting a plane wave, which excites all available transmission channels very broadly, the averaged intensity diminishes linearly with depth, as predicted by Ohm's law. When the wavefront corresponds to injecting a closed channel,  the decay is much faster and exponential, while when injecting an open channel the  intensity first increases with depth until the center of the slab, and only diminishes thereafter. As a result, the intensity is almost symmetric with a maximum in the center of the slab, as necessary in order to transmit a significant amount of energy through the medium. Analytical expressions for these distribution functions have been proposed in \cite{davy_universal_2015}.

Note that the above description considers a simplified system, which serves as a good starting point to understand the difficulty to measure this distribution and inject the corresponding modes in practice: First, it is assumed that all modes are accessible (from both sides of the slab, including also the polarization degrees of freedom), and that the system is two-dimensional only (as in section \ref{section2} for the waveguide geometry). First experiments which could satisfy these demanding requirements were reported in acoustics \cite{gerardin_full_2014} and in optics \cite{sarma_control_2016}. In both setups the dramatic change in the internal energy distribution (see Fig.~\ref{fig:choibimodal}) could, indeed, be observed. In most experiments, however,  these conditions are difficult to meet in practice.

\begin{figure}
  \centering
  \includegraphics[width=0.9\linewidth]{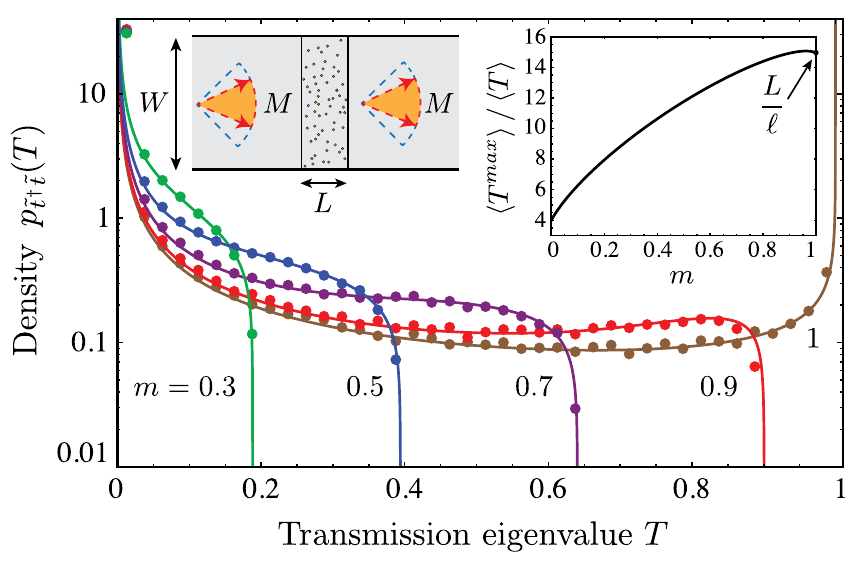}
  \caption{ (color online). Transmission eigenvalue density of a disordered slab placed in a waveguide with $N=485$ channels (length $L=150/k$, width $W=900/k$), for different fractions of controlled channels $m=M/N$. Numerical results (dots) are obtained from solving the wave equation for 120 realizations of the slab for fixed disorder strength. The solid lines are the theoretical prediction based on free probability theory. The inset shows the maximal transmission enhancement possible for a given $m$. (Figure adapted from \cite{goetschy_filtering_2013}.)}
  \label{fig:goetschybimodal}
\end{figure}

An analytical model and numerical simulations in the waveguide geometry \cite{goetschy_filtering_2013} were dedicated to what happens to the distribution of measured transmission eigenvalues in case of partial channel access (considering both control and detection), and how it affects the maximal transmission $T_{\rm max}$  that can be achieved by wavefront shaping when injecting the maximally open channel that can be measured. For this purpose the  control  parameter $m=M/N$ was introduced, where $M$ is the number of channels controlled and $N$ the total number of channels. The most striking result is that even a small degree of imperfect control ($m\lesssim 1$) abruptly suppresses the mode of unit transmission, and the measured distribution rapidly deviates from the bimodal one, with the disappearance of the peaked distribution around unity (see Fig.~\ref{fig:goetschybimodal}). An increase of the transmission relatively to the mean transmission (i.e., $T_{max}\gg \overline{T} $) can nonetheless be achieved with partial control, until the distribution converges to the Mar\v{c}enko-Pastur distribution (as in Eq.~(\ref{eq:Marcenkopastur}) and in Fig.~\ref{fig:marcenkopastur}). Interestingly, the crossover to uncorrelated Gaussian matrices occurs when the number of modes controlled is lower than the 
total transmission $T$.  Similar results are derived in reflection, where the perfect reflection expected when injecting closed modes is suppressed with imperfect control and detection. In particular, detection or control of a single polarization immediately results in loss of half of the modes. The result by Goetschy and Stone \cite{goetschy_filtering_2013} was extended in \cite{popoff_coherent_2014} to include not only the waveguide geometry, but also the slab geometry with partial illumination. In particular, the effective control parameter $m$ is extended for the case of an illumination zone $D$ smaller than the thickness $L$ of the slab. Due to the fact that the area of the diffusive halo on the far side is larger than the injection area,  the number of modes at the output is automatically larger than the number of input modes, which results in a diminution of the maximal  transmission achievable.

\subsubsection{Unraveling and exploiting open and closed channels}\label{subsection5.2.2}

Despite these difficulties in accessing the full bimodal nature of the transmission eigenchannels of a disordered slab, several experiments have managed to reveal some features of bimodality by means of wavefront shaping.

\begin{figure}
  \centering
  \includegraphics[width=1\linewidth]{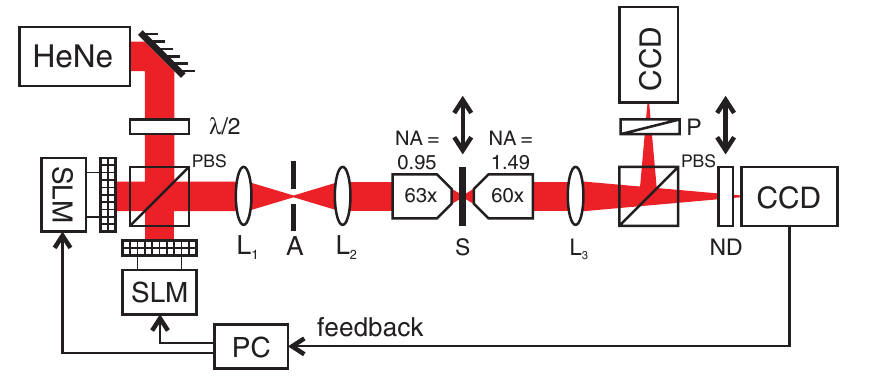}  
  \includegraphics[width=1\linewidth]{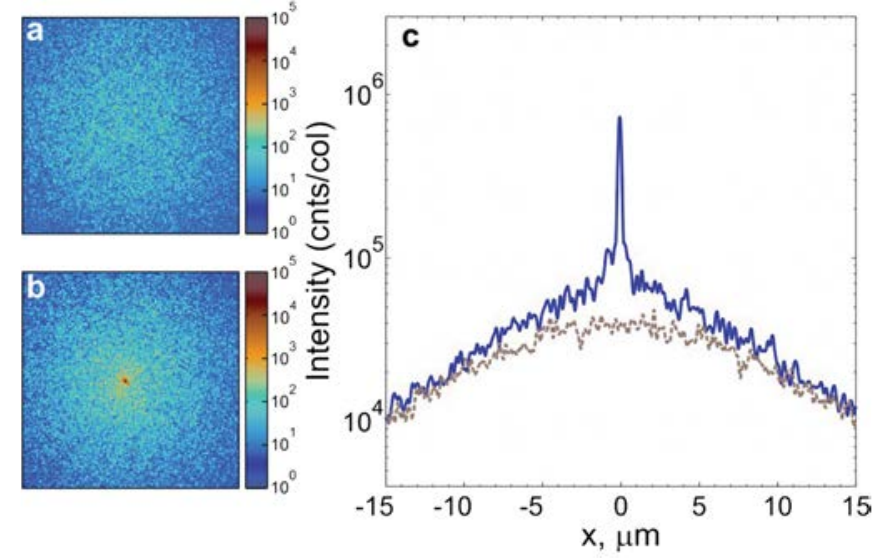}
  \caption{ (color online). Coupling to open channels by wavefront optimization to a focus spot. (top) Schematic of the experiment: two SLMs are used to control both polarization directions; high numerical aperture objectives ensure maximal coverage of incident angles; both polarizations are detected on two CCD cameras on the far side. (bottom) Intensity of the transmitted speckle figures at the output plane (a) before and (b) after optimization of the wavefront. (c) Intensity summed over the $y$ direction to average the speckle: dashed curve for unoptimized and solid curve for optimized wavefront. The total transmission is increased by $35\%$. (Figure adapted from \cite{vellekoop_universal_2008}.)
  }
  \label{fig:vellekoopuniversal}
\end{figure}

Historically, the first experimental result was reported by Vellekoop and Mosk as early as 2008  \cite{vellekoop_universal_2008}, where an optimization  through a slab was performed in the limit where a noticeable fraction of the modes is controlled (up to approximately $30\%$ at the input). Experimentally, this was achieved  in two ways; first by designing a relatively thin multiply-scattering sample  (down to $5.7$ $\mu$m) to minimize the number of modes to be controlled; secondly by controlling and detecting both polarizations using polarization separators, and using high numerical aperture objectives to access high angles of incidence. The result, when performing the same point optimization as in \cite{vellekoop_focusing_2007} was a spectacular deviation from the opaque lens predictions (see Fig.~\ref{fig:vellekoopuniversal}). While optimizing a single speckle spot, an increase of the  overall transmitted intensity was observed, not only in the focus, but also in the surrounding speckle, which meant that strong spatial correlations must be present in the speckle, due to the fact that only a few  modes contribute to the transmitted speckle.  The increased transmission was compared with RMT predictions, and could be well-interpreted as a redistribution of the input energy from closed to open channels. 

The authors further derive that perfect optimization to a single point should, on average, increase the total transmission to the universal value of $2/3$. This value of 2/3 is directly linked to the electronic quantum shot noise for the case of a bimodal distribution of transmission eigenvalues \cite{beenakker_applications_2011}, which we have found to be characterized by a sub-Poissonian shot noise Fano factor $F= {{\rm tr}[\bf{t} \bf{t}^\dagger (1-\bf{t} \bf{t}^\dagger)]}/{{\rm tr}[\bf{t} \bf{t}^\dagger]}=1/3$ in the diffusive limit (see section \ref{subsection2.2}). Indeed, when aiming to maximize the intensity to a given position, Eq.~(\ref{eq:phaseconjugation}) tells us that this can be achieved through phase conjugation ${\bf E}^{\rm in}={\bf t}^\dagger\,{\bf E}^{\rm target}$ (we neglect here for the moment that in the experiment only $\tilde{\bf t}$, i.e., the partial transmission matrix restricted to the measured and controlled modes is available). 
The output field, in turn, is given as follows, ${\bf E}^{\rm out}= {\bf t}{ \bf t}^\dagger {\bf  E}^{\rm target}$, and the total intensity is $I^{\rm out}={\vert {\bf E}^{\rm out}}\vert^2$. When averaging over realizations or over target positions, one can show that the average transmission is ${\overline{T}}={I_{out}}/{I_{in}}={{\rm tr}(\bf t \bf t^\dagger t t^\dagger) }/{{\rm tr}(\bf t t^\dagger)}$. Since this relation  is directly related to the shot noise Fano factor by $\overline{T}=1-F$, we simply get the result  $\overline{T}=2/3$.  This  result was numerically confirmed in \cite{choi_transmission_2011}, where it was further noted that this result is connected to the fact that the contribution of each eigenchannel $n$ to the optimized wavefront was, on average, proportional to its eigenvalue $\vert\tau_n\vert^2$, a well-known property of the time reversal operator $\bf t t^\dagger$ \cite{tanter_time_2000}. This theoretical result was later verified experimentally in \cite{kim_relation_2013}. Of course, imperfect control of the wavefront leads to a reduced transmission compared to the ideal case. The authors introduce in \cite{vellekoop_universal_2008} a parameter $\gamma$ that represents the overlap between the injected mode and the perfect optimized mode, the difference coming both from imperfect channel control, phase-only operation, and noise in the optimization process. The expected total transmission is $T_c=\vert \gamma^2\vert \overline{T}$, and excellent agreement in different experimental conditions is found. 

In contrast to the opaque lens case, optimization and imaging through a medium gives quite different results if the number of open channels becomes lower than the number of degrees of freedom that one has access to. As shown in \cite{vellekoop_universal_2008}, when optimizing a single point, the background increases, which means that the signal to background ratio is lower than expected from the opaque lens analysis.  This effect has been discussed in  \cite{davy_focusing_2012}, in the context of microwaves but the result remains valid for optics. In essence, the signal to background in a point optimization experiment is bounded by the number of transmitting modes and can be down to one (no optimization) in the single open channel regime \cite{pena_single-channel_2014}. Of course, in this limiting case, the intensity at the focus has been indeed increased, but since there is only a single mode that dominates the transmission, the background has also increased correspondingly. 

Several works have also reported on measuring a TM and subsequently injecting the mode with the largest transmission.  In \cite{kim_maximal_2012}, the  limiting case of very sparse measurement was explored: A square transmission matrix was measured over a single polarization and a very limited angular view (covering a numerical aperture of $NA \simeq 0.32)$. In this configuration, the quarter circle law, that corresponds to the Mar\v{c}enko-Pastur distribution for a square random matrix, is expected (see Eq.~\ref{eq:Marcenkopastur}). While this distribution had already been  measured in optics in \cite{popoff_measuring_2010}, the authors show that when sending the input vector corresponding to the highest transmission mode given by the measured transmission matrix, they recover a higher transmission by a factor $3.99$ relative to the mean transmission within the detection angle, in good agreement with the fact that the distribution of eigenvalues is bounded to twice the mean amplitude transmission in the  case of a square random matrix (corresponding to a factor four in intensity).  In \cite{kim_implementing_2014}, the authors reported on the promising use of a  binary amplitude modulator to measure the TM and to inject eigenchannels. They show that such a binary modulator is able to match the calculated mode  for single mode injection with $40\%$ fidelity, and demonstrate a two-fold increase of transmission over the mean transmission in the detection angle of their apparatus.  Measurements of very large TMs in strongly scattering media to approach the bimodal distribution are reported in \cite{yu_measuring_2013} and \cite{akbulut_optical_2016}. In both studies, deviations from the Mar\v{c}enko-Pastur distribution were observed, and could be qualitatively modeled, but due to limited control, the direct observation and injection of open channels in optics has so far not yet been reported in the literature (see \cite{gerardin_full_2014} for a first realization of open channels in  acoustics).

Another possibility of high practical interest, is to measure the reflection matrix. When sending the mode with the lowest reflection, a high transmission should be obtained. This idea was experimentally implemented in \cite{kim_exploring_2015} where a three-fold increase in transmission is reported when injecting the mode with the smallest eigenvalue in reflection.

\begin{figure}
  \centering
  \includegraphics[width=1\linewidth]{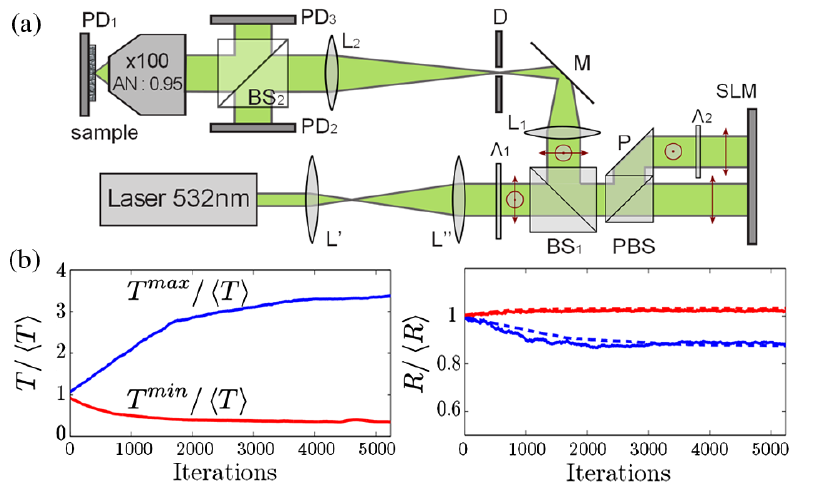}  
  \caption{(color online). Experimental setup and results for total transmission optimization. (a) Schematic of the experiment for the control of total transmission. The two polarizations of the laser are modulated by two different areas of a phase-only SLM, with 1740 macropixels controlled. The scattering sample is placed at the focal plane of the objective. Three photodetectors PD1, PD2, and PD3 measure, respectively, the intensities of transmitted, incident, and reflected light, and the optimization is performed on the total transmission $T$, measured as the intensity on detector PD1 normalized by the incident intensity measured on detector PD2. (b) Measured ${T}/{\langle T\rangle}$ (left panel) and ${R}/{\langle R\rangle}$ (right panel) versus the step number of optimization for enhancement (increasing blue curve) and reduction (decreasing red curve) of the total transmission. The sample is a 20 m thick ZnO layer, and the average transmission is $\langle T\rangle \sim 5\%$. The dotted line represents the reflection estimated from the transmission using ${R}/{\langle R\rangle}=(1-T)/(1-\langle T \rangle)$. (Figure adapted from \cite{popoff_coherent_2014}.)}
  \label{fig:maxtransmission}
\end{figure}

An  alternative to single-point optimization was reported experimentally in \cite{popoff_coherent_2014}, see
Fig.~\ref{fig:maxtransmission} for a corresponding illustration. Instead of measuring a single position and observing an increase of the  transmission, here the total transmitted intensity was optimized directly. In order to detect all modes at the output and not to be limited by collection optics,  a large-area photodetector is directly placed on the backside of the sample, thus collecting all modes with the same efficiency. This is in contrast with previous results where a limited collection angle was inherent to all implementations, and therefore only partial transmission enhancements were reported, from which changes in total transmission could only be inferred. When optimizing the wavefront to maximize or minimize the total transmitted intensity, an approximately $3.6$-fold enhancement, and a $3.1$-fold reduction in transmission relatively to the average could be found, respectively. To confirm that this effect was, indeed, a mesoscopic effect, the reflection was also monitored and its increase or decrease was found to be anticorrelated with a corresponding change in transmission. Yet other approaches have been proposed, e.g., optimization algorithms to maximize transmission by analyzing only the backscattered light \cite{jin_backscatter_2014,jin_iterative_2013}, or a photocurrent \cite{liew_coherent_2016}. As was shown both experimentally and theoretically \cite{hsu_correlation-enhanced_2017} long-range-correlation effects as discussed in section \ref{subsubsection3.1.1} and in \cite{scheffold_observation_1997,garcia-martin_finite-size_2002} increase the dynamic range of light-delivery into regions containing multiple speckles. 

\subsection{Time delay eigenstates}\label{subsection5b.1}

In this section we will discuss the interesting properties of the eigenstates of 
the time-delay operator or matrix ${\bf Q}=-i{\bf S}^\dagger (\partial {\bf S})/(\partial \omega)$
defined in sec.~\ref{subsection2.4}, see Eq.~(\ref{eq:timedelayop}). Thanks to the techniques of wavefront shaping discussed in section \ref{section4}, it is now within reach not only to measure the time delay matrix of a system, but also to excite an eigenvector at the input, thus generating a time-delay eigenstate. We will first describe these states in all their generality, then discuss their applications in two specific cases relevant for optics: principal states and particle-like states (see sections \ref{subsubsection5b.2} and  \ref{subsubsection5b.1a}). The eigenstates ${\bf q}_i$ of ${\bf Q}$ are
defined as follows, ${\bf Q\,q}_i=q_i{\bf q}_i$, and they are associated with a well-defined scattering time delay
$q_i$ (also called proper delay time), which measures the time accumulated between entering and
exiting the scattering region. When the scattering matrix $\bf S$ of a problem is unitary, the associated
time delay operator $\bf Q$ is Hermitian, such that the proper delay times are real and the time delay eigenstates ${\bf q}_i$ form a
complete and orthogonal set at the input to the scattering region.

The question we address in the following is, which practical consequences can be associated
with the above definitions and how such states can be determined and  generated experimentally. Generally speaking, a wave that enters a disordered slab will have
components that exit the slab rather quickly, while other components will stay inside the slab for longer (as discussed in section \ref{subsubsection4.3.2}). This is different when injecting a state defined by the time-delay coefficient vector ${\bf q}_i$ into the slab (e.g., through an SLM), 
since this state is characterized by just a single and well-defined time, 
i.e., its proper delay time $q_i$.  This feature leads to very advantageous properties related to the fact that
a well-defined time-delay can be linked to a suppression of frequency dispersion and to a strong
collimation of ballistic scattering states.  
 
\subsubsection{Principal modes in a fiber}\label{subsubsection5b.2}
Consider a multi-mode fiber (MMF) which transmits light almost perfectly, i.e., it has very little reflection and absorption. In this case
the transmission matrix $\bf t$ associated with this fiber is close to unitary such that all the transmission eigenvalues $\tau_n$ are 
near unity. Correspondingly, the associated transmission eigenchannels studied in section \ref{subsection2.2} will not be in any way special, since the massive degeneracy in the 
linear subspace associated with $\tau\approx 1$ will mix all these channels indiscriminately.
The question thus arises whether this degeneracy can be lifted by a clever choice for
a different basis that has advantageous properties, e.g., for the transfer of information through the fiber. One might think that such a
suitable basis is given by the fiber modes themselves, as determined by the boundary conditions in the fiber cross-section. In particular, since
the different transverse mode profiles also result in mode-specific longitudinal velocities it is tempting to think that the fiber modes are the
suitable modes for avoiding dispersion in the fiber. It turns out, however, that this is generally not the case and that such a dispersion-free basis is rather given through the eigenbasis of the time-delay operator 
\cite{fan_principal_2005,poole_phenomenological_1986}.

Specifically, consider an input coefficient vector ${\bf v}$, which is transmitted by the fiber to an output vector ${\bf u}={\bf t}{\bf v}$, where we will assume the vectors and the matrix
to be given in the mode basis (polarization degrees of freedom will be neglected). 
If we now demand that the transverse profile at
the fiber output should be dispersion-free this means that the output vector ${\bf u}$ should not change, when changing the input frequency $\omega$
slightly while keeping the input vector ${\bf v}$ the same 
\cite{fan_principal_2005}. To be more specific, we demand that the 
orientation of the output vector ${\bf u}$ stays invariant (which is equivalent to demanding that 
the output field distribution stays unchanged up to a prefactor).
Decomposing ${\bf u}$ into an amplitude and the corresponding unit vector which contains this orientation
${\bf u}=u\hat{\bf u}$, we obtain the following relations,
\begin{equation}
\frac{d {\bf u}}{d \omega}=\frac{d u}{d \omega} \hat{\bf u}+u\frac{d \hat{\bf u}}{d \omega} =
\frac{d u}{d \omega}u^{-1}{\bf t}{\bf v}+u\frac{d \hat{\bf u}}{d \omega}\equiv
\frac{d {\bf t}}{d \omega} {\bf v}\,.
\label{eq:pms1}
\end{equation}
Requiring that $\hat{\bf u}$ is dispersion free, $d \hat{\bf u}/d\omega\equiv 0$,
and multiplying from the left with $-i{\bf t}^{-1}$, we end up with the following relation,
\begin{equation}
-i{\bf t}^{-1}\frac{d {\bf t}}{d\omega}\,{\bf v}=-iu^{-1}\frac{d u}{d\omega}\,{\bf v}\,.
\label{eq:pms3}
\end{equation}
which tells us that those input states ${\bf v}$ which are transmitted without transverse dispersion (to first order) are eigenstates
of the matrix $\tilde{\bf Q}=-i{\bf t}^{-1}d {\bf t}/d\omega$. 
For unitary transmission matrices, for which ${\bf t}^{-1}={\bf t}^{\dagger}$, 
this expression for $\tilde{\bf Q}$ is perfectly equivalent to the expression for the Wigner-Smith time-delay operator ${\bf Q}$ which we had found before, see Eq.~(\ref{eq:timedelayop}).
For non-unitary transmission matrices as for fibers with finite reflection or loss, one can further modify 
the right hand side of Eq.~(\ref{eq:pms3}) [using $u=|u|\exp(i\phi)$] 
to find the following expression for the corresponding eigenvalue of this new operator,
\begin{equation}
q=-i \,\frac{d \ln |{\bf t v}|}{d\omega}+ \frac{d \arg ({\bf t v})}{d\omega}\,.
\label{eq:pms4}
\end{equation}
The first term on the right-hand side of Eq.~(\ref{eq:pms4}) is a measure for the losses due to reflection
or outcoupling that depends only on the norm of the output; the second term is the
derivative of the scattering phase at the output, i.e., the time delay. 

\begin{figure}
  \centering
  \includegraphics[width=\linewidth]{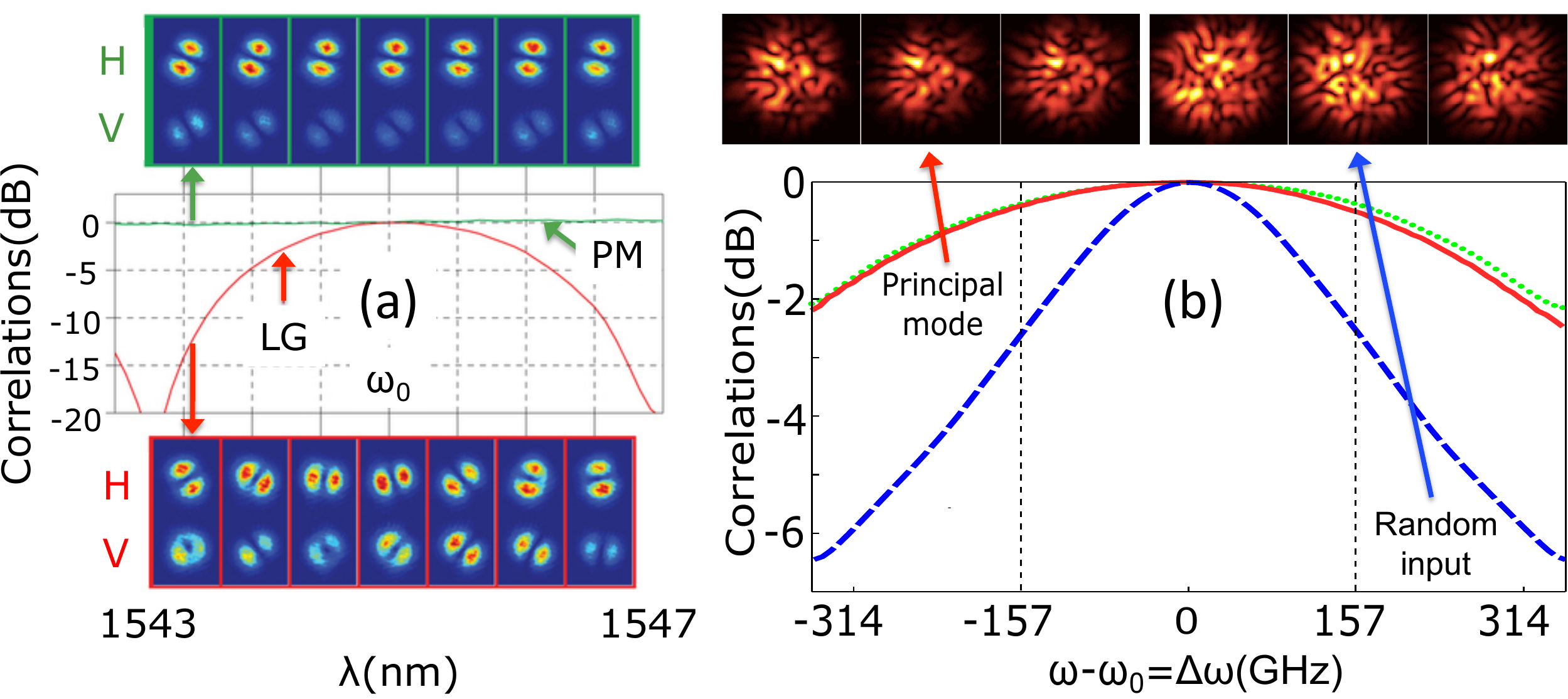}
  \caption{(color online). Experimental data on principal modes (PMs) in fibers with (a) weak and (b) strong mode mixing, adapted from \cite{carpenter_observation_2015} for (a) and from \cite{xiong_spatiotemporal_2016} for (b).
In both figure parts the main panel shows the spectral correlation of the output field pattern measured relative to the center frequency for a PM as compared to a Laguerre-Gaussian (LG) mode in (a) and as compared to a random superposition of LP modes in (b). 
In both cases the PM features a considerably increased stability of the field configuration at the fiber output. This output is shown in the color images: (a) Top: PM, bottom: LG mode. Horizontal (H) and veritcal (V) polarization directions are shown separately. (b) Top left: PM, top right: random input. Images recorded at $\Delta \omega=-$157, 0, 157 GHz (from left to right). The green dashed line in (b) shows the calculated correlation function based on the experimentally obtained fiber transmission matrix, which is in excellent agreement with the experimentally determined correlation function (red).}
  \label{fig:principalmodes}
\end{figure}

Due to their superior properties, the modes associated with the eigenvectors of $\tilde{\bf Q}$ have been termed ``principal modes (PMs)'' \cite{fan_principal_2005}.
Note in this context, that for a perfectly straight fiber without mode-coupling, 
the PMs and the fiber
modes become the same (in the absence of degeneracies). In this sense, the advantages of the PMs assert themselves fully in the presence of a finite
crosstalk between the ideal fiber modes \cite{ho_linear_2014}. 
First observations of PMs in MMFs have been reported in \cite{xiong_spatiotemporal_2016,carpenter_observation_2015,carpenter_first_2014}.
In Fig.~\ref{fig:principalmodes}a we show data from a measurement on fibers with weak mode coupling, which demonstrates the
increased stability of PMs as compared to conventional linearly polarized (LP) 
fiber modes. Whereas PMs feature already by design a frequency-stability to first order, 
the weak coupling of modes enhances their stability further.
In the regime of strong mode coupling the frequency stability of PMs is reduced, but still far superior 
compared to arbitrary input configurations (see Fig.~\ref{fig:principalmodes}b).
The same can be expected for a disordered slab geometry, for which PMs can also be constructed,
but for which case no experiments have been reported so far. We will see in the following
that also another class of time-delay eigenstates can be found in ballistic or quasi-ballistic scattering
structures with a frequency robustness that goes  beyond the above first-order stability. 
 
\subsubsection{Particle-like scattering states}\label{subsubsection5b.1a}
\begin{figure}
  \centering
  \includegraphics[width=1\linewidth]{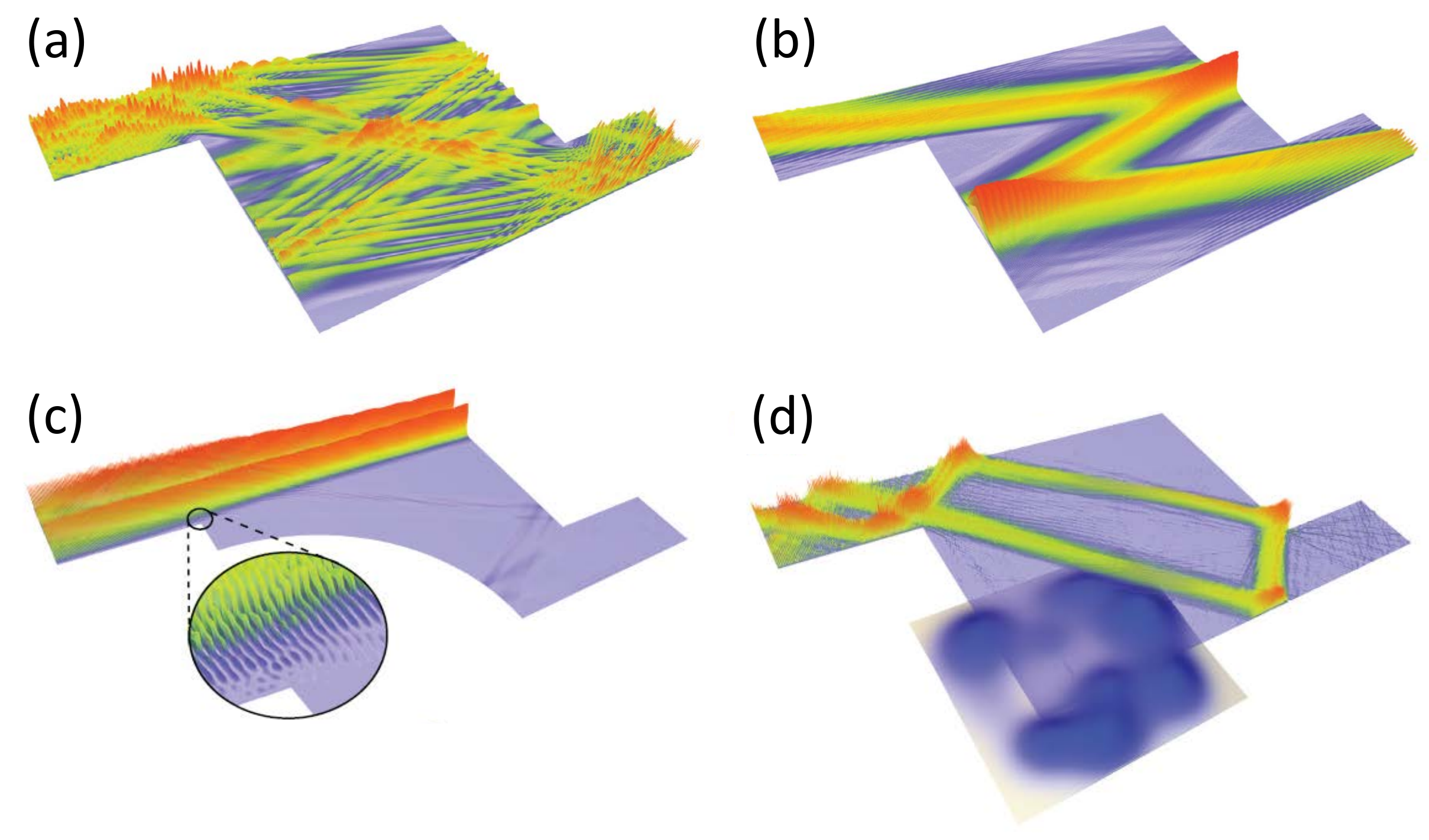}
  \caption{(color online). (a) Wave intensity of a transmission eigenchannel with transmission close to unity in scattering through a resonator with two waveguides attached on the left and right (the flux is incoming from the left). Whereas transmission eigenchannels typically lead to highly complex interference patterns inside the resonator (as in wave optics), the particle-like states shown in (b,c,d) follow a classical bouncing pattern throughout the entire scattering process (as in geometric optics). (b) Transmitted particle-like state in a clean rectangular resonator. (c,d) Reflected particle-like states in a geometry of the same dimensions as in (a,b), from which, however, a quarter-circular piece was removed in (c) and a  smooth and weak disorder potential was added in (d) (see bottom part of this panel). (Figures partially adapted from \cite{rotter_generating_2011}.)}
  \label{fig:particlelike}
\end{figure}

Consider the simple case of a resonator geometry, which, for reasons of simplicity, is
assumed to be just 
two-dimensional. The scalar waves, which are injected through a waveguide on the left, can be reflected
through the same waveguide or transmitted through a second waveguide attached on the right 
(see illustration in Fig.~\ref{fig:particlelike}). One can now try to 
steer waves through the resonator such that they will
follow the path of a classical trajectory throughout the entire scattering process rather than being diffractively scattered. To select the ``geometric optics'' states from the 
full set of scattering states that ``wave optics'' will produce in this setup, we further assume that only the scattering matrix $\bf S$ of the system is available (but no information on its interior scattering landscape). 
The presence of such ballistic scattering states
leads to non-universal contributions to the distribution of
transmission eigenvalues $P(\tau)$ at the values $\tau\approx 0$ and $\tau\approx 1$, corresponding
to fully closed and open transmission channels, respectively.
In the mesoscopic regime (see section \ref{subsection2.2}) it was exactly such
system-specific contributions which were responsible for the suppression of electronic shot-noise below the universal limit \cite{jacquod_breakdown_2004,sukhorukov_quantum--classical_2005,agam_shot_2000,aigner_shot_2005}, which itself
was already reduced below the Poissonian limit by Schottky \cite{beenakker_quantum_2003,schottky_uber_1918}.

In analogy to the situation found for the MMF in the previous 
section \ref{subsubsection5b.2}, all the fully  transmitted (reflected) waves are 
completely mixed in the degenerate
subspace corresponding to $\tau\approx 1$ ($\tau\approx 0$). Correspondingly, exciting transmission or reflection eigenchannels will not yield states that follow individual classical 
bouncing patterns in the system, but several of them simultaneously, see Fig.~\ref{fig:particlelike}(a). 
To unmix these contributions one can now make use of the Wigner-Smith time-delay operator $\bf Q=-i{\bf S}^\dagger\partial {\bf S}/(\partial \omega)$ introduced in section \ref{subsection2.4}; this is because its eigenvalues, i.e., the ``proper delay times'' allow one to sort all the different 
ballistic scattering contributions by way of their different time-delays 
\cite{rotter_generating_2011}. 
Specifically, one determines those eigenstates $\bf q$ of $\bf Q$ that only have incoming components in the left
wave guide, i.e., ${\bf q}=({\bf q}_l,{\bf 0})^{\rm T}$. Writing the time-delay operator with its four sub-blocks (in correspondence
to the subdivision of the scattering matrix itself, see Eq.~(\ref{eq:smatrixblock})), one finally obtains the following eigenvalue problem, 
\begin{equation}
  \left( \begin{array}{cc}
      {\bf Q}_{11}\phantom{\Big|}\! & {\bf Q}_{12}  \\
      {\bf Q}_{21}\phantom{\Big |}\! & {\bf Q}_{22} \end{array}
  \right)\left( \begin{array}{c}
      \!\!\phantom{\Big |}{\bf q}_{l}^{\,\,{\rm}}\phantom{\Big |}\!\!
      \\\!\!\phantom{\Big|}{\bf 0}\phantom{\Big|}\!\!
    \end{array} \right)=
  \left( \begin{array}{c}
      \!\!\phantom{\Big|} {\bf Q}^{\phantom{i}}_{11}\,{\bf q}_{l}^{\,\,{\rm }}\phantom{\Big|}\!\!
      \\\!\!\phantom{\Big|}{\bf Q}^{\phantom{i}}_{21}\,{\bf q}_{l}^{\,\,{\rm }}\phantom{\Big|}\!\!
\end{array} \right)=
  q \left( \begin{array}{c}
      \!\!\phantom{\Big |}{\bf q}_{l}^{\,\,{\rm }}\phantom{\Big |}\!\!
      \\\!\!\phantom{\Big|}{\bf 0}\phantom{\Big|}\!\!
\end{array} \right)
\,.\label{eq:q2}
\end{equation}
From the last equality the following two conditions can be deduced: 
(i) ${\bf Q}^{\phantom{i}}_{11}\,{\bf q}_{l}^{\,\,{\rm
    }}=q\,{\bf q}_{l}^{\,\,{\rm }}$ and (ii)
${\bf Q}^{\phantom{i}}_{21}\,{\bf q}_{l}^{\,\,{\rm }}=\vec{0}$.  
Since ${\bf Q}_{11}$ is a Hermitian matrix (${\bf Q}_{12}$ and ${\bf Q}_{21}$ are not), 
condition (i) yields an orthogonal and complete set of eigenstates in the incoming waveguide. Out
of this set of states, those which, according to condition (ii), lie in the null-space (kernel) of 
${\bf Q}_{21}$ are the desired time-delay eigenstates with a well-defined input port. One can show 
\cite{rotter_generating_2011} 
that both conditions can only be fulfilled by waves which are either fully transmitted or reflected; in other
words these states are simultaneously eigenstates of ${\bf Q}$ and of ${\bf t}^\dagger {\bf t}$ with deterministic transmission
eigenvalues $\tau$ close to zero or one. Practically, the degree to which condition (ii) is fulfilled 
can be evaluated by a measure $\chi=\|{\bf Q}_{21}\,{\bf q}_{l}^{\,\,{\rm }}\|$ which should be the closer to zero
the better condition (ii) is fulfilled. This measure can thus be conveniently used for assessing how well
a given state will be able to follow a classical bouncing pattern. Typically those states with a small 
value of $\chi$ are also those with a small time-delay eigenvalue $q$, in agreement with the expectation that
only states which stay inside the scattering region for a time shorter than the Ehrenfest time $q<\tau_E$ will
be able to behave ``particle-like'' \cite{agam_shot_2000}. 
In Fig.~(\ref{fig:particlelike})b-d three examples of states are shown which feature very small 
values of $\chi$ for different resonator geometries. It can clearly be seen how these
states tend to follow a (short) ray from geometric optics that avoids any diffractive scattering throughout its propagation (as, e.g., at the sharp corners of the input and output facets). 
A first experimental demonstration of particle-like scattering states as well as of  
corresponding wave packets has been reported  
based on acoustic wave scattering in a metal plate studied by laser interferometry
\cite{gerardin_particlelike_2016}. 

When comparing these particle-like states to the PMs from the previous section \ref{subsubsection5b.2}, the following comments can be made: PMs can be 
constructed for arbitrary scattering media (including strongly disordered samples) and their
frequency stability to first order will be assured in all of these systems by construction. Particle-like
states, on the other hand, can only be found in systems where waves can propagate along sufficiently
stable ballistic scattering pathways. Due to their collimation on these pathways, particle-like states 
feature, in turn, a much higher frequency stability than principle modes (in a similar way as geometric 
optics states are frequency-independent by default). Both sets of states have very advantageous
properties for communication purposes, like the dispersion-free transmission as well as the high 
directionality of the particle-like states that seems well suited for steering a signal to
a well-defined target. In section \ref{subsection5b.4} we will also see that the  time-delay eigenstates
optimally avoid or enhance the effect of dissipation in a medium. 

\subsection{Wavefront shaping in media with gain or loss}\label{subsection5b.3}
In this section we will discuss the application of wave front shaping techniques in systems with gain or loss,
with a focus on disordered media. 

\subsubsection{Absorbing media}\label{subsection5b.4}
Consider  the case of an absorbing disordered medium, which, for simplicity, we
will assume to be uniformly absorbing (i.e., the absorption rate is independent of the spatial position in the medium). 
For this situation we know from our analysis in section \ref{subsection2.4} 
that the absorption is directly proportional to the time spent inside the absorbing medium.
Since, in turn, the time associated with a stationary scattering state can be measured through the dwell time operator ${\bf Q}_d$, minimal or maximal absorption of waves in a medium can be achieved by injecting those eigenvectors of ${\bf Q}_d$, 
which are associated with the smallest or largest eigenvalue, respectively. (We recall here the result from section \ref{subsection2.4} that in the limit of vanishing absorption the dwell time operator ${\bf Q}_d$ and the time-delay operator ${\bf Q}$ coincide up to mostly negligible self-interference terms.)
The procedure to obtain the states with minimal absorption is thus equivalent to the approach we presented 
in section \ref{subsubsection5b.1a} for the generation of particle-like scattering states, which are associated with 
the smallest values of the time-delay. The states with maximal absorption have been explicitly studied in 
\cite{chong_hidden_2011}, where it was shown how in a weakly scattering medium a suitably chosen input wave front
can increase the degree of absorption from a few percent to more than $99\%$. Such a coherent enhancement of absorption (CEA) can, in
principle, be realized at any frequency for which the input wave is shaped appropriately.

An interesting point to observe
here is that in the theoretical approach put forward in \cite{chong_hidden_2011} these maximally absorbed states were
not identified through the help of the dwell-time operator, but rather as those states, which are minimally reflected from an
absorbing disordered medium. In the considered single-port systems, where the reflection matrix is equivalent to the scattering matrix, 
we know, however, from Eq.~(\ref{eq:unitarydeficit}), that the dwell time operator is equivalent to the unitary deficit of
the scattering matrix such that these two different concepts to evaluate maximum absorption perfectly coincide. 
In \cite{chong_hidden_2011} the analysis was also extended to the case of a 
spatially localized absorber buried behind a layer of lossless scattering medium -- a situation that
has also been studied experimentally by Vellekoop et al.~in an attempt to focus light on a fluorescent nanoscopic bead inside a disordered
medium to increase the fluorescence \cite{vellekoop_demixing_2008}. 
In this case the degree of optimal absorption was found  
to be more strongly bounded as compared to the case where the entire medium is absorbing. Further work
also shows how the long-range spectral correlations inherent in the transmission and reflection matrices can help to achieve enhanced absorption in a broadband frequency range \cite{hsu_broadband_2015}.
A first experimental realization of a variant of CEA has been reported in \cite{liew_coherent_2016}.  

In another numerical study the effect of absorption was investigated in scattering systems with more than one port, like a 2D 
disordered wave guide connected to two perfect leads on the left and right 
\cite{liew_transmission_2014}. In such systems both the
maximally transmitted and the minimally reflected channels were studied.
For weak absorption these two types of channels were found to be dominated by diffusive transport and to be
equivalent (as following from the connection between the transmission and reflection matrices in unitary systems, 
see section \ref{subsubsection2.1.4}). For increasing absorption, however, the behavior of these two different channels decouples, as the
reflection can then be minimized not only by increased transmission, but also by enhanced absorption: at a given 
absorption strength, the maximum transmission channel was found to display a sharp transition  
to a quasi-ballistic transport regime. This transition does not occur for the minimal reflection channel, which gets
increasingly dominated by the absorption when the absorption strength is increased. A very interesting
aspect that was also found in this context is that the shape of the transmission and reflection eigenvalue distributions in disordered and dissipative media depends on the confinement geometry \cite{yamilov_shape_2016} - a fact that may be used for controlling this distribution at will.

Whereas the above concepts relating to coherent enhancement of absorption (CEA) can be implemented at any input frequency, it was shown in \cite{chong_coherent_2010} 
that at well-defined frequencies and at a carefully chosen amount of dissipation, certain incoming
channels of light can be fully absorbed. Such a coherent perfect absorber (CPA) of light corresponds to the multi-mode generalization
of a critically coupled oscillator, with the difference that at least two input beams are required, which have to have the correct
amplitude and phase to interfere appropriately. 
As a result, the relative phase between 
the input beams can be used to sensitively tune the degree of absorption, as was done in the first CPA experiment
\cite{wan_time-reversed_2011}. From the conceptual point of
view a CPA is a time-reversed laser, in the sense that a gain medium at its first lasing threshold will emit coherent radiation 
at a well-defined frequency and with a well-defined phase relationship, e.g., between beams emitted on either side of the laser. The time-reversed process corresponds to an absorbing medium which perfectly absorbs the coherent illumination
which impinges on it. If one considers a simple 1D edge-emitting laser that emits to the left and right, 
the coherently absorbed 
light field of the corresponding CPA features two beams (incoming from the left and right), which share a specific phase relationship to each other. At the points where this phase relationship is satisfied, maximal absorption occurs.  It is interesting that, prior to these theoretical and experimental developments, a seminal experiment in the field of plasmonics showed extraordinary absorption for a gold grating under  specific incident illumination \cite{Hutley_total_1976}, which was explained using a similar reasoning.  

The concept of a CPA can also be extended to other systems \cite{noh_perfect_2012,zanotto_perfect_2014}, to
2D or 3D as well as to disordered media. In the latter case the CPA would be the time-reverse of a
random laser, corresponding to an absorbing random medium, which sucks in 
incoming waves that have exactly the same complex wave front as the emission profile of the
random laser. To generate such a complex wave pattern one would of course have to resort
to wave front shaping techniques using SLMs or equivalent tools. 
     
Generally speaking, the theoretical concept behind CPAs builds on the analytical properties of the scattering matrix ${\bf S}(\omega)$ (see section \ref{subsubsection2.1.4} for more information). 
In the absence of loss or gain, this matrix is unitary and 
features poles (zeros) at complex frequencies with negative (positive) imaginary parts, respectively, located in the complex plane as mirror-symmetric pairs with respect to the real axis. 
 When adding gain to the system the poles and zeros move upwards in the complex plane 
 until the point where the first pole reaches the real axis and lasing sets in. Alternatively, when adding loss to the system, the poles and zeros move downwards until the first zero hits the real axis, at which point
coherent perfect absorption can be realized \cite{chong_coherent_2010}. Adding even more loss drags additional zeros across the real axis, creating a CPA at each 
new intersection. Subsequent work also demonstrated how a laser and a CPA can be combined in a single device \cite{longhi_pt-symmetric_2010,chong_pt-symmetry_2011}.
Such a laser-absorber (or CPA-laser) can be realized based on the concept of $\mathcal{PT}$-symmetric optical systems \cite{ruter_observation_2010,makris_beam_2008,el-ganainy_theory_2007,bender_real_1998} in which gain and loss
are carefully balanced and poles and zeros of the scattering matrix can be
brought to meet on the real axis. Realizing such concepts in the optical experiments is challenging as the CPA-lasing points at the pole-zero crossing are affected by the noise due to amplified spontaneous emission. The first realization of a CPA with a $\mathcal{PT}$-phase transition has
been reported with a pair of coupled resonators coupled to a microwave transmission line \cite{sun_experimental_2014}, followed by the first successful demonstration of lasing and anti-lasing in the same $\mathcal{PT}$-symmetric  device \cite{wong_lasing_2016}
.
  
\subsubsection{Amplifying media}\label{subsection5b.5}
In the previous section we discussed how waves that are injected into a certain disordered
medium with absorption can be shaped such as to be maximally or minimally absorbed. 
Such an approach can, of course, also be considered with an amplifying medium, where 
one is naturally concerned with maximal or minimal amplification. Work in this direction 
has, e.g., dealt with the non-trivial transient dynamics in photonic waveguide structures  
composed of a combination of materials with both loss and gain. Contrary to 
conventional expectation, specific initial conditions for the incoming wave can lead to 
power amplification by several orders of magnitude 
even if the waveguide is, on average, lossy \cite{makris_anomalous_2014}.   
Systems with gain and loss  have also been proposed for the realization of a 
special family of waves that have the curious property of featuring a constant intensity 
in the presence of a non-homogeneous scattering landscape \cite{makris_constant-intensity_2015} -- 
a feature that can not be realized with Hermitian scattering potentials. Extending this concept allows one to achieve perfect transmission even 
through strongly scattering disorder \cite{makris_wave_2016}.
A realization of these curious wave solutions requires a careful shaping of the incoming wave 
front and of the medium's gain-loss profile. 
 
As  a medium with a sufficient amount of gain can emit coherent radiation 
on its own when crossing the laser threshold, work on amplifying media has also always had
a very strong focus on engineering the gain for a desired lasing action. In principle, 
optimizing the gain profile for a medium is a well-studied problem. Consider here, 
e.g., the case of a distributed feedback laser in which a so-called ``gain grating'', consisting of
a periodic arrangement of purely amplifying components, can efficiently pump lasing modes
with the same periodicity as the grating  
\cite{carroll_distributed_1998}. Whereas these concepts for
 quasi-1D laser structures (like ridge or ring lasers) have meanwhile 
 reached the level of industrial applications, more advanced concepts on lasers with a
 quasi-2D (planar) geometry have just been explored very recently. Here, the main focus
 was on reducing the laser threshold of specific modes by increasing the spatial overlap 
 between these selected modes and an externally applied pump profile. Practical implementations
 of this concept include, e.g., electrically pumped devices 
for which the electrodes were patterned appropriately \cite{shinohara_chaos-assisted_2010,fukushima_ring_2002,kneissl_current-injection_2004}. 
All of these implementations require, however, the prior knowledge of the spatial pattern of
the selected mode. 

\begin{figure}
  \centering
  \includegraphics[width=1\linewidth]{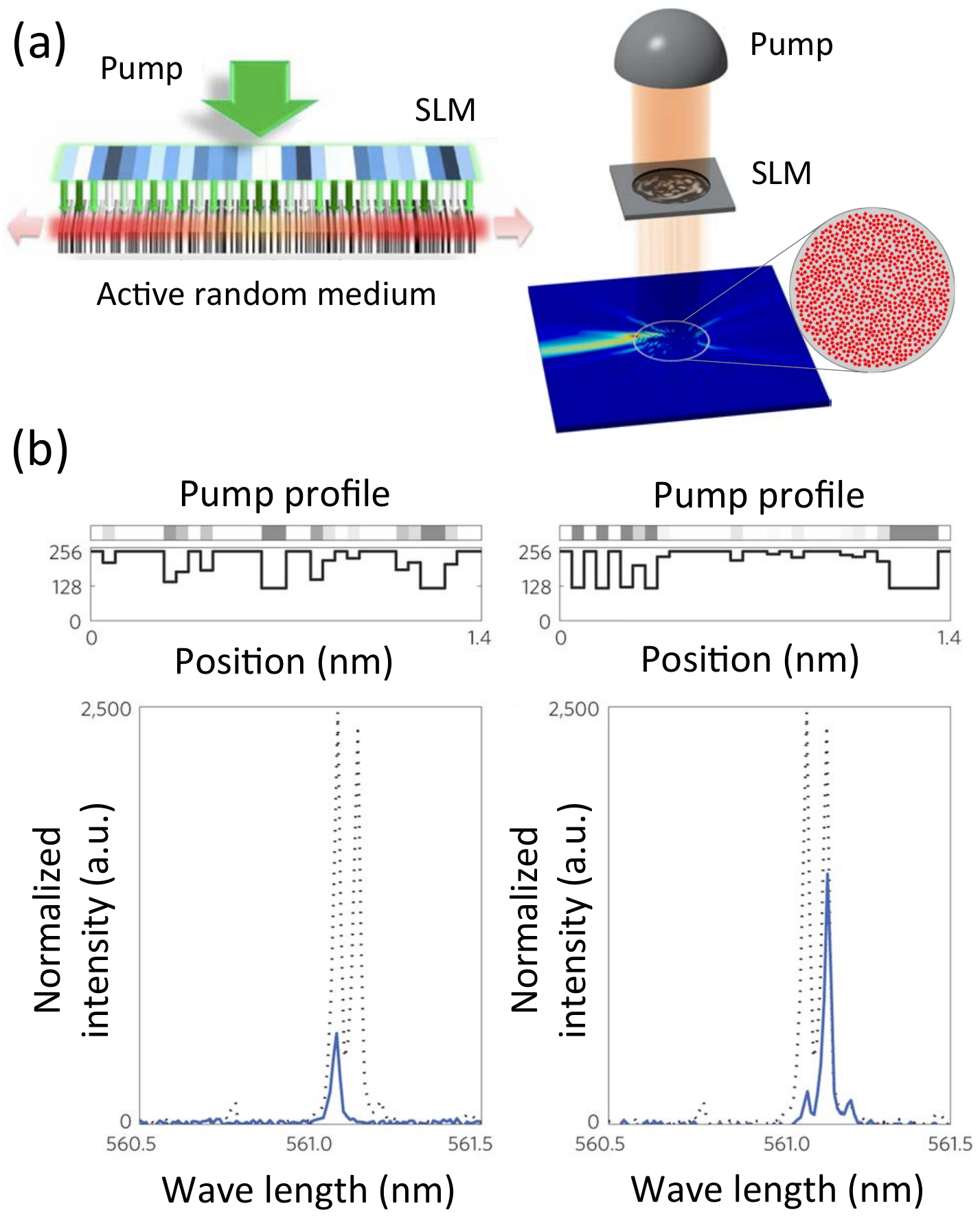}
  \caption{(color online). (a) (left) Schematic to control the emission spectrum of a random laser consisting of a quasi-one-dimensional sequence of different dielectric layers. Tuning the incident pump beam through the spatial light modulator allows here to change how many lasing modes are active and at which frequencies they emit. (right) A corresponding setup proposed to control the directionality of the emission. (b) The spectral control of a random laser (left panel in (a), was realized in an experiment using opto-fluidic random lasers \cite{bachelard_adaptive_2014}, where the laser emitted in two modes for uniform pumping (see dotted lines in lower panels). By shaping the pump profile in the way shown in the top panel, single-mode emission through either one of these two modes was achieved (see solid lines in lower panels). (Figures adapted from \cite{bachelard_taming_2012} for (a) (left panel), from \cite{hisch_pump-controlled_2013} for (a) (right panel) and from \cite{bachelard_adaptive_2014} for (b).)}  
  \label{fig:pump_shaping_figure}
\end{figure}

With the availability of wave front shaping tools, the pump profile
as exerted on an optically pumped laser can, however, be tuned in a manner which is
flexible enough to select a given mode based on a simple feedback loop. This feedback 
can be set up between the pump profile (as determined by the pixel configuration in an SLM)
and, e.g., the light spectrum of the laser pumped with this profile. Combining the feedback with 
an optimization algorithm has, e.g., been suggested as a means to 
make the multi-mode emission spectrum of a random laser single-moded
\cite{bachelard_taming_2012}. In an experimental realization,
realized shortly after the theoretical proposal 
\cite{bachelard_adaptive_2014} a mode-specific 
pump-selection and a corresponding single-mode operation could be successfully 
demonstrated for the challenging case of a 
weakly scattering random laser, see Fig.~\ref{fig:pump_shaping_figure}. 
For such a system no a-priori knowledge of the lasing mode is available and finding the appropriate 
pump-grating is thus only possible through optimization. 
As in this case the laser modes are also strongly overlapping
both spectrally and spatially, the pump profiles obtained as a result of the optimization process
do not just follow the intensity of the laser modes, but instead display a highly complex
pattern only remotely related to the mode profiles. The connection between the
pump profiles and the modes they select was elucidated in subsequent theoretical work 
\cite{bachelard__2014,cerjan_controlling_2016} based on a non-Hermitian perturbation theory analysis. Alternatively, 
a mode-selection approach was proposed based on insights from gain saturation of 
interacting laser modes \cite{ge_selective_2015}.

A remarkable feature of the above feedback-based pump optimization is its flexibility in 
terms of the optimization goals that it can be employed for.  Specifically, it has been proposed that
not only the multi-mode spectrum of a random laser can be ``tamed'' with it, but also its
highly irregular emission profile, see Fig.~\ref{fig:pump_shaping_figure}a (right panel)  
\cite{hisch_pump-controlled_2013}. A corresponding pump-shaping strategy for tuning the emission profile 
of a laser has meanwhile been successfully implemented for micro-cavity 
lasers \cite{liew_active_2014} (for random lasers a collimated output beam was achieved with other means
\cite{schonhuber_random_2016}). A next step could be to use the spatial control of the
applied pump to enhance the power efficiency of lasers \cite{ge_enhancement_2014}.

\section{Conclusions and outlook}\label{section6}
Wave front shaping techniques will become faster,
more accurate and will involve an increasing number of controlled pixels. Loosely speaking, this
development can be connected to the exponential increase in efficiency of computer hardware (known
as Moore's law). With this projection in mind, one can foresee that experiments using 
light modulation technology will soon be able to do much more than they can do already now. 
In this last section 
a few ideas will be provided on where the further technological developments could bring us or on what we believe could be 
promising research topics in the next few years.

\subsection{Wavefront shaping for unraveling mesoscopic phenomena}\label{subsection6.1}
Whereas our review has already highlighted quite a few mesoscopic effects that were brought to light
with wavefont shaping tools, we believe that many more fundamental phenomena can still be explored with this new technology. 
Prominent examples could be, e.g., an unambiguous proof for Anderson localization 
in a three-dimensional 
medium \cite{scheffold_inelastic_2013,sperling_direct_2013,skipetrov_absence_2014,segev_anderson_2013,lagendijk_fifty_2009} or the direct experimental demonstration of the bimodal distribution of transmission eigenvalues
in optical scattering through a disordered medium (for which several indications have already been
put forward \cite{popoff_coherent_2014,vellekoop_universal_2008,goetschy_filtering_2013}). 
In this context it would also be of high interest to
directly inject the fully open transmission eigenchannels 
and to observe their superior transmission characteristics, in a similar way to what 
has been done
in acoustics \cite{gerardin_full_2014}. Likewise, one could try to directly observe 
the propagation of time-delay eigenstates (such as principal modes \cite{fan_principal_2005} 
or particle-like states \cite{rotter_generating_2011}) through a weakly or strongly disordered 
optical medium, similar to what as recently achieved in optical fibers \cite{xiong_spatiotemporal_2016,carpenter_observation_2015} and in acoustic waveguides \cite{gerardin_particlelike_2016}.
An experimental demonstration that also still remains to be done is that of a multi-channel
coherent perfect absorber \cite{chong_coherent_2010}, 
which could go as far as to demonstrate the  time-reverse process of random lasing.
 Also other exotic effects, like ``rogue waves'' \cite{solli_optical_2007, liu_triggering_2015} or 
``branched flow'' \cite{topinka_coherent_2001,metzger_universal_2010}, 
that have also been studied in the context of ocean acoustics, could now
be enhanced and tuned in various ways through wave front shaping.
Optical scattering of course also brings in many new aspects as compared to its electronic counterpart, 
in particular through non-linearities \cite{wellens_nonlinear_2008,bortolozzo_experimental_2011} 
and the breaking of reciprocity \cite{muskens_partial_2012,peng_parity-time-symmetric_2014},
which features have not yet been fully explored with wavefront shaping tools.

\subsection{New systems}\label{subsection6.2}
We have seen that mesoscopic physics theory and experiments have mostly  been focused on a handful of canonical systems in electronics: the disordered wire, the quantum billiard, the quantum point contact, etc. In the optical domain, a large fraction of the theoretical and experimental work was focused on scattering in three-dimensional bulk disorder, and restricted to a few geometries as well. Here, we want to comment on the opportunities of new optical systems to be studied with the mesoscopic physics concepts already developed. 

Two systems  have been discussed extensively already in sections \ref{section4} and \ref{section5}: biological tissues and multimode fibers. We have tried to highlight their specific features in terms of scattering and in which way these can be appropriately described by adapting the existing mescoscopic physics concepts. We have also tried to highlight how these system-specific aspects could lead to new opportunities, e.g., for imaging (see here, in particular, the discussion on the memory effect in chapters \ref{section2} and \ref{section5}). Plasmonic systems are also an interesting playground to study the effect of disorder and of wavefront shaping,  in particular due to the capability of the metal to localize light well-below the diffraction limit. This includes not only metallic hole arrays with disorder \cite{gjonaj_active_2011, gjonaj_focusing_2013, seo_far-field_2014}, but also metal-dielectric fractal structures \cite{gaio_percolating_2015,bondareff_Probing_2015}.

New optical systems are meanwhile emerging due to the exciting possibility in photonics to tailor the propagation medium, e.g., to vary the amount of order and disorder, or to change the dimensionality of the problem.  Light transport has been studied in 1D stacks \cite{bertolotti_optical_2005}, in 1D waveguides \cite{topolancik_random_2007, sapienza_cavity_2010}, and in 2D disordered structures \cite{garcia_nonuniversal_2012, riboli_anderson_2011}. 
A whole community works on photonic crystals, trying to obtain perfectly regular structures, but the very small amount of residual disorder has been known to strongly affect transport within these structures. Careful engineering of the amount of disorder can, in turn, allow to control the transport properties of the system \cite{topolancik_random_2007, garcia_quantifying_2013}. In 3D, self-organization allows the fabrication of near-perfect 3D photonic crystals \cite{galisteo-lopez_self-assembled_2011}, for which the amount of disorder and its effect on the transport properties of light can again be controlled. Another potential revolution has been initiated by the possibility of engraving on a medium a completely designed refractive index distribution via direct laser writing \cite{kawata_finer_2001, deubel_direct_2004}, which meanwhile allows to create new kinds of disorder, such as hyper-uniform structures \cite{haberko_direct_2013, muller_silicon_2014, florescu_designer_2009,froufe-perez_role_2016}.  Finally, we also mention optically reconfigurable structures where a refractive index distribution is engraved from an intensity pattern, such as photorefractive crystals \cite{levi_disorder-enhanced_2011, schwartz_transport_2007}, an optical valve \cite{bortolozzo_experimental_2011} or integrated silicon-on-insulator multi-mode interference devices \cite{bruck_all-optical_2016,bruck_device-level_2015}. These structures are promising platforms on which not only the wavefront, but also the disorder can be controlled using a spatial light modulator. 

\subsection{Applications of mesoscopic concepts in optics}\label{subsection6.3}
A field where wavefront shaping concepts have already had a large impact is the field of optical imaging in turbid tissues (see also section \ref{section4}). Several proof-of-concept experiments have shown that the conventional paradigm of ballistic imaging could be extended to the
deep multiple scattering regime (see section \ref{subsubsection4.5.2}). While the multiscale nature and variability of these media makes them very difficult to model, they are also a challenge for wavefront shaping due to a potentially very short decorrelation time, inhomogeneous absorption, and due to the fact that one typically has access to one side of them only. In this respect, the exploitation of the memory effect  (see section \ref{subsection5.1}) has already overcome this limitation, by providing an order of magnitude increase in speed for scanning a point \cite{tang_superpenetration_2012} or even for single shot imaging \cite{katz_noninvasive_2014}. Other mesoscopic concepts, such as coherent perfect absorption or the generation of time-delay eigenstates (see section \ref{subsection5b.1} and \ref{subsection5b.3}), could be exploited in the future for a further improvement of imaging or light delivery, e.g., to avoid or to address regions in a tissue that are absorbing or  where movements induce decorrelations. Due to the enormous complexity involved, calculating the transmission matrix of a medium from the characterization of its three-dimensional shape, or worse, recovering the shape of a medium from its transmission matrix, currently remains out of reach for disordered systems. Only in very simple cases such demanding tasks can be achieved, such as, e.g., in short straight multimode fibers \cite{ploschner_seeing_2015}.

The perspective of better controlling the transmission through a medium has tremendous potential in communication technology, not only in optics but also in the radio frequency domain, where 
increased transmission and bandwith could be realized as well as more secure and well-isolated channels \cite{kaina_shaping_2014}. In this context, the direct access to open channels or time-delay eigenstates would be particularly useful (see section \ref{section5}).  

Wavefront shaping has also moved multimode fibers (MMFs) into the focus of attention, both for imaging and for telecommunication purposes. Imaging through an MMF is meanwhile a direct competitor to the bulkier fiber bundles \cite{gigan_endoscopy_2012}. In fiber communications, where the 
spatial degrees of freedom in MMFs are the last ones remaining to be exploited for higher data rate \cite{richardson_space-division_2013}, wavefront shaping has already opened the possibility to physically decouple the transmission modes, rather than unmixing the transmitted information a posteriori using multiple-input multiple-output technology. 

Another domain of application is nanophotonics and quantum optics, where complex systems have increasingly been considered as a platform for light matter interaction. In a first step, a modification of the emission properties of isolated single emitters has been discussed and was connected to the local density of states \cite{sapienza_cavity_2010,birowosuto_observation_2010,krachmalnicoff_fluctuations_2010,sapienza_long-tail_2011}, which in turn can be linked to the modal structure of the medium as described by mesoscopic theory. More recently, the concepts of quantum networks have been studied, where multiple emitters are distributed and connected for quantum computing or simulations \cite{vedral_quantum_1996,plenio_dephasing-assisted_2008} as well as to understand quantum phenomena in biology such as photosynthesis \cite{hildner_quantum_2013}. In a disordered system,  the connections between emitters can again be understood through the underlying modal structure and the associated Green's function of the medium \cite{caze_spatial_2013}. If mesoscopic theory can help to understand and better design such a network, wavefront shaping can also be a particularly useful tool to interrogate such a system for computation or simulation. Even a linear disordered system can be  interesting in the context of quantum random walks of single or multiple photons \cite{ott_quantum_2010,huisman_controlling_2014,goorden_quantum-secure_2014,defienne_nonclassical_2014, defienne_two-photon_2016, wolterink_programmable_2016}. Mesoscopic effects such as Anderson localization have been studied with non-classical states in waveguide arrays 
\cite{schreiber_decoherence_2011,crespi_anderson_2013} and it would be very interesting to study these effects in genuine disordered systems. Again, wavefront shaping could serve here as an indispensable ingredient to make these systems useful for applications.

Finally, we mention the following proofs of concept that have already been given for new devices, be it for compact spectrometers \cite{redding_compact_2013}, ultrafast switches \cite{strudley_ultrafast_2014}, tunable random lasers \cite{bachelard_adaptive_2014,hisch_pump-controlled_2013} as well as for light harvesting \cite{vynck_photon_2012, riboli_engineering_2014}. We expect this list to be significantly extended in the very near future.

\section*{Acknowledgments}
  The authors would like to thank all members of the community that were kind enough to share the rights to reprint their figures and the following colleagues for very
  fruitful discussions: Philipp Ambichl, Alexandre Aubry, Nicolas Bachelard, Hui Cao, R{\'e}mi Carminati, Adrian Girschik, Michel Gross, Thomas Hisch, Ori Katz, Florian Libisch, Matthias Liertzer, Stefan Nagele, Renate Pazourek, Romain Pierrat, and Patrick Sebbah. 
  SR gratefully acknowledges the generous support of Institut Langevin and Ecole Normale Sup{\'e}rieure through invited professor positions, 
  during which part of this review was written. The authors' own research presented in this review was 
  supported by the following funding agencies: Austrian Science Fund (FWF) projects SFB-ADLIS F16, P17359, SFB-IR-ON F25, SFB-NextLite F49 and I1142-N27 (GePartWave); Vienna Science Fund (WWTF) project MA09-030 (LICOTOLI); European Research Council (ERC) grant 278025. 

\bibliography{rmprefs}
\end{document}